{}

\documentclass[11pt,a4paper]{article}

\newif\ifpublic\publictrue
\newif\iffancy\fancytrue

\usepackage[a4paper,text={160mm,247mm},centering]{geometry}
\usepackage{setspace}
\usepackage{rotating}
\usepackage{amsmath,amssymb, amsfonts}
\usepackage{dsfont}

\usepackage[compile=false,fonts]{mpostinl}

\usepackage[usenames,dvipsnames,table]{xcolor}
\definecolor{mygreen}{rgb}{0,0.4,0}
\definecolor{myblue}{rgb}{0,0.0,0.4}
\definecolor{refrcolor}{rgb}{0,0.4,0}
\definecolor{cgreen}{rgb}{0,0.7,0}
\definecolor{ecolor}{rgb}{.52,.03,.06}
\definecolor{bgcolor}{rgb}{.96,.95,.80}
\definecolor{bgcolordark}{rgb}{.80,.80,.67}
\definecolor{faint}{rgb}{.80,.80,.80}

\def\beq{\begin{equation}}
\def\eeq{\end{equation}}

\usepackage{enumitem}

\begin{mpostdef}
pair vpos[];
pair epos[];
pair ext[];
pair exta[];
path paths[];
picture pic;
picture pic[];
picture savepic;
path shuffle;
path cpath;
path hpath;
path fs;
xu:=1cm;
yu:=1cm;
\end{mpostdef}

\begin{mpostdef}
	shuffle:= (0xu,1.414 xu)--(0xu,0xu)--(1xu,0xu)--(1xu,1.414xu)--(1xu,0xu)--(2xu,0xu)--(2xu,1.414 xu);
\end{mpostdef}

\begin{mpostdef}
def pensize(expr s)=withpen pencircle scaled s enddef;
def fillshape(expr p,ci,tb,cb)=
  fill p withcolor ci;
  draw p pensize(tb) withcolor cb;
enddef;
def filldot(expr z,s,ci)=
  fillshape(fullcircle scaled s shifted z, ci, 0.5pt, 0.0white);
enddef;
def filltrig(expr z,s,r,ci)=
  fillshape(((dir 60)--(dir 180)--(dir 300)--cycle) scaled s rotated r shifted z, ci, 0.5pt, 0.0white);
enddef;
def fillsqr(expr z,s,r,ci)=
  fillshape(((dir 45)--(dir 135)--(dir 225)--(dir 315)--cycle) scaled s rotated r shifted z, ci, 0.5pt, 0.0white);
enddef;
def drawcross(expr z,s,r,t,c)=
  draw ((-0.5,-0.5)--(+0.5,+0.5)) scaled s rotated r shifted z pensize(t) withcolor c;
  draw ((+0.5,-0.5)--(-0.5,+0.5)) scaled s rotated r shifted z pensize(t) withcolor c;
enddef;

def midarrow (expr p, t) =
  fill arrowhead subpath(0,arctime(arclength(subpath (0,t) of p)+0.5ahlength) of p) of p;
enddef;

def drawprop expr p=draw p pensize(1.5pt) enddef;
def drawline expr p=draw p pensize(0.5pt) enddef;
def drawpos expr p=drawcross(p, 5pt, 0, 1.0pt, 0.0white) enddef;
def drawvertex expr p=filldot(p, 5pt, 0.5white+0.5red) enddef;
def drawblackdotscale expr p=filldot(p, 0.4*xu, black) enddef;
def drawreddotscale expr p=filldot(p, 0.4*xu, red) enddef;
def drawwhitedotscale expr p=filldot(p, 0.4*xu, white) enddef;
def drawblackdot expr p=filldot(p, 8pt, black) enddef;
def drawwhitedot expr p=filldot(p, 8pt, white) enddef;
def drawlinearrow expr p=draw p pensize(0.5pt); midarrow (p,0.5); enddef;
def drawproparrow expr p=draw p pensize(1.5pt); midarrow (p,0.5); enddef;
def drawproparrowdouble expr p=draw p pensize(1.5pt); midarrow (p,0.65); midarrow (reverse p,0.65); enddef;
def drawwhitedotmr (expr p, sz)=filldot(p, sz*xu, white) enddef;
def drawblackdotmr (expr p, sz)=filldot(p, sz*xu, black) enddef;
def trdrawarrow expr p=draw p pensize(0.5pt); midarrow (p,0.8); enddef;
\end{mpostdef}
\begin{mpostdef}
u:=xu;									
a=2u;									
l=8u;									
u=12.5v;								
del=23.44;                             	
pair w$;								

let vector=color;                       
let xc=redpart; let yc=greenpart; let zc=bluepart;  
def dot(expr u,v)=
    (xc(u)*xc(v)+yc(u)*yc(v)+zc(u)*zc(v))
enddef;
def g(expr phi,lambda)=                 
    (cosd(phi)*cosd(lambda),cosd(phi)*sind(lambda),sind(phi))
enddef;
def G(expr omega,incl,Omega)=           
    (cosd(omega)*cosd(Omega)-sind(omega)*cosd(incl)*sind(Omega),
     cosd(omega)*sind(Omega)+sind(omega)*cosd(incl)*cosd(Omega),sind(omega)*sind(incl))
enddef;
def ellipse(expr ra,rb,an)=
    (fullcircle xscaled 2ra yscaled 2rb rotated an)
enddef;
def halfell(expr ra,rb,an)=
    (halfcircle xscaled 2ra yscaled 2rb rotated an)
enddef;

def circle(expr u,psi,a,f)=
    numeric s, c, e, c[]; path p[]; pair q[];

    c0=dot(u,g(phi,lambda));
    c1=dot(u,g(0,lambda+90));
    c2=dot(u,g(phi+90,lambda));

    e  = 1+-+c0;                        
    s  = sind(psi); c  = cosd(psi);     %
    q1 = (a,0); q2 = (0,a);             
    w$:=q0=c1*q1 + c2*q2;               
    p0 = origin--q0;                    
    p9 = ellipse(c*a,abs(c0)*c*a,angle(c2,-c1)) shifted (s*q0);

    if (psi=0):                         
        p1 = halfell(a,c0*a,angle(-c2,c1)); 
        p2 = halfell(a,c0*a,angle(c2,-c1)); 
    elseif (abs(c0)>c):                 
        if (s*c0>0):                    
            p1=p9;
            p2=origin;
        else:                           
            p1=origin;
            p2=p9;
        fi
    else:                               
        c5=e+-+s; c6=e*e;
        q3=((c1*s+c2*c5)*a/c6,(c2*s-c1*c5)*a/c6);
        q4=((c1*s-c2*c5)*a/c6,(c2*s+c1*c5)*a/c6);
        c3=xpart(p9 intersectiontimes ((s*q0)--1.1q3));
        c4=xpart(p9 intersectiontimes ((s*q0)--1.1q4));
        if (psi>0):
            p3 = subpath(c4,8) of p9 & subpath(0,c3) of p9;
            p4 = subpath(c3,c4) of p9;
        else:
            p3 = subpath(c4,c3) of p9;
            p4 = subpath(c3,8) of p9 & subpath(0,c4) of p9;
        fi

        if (c0>0):
            p1=p3; p2=p4;
        else:                           
            p1=p4; p2=p3;
        fi
    fi

    draw p1;
    if (f>1):
        draw p2 dashed evenly;
    fi
    if (f=1) or (f=3):
        drawarrow p0 dashed evenly;
    fi

enddef;
\end{mpostdef}
\begin{mpostdef}
input cmarrows;
setup_cmarrows(
  brace_name     = "bigbrace";
  parameter_file = "cmr10.mf";
  macro_name     = "bracea");
setup_cmarrows(
  brace_name     = "Biggbrace";
  parameter_file = "cmr12.mf";
  macro_name     = "braceb");
setup_cmarrows(
  brace_name     = "extensiblebrace";
  parameter_file = "cmr9.mf";
  macro_name     = "bracec");

pair vtpos[];
pair lnpos[];
trds:=4pt; 
string s;

def trbaseline(expr xlength)=
fill (-1xu,-0.2xu) -- (-1xu + (xlength)*xu,-0.2xu) -- (-1xu + (xlength)*xu,0.2xu) -- (-1xu,0.2xu) -- cycle withcolor .84 white;
enddef;

def labeltranslate(expr code,pos)=
if code=  "l0" : label.rt(btex \small $0$ etex,pos); fi;
if code=  "l1" : label.rt(btex \small $1$ etex,pos); fi;
if code=  "l2" : label.rt(btex \small $2$ etex,pos); fi;
if code=  "l3" : label.rt(btex \small $3$ etex,pos); fi;
if code=  "l4" : label.rt(btex \small $4$ etex,pos); fi;
if code=  "l5" : label.rt(btex \small $5$ etex,pos); fi;
if code=  "l6" : label.rt(btex \small $6$ etex,pos); fi;
if code=  "l7" : label.rt(btex \small $7$ etex,pos); fi;
if code=  "l11" : label.rt(btex \small $a_1$ etex,pos); fi;
if code=  "l12" : label.rt(btex \small $a_2$ etex,pos); fi;
if code=  "l13" : label.rt(btex \small $a_{i-1}$ etex,pos); fi;
if code=  "l14" : label.rt(btex \small $a_{i}$ etex,pos); fi;
if code=  "l15" : label.rt(btex \small $a_{i+1}$ etex,pos); fi;
if code=  "l16" : label.rt(btex \small $a_{k-1}$ etex,pos); fi;
if code=  "l17" : label.rt(btex \small $a_{k}$ etex,pos); fi;
if code=  "l18" : label.rt(btex \small $a_{k+1}$ etex,pos); fi;
if code=  "l19" : label.rt(btex \small $a_p$ etex,pos); fi;
if code=  "l21" : label.rt(btex \small $b_1$ etex,pos); fi;
if code=  "l22" : label.rt(btex \small $b_2$ etex,pos); fi;
if code=  "l23" : label.rt(btex \small $b_{l-1}$ etex,pos); fi;
if code=  "l24" : label.rt(btex \small $b_{l}$ etex,pos); fi;
if code=  "l25" : label.rt(btex \small $b_{l+1}$ etex,pos); fi;
if code=  "l26" : label.rt(btex \small $b_{j-1}$ etex,pos); fi;
if code=  "l27" : label.rt(btex \small $b_{j}$ etex,pos); fi;
if code=  "l28" : label.rt(btex \small $b_{j+1}$ etex,pos); fi;
if code=  "l29" : label.rt(btex \small $b_q$ etex,pos); fi;
if code=  "l61" : label.rt(btex \small $c_1$ etex,pos); fi;
if code=  "l62" : label.rt(btex \small $c_2$ etex,pos); fi;
if code=  "l68" : label.rt(btex \small $c_{m-1}$ etex,pos); fi;
if code=  "l69" : label.rt(btex \small $c_m$ etex,pos); fi;
if code=  "l71" : label.rt(btex \small $d_1$ etex,pos); fi;
if code=  "l72" : label.rt(btex \small $d_2$ etex,pos); fi;
if code=  "l78" : label.rt(btex \small $d_{l-1}$ etex,pos); fi;
if code=  "l79" : label.rt(btex \small $d_l$ etex,pos); fi;
if code=  "l40" : label.rt(btex \small $r$ etex,pos); fi;
if code=  "l41" : label.rt(btex \small $r$ etex,pos); fi;
if code=  "l48" : label.rt(btex \small $r_0$ etex,pos); fi;
if code=  "l49" : label.rt(btex \small $r_1$ etex,pos); fi;
if code=  "l51" : label.rt(btex \small $i$ etex,pos); fi;
if code=  "l52" : label.rt(btex \small $j$ etex,pos); fi;
if code=  "l53" : label.rt(btex \small $k$ etex,pos); fi;
if code=  "xi" : label.rt(btex \small $\xi$ etex,pos); fi;
if code=  "eta" : label.rt(btex \small $\eta$ etex,pos); fi;
if code=  "e2" : label.rt(btex \small $\eta_2$ etex,pos); fi;
if code=  "e3" : label.rt(btex \small $\eta_3$ etex,pos); fi;
if code=  "e4" : label.rt(btex \small $\eta_4$ etex,pos); fi;
if code=  "e5" : label.rt(btex \small $\eta_5$ etex,pos); fi;
if code= "e23" : label.rt(btex \small $\eta_{23}$ etex,pos); fi;
if code= "e24" : label.rt(btex \small $\eta_{24}$ etex,pos); fi;
if code= "e25" : label.rt(btex \small $\eta_{25}$ etex,pos); fi;
if code= "e34" : label.rt(btex \small $\eta_{34}$ etex,pos); fi;
if code= "e35" : label.rt(btex \small $\eta_{35}$ etex,pos); fi;
if code= "e45" : label.rt(btex \small $\eta_{45}$ etex,pos); fi;
if code="e234" : label.rt(btex \small $\eta_{234}$ etex,pos); fi;
if code="e23n" : label.rt(btex \small $\eta_{23\dots n}$ etex,pos); fi;
if code=  "er" : label.rt(btex \small $\eta_{r}$ etex,pos); fi;
if code=  "eB" : label.rt(btex \small $\eta_{B}$ etex,pos); fi;
if code=  "eBxi" : label.rt(btex \small $\eta_{B}+\xi$ etex,pos); fi;
if code=  "erB" : label.rt(btex \small $\eta_r+\eta_{B}$ etex,pos); fi;
if code=  "ea" : label.rt(btex \small $\eta_{a}$ etex,pos); fi;
if code=  "eab" : label.rt(btex \small $\eta_{ab}$ etex,pos); fi;
if code=  "eamxi" : label.rt(btex \small $\eta_{a}-\xi$ etex,pos); fi;
if code=  "ebpxi" : label.rt(btex \small $\eta_{b}+\xi$ etex,pos); fi;
if code=  "eA" : label.rt(btex \small $\eta_{A}$ etex,pos); fi;
if code=  "eBr" : label.rt(btex \small $\eta_{B}+\eta_r$ etex,pos); fi;
if code=  "eAxi" : label.rt(btex \small $\eta_{A}-\xi$ etex,pos); fi;
if code=  "ea1im1" : label.rt(btex \small $\eta_{a_{1 \dots i-1}}$ etex,pos); fi;
if code=  "eaikp1" : label.rt(btex \small $\eta_{a_{i \dots k-1}}$ etex,pos); fi;
if code=  "eaip" : label.rt(btex \small $\eta_{a_{i \dots p}}$ etex,pos); fi;
if code=  "meaip" : label.rt(btex \small $-\eta_{a_{i \dots p}}$ etex,pos); fi;
if code=  "eakp" : label.rt(btex \small $\eta_{a_{k \dots p}}$ etex,pos); fi;
if code=  "eakp1p" : label.rt(btex \small $\eta_{a_{k+1 \dots p}}$ etex,pos); fi;
if code="ea23p" : label.rt(btex \small $\eta_{a_{23\dots p}}$ etex,pos); fi;
if code="ea23pxi" : label.rt(btex \small $\eta_{a_{23\dots p}}-\xi$ etex,pos); fi;
if code="ea23pereB" : label.rt(btex \small $\eta_{a_{23\dots p}}+\eta_r+\eta_B$ etex,pos); fi;
if code="eaipereB" : label.rt(btex \small $\eta_{a_{i\dots p}}+\eta_r+\eta_B$ etex,pos); fi;
if code=  "mea" : label.rt(btex \small $-\eta_{a}$ etex,pos); fi;
if code=  "meakp" : label.rt(btex \small $-\eta_{A_{k\dots p+1}}$ etex,pos); fi;
if code=  "eakpmxi" : label.rt(btex \small $\eta_{A_{k\dots p+1}}{-}\xi$ etex,pos); fi;
if code=  "eb" : label.rt(btex \small $\eta_{b}$ etex,pos); fi;
if code=  "eab" : label.rt(btex \small $\eta_{a b}$ etex,pos); fi;
if code=  "ebjq" : label.rt(btex \small $\eta_{b_{j\dots q}}$ etex,pos); fi;
if code=  "eblq" : label.rt(btex \small $\eta_{b_{l\dots q}}$ etex,pos); fi;
if code=  "eblqxi" : label.rt(btex \small $\eta_{b_{l\dots q}}{+}\xi$ etex,pos); fi;
if code=  "ebjp1q" : label.rt(btex \small $\eta_{b_{j+1\dots q}}$ etex,pos); fi;
if code=  "ebjxi" : label.rt(btex \small $\eta_{b_{j\dots q}}{+}\xi$ etex,pos); fi;
if code=  "eakpblq" : label.rt(btex \small $\eta_{a_{k\dots p}}{+}\eta_{b_{l\dots q}}$ etex,pos); fi;
if code=  "eaipblq" : label.rt(btex \small $\eta_{a_{i\dots p}}{+}\eta_{b_{l\dots q}}$ etex,pos); fi;
if code=  "eakpxi" : label.rt(btex \small $\eta_{a_{k\dots p}}{-}\xi$ etex,pos); fi;
if code=  "eaipxi" : label.rt(btex \small $\eta_{a_{i\dots p}}{-}\xi$ etex,pos); fi;
if code="eb23q" : label.rt(btex \small $\eta_{b_{23\dots q}}$ etex,pos); fi;
if code="eb23qxi" : label.rt(btex \small $\eta_{b_{23\dots q}}+ \xi$ etex,pos); fi;
if code=   "A" : label.lft(btex \small $A$ etex,pos); fi;
if code=   "B" : label.lft(btex \small $B$ etex,pos); fi;
if code=   "shuffle" : label.lft(btex \small $s$ etex,pos); fi;
enddef;

def trbasevert(expr vert, xpos)=
   vtpos[vert]:=(xpos*xu,0);
   draw vtpos[vert] pensize(trds);
   labeltranslate("l" & decimal(vert),(xpos*xu,0));
enddef;

def trvert(expr vert, pos)=
   draw pos pensize(trds);
   labeltranslate("l" & decimal(vert),pos);
enddef;

def trbaseset(expr vert, xpos, setname)=
   vtpos[vert]:=(xpos*xu,0);
   vtpos[vert+1]:=(xpos*xu,1.6*xu);
   draw vtpos[vert] pensize(trds);
   draw vtpos[vert+1] pensize(trds);
   draw vtpos[vert]--vtpos[vert+1] dashed withdots scaled 0.5 pensize (1pt);
   labeltranslate(setname,1/2*(vtpos[vert]+vtpos[vert+1])-(0.2xu,0));
   bracea vtpos[vert] shifted (-0.1xu,0)--vtpos[vert+1] shifted (-0.1xu,0) withcolor 0.2*white;
enddef;

def trset(expr startlab,setlab,linklab,setname)=
   vtpos[setlab]:=vtpos[startlab]+dir(90)*xu;
   vtpos[setlab+1]:=vtpos[startlab]+2.6*dir(90)*xu;
   draw vtpos[setlab]--vtpos[setlab+1] dashed withdots scaled 0.5 pensize (1pt);
   draw vtpos[startlab]--vtpos[setlab];
   draw vtpos[setlab] pensize(trds);
   draw vtpos[setlab+1] pensize(trds);
   labeltranslate(setname,1/2*(vtpos[setlab]+vtpos[setlab+1])-(0.2xu,0));
   bracea vtpos[setlab] shifted (-0.1xu,0)--vtpos[setlab+1] shifted (-0.1xu,0) withcolor 0.2*white;
   lnpos[setlab]:=vtpos[startlab]+1/2*dir(90)*xu; 
   labeltranslate(linklab,lnpos[setlab]-(0.05xu,0));
enddef;

def trlink(expr startlab,direction,endlab,linklab)=
  if direction="l" : 
    vtpos[endlab]:=vtpos[startlab]+dir(120)*xu; lnpos[endlab]:=vtpos[startlab]+1/2*dir(120)*xu; trdrawarrow vtpos[endlab]--vtpos[startlab]; trvert(endlab,vtpos[endlab]); labeltranslate(linklab,lnpos[endlab]-(0.07xu,0)); fi;
  if direction="m" : 
  vtpos[endlab]:=vtpos[startlab]+dir(90)*xu; lnpos[endlab]:=vtpos[startlab]+1/2*dir(90)*xu; trdrawarrow vtpos[endlab]--vtpos[startlab]; trvert(endlab,vtpos[endlab]); labeltranslate(linklab,lnpos[endlab]-(0.05xu,0)); fi;
  if direction="r" : 
    vtpos[endlab]:=vtpos[startlab]+dir(60)*xu; lnpos[endlab]:=vtpos[startlab]+1/2*dir(60)*xu; trdrawarrow vtpos[endlab]--vtpos[startlab]; trvert(endlab,vtpos[endlab]); labeltranslate(linklab,lnpos[endlab]); fi;
enddef;

def trbasevertleft(expr vert, xpos, ypos)=
   vtpos[vert]:=(xpos*xu,ypos*yu);
   draw vtpos[vert] pensize(trds);
   labeltranslate("l" & decimal(vert),(xpos*xu-0.5*xu,ypos*yu));
enddef;

def trbasevertright(expr vert, xpos, ypos)=
   vtpos[vert]:=(xpos*xu,ypos*yu);
   draw vtpos[vert] pensize(trds);
   labeltranslate("l" & decimal(vert),(xpos*xu+0.1*xu,ypos*yu));
enddef;

def trvertleft(expr vert, pos)=
   draw pos pensize(trds);
   labeltranslate("l" & decimal(vert),pos-(0.5xu,0));
enddef;

def trvertright(expr vert, pos)=
   draw pos pensize(trds);
   labeltranslate("l" & decimal(vert),pos+(0.1xu,0));
enddef;

def trlinkleft(expr startlab,direction,endlab,linklab)=
  if direction="l" : 
    vtpos[endlab]:=vtpos[startlab]+dir(120)*xu; lnpos[endlab]:=vtpos[startlab]+1/2*dir(120)*xu; trdrawarrow vtpos[endlab]--vtpos[startlab]; trvertleft(endlab,vtpos[endlab]); labeltranslate(linklab,lnpos[endlab]-(0.07xu,0)); fi;
  if direction="m" : 
  vtpos[endlab]:=vtpos[startlab]+dir(90)*xu; lnpos[endlab]:=vtpos[startlab]+1/2*dir(90)*xu; trdrawarrow vtpos[endlab]--vtpos[startlab]; trvertleft(endlab,vtpos[endlab]); labeltranslate(linklab,lnpos[endlab]-(0.05xu,0)); fi;
  if direction="r" : 
    vtpos[endlab]:=vtpos[startlab]+dir(60)*xu; lnpos[endlab]:=vtpos[startlab]+1/2*dir(60)*xu; trdrawarrow vtpos[endlab]--vtpos[startlab]; trvertleft(endlab,vtpos[endlab]); labeltranslate(linklab,lnpos[endlab]); fi;
enddef;

def trlinkleftdotted(expr startlab,direction,endlab,linklab)=
  if direction="l" : 
    vtpos[endlab]:=vtpos[startlab]+dir(120)*xu; lnpos[endlab]:=vtpos[startlab]+1/2*dir(120)*xu; draw vtpos[endlab]--vtpos[startlab] dashed withdots scaled 0.5 pensize (1pt); trvertleft(endlab,vtpos[endlab]); labeltranslate(linklab,lnpos[endlab]-(0.07xu,0)); fi;
  if direction="m" : 
  vtpos[endlab]:=vtpos[startlab]+dir(90)*xu; lnpos[endlab]:=vtpos[startlab]+1/2*dir(90)*xu; draw vtpos[endlab]--vtpos[startlab] dashed withdots scaled 0.5 pensize (1pt); trvertleft(endlab,vtpos[endlab]); labeltranslate(linklab,lnpos[endlab]-(0.05xu,0)); fi;
  if direction="r" : 
    vtpos[endlab]:=vtpos[startlab]+dir(60)*xu; lnpos[endlab]:=vtpos[startlab]+1/2*dir(60)*xu; draw vtpos[endlab]--vtpos[startlab] dashed withdots scaled 0.5 pensize (1pt); trvertleft(endlab,vtpos[endlab]); labeltranslate(linklab,lnpos[endlab]); fi;
enddef;

def trlinkright(expr startlab,direction,endlab,linklab)=
  if direction="l" : 
    vtpos[endlab]:=vtpos[startlab]+dir(120)*xu; lnpos[endlab]:=vtpos[startlab]+1/2*dir(120)*xu; trdrawarrow vtpos[endlab]--vtpos[startlab]; trvertright(endlab,vtpos[endlab]); labeltranslate(linklab,lnpos[endlab]-(0.07xu,0)); fi;
  if direction="m" : 
  vtpos[endlab]:=vtpos[startlab]+dir(90)*xu; lnpos[endlab]:=vtpos[startlab]+1/2*dir(90)*xu; trdrawarrow vtpos[endlab]--vtpos[startlab]; trvertright(endlab,vtpos[endlab]); labeltranslate(linklab,lnpos[endlab]-(0.05xu,0)); fi;
  if direction="r" : 
    vtpos[endlab]:=vtpos[startlab]+dir(60)*xu; lnpos[endlab]:=vtpos[startlab]+1/2*dir(60)*xu; trdrawarrow vtpos[endlab]--vtpos[startlab]; trvertright(endlab,vtpos[endlab]); labeltranslate(linklab,lnpos[endlab]); fi;
enddef;

def trlinkrightspecial(expr startlab,direction,endlab,linklab)=
  if direction="m" : 
  vtpos[endlab]:=vtpos[startlab]+dir(90)*xu; lnpos[endlab]:=vtpos[startlab]+1/2*dir(90)*xu; trdrawarrow vtpos[endlab]--vtpos[startlab]; trvertright(endlab,vtpos[endlab]); labeltranslate(linklab,lnpos[endlab]-(2.05xu,0)); fi;
enddef;

def trlinkrightdotted(expr startlab,direction,endlab,linklab)=
  if direction="l" : 
    vtpos[endlab]:=vtpos[startlab]+dir(120)*xu; lnpos[endlab]:=vtpos[startlab]+1/2*dir(120)*xu; draw vtpos[endlab]--vtpos[startlab] dashed withdots scaled 0.5 pensize (1pt); trvertright(endlab,vtpos[endlab]); labeltranslate(linklab,lnpos[endlab]-(0.07xu,0)); fi;
  if direction="m" : 
  vtpos[endlab]:=vtpos[startlab]+dir(90)*xu; lnpos[endlab]:=vtpos[startlab]+1/2*dir(90)*xu; draw vtpos[endlab]--vtpos[startlab] dashed withdots scaled 0.5 pensize (1pt); trvertright(endlab,vtpos[endlab]); labeltranslate(linklab,lnpos[endlab]-(0.05xu,0)); fi;
  if direction="r" : 
    vtpos[endlab]:=vtpos[startlab]+dir(60)*xu; lnpos[endlab]:=vtpos[startlab]+1/2*dir(60)*xu; draw vtpos[endlab]--vtpos[startlab] dashed withdots scaled 0.5 pensize (1pt); trvertright(endlab,vtpos[endlab]); labeltranslate(linklab,lnpos[endlab]); fi;
enddef;

def trlinkvertices(expr startlab,endlab,linklab,linklabxpos,linklabypos)=
  trdrawarrow vtpos[endlab]--vtpos[startlab]; 
  labeltranslate(linklab,(vtpos[startlab]+vtpos[endlab])*0.5+(-0.2xu+linklabxpos*xu,0.2yu+linklabypos*yu)); 
enddef;

def trlinkverticesdotted(expr startlab,endlab,linklab,linklabxpos,linklabypos)=
  draw vtpos[endlab]--vtpos[startlab] dashed withdots scaled 0.5 pensize (1pt); labeltranslate(linklab,(vtpos[startlab]+vtpos[endlab])*0.5+(-0.2xu+linklabxpos*xu,0.2yu+linklabypos*yu)); 
enddef;

def trbasesetright(expr vert, xpos, setname)=
   vtpos[vert]:=(xpos*xu,0);
   vtpos[vert+1]:=(xpos*xu,1.6*xu);
   draw vtpos[vert] pensize(trds);
   draw vtpos[vert+1] pensize(trds);
   draw vtpos[vert]--vtpos[vert+1] dashed withdots scaled 0.5 pensize (1pt);
   labeltranslate(setname,1/2*(vtpos[vert]+vtpos[vert+1])-(0.2xu,0));
   bracea vtpos[vert] shifted (-0.1xu,0)--vtpos[vert+1] shifted (-0.1xu,0) withcolor 0.2*white;
enddef;
\end{mpostdef}

\begin{mpostfig}[label=example1]
  trbasevert(1,0);
  trlink(1,"m",2,"e234");
  trlink(2,"l",3,"e3");
  trlink(2,"r",4,"e4");
  trbasevert(0,1.5);
  trlink(0,"m",5,"e5");
\end{mpostfig}

\begin{mpostfig}[label=example1RHS1]
	trbasevert(1,0);
	trlink(1,"m",2,"e234");
	trlink(2,"m",3,"e34");
	trlink(3,"m",4,"e4");
	trbasevert(0,1.5);
	trlink(0,"m",5,"e5");
\end{mpostfig}

\begin{mpostfig}[label=example4Pt]
	trbasevert(1,0);
	trlink(1,"m",2,"e234");
	trlink(2,"m",3,"e34");
	trlink(3,"m",4,"e4");
\end{mpostfig}

\begin{mpostfig}[label=example1RHS2]
	trbasevert(1,0);
    trlink(1,"m",2,"e234");
    trlink(2,"m",4,"e34");
    trlink(4,"m",3,"e3");
    trbasevert(0,1.5);
    trlink(0,"m",5,"e5");
\end{mpostfig}

\begin{mpostfig}[label=concatenationLHS]
	trbasevertright(11,0,0);
    trlinkright(11,"m",12,"ea23pereB");
    trlinkrightdotted(12,"m",13," ");
    trlinkright(13,"m",14,"eaipereB");
    trlinkrightdotted(14,"m",19," ");
    trlinkright(19,"m",40,"erB");
	trbasevertright(40,3,0);
	trlinkright(40,"m",21," ");
	trlinkright(21,"m",22," ");
	trlinkrightdotted(22,"m",29," ");
\end{mpostfig}
\begin{mpostfig}[label=concatenationRHS]
	trbasevertright(11,0,0);
	trlinkright(11,"m",12,"");
	trlinkrightdotted(12,"m",19," ");
	trlinkright(19,"m",40,"");
	trlinkright(40,"m",21," ");
	trlinkright(21,"m",22," ");
	trlinkrightdotted(22,"m",29," ");
\end{mpostfig}

\begin{mpostfig}[label=shuffleSameEndingLHS]
	trbasevertright(40,0,0);
	trlinkleft(40,"l",11," ");
	trlinkleft(11,"m",12," ");
	trlinkleftdotted(12,"m",19," ");
	trlinkright(40,"r",21," ");
    trlinkright(21,"m",22," ");
    trlinkrightdotted(22,"m",29," ");
\end{mpostfig}
\begin{mpostfig}[label=shuffleSameEndingRHS]
	trbasevertright(40,0,0);
	trlinkleft(40,"l",11," ");
	trlinkleft(11,"m",12," ");
	trlinkleftdotted(12,"m",19," ");
	trlinkright(40,"r",21," ");
	trlinkright(21,"m",22," ");
	trlinkrightdotted(22,"m",29," ");
	draw shuffle scaled 0.1 shifted ((vtpos[11]+vtpos[21])*0.5-(0.08xu,0.45xu)) pensize(0.4pt);
\end{mpostfig}

\begin{mpostfig}[label=shuffleSameBeginningLHS]
	trbasevertright(40,0,0);
	trbasevertleft(69,-0.5,-1);
	trlinkvertices(69,40,"eA",-0.5,0);
	trbasevertleft(62,-0.5,-2);
	trlinkverticesdotted(62,69,"",0,0);
	trbasevertleft(61,-0.5,-3);
	trlinkvertices(61,62,"",0,0);
	trbasevertright(79,0.5,-1);
	trlinkvertices(79,40,"eBr",0.2,0);
	trbasevertright(72,0.5,-2);
	trlinkverticesdotted(72,79,"",0,0);
	trbasevertright(71,0.5,-3);
	trlinkvertices(71,72,"",0,0);

\end{mpostfig}
\begin{mpostfig}[label=shuffleSameBeginningRHS]
	trbasevertright(40,0,0);
	trbasevertleft(69,-0.5,-1);
	trlinkvertices(69,40,"eA",-0.5,0);
	trbasevertleft(62,-0.5,-2);
	trlinkverticesdotted(62,69,"",0,0);
	trbasevertleft(61,-0.5,-3);
	trlinkvertices(61,62,"",0,0);
	trbasevertright(79,0.5,-1);
	trlinkvertices(79,40,"eBr",0.2,0);
	trbasevertright(72,0.5,-2);
	trlinkverticesdotted(72,79,"",0,0);
	trbasevertright(71,0.5,-3);
	trlinkvertices(71,72,"",0,0);
        draw shuffle scaled 0.1 shifted ((vtpos[69]+vtpos[79])*0.5-(0.1xu,0.04xu)) pensize(0.4pt);
\end{mpostfig}

\begin{mpostfig}[label=reflectionLHS]
	trbasevertright(40,0,0);
	trbasevertright(11,0,1);
	trlinkvertices(40,11,"",0,0);
	trbasevertright(12,0,2);
	trlinkvertices(11,12,"",0,0);
	trbasevertright(19,0,3);
	trlinkverticesdotted(12,19,"",0,0);
	
\end{mpostfig}
\begin{mpostfig}[label=reflectionRHS]
	trbasevertright(40,0,0);
	trbasevertright(11,0,1);
	trlinkvertices(11,40,"",0,0);
	trbasevertright(12,0,2);
	trlinkvertices(12,11,"",0,0);
	trbasevertright(19,0,3);
	trlinkverticesdotted(19,12,"",0,0);
\end{mpostfig}

\begin{mpostfig}[label=phir0Ar1BLHS]
	trbasevertleft(48,0,0);
	trbasevertleft(11,0,0.5);
	trlinkvertices(48,11,"",0,0);
	trbasevertleft(19,0,1.5);
	trlinkverticesdotted(11,19,"",0,0);
	trbasevertleft(49,0,2);
	trlinkvertices(19,49,"",0,0);
	trbasevertleft(21,0,2.5);
	trlinkvertices(49,21,"",0,0);
	trbasevertleft(29,0,3.5);
	trlinkverticesdotted(21,29,"",0,0);
\end{mpostfig}

\begin{mpostfig}[label=phir0Ar1BRHS]
	trbasevertleft(48,0,0);
	trlinkleft(48,"l",11," ");
	trlinkrightdotted(11,"l",13," ");
	trlinkleft(48,"r",49,"");
	trlinkright(49,"l",19," ");
	trlinkrightdotted(19,"m",14," ");
	trlinkright(49,"r",21," ");
	trlinkrightdotted(21,"m",29," ");
        draw shuffle scaled 0.1 shifted ((vtpos[19]+vtpos[21])*0.5-(0.08xu,0.45xu)) pensize(0.4pt); 	
\end{mpostfig}

\begin{mpostfig}[label=FayLHS]
	trbasevertright(53,0.5,0);
	trbasevertleft(51,0,0.75);
	trbasevertright(52,1,0.75);
	trlinkvertices(53,51,"ea",-0.2,-0.3);
	trlinkvertices(53,52,"eb",0.2,-0.3);
\end{mpostfig}

\begin{mpostfig}[label=FayRHS1]
	trbasevertright(53,0.5,0);
	trbasevertleft(51,0,0.75);
	trbasevertright(52,1,0.75);
	trlinkvertices(53,51,"eab",-0.3,-0.3);
	trlinkvertices(51,52,"eb",0,0);
\end{mpostfig}

\begin{mpostfig}[label=FayRHS2]
	trbasevertright(53,0.5,0);
	trbasevertleft(51,0,0.75);
	trbasevertright(52,1,0.75);
	trlinkvertices(52,51,"",0,0);
	trlinkvertices(53,52,"",0.2,-0.3);
\end{mpostfig}

\begin{mpostfig}[label=PFLHS]
	trbasevertright(53,0.5,0);
	trbasevertleft(51,0,0.75);
	trbasevertright(52,1,0.75);
	trlinkvertices(53,51,"",-0.2,-0.3);
	trlinkvertices(53,52,"",0.2,-0.3);
\end{mpostfig}

\begin{mpostfig}[label=PFRHS1]
	trbasevertright(53,0.5,0);
	trbasevertleft(51,0,0.75);
	trbasevertright(52,1,0.75);
	trlinkvertices(53,51,"",-0.3,-0.3);
	trlinkvertices(51,52,"",0,0);
\end{mpostfig}

\begin{mpostfig}[label=PFRHS2]
	trbasevertright(53,0.5,0);
	trbasevertleft(51,0,0.75);
	trbasevertright(52,1,0.75);
	trlinkvertices(52,51,"ea",0,0);
	trlinkvertices(53,52,"eab",0.2,-0.3);
\end{mpostfig}

\begin{mpostfig}[label=EKSeries]
	trbasevertleft(51,0,0);
	trbasevertright(52,1,0);
	trlinkvertices(51,52,"eta",0,0);
\end{mpostfig}

\begin{mpostfig}[label=EKSeriesz]
	trbasevertleft(51,0,0);
	trbasevertright(52,1,0);
	trlinkvertices(51,52,"",0,0);
\end{mpostfig}

\begin{mpostfig}[label=phiChainA]
	trbasevertleft(11,0,0);
    trlinkleft(11,"m",12," ");
    trlinkleftdotted(12,"m",19," ");
\end{mpostfig}

\begin{mpostfig}[label=phiChainAFactors]
	trbasevertright(13,0,0);
	trlinkright(13,"m",14,"eaip");
\end{mpostfig}

\begin{mpostfig}[label=ExampleTreeWeights]
	trbasevertright(1,0,0);
	trlinkright(1,"m",2,"e234");
	trlinkright(2,"l",3,"e3");
	trlinkright(2,"r",4,"e4");
\end{mpostfig}

\begin{mpostfig}[label=ExampleTree]
	trbasevertright(1,0,0);
	trlinkright(1,"m",2," ");
	trlinkright(2,"l",3," ");
	trlinkright(2,"r",4," ");
\end{mpostfig}

\begin{mpostfig}[label=startzzero]
  trbasevertleft(1,0,0);
  trlinkleft(1,"m",11,"eA");
  trlinkleft(11,"m",12,"ea23p");
  trlinkleftdotted(12,"m",17," ");
  trlinkleftdotted(17,"m",19," ");
  trbasevertright(0,2,0);
  trlinkright(0,"m",21,"eB");
  trlinkright(21,"m",22,"eb23q");
  trlinkrightdotted(22,"m",27," ");
  trlinkrightdotted(27,"m",29," ");
  trlinkvertices(17,27,"xi",0,0);
\end{mpostfig}

\begin{mpostfig}[label=startzzeroXiDown]
	trbasevertleft(1,0,0);
	trlinkright(1,"m",11,"eAxi");
	trlinkrightdotted(11,"m",16," ");
	trlinkright(16,"m",17,"eakpxi");
	trlinkright(17,"m",18,"eakp1p");
	trlinkrightdotted(18,"m",19," ");
	trbasevertright(0,2,0);
	trlinkright(0,"m",21,"eBxi");
	trlinkrightdotted(21,"m",26," ");
	trlinkright(26,"m",27,"ebjxi");
	trlinkright(27,"m",28,"ebjp1q");
	trlinkrightdotted(28,"m",29," ");
	trlinkvertices(1,0,"xi",0,-0.5);
\end{mpostfig}

\begin{mpostfig}[label=startzzeroXiDownj0]
	trbasevertleft(1,0,0);
	trlinkright(1,"m",11,"eAxi");
	trlinkrightdotted(11,"m",16," ");
	trlinkright(16,"m",17,"eakpxi");
	trlinkright(17,"m",18,"eakp1p");
	trlinkrightdotted(18,"m",19," ");
	trbasevertright(0,2,0);
	trlinkright(0,"m",21," ");
	trlinkright(21,"m",22," ");
	trlinkrightdotted(22,"m",29," ");
	trlinkvertices(1,0,"xi",0,-0.5);
\end{mpostfig}

\begin{mpostfig}[label=startzzeroXiDownk0]
	trbasevertleft(1,0,0);
	trlinkright(1,"m",11,"");
	trlinkright(11,"m",12,"");
	trlinkrightdotted(12,"m",19," ");
	trbasevertright(0,2,0);
	trlinkright(0,"m",21,"eBxi");
	trlinkrightdotted(21,"m",26," ");
	trlinkright(26,"m",27,"ebjxi");
	trlinkright(27,"m",28,"ebjp1q");
	trlinkrightdotted(28,"m",29," ");
	trlinkvertices(1,0,"xi",0,-0.5);
\end{mpostfig}

\begin{mpostfig}[label=fundamentalIdentityLHS]
  trbasevertleft(1,0,0);
  trlinkleft(1,"m",17,"ea");
  trbasevertright(0,1.5,0);
  trlinkright(0,"m",27,"eb");
  trlinkvertices(17,27,"xi",0,0);
\end{mpostfig}

\begin{mpostfig}[label=fundamentalIdentityRHS1]
  trbasevertleft(1,0,0);
  trlinkleft(1,"m",17,"eab");
  trbasevertright(0,1.5,0);
  trbasevertright(27,1.5,1);	
  trlinkvertices(0,1,"eb",0,0);
  trlinkvertices(17,27,"xi",0,0);
\end{mpostfig}

\begin{mpostfig}[label=fundamentalIdentityRHS2]
  trbasevertleft(1,0,0);
  trbasevertleft(17,0,1);
  trbasevertright(0,1.5,0);
  trlinkright(0,"m",27,"eab");	
  trlinkvertices(0,1,"mea",-0.3,0);
  trlinkvertices(27,17,"xi",0,0);
\end{mpostfig}

\begin{mpostfig}[label=fundamentalIdentityRHS3]
  trbasevertleft(1,0,0);
  trlinkleft(1,"m",17,"ea");
  trbasevertright(0,1.5,0);
  trlinkleft(0,"m",27,"eb");
  trlinkvertices(1,0,"xi",0,0);
\end{mpostfig}

\begin{mpostfig}[label=fundamentalIdentityRHS1FullFactor]
	trbasevertleft(1,0,0);
	trlinkleft(1,"m",17,"eaipblq");
	trbasevertright(0,1.5,0);
	trbasevertright(27,1.5,1);	
	trlinkvertices(0,1,"eblq",-0.125,-0.5);
	trlinkvertices(17,27,"eblqxi",-0.5,0);
\end{mpostfig}

\begin{mpostfig}[label=fundamentalIdentityRHS2FullFactor]
	trbasevertleft(1,0,0);
	trbasevertleft(17,0,1);
	trbasevertright(0,1.5,0);
	trlinkrightspecial(0,"m",27,"eaipblq");	
	trlinkvertices(0,1,"meaip",-0.5,-0.5);
	trlinkvertices(27,17,"eaipxi",-0.5,0);
\end{mpostfig}

\begin{mpostfig}[label=fundamentalIdentityRHS3FullFactor]
	trbasevertleft(1,0,0);
	trlinkleft(1,"m",17,"eaip");
	trbasevertright(0,1.5,0);
	trlinkleft(0,"m",27,"eblq");
	trlinkvertices(1,0,"xi",0,-0.5);
\end{mpostfig}

\begin{mpostfig}[label=fundamentalIdentityJLHS]
	trbasevertleft(1,0,0);
	trlinkright(1,"m",17,"ea");
	trbasevertright(0,1.5,0);
	trlinkvertices(17,0,"xi",0,0);
\end{mpostfig}

\begin{mpostfig}[label=fundamentalIdentityJRHS2]
	trbasevertleft(1,0,0);
	trlinkright(1,"m",17,"eamxi");
	trbasevertright(0,1.5,0);
	trlinkvertices(1,0,"xi",0,-0.015*yu);
\end{mpostfig}

\begin{mpostfig}[label=fundamentalIdentityJRHS1]
	trbasevertleft(1,0,0);
	trbasevertright(17,0,1);
	trbasevertright(0,1.5,0);
	trlinkvertices(0,1,"mea",0,-0.015*yu);
	trlinkvertices(0,17,"eamxi",0,0);
\end{mpostfig}

\begin{mpostfig}[label=fundamentalIdentityKLHS]
	trbasevertleft(1,0,0);
	trbasevertleft(0,1.5,0);
	trlinkright(0,"m",27,"eb");
	trlinkvertices(1,27,"xi",0,0);
\end{mpostfig}

\begin{mpostfig}[label=fundamentalIdentityKRHS2]
	trbasevertleft(1,0,0);
	trbasevertright(0,1.5,0);
	trlinkright(0,"m",27,"ebpxi");
	trlinkvertices(1,0,"xi",0,-0.015*yu);
\end{mpostfig}

\begin{mpostfig}[label=fundamentalIdentityKRHS1]
	trbasevertleft(1,0,0);
	trbasevertright(27,1.5,1);
	trbasevertright(0,1.5,0);
	trlinkvertices(0,1,"eb",0,-0.015*yu);
	trlinkvertices(1,27,"ebpxi",0.015*xu,-0.005*yu);
\end{mpostfig}

\begin{mpostfig}[label=phiA]
  trbasevertleft(1,0,0);
  trlinkleft(1,"m",11," ");
  trlinkleft(11,"m",12," ");
  trlinkleftdotted(12,"m",17," ");
  trlinkleftdotted(17,"m",19," ");
\end{mpostfig}

\begin{mpostfig}[label=phi1ak]
  trbasevertleft(1,0,0);
  trlinkleft(1,"l",11," ");
  trlinkrightdotted(11,"l",13," ");
  trlinkleft(1,"r",17," ");
  trlinkright(17,"l",16," ");
  trlinkrightdotted(16,"m",14," ");
  trlinkright(17,"r",18," ");
  trlinkrightdotted(18,"m",19," ");
  draw shuffle scaled 0.1 shifted ((vtpos[16]+vtpos[18])*0.5-(0.08xu,0.45xu)) pensize(0.4pt);
\end{mpostfig}

\begin{mpostfig}[label=phi1akFullFactor]
	trbasevertleft(1,0,0);
	trlinkleft(1,"l",11," ");
	trlinkrightdotted(11,"l",13," ");
    trbasevertright(17,0.5,0.866);
	trlinkright(17,"l",16," ");
	trlinkrightdotted(16,"m",14," ");
	trlinkright(17,"r",18," ");
	trlinkrightdotted(18,"m",19," ");
        draw shuffle scaled 0.1 shifted ((vtpos[16]+vtpos[18])*0.5-(0.08xu,0.45xu)) pensize(0.4pt);
\end{mpostfig}

\begin{mpostfig}[label=phiB]
  trbasevertleft(0,0,0);
  trlinkleft(0,"m",21," ");
  trlinkleft(21,"m",22," ");
  trlinkleftdotted(22,"m",27," ");
  trlinkleftdotted(27,"m",29," ");
\end{mpostfig}

\begin{mpostfig}[label=phi0bj]
  trbasevertleft(0,0,0);
  trlinkright(0,"r",21," ");
  trlinkrightdotted(21,"r",23," ");
  trlinkright(0,"l",27," ");
  trlinkright(27,"r",26," ");
  trlinkrightdotted(26,"m",24," ");
  trlinkright(27,"l",28," ");
  trlinkrightdotted(28,"m",29," ");
  draw shuffle scaled 0.1 shifted ((vtpos[26]+vtpos[28])*0.5-(0.08xu,0.45xu)) pensize(0.4pt);
\end{mpostfig}

\begin{mpostfig}[label=phi0bjFullFactor]
	trbasevertright(0,0,0);
	trlinkright(0,"r",21," ");
	trlinkrightdotted(21,"r",23," ");
	trbasevertright(27,-0.5,0.866);
	trlinkright(27,"r",26," ");
	trlinkrightdotted(26,"m",24," ");
	trlinkright(27,"l",28," ");
	trlinkrightdotted(28,"m",29," ");
        draw shuffle scaled 0.1 shifted ((vtpos[26]+vtpos[28])*0.5-(0.08xu,0.45xu)) pensize(0.4pt);
\end{mpostfig}

\begin{mpostfig}[label=phi1ak0bj]
  trbasevertleft(1,0,0);
  trlinkleft(1,"l",11," ");
  trlinkrightdotted(11,"l",13," ");
  trlinkleft(1,"r",17,"eaip");
  trlinkright(17,"l",16," ");
  trlinkrightdotted(16,"m",14," ");
  trlinkright(17,"r",18," ");
  trlinkrightdotted(18,"m",19," ");
  trbasevertleft(0,3,0);
  trlinkright(0,"r",21," ");
  trlinkrightdotted(21,"r",23," ");
  trlinkright(0,"l",27," ");
  trlinkright(27,"r",26," ");
  trlinkrightdotted(26,"m",24," ");
  trlinkright(27,"l",28," ");
  trlinkrightdotted(28,"m",29," ");
  trlinkvertices(17,27,"xi",0,0);
  draw shuffle scaled 0.1 shifted ((vtpos[16]+vtpos[18])*0.5-(0.08xu,0.45xu)) pensize(0.4pt);
  draw shuffle scaled 0.1 shifted ((vtpos[26]+vtpos[28])*0.5-(0.08xu,0.45xu)) pensize(0.4pt);
\end{mpostfig}

\begin{mpostfig}[label=phi1ak0bj1]
	trbasevertleft(1,0,0);
	trlinkleft(1,"l",11," ");
	trlinkrightdotted(11,"l",13," ");
	trlinkleft(1,"r",17,"eaipblq");
	trlinkright(17,"l",16," ");
	trlinkrightdotted(16,"m",14," ");
	trlinkright(17,"r",18," ");
	trlinkrightdotted(18,"m",19," ");
	trbasevertright(0,3,0);
	trlinkright(0,"r",21," ");
	trlinkrightdotted(21,"r",23," ");
	trbasevertright(27,2.5,0.866);
	trlinkvertices(0,1,"eblq",0,-0.5);
	trlinkright(27,"r",26," ");
	trlinkrightdotted(26,"m",24," ");
	trlinkright(27,"l",28," ");
	trlinkrightdotted(28,"m",29," ");
	trlinkvertices(17,27,"eblqxi",-0.5,0);
        draw shuffle scaled 0.1 shifted ((vtpos[16]+vtpos[18])*0.5-(0.08xu,0.45xu)) pensize(0.4pt);
        draw shuffle scaled 0.1 shifted ((vtpos[26]+vtpos[28])*0.5-(0.08xu,0.45xu)) pensize(0.4pt);
\end{mpostfig}

\begin{mpostfig}[label=phi1ak0bj1Chain]
	trbasevertleft(1,0,0);
	trlinkleft(1,"l",11," ");
	trlinkrightdotted(11,"l",13," ");
	trlinkleft(1,"r",17," ");
	trlinkright(17,"l",16," ");
	trlinkrightdotted(16,"m",14," ");
	trlinkright(17,"r",18," ");
	trlinkrightdotted(18,"m",19," ");
	trbasevertright(0,3,0);
	trlinkright(0,"r",21," ");
	trlinkrightdotted(21,"r",23," ");
	trbasevertright(27,2.5,0.866);
	trlinkvertices(0,1,"eblq",0,-0.5);
	trlinkright(27,"r",26," ");
	trlinkrightdotted(26,"m",24," ");
	trlinkright(27,"l",28," ");
	trlinkrightdotted(28,"m",29," ");
	trlinkvertices(17,27," ",-0.5,0);
        draw shuffle scaled 0.1 shifted ((vtpos[16]+vtpos[18])*0.5-(0.08xu,0.45xu)) pensize(0.4pt);
        draw shuffle scaled 0.1 shifted ((vtpos[26]+vtpos[28])*0.5-(0.08xu,0.45xu)) pensize(0.4pt);
        draw shuffle scaled 0.1 shifted ((vtpos[18]+vtpos[27])*0.5-(0.94xu,0.22xu)) pensize(0.4pt);
        draw shuffle scaled 0.1 shifted ((vtpos[11]+vtpos[17])*0.5-(0.08xu,0.42yu)) pensize(0.4pt);
\end{mpostfig}

\begin{mpostfig}[label=phi1ak0bj1SMap]
	trbasevertleft(1,0,0);
	trlinkleft(1,"l",11," ");
	trbasevertright(16,0,2*0.87);
	trlinkverticesdotted(11,16," ",0,0);
	trbasevertleft(17,0.5,0.87);
	trlinkvertices(16,17," ",0,0);
	trbasevertright(18,1,2*0.87);
	trlinkvertices(18,17," ",0,0);
	trbasevertright(19,1,2*0.87+1);
	trlinkverticesdotted(19,18," ",0,0);
	trbasevertright(0,3,0);
	trlinkright(0,"r",21," ");
	trlinkrightdotted(21,"r",23," ");
	trbasevertright(27,2.5,0.866);
	trlinkvertices(0,1,"eblq",0,-0.5);
	trlinkright(27,"r",26," ");
	trlinkrightdotted(26,"m",24," ");
	trlinkright(27,"l",28," ");
	trlinkrightdotted(28,"m",29," ");
	trlinkvertices(17,27,"eblqxi",-0.5,0);
        draw shuffle scaled 0.1 shifted ((vtpos[16]+vtpos[18])*0.5-(0.08xu,0.45xu)) pensize(0.4pt);
        draw shuffle scaled 0.1 shifted ((vtpos[26]+vtpos[28])*0.5-(0.08xu,0.45xu)) pensize(0.4pt);
\end{mpostfig}

\begin{mpostfig}[label=phi1ak0bj2]
	trbasevertleft(1,0,0);
	trlinkleft(1,"l",11," ");
	trlinkrightdotted(11,"l",13," ");
	trbasevertleft(17,0.5,0.866);
	trlinkright(17,"l",16," ");
	trlinkrightdotted(16,"m",14," ");
	trlinkright(17,"r",18," ");
	trlinkrightdotted(18,"m",19," ");
	trbasevertright(0,3,0);
	trlinkright(0,"r",21," ");
	trlinkrightdotted(21,"r",23," ");
	trbasevertright(27,2.5,0.866);
	trlinkvertices(0,1,"meaip",0,-0.5);
	trlinkright(27,"r",26," ");
	trlinkrightdotted(26,"m",24," ");
	trlinkright(27,"l",28," ");
	trlinkrightdotted(28,"m",29," ");
	trlinkvertices(27,17,"eaipxi",-0.55,0);
	trlinkvertices(0,27,"eaipblq",-2,-0.2);
        draw shuffle scaled 0.1 shifted ((vtpos[16]+vtpos[18])*0.5-(0.08xu,0.45xu)) pensize(0.4pt);
        draw shuffle scaled 0.1 shifted ((vtpos[26]+vtpos[28])*0.5-(0.08xu,0.45xu)) pensize(0.4pt);
\end{mpostfig}

\begin{mpostfig}[label=phi1ak0bj2Chain]
	trbasevertleft(1,0,0);
	trlinkleft(1,"l",11," ");
	trlinkrightdotted(11,"l",13," ");
	trbasevertleft(17,0.5,0.866);
	trlinkright(17,"l",16," ");
	trlinkrightdotted(16,"m",14," ");
	trlinkright(17,"r",18," ");
	trlinkrightdotted(18,"m",19," ");
	trbasevertright(0,3,0);
	trlinkright(0,"r",21," ");
	trlinkrightdotted(21,"r",23," ");
	trbasevertright(27,2.5,0.866);
	trlinkvertices(0,1,"meaip",0,-0.5);
	trlinkright(27,"r",26," ");
	trlinkrightdotted(26,"m",24," ");
	trlinkright(27,"l",28," ");
	trlinkrightdotted(28,"m",29," ");
	trlinkvertices(27,17," ",-0.55,0);
	trlinkvertices(0,27," ",-2,-0.2);
        draw shuffle scaled 0.1 shifted ((vtpos[16]+vtpos[18])*0.5-(0.08xu,0.45xu)) pensize(0.4pt);
        draw shuffle scaled 0.1 shifted ((vtpos[26]+vtpos[28])*0.5-(0.08xu,0.45xu)) pensize(0.4pt);
        draw shuffle scaled 0.1 shifted ((vtpos[17]+vtpos[28])*0.5-(-0.68xu,0.22xu)) pensize(0.4pt);
        draw shuffle scaled 0.1 shifted ((vtpos[27]+vtpos[21])*0.5-(0.08xu,0.42yu)) pensize(0.4pt);
\end{mpostfig}

\begin{mpostfig}[label=phi1ak0bj2SMap]
	trbasevertleft(1,0,0);
	trlinkleft(1,"l",11," ");
	trlinkrightdotted(11,"l",13," ");
	trbasevertleft(17,0.5,0.866);
	trlinkright(17,"l",16," ");
	trlinkrightdotted(16,"m",14," ");
	trlinkright(17,"r",18," ");
	trlinkrightdotted(18,"m",19," ");
	trbasevertright(0,3,0);
	trlinkright(0,"r",21," ");
	trbasevertright(27,2.5,0.866);
	trlinkvertices(0,1,"meaip",0,-0.5);
	trbasevertright(26,3,2*0.866);
	trlinkvertices(26,27," ",0,0);
	trlinkverticesdotted(21,26," ",0,0);
	trbasevertright(28,2,2*0.866);
	trlinkvertices(28,27," ",0,0);
	trlinkrightdotted(28,"m",29," ");
	trlinkvertices(27,17,"eaipxi",-0.55,0);
        draw shuffle scaled 0.1 shifted ((vtpos[16]+vtpos[18])*0.5-(0.08xu,0.45xu)) pensize(0.4pt);
        draw shuffle scaled 0.1 shifted ((vtpos[26]+vtpos[28])*0.5-(0.08xu,0.45xu)) pensize(0.4pt);
\end{mpostfig}

\begin{mpostfig}[label=phiAphiBxiJ0]
	trbasevertleft(1,0,0);
	trlinkleft(1,"m",11,"eA");
	trlinkleft(11,"m",12,"ea23p");
	trlinkleftdotted(12,"m",17," ");
	trlinkleftdotted(17,"m",19," ");
	trbasevertright(0,2,0);
    trlinkright(0,"m",21," ");
    trlinkright(21,"m",22," ");
    trlinkrightdotted(22,"m",29," ");
	trlinkvertices(17,0,"xi",0.005xu,0);
\end{mpostfig}

\begin{mpostfig}[label=phi1ak0]
	trbasevertleft(1,0,0);
	trlinkleft(1,"l",11," ");
	trlinkrightdotted(11,"l",13," ");
	trlinkleft(1,"r",17,"eaip");
	trlinkright(17,"l",16," ");
	trlinkrightdotted(16,"m",14," ");
	trlinkright(17,"r",18," ");
	trlinkrightdotted(18,"m",19," ");
	trbasevertright(0,2,0);
    trlinkright(0,"m",21," ");
    trlinkright(21,"m",22," ");
    trlinkrightdotted(22,"m",29," ");
	trlinkvertices(17,0,"xi",0,0);
        draw shuffle scaled 0.1 shifted ((vtpos[16]+vtpos[18])*0.5-(0.08xu,0.45xu)) pensize(0.4pt);
\end{mpostfig}

\begin{mpostfig}[label=phiak01]
	trbasevertleft(1,0,0);
	trlinkleft(1,"l",11," ");
	trlinkrightdotted(11,"l",13," ");
	trbasevertright(0,2,0);
	trlinkvertices(0,1,"meaip",-0.5,-0.5);
	trbasevertright(17,0.5,0.87);
	trlinkvertices(0,17,"eaipxi",-1.2,-0.4);
	trlinkright(17,"l",16," ");
	trlinkrightdotted(16,"m",14," ");
	trlinkright(17,"r",18," ");
	trlinkrightdotted(18,"m",19," ");
	trlinkright(0,"m",21," ");
	trlinkright(21,"m",22," ");
	trlinkrightdotted(22,"m",29," ");
        draw shuffle scaled 0.1 shifted ((vtpos[16]+vtpos[18])*0.5-(0.08xu,0.45xu)) pensize(0.4pt);
\end{mpostfig}

\begin{mpostfig}[label=phiak01Chain]
	trbasevertleft(1,0,0);
	trlinkleft(1,"l",11," ");
	trlinkrightdotted(11,"l",13," ");
	trbasevertright(0,2,0);
	trlinkvertices(0,1,"meaip",-0.5,-0.5);
	trbasevertright(17,0.5,0.87);
	trlinkvertices(0,17,"eaipxi",-1.2,-0.4);
	trlinkright(17,"l",16," ");
	trlinkrightdotted(16,"m",14," ");
	trlinkright(17,"r",18," ");
	trlinkrightdotted(18,"m",19," ");
	trlinkright(0,"m",21," ");
	trlinkright(21,"m",22," ");
	trlinkrightdotted(22,"m",29," ");
        draw shuffle scaled 0.1 shifted ((vtpos[16]+vtpos[18])*0.5-(0.08xu,0.45xu)) pensize(0.4pt);
        draw shuffle scaled 0.1 shifted ((vtpos[17]+vtpos[21])*0.5-(-0.3xu,0.45xu)) pensize(0.4pt);
\end{mpostfig}

\begin{mpostfig}[label=phiak01SMap]
	trbasevertleft(1,0,0);
	trlinkleft(1,"l",11," ");
	trlinkrightdotted(11,"l",13," ");
	trbasevertright(0,2,0);
	trlinkvertices(0,1,"meaip",-0.5,-0.5);
	trbasevertright(17,0.5,0.87);
	trlinkvertices(0,17,"eaipxi",-1.2,-0.4);
	trlinkright(17,"l",16," ");
	trlinkrightdotted(16,"m",14," ");
	trlinkright(17,"r",18," ");
	trlinkrightdotted(18,"m",19," ");
	trbasevertright(21,2,1);
	trbasevertright(22,2,2);
	trbasevertright(29,2,3);
	trlinkvertices(21,0,"",0,0);
	trlinkvertices(22,21,"",0,0);
	trlinkverticesdotted(29,21,"",0,0);
        draw shuffle scaled 0.1 shifted ((vtpos[16]+vtpos[18])*0.5-(0.08xu,0.45xu)) pensize(0.4pt);
\end{mpostfig}

\begin{mpostfig}[label=phik0bj]
	trbasevertleft(1,0,0);
    trlinkleft(1,"m",11," ");
    trlinkleft(11,"m",12," ");
    trlinkleftdotted(12,"m",19," ");	
    trbasevertright(0,2,0);
	trlinkright(0,"r",21," ");
	trlinkrightdotted(21,"r",23," ");
	trbasevertright(27,1.5,0.87);
	trlinkright(27,"r",26," ");
	trlinkrightdotted(26,"m",24," ");
	trlinkright(27,"l",28," ");
	trlinkrightdotted(28,"m",29," ");
	trlinkvertices(1,27,"eblqxi",0,-0.35);
	trlinkvertices(0,1,"eblq",0,-0.5);
        draw shuffle scaled 0.1 shifted ((vtpos[26]+vtpos[28])*0.5-(0.08xu,0.45xu)) pensize(0.4pt);
\end{mpostfig}

\begin{mpostfig}[label=phik0bjChain]
	trbasevertleft(1,0,0);
	trlinkleft(1,"m",11," ");
	trlinkleft(11,"m",12," ");
	trlinkleftdotted(12,"m",19," ");	
	trbasevertright(0,2,0);
	trlinkright(0,"r",21," ");
	trlinkrightdotted(21,"r",23," ");
	trbasevertright(27,1.5,0.87);
	trlinkright(27,"r",26," ");
	trlinkrightdotted(26,"m",24," ");
	trlinkright(27,"l",28," ");
	trlinkrightdotted(28,"m",29," ");
	trlinkvertices(1,27,"eblqxi",0,-0.35);
	trlinkvertices(0,1,"eblq",0,-0.5);
        draw shuffle scaled 0.1 shifted ((vtpos[26]+vtpos[28])*0.5-(0.08xu,0.45xu)) pensize(0.4pt);
        draw shuffle scaled 0.1 shifted ((vtpos[11]+vtpos[27])*0.5-(0.58xu,0.5xu)) pensize(0.4pt);
\end{mpostfig}

\begin{mpostfig}[label=phik0bjSMap]
	trbasevertleft(1,0,0);
	trbasevertleft(11,0,1);
	trbasevertleft(12,0,2);
	trbasevertleft(19,0,3);
	trlinkvertices(11,1,"",0,0);
	trlinkvertices(12,11,"",0,0);
	trlinkverticesdotted(19,12,"",0,0);
	trbasevertright(0,2,0);
	trlinkright(0,"r",21," ");
	trlinkrightdotted(21,"r",23," ");
	trbasevertright(27,1.5,0.87);
	trlinkright(27,"r",26," ");
	trlinkrightdotted(26,"m",24," ");
	trlinkright(27,"l",28," ");
	trlinkrightdotted(28,"m",29," ");
	trlinkvertices(1,27,"eblqxi",0,-0.35);
	trlinkvertices(0,1,"eblq",0,-0.5);
        draw shuffle scaled 0.1 shifted ((vtpos[26]+vtpos[28])*0.5-(0.08xu,0.45xu)) pensize(0.4pt);
\end{mpostfig}

\begin{mpostfig}[label=phiAphiBk0]
	trbasevertleft(1,0,0);
	trlinkleft(1,"m",11,"");
	trlinkleft(11,"m",12," ");
	trlinkleftdotted(12,"m",19," ");
	trbasevertright(0,2,0);
	trlinkright(0,"m",21,"eB");
	trlinkright(21,"m",22,"eb23q");
	trlinkrightdotted(22,"m",27," ");
	trlinkrightdotted(27,"m",29," ");
	trlinkvertices(1,27,"xi",-0.005xu,0);
\end{mpostfig}

\begin{mpostdef}
color ecolor; ecolor:=(.72,.03,.06);
color bcolor; bcolor:=(.07,.05,.66);
color gcolor; 
gcolor:=(.0,.44,.0);
color faint; faint:=(.7,0.7,0.7);
vsize := 0.16xu;
rad:= 2xu;
cpath:= for i=0 upto 35: 
  dir (i*10) scaled rad ..
endfor cycle;
def vert(expr pos)=
  fill fullcircle scaled vsize shifted point pos of cpath withcolor white;
  draw halfcircle scaled vsize rotated (90+10*pos) shifted point pos of cpath withcolor ecolor pensize(1.3pt);
enddef;
def vertthick(expr pos)=
  fill fullcircle scaled vsize shifted point pos of cpath withcolor white;
  draw halfcircle scaled vsize rotated (90+10*pos) shifted point pos of cpath withcolor ecolor pensize(2.0pt);
enddef;
def vertcol(expr pos,col)=
  fill fullcircle scaled vsize shifted point pos of cpath withcolor white;
  draw halfcircle scaled vsize rotated (90+10*pos) shifted point pos of cpath withcolor col pensize(1.3pt);
enddef;
def drawbdry(expr startpos,endpos)=
  draw subpath(startpos,endpos) of cpath withcolor ecolor pensize(1.3pt);
enddef;
def drawbdryblack(expr startpos,endpos)=
  draw subpath(startpos,endpos) of cpath withcolor black pensize(1.3pt);
enddef;
def drawbdrythick(expr startpos,endpos)=
  draw subpath(startpos,endpos) of cpath withcolor ecolor pensize(2.0pt);
enddef;
def drawbdrythickblack(expr startpos,endpos)=
  draw subpath(startpos,endpos) of cpath withcolor black pensize(2.0pt);
enddef;
def drawbdrydashed(expr startpos,endpos)=
  draw subpath(startpos,endpos) of cpath withcolor ecolor dashed withdots scaled 0.5 pensize(1.3pt);
enddef;
def drawbdrydashedthick(expr startpos,endpos)=
  draw subpath(startpos,endpos) of cpath withcolor ecolor dashed withdots scaled 0.5 pensize(2.0pt);
enddef;
def vlabel(expr pos, vscal, tt)=
vert(pos);
label(tt,point pos of cpath scaled vscal);
enddef;

def vlabelcol(expr pos, vscal, tcol, vcol, tt)=
vertcol(pos, vcol);
label(tt,point pos of cpath scaled vscal) withcolor tcol;
enddef;

def ppath(expr p,spos,epos)=
subpath(arctime(spos*arclength(p))of p,arctime(epos*arclength(p)) of p) of p
enddef;
\end{mpostdef}

\begin{mpostfig}[label=fundamentaldomain]
  picture pic;
  draw (-2xu,0xu) ..controls (-2xu,0.9xu) and (2xu,0.9xu) .. ( 2xu,0xu) dashed evenly pensize (0.8pt) withcolor ecolor;
  draw ((-2xu,0xu) ..controls (-2xu,0.9xu) and (2xu,0.9xu) .. ( 2xu,0xu)) scaled 0.47 pensize (0.8pt) withcolor gcolor;
  draw (0xu,-0.15xu) .. controls (0.2xu,-0.18xu) and (0.2xu,-0.92xu) .. ( 0xu,-0.95xu) dashed evenly pensize (0.8pt) withcolor bcolor;
  draw fullcircle xscaled 4xu yscaled 1.9xu pensize (1pt);
  draw (-1xu,0.03xu){dir -20}..(1xu,0.03xu) pensize (0.8pt);
  draw (-0.8xu,-0.02xu){dir 25}..(0.8xu,-0.02xu) pensize (0.8pt);
  draw (-2xu,0xu) ..controls (-2xu,-0.9xu) and (2xu,-0.9xu) .. ( 2xu,0xu) pensize (0.8pt) withcolor ecolor;
  draw ((-2xu,0xu) ..controls (-2xu,-0.9xu) and (2xu,-0.9xu) .. ( 2xu,0xu)) scaled 0.47 pensize (0.8pt) dashed evenly withcolor gcolor;
  draw (0xu,-0.15xu) .. controls (-0.2xu,-0.18xu) and (-0.2xu,-0.92xu) .. ( 0xu,-0.95xu) pensize (0.8pt) withcolor bcolor;
  pic:=currentpicture;
  currentpicture := nullpicture;
  pickup pencircle scaled 1pt;
  drawarrow (-0.15xu,0xu)--(5.2xu,0xu) withcolor black;
  drawarrow (0xu,-0.15xu)--(0xu,2.6xu) withcolor black;
  draw (0xu,0xu)--(3.9xu, 0xu) withcolor ecolor;
  draw (4.9 xu, 2 xu)--(1 xu, 2 xu) withcolor ecolor;
  draw (4.4 xu, 1 xu)--(0.5 xu, 1 xu) withcolor gcolor;
  draw (3.9 xu, 0 xu)--(4.9 xu, 2 xu) withcolor bcolor;
  draw (0xu, 0xu)--(1 xu, 2 xu) withcolor bcolor;
  label.bot (btex $0$ etex, (0,-.1xu));
  label.top (btex $\tau$ etex, (0.9xu,2xu));
  label.top (btex $\tau+1$ etex, (4.8xu,2xu));
  label.bot(btex $1$ etex, (3.9xu,-.1xu));
  label.lft(btex Im$(z)$ etex, (0xu,2.6xu));
  label.top(btex Re$(z)$ etex, (5.2xu,0xu));
  draw pic shifted (-4xu,0.8xu) scaled (1.2);

\end{mpostfig}

\begin{mpostfig}[label=fundamentaldomain2]
  fs := (0.0xu,0.0xu)--(4.0xu, 0.0xu)--(4.0 xu, 1 xu)--(0.0 xu, 1 xu)--cycle;
  fillshape(fs,0.8*white,0pt,0.8*white);
  pickup pencircle scaled 1pt;
  drawarrow (-0.15xu,0xu)--(5.2xu,0xu) withcolor black;
  drawarrow (0xu,-0.15xu)--(0xu,2.6xu) withcolor black;
  draw (0xu,0xu)--(4.0xu, 0xu) withcolor ecolor;
  draw (4.0 xu, 2 xu)--(0 xu, 2 xu) withcolor ecolor;
  draw (4.0 xu, 0 xu)--(4.0 xu, 2 xu) withcolor bcolor;
  draw (0.0 xu, 1 xu)--(4.0 xu, 1 xu) withcolor gcolor;
  draw (0xu, 0xu)--(0 xu, 2 xu) withcolor bcolor;
  label.lft (btex $\tau$ etex, (-0.1xu,2xu));
  label.top (btex $\tau+1$ etex, (4.0xu,2xu));
  label.bot(btex $1\cong z_1$ etex, (4.2xu,-.03xu));
  label.lft(btex Im$(z)$ etex, (0xu,2.6xu));
  label.top(btex Re$(z)$ etex, (5.2xu,0xu));
  pic := currentpicture;
  currentpicture:=nullpicture;

  draw pic;
  for i=0 upto 2:
  filldot((0.78*i*xu,0xu),4pt,white)
  endfor;
  filldot((0.78*4*xu,0xu),4pt,white)
  filldot((0.78*5*xu,0xu),4pt,white)
  label.bot (btex $z_1$ etex, (0xu,-.1xu));
  label.bot (btex $z_2$ etex, (0.78xu,-.1xu));
  label.bot (btex $z_3$ etex, (1.56xu,-.1xu));
  label.bot (btex $\cdots$ etex, (2.34xu,-.1xu));
  label.bot (btex $z_n$ etex, (3.14xu,-.1xu));
  draw pic shifted (8xu,0xu);
  for i=0 upto 2:
  filldot(((8+0.78*i)*xu,0xu),4pt,white)
  endfor;
  filldot(((8+0.78*4)*xu,0xu),4pt,white)
  for i=1 upto 2:
    filldot(((8.1+0.78*i)*xu,1xu),4pt,white)
  endfor;
  filldot(((8.1+0.78*4)*xu,1xu),4pt,white)
  label.bot (btex $z_1$ etex, (8xu,-.1xu));
  label.bot (btex $z_2$ etex, (8.78xu,-.1xu));
  label.bot (btex $z_3$ etex, (9.56xu,-.1xu));
  label.bot (btex $\cdots$ etex, (10.34xu,-.1xu));
  label.bot (btex $z_k$ etex, (11.14xu,-.1xu));
  label.top (btex $z_{k+1}$ etex, (8.88xu,1.1xu));
  label.top (btex $z_{k+2}$ etex, (9.66xu,1.1xu));
  label.top (btex $\cdots$ etex, (10.44xu,1.1xu));
  label.top (btex $z_n$ etex, (11.32xu,1.1xu));
\end{mpostfig}

\makeatletter
\providecommand*{\shuffle}{%
  \mathbin{\mathpalette\shuffle@{}}%
}
\newcommand*{\shuffle@}[2]{%
  \sbox0{$#1\vcenter{}$}%
  \kern .15\ht0 
  \rlap{\vrule height .25\ht0 depth 0pt width 2.5\ht0}%
  \raise.1\ht0\hbox to 2.5\ht0{%
    \vrule height 1.75\ht0 depth -.1\ht0 width .17\ht0 %
    \hfill
    \vrule height 1.75\ht0 depth -.1\ht0 width .17\ht0 %
    \hfill
    \vrule height 1.75\ht0 depth -.1\ht0 width .17\ht0 %
  }%
  \kern .15\ht0 
}
\makeatother

\makeatletter
\providecommand*{\boldshuffle}{%
  \mathbin{\mathpalette\boldshuffle@{}}%
}
\newcommand*{\boldshuffle@}[2]{%
  \sbox0{$#1\vcenter{}$}%
  \kern .15\ht0 
  \rlap{\vrule height .25\ht0 depth 0.25pt width 2.5\ht0}%
  \raise.1\ht0\hbox to 2.5\ht0{%
    \vrule height 1.75\ht0 depth -.1\ht0 width .31\ht0 %
    \hfill
    \vrule height 1.75\ht0 depth -.1\ht0 width .31\ht0 %
    \hfill
    \vrule height 1.75\ht0 depth -.1\ht0 width .31\ht0 %
  }%
  \kern .15\ht0 
}
\makeatother

\makeatletter
\g@addto@macro\bfseries{\boldmath}
\makeatother

\usepackage{xparse}

\ExplSyntaxOn
\NewDocumentCommand{\Gtargz}{m m}
{
 \Gt\left(\begin{smallmatrix}
 \Gtargz_print:n {#1} \\
 \Gtargz_print:n {#2}
 \end{smallmatrix};z\right)
}
\seq_new:N \l_Gtargz_list_seq
\cs_new_protected:Npn \Gtargz_print:n #1
{
  \seq_set_split:Nnn \l_Gtargz_list_seq { , } { #1 }
  \seq_use:Nn \l_Gtargz_list_seq { , & }
}
\ExplSyntaxOff

\ExplSyntaxOn
\NewDocumentCommand{\Gtargzt}{m m}
{
 \Gt\left(\begin{smallmatrix}
 \Gtargzt_print:n {#1} \\
 \Gtargzt_print:n {#2}
 \end{smallmatrix};z,\tau\right)
}
\seq_new:N \l_Gtargzt_list_seq
\cs_new_protected:Npn \Gtargzt_print:n #1
{
  \seq_set_split:Nnn \l_Gtargzt_list_seq { , } { #1 }
  \seq_use:Nn \l_Gtargzt_list_seq { , & }
}
\ExplSyntaxOff

\newcommand{\SI}[1]{\Sel[#1]}

\ExplSyntaxOn
\NewDocumentCommand{\SIE}{m m}
{
\SelEn\!\Big[\begin{smallmatrix}
 \SI_print:n {#1} \\
 \SI_print:n {#2}
 \end{smallmatrix}\Big]
}
\seq_new:N \l_SI_list_seq
\cs_new_protected:Npn \SI_print:n #1
{
  \seq_set_split:Nnn \l_SI_list_seq { , } { #1 }
  \seq_use:Nn \l_SI_list_seq { , & }
}
\ExplSyntaxOff

\ExplSyntaxOn
\NewDocumentCommand{\SIEzwei}{m m}
{
\SelEz\!\Big[\begin{smallmatrix}
 \SI_print:n {#1} \\
 \SI_print:n {#2}
 \end{smallmatrix}\Big]
}
\seq_new:N \l_SIz_list_seq
\cs_new_protected:Npn \SIz_print:n #1
{
  \seq_set_split:Nnn \l_SIz_list_seq { , } { #1 }
  \seq_use:Nn \l_SIz_list_seq { , & }
}
\ExplSyntaxOff

\usepackage[pdfencoding=auto,bookmarks=true,hyperfigures=true]{hyperref}%
\PassOptionsToPackage{unicode}{hyperref}
\usepackage{graphbox}
\usepackage{float}
\usepackage[nosort]{cite}
\usepackage{mcite}
\usepackage{lmodern}
\usepackage{amsbsy}
\usepackage[utf8]{inputenc}
\usepackage[font=small,labelfont=bf]{caption}

\iffancy

\def\showkeysrefformat#1{{\normalfont\tiny\ttfamily#1}}
\makeatletter
\def\SK@@ref#1>#2\SK@{%
 {\@inlabelfalse\leavevmode\vbox to\z@{%
 \vss\SK@refcolor\rlap{\vrule\raise .75em%
  \hbox{\showkeysrefformat{#2}}}}}}
\makeatother
\fi

\numberwithin{equation}{section}

\newcommand{\eqn}[1]{eq.~\eqref{#1}}

\newcommand{\eqns}[2]{eqs.~\eqref{#1} and~\eqref{#2}}

\newcommand{\rcite}[1]{ref.~\cite{#1}}
\newcommand{\rcites}[1]{refs.~\cite{#1}}

\providecommand{\href}[2]{#2}

\makeatletter
\def\mr@ignsp#1 {\ifx\:#1\@empty\else #1\expandafter\mr@ignsp\fi}%
\newcommand{\multiref}[1]{\begingroup
\xdef\mr@no@sparg{\expandafter\mr@ignsp#1 \: }%
\def\mr@comma{}%
\@for\mr@refs:=\mr@no@sparg\do{\mr@comma\def\mr@comma{,}\ref{\mr@refs}}%
\endgroup}
\renewcommand{\eqref}[1]{(\multiref{#1})}
\makeatother

\makeatletter
\newcommand{\namedref}[2]{#1~\hyperref[#2]{\ref*{#2}}}
\newcommand{\secref}{\@ifstar{\namedref{Section}}{\namedref{section}}}
\newcommand{\subsecref}{\@ifstar{\namedref{Subsection}}{\namedref{subsection}}}
\newcommand{\appref}{\@ifstar{\namedref{Appendix}}{\namedref{appendix}}}
\newcommand{\tabref}{\@ifstar{\namedref{Table}}{\namedref{table}}}
\newcommand{\figref}{\@ifstar{\namedref{Figure}}{\namedref{figure}}}
\makeatother

\providecommand{\hypersetup}[1]{}
\providecommand{\texorpdfstring}[2]{#1}

\hypersetup{plainpages=false}
\hypersetup{pdfpagemode=UseNone}
\hypersetup{bookmarksnumbered=true}
\hypersetup{pdfstartview=FitH}
\hypersetup{colorlinks=false}
\hypersetup{citebordercolor={0.7 0.7 1}}
\hypersetup{urlbordercolor={.4 .8 1}}
\hypersetup{linkbordercolor={1 .8 .6}}
\hypersetup{colorlinks=true, urlcolor=[rgb]{0.13,0.30,0.45}, linkcolor=[rgb]{0.13,0.30,0.45}, citecolor=[rgb]{0.55,0.0,0.05}}
\makeatletter
\let\@keywords\@empty
\let\@subject\@empty
\providecommand{\keywords}[1]{\gdef\@keywords{#1}}
\providecommand{\subject}[1]{\gdef\@subject{#1}}
\def\thetitle{\@title}
\def\theauthor{\@author}
\def\thesubject{\@subject}
\def\thedate{\@date}
\def\thekeywords{\@keywords}
\makeatother
\AtBeginDocument{
\hypersetup{pdftitle={\thetitle}}%
\hypersetup{pdfauthor={\theauthor}}%
\hypersetup{pdfsubject={\thesubject}}%
\hypersetup{pdfkeywords={\thekeywords}}%
}

\newif\ifnote 
\notetrue

\allowdisplaybreaks

\newcommand{\smat}[1]{\begin{smallmatrix}#1\end{smallmatrix}}

\let\Re\relax\DeclareMathOperator{\Re}{Re}
\let\Im\relax\DeclareMathOperator{\Im}{Im}

\newcommand{\pd}{\partial}

\newcommand{\XXX}{X^\tau_{0,n}}
\newcommand{\DDD}{D^\tau_{0,n}}
\newcommand{\BBB}{B^\tau_{0,n}}
\newcommand{\XXXn}[1]{X^\tau_{0,#1}}

\newcommand{\eps}{\epsilon}
\newcommand{\ep}{\epsilon}

\newcommand{\s}{\sigma}

\newcommand{\om}{\omega}

\newcommand{\SL}{\mathrm{SL}}

\newcommand{\te}{\textrm}

\newcommand{\dd}{\mathrm{d}}

\newcommand{\ap}{\alpha'}

\newcommand{\vecb}{\left(\begin{array}{c}}
\newcommand{\vece}{\end{array}\right)}
\newcommand{\ccb}{\left(\begin{array}{cc}}
\newcommand{\cce}{\end{array}\right)}
\newcommand{\cccb}{\left(\begin{array}{ccc}}
\newcommand{\ccce}{\end{array}\right)}
\newcommand{\ccccb}{\left(\begin{array}{cccc}}
\newcommand{\cccce}{\end{array}\right)}
\newcommand{\cccccb}{\left(\begin{array}{ccccc}}
\newcommand{\ccccce}{\end{array}\right)}
\newcommand{\ccccccb}{\left(\begin{array}{cccccc}}
\newcommand{\cccccce}{\end{array}\right)}

\newcommand{\ZC}{\mathbb C}

\newcommand{\ZR}{\mathbb R}
\newcommand{\ZZ}{\mathbb Z}

\newcommand{\CG}{\mathcal{G}}

\newcommand{\CO}{\mathcal{O}}      
\newcommand{\CP}{\mathcal{P}}

\newcommand{\BT}{\boldsymbol{\mathrm{T}}}

\def\tree{\text{tree}}

\def\BCJ{\textrm{BCJ}}

\let\emptyset\varnothing

\newcommand{\nnl}{\nonumber\\}

\DeclareMathOperator{\GL}{\Gamma}
\DeclareMathOperator{\Gt}{\tilde{\Gamma}}
\newcommand{\KN}{\mathrm{KN}}
\newcommand{\KNt}{\mathrm{KN}^\tau}

\DeclareMathOperator{\Sel}{S}

\DeclareMathOperator{\SelEn}{S^E_\mathit{n}}
\DeclareMathOperator{\SelEz}{S^E_2}
\DeclareMathOperator{\Selbld}{\mathbf{S}}
\DeclareMathOperator{\SelbldE}{\mathbf{S}^E}
\DeclareMathOperator{\SelbldEn}{\mathbf{S}^E_\mathit{n}}

\DeclareMathOperator{\BC}{\mathbf{C}}
\DeclareMathOperator{\gm}{\gamma}
\DeclareMathOperator{\gmz}{\gamma_0}

\DeclareMathOperator{\zm}{\zeta}

\DeclareMathOperator{\dlog}{\mathrm{dlog}}

\newcommand{\El}{\text{E}}

\DeclareMathOperator{\GGs}{G}
\newcommand{\GGz}[1]{\GGs^0_{#1}}
\newcommand{\GG}[1]{\GGs_{#1}}
\newcommand{\GGn}[1]{\GGs^0_{#1}}

\usepackage{nicefrac}

\DeclareMathOperator{\E}{E}

\DeclareMathOperator{\BZ}{\boldsymbol{\mathrm{Z}}}

\newcommand{\Zt}{Z^{\tau}}
\newcommand{\Ztn}[1]{Z^{\tau}_#1}

\newcommand{\Ztzn}[1]{Z^{\tau}_{0,#1}}
\newcommand{\BZtzn}[1]{\BZ^{\tau}_{0,#1}}
\newcommand{\Ztree}{Z^{\tree}}
\newcommand{\phiChain}{\varphi^\tau}

\title{\textbf{
    Two dialects for KZB equations: generating one-loop open-string integrals 
    }}
    \author{Johannes Broedel\texorpdfstring{$^{\,\textit{a}}$}{},
            Andr\'e Kaderli\texorpdfstring{$^{\,\textit{a},\textit{b}}$}{},
            Oliver Schlotterer\texorpdfstring{$^{\,\textit{c}}$}{}
}
\date{\today}

\begin{document}
\pdfbookmark[1]{Title Page}{title} \thispagestyle{empty}
\begin{flushright}
  \verb!HU-EP-20/14!\\
  \verb!HU-Mathematik-2020-04!\\
  \verb!UUITP-22/20!
\end{flushright}
\vspace*{0.4cm}
\begin{center}%
  \begingroup\LARGE\bfseries\thetitle\par\endgroup
\vspace{1.0cm}

\begingroup\large\theauthor\par\endgroup
\vspace{9mm}
\begingroup\itshape
$^{\te{a}}$Institut f\"ur Mathematik und Institut f\"ur Physik, Humboldt-Universit\"at zu Berlin\\
IRIS Adlershof, Zum Gro\ss{}en Windkanal 6, 12489 Berlin, Germany
\par\endgroup
\vspace{3mm}
\begingroup\itshape
$^{\te{b}}$Max-Planck-Institut f\"ur Gravitationsphysik, Albert-Einstein-Institut\\
Am M\"uhlenberg 1, 14476 Potsdam, Germany
\par\endgroup
\vspace{3mm}
\begingroup\itshape
$^{\te{c}}$Department of Physics and Astronomy, Uppsala University\\
75108 Uppsala, Sweden
\par\endgroup
\vspace{3mm}

\vspace{0.4cm}

\begingroup\ttfamily
jbroedel@physik.hu-berlin.de, kaderlia@physik.hu-berlin.de, oliver.schlotterer@physics.uu.se
\par\endgroup

\vspace{1.2cm}

\bigskip

\textbf{Abstract}\vspace{5mm}

\begin{minipage}{14.4cm}
Two different constructions generating the low-energy expansion of genus-one configuration-space integrals appearing in one-loop open-string amplitudes have been put forward in \rcites{Mafra:2019xms, *Mafra:2019ddf, Broedel:2019gba}. We are going to show that both approaches can be traced back to an elliptic system of Knizhnik--Zamolodchikov--Bernard(KZB) type on the twice-punctured torus. 

We derive an explicit all-multiplicity representation of the elliptic KZB system for a vector of iterated integrals with an extra marked point and explore compatibility conditions for the two sets of algebra generators appearing in the two differential equations.  
\end{minipage}

\vspace*{4cm}

\end{center}

\newpage

\tableofcontents

\vspace*{20pt}

\section{Introduction and summary}

During the last years we have been experiencing a significant growth in understanding the mathematical concepts leading to recursion relations for scattering amplitudes in quantum field and string theory.  A multitude of languages and approaches is available for various quantum field theories, see for instance \cite{Berends:1987me,Cachazo:2004kj,Britto:2005fq, CaronHuot:2011kk, Baadsgaard:2015twa, Cachazo:2015aol, Arkani-Hamed:2016byb, Jurco:2019yfd} and references therein. Recent progress on string amplitudes in turn was driven by disentangling their polarization degrees of freedom from moduli-space integrals over punctured worldsheets and finding separate recursions for both types of building blocks.  
The low-energy expansion of string amplitudes exposed by such recursions at tree and loop level contains a wealth of information on relations between gauge theories and gravity, string dualities and counterterms including their non-renormalization theorems.
For the moduli-space integrals in open-string tree-level amplitudes, a recursion based on the Knizhnik-Zamolodchikov equation was already identified in \rcite{Broedel:2013aza} based on \rcites{Aomoto, Terasoma} and later complemented by other methods put forward in \rcites{Mafra:2016mcc, Puhlfuerst:2015gta}. 

The problem of finding a one-loop (or genus-one) analogue of the open-string tree-level recursions was long-standing. A first mathematical challenge was to thoroughly understand iterated integrals on the elliptic curve and their associated special values, elliptic multiple zeta values \cite{Levin, BrownLev, Enriquez:Emzv, Broedel:2015hia}. Then, the cooperation of mathematicians and physicists was instrumental to investigate and understand the relation of those iterated integrals to one-loop open-string amplitudes and their differential equations \cite{Broedel:2014vla,Broedel:2017jdo, Broedel:2019vjc}. The closed-string counterparts of these genus-one integrals lead to an intriguing system of non-holomorphic modular forms \cite{DHoker:2015wxz, DHoker:2016mwo} that inspired mathematical research lines including \rcites{Zerbini:2015rss, Brown:2017qwo, Brown:2017qwo2, DHoker:2019xef, Zagier:2019eus}.

These structural considerations paved the way for two recent methods \cite{Mafra:2019xms, Broedel:2019gba} to systematically evaluate the integrals over punctures on the boundary of a genus-one surface order by order in the inverse string tension $\alpha'$. These integrals to be referred to as genus-one configuration-space integrals\footnote{We distinguish {\it moduli-space integrals} over both the punctures $z_i$ and the modular parameter $\tau$ of a genus-one surface from the {\it configuration-space integrals} over the $z_i$ which are still functions of $\tau$.} form the backbone of one-loop open-string amplitudes. Both algorithms rely on differential equations of Knizhnik--Zamolodchikov--Bernard(KZB) type on a genus-one surface with boundaries. 
\begin{itemize}
	\item In \rcite{Mafra:2019xms}, a KZB-type differential equation with respect to the modular parameter $\tau$, which encodes the geometry of genus-one surfaces, was established. Acting on a vector of generating functions for one-loop configuration-space integrals, the $\tau$-derivative can be expressed as a linear operator that mixes the components in different vector entries. In particular, this exposes finite-dimensional conjectural matrix representations of a special derivation algebra with corresponding generators $\eps_k$. Using Picard iteration, the equation can be solved starting from a particular value which is conveniently chosen as the limit $\tau\to i\infty$ where the genus-one configuration-space integrals degenerate to their genus-zero counterparts with two additional legs.
	\item In \rcite{Broedel:2019gba}, a KZB-type differential equation with respect to the position of an auxiliary point $z_0$ was identified. Facilitating a vector of configuration-space integrals with the auxiliary point (genus-one Selberg integrals), a solution can be obtained using the KZB associator: it relates two regularized boundary values, which emerge when sending the auxiliary point to the poles of the differential equation in two distinct ways. At one boundary value one obtains the one-loop configuration-space integrals, while at the other boundary one recovers again genus-zero configuration-space integrals with two additional legs. The main players in the construction are infinite-dimensional matrix representations of an algebra with generators $x_k$, which can be cut off to finite size when calculating up to a certain order in the $\alpha'$-expansion of the string amplitudes. 
\end{itemize}
The two algorithms relate open-string tree-level and one-loop amplitudes in the same way: both are capable of determining the $n$-point configuration-space integrals at genus one from $(n{+}2)$-point configuration-space integrals at genus zero. On the contrary, the representations of the KZB equations and underlying algebra generators are quite distinct. The relation between the two approaches can be best understood and investigated by considering a formalism combining the advantages of each of the previous methods: the central object to be considered in this article is a length-$n!$ vector of generating functions for planar $n$-point one-loop configuration-space integrals to be denoted by $\BZ^\tau_{0,n}$ with an auxiliary point $z_0$: In particular
\begin{enumerate}[label=\alph*)]
	\item we will find an all-multiplicity expression for the $\tau$-derivative of $\BZtzn{n}$ in order to connect with the approach in \rcite{Mafra:2019xms}. This will be an equation of the form
	\begin{align}\label{eqn:tauderivative}
	2\pi i \partial_{\tau} \BZtzn{n}&=
	\left(\DDD(\{\eps_k\})+\BBB(\{x_{j}\})\right)
	\BZtzn{n}\, ,\quad k=0,4,6,8,\dots,\quad j=1,2,3,\dots\,,
	\end{align}
	where the operators $ \DDD$ and $\BBB$ are $n!{\times}n!$ matrices with entries proportional to $\alpha'$.
	\item we will rewrite the formalism of \rcite{Broedel:2019gba} in terms of the vector of generating series $\BZ^\tau_{0,n}$, leading to finite-size matrix representations and an all-multiplicity expression for the \mbox{$z_0$-derivative} of $ \BZtzn{n}$ of the form
	\begin{align}\label{eqn:z0derivative}
	\partial_{z_0} \BZtzn{n}&= \XXX(\{x_k\}) \BZtzn{n}\,,\quad k=0,1,2,\dots\,,
	\end{align}
where $\XXX$ is a $n!{\times} n!$ matrix proportional to $\alpha'$ as well. The constituents $x_k$ are related to the braid matrices that govern the genus-zero counterparts of $\BZtzn{n}$ \cite{Mizera:2019gea}.
\end{enumerate}
Hence, the generating functions $\BZtzn{n}$ of genus-one configuration-space integrals to be introduced
in this work furnish integral representations for solutions to the elliptic KZB system.
Having two differential equations (\ref{eqn:tauderivative}) and (\ref{eqn:z0derivative}) at our disposal, we can demand commutativity of the two derivatives. This implies consistency conditions for the two classes of algebra generators involved. We have checked on a case-by-case basis that our realizations of the generators
satisfy these relations.

In \secref{sec:review} we are going to provide the mathematical and physical setting: we will discuss genus-zero and genus-one configuration-space integrals contributing to tree-level and one-loop open-string scattering amplitudes, respectively. This will set our conventions and incorporate a review of general properties of configuration-space integrals, iterated integrals and (elliptic) multiple zeta values. \secref*{sec:DGLs} is devoted to the discussion of several types of differential equations allowing for recursive solutions: in \subsecref{ssec:z0tree} we review the genus-zero recursion from \rcite{Broedel:2013aza} and bring it into the context of the later genus-one results.
In subsections \ref{ssec:taulang} and \ref{ssec:z0lang} we discuss the \mbox{$\tau$-based} and \mbox{$z_0$-based} genus-one recursions from \rcite{Mafra:2019xms} and \rcite{Broedel:2019gba}, respectively.  The central object to be discussed in \secref{sec:derivatives} is the vector of configuration-space integrals $\BZ^\tau_{0,n}$ with an auxiliary point. After introducing the vector, we will perform the two steps a) and b) lined out above, resulting in an all-multiplicity representation of the elliptic KZB system on the twice-punctured torus.  By considering the regularized boundary values for the elliptic KZB system, we will relate the different approaches in \secref{sec:translation}, before we conclude in \secref{sec:conclusion}. 


\section{Open-string scattering amplitudes and configuration-space integrals}
\label{sec:review}

In this review section we will introduce several mathematical objects and concepts necessary for the description of open-string scattering amplitudes at genus zero and genus one. Rather than providing yet another thorough and detailed introduction, we will just mention and collect the key concepts here and provide numerous links to elaborate discussions.  

The structure of scattering amplitudes in open-string theories can be most easily captured and understood when disentangling the results from evaluation of a conformal worldsheet correlator: the latter depends on the external polarizations through a \textit{kinematical part} which we will separate from the \textit{moduli-space integrals} that encode string corrections to field-theory amplitudes through their series expansion in $\ap$. Moduli-space integrals are dimensionless as they depend on dimensionless Mandelstam variables
\begin{equation}
	\label{eqn:Mandelstam}
	s_{i_1 i_2 \dots i_r}=-\alpha'(k_{i_1}+k_{i_2}+\dots +k_{i_r})^2,\qquad 1\leq i_k \leq n\, ,
\end{equation}
where $n$ denotes the number of external particles. Their integrands are calculated as conformal correlators of vertex positions $z_i$ on Riemann surfaces, whose genus refers to the loop order in question. 
In the next two subsections, we are going to collect the basic formalism for the integration over 
open-string punctures at genus zero and one: tree level and one loop, respectively. Since we
do not perform the integral over the modular parameter $\tau$ of genus-one surfaces in this work, the integrals 
over the $z_i$ will be referred to as {\it configuration-space integrals} in contradistinction to the full {\it moduli-space integrals}
entering one-loop string amplitudes.

The Mandelstam variables defined above in \eqn{eqn:Mandelstam} are going to take a role as (complex) parameters in the configuration-space integrals to be considered in this article. Naturally, the convergence behavior of those integrals depends on the values of the Mandelstam variables. Convergent integrals are obtained, when the Mandelstam variables are taken to satisfy the conditions listed below, though one can analytically continue to different regions. The conditions are formulated in terms of Mandelstam variables whose indices are related to \textit{consecutive} insertion points on the disk or cylinder boundary. 

\begin{itemize}
	\item For genus-zero configuration-space integrals, the issue of convergence was discussed at various places, see e.g.\ \rcites{Mandelstam:1974fq, Brown:2019wna}: tree-level configuration-space integrals converge, if 
\begin{equation}
	\Re(s_{i_1i_2\ldots i_r})<0\text{ for all consecutive labels }i_1,i_2,\ldots,i_r
	\,,
	\label{assump}
\end{equation}
unless $s_{i_1 i_2 \ldots i_r}$ vanishes by momentum conservation.
	\item For the augmented genus-one configuration-space integrals $\BZ^\tau_{0,n}$, we relax momentum conservation and consider all the $s_{ij}$ with $i<j$ as independent. Still, the above condition \eqref{assump} for convergence
carries over to the genus-one configuration-space integrals, where the notion of consecutive insertion points is adapted
to an auxiliary puncture $z_0$ between $z_n$ and $z_1$. The associated auxiliary Mandelstam invariant $s_{01}$ is furthermore taken to obey
	\begin{equation}
	\Re(s_{i_1i_2\ldots i_r})<\Re(s_{01})<0\text{ for all consecutive labels }(i_1,i_2,\ldots,i_r)\neq (0,1)
	\,,
	\label{assumpGenusOne}
	\end{equation}
which is no restriction in the applications to one-loop open-string amplitudes since $s_{01}$ will
drop out from the final results.
In the context of one-loop open-string amplitudes, integrals of the type in $\BZ^\tau_{0,n}$ are
analytically continued from their region of convergence to physically sensible situations.
The resulting singularities in the form of poles and branch cuts
have been for example explored in a closed-string context in \rcite{DHoker:1994gnm}. 
\end{itemize}

\subsection{Tree level: genus zero}
Calculating open-string amplitudes at tree level amounts to the evaluation of configuration-space integrals on a genus-zero surface with boundary. The corresponding genus-zero Green's function is a plain logarithm
\begin{equation}
	\CG^{\rm tree}_{ij}=\log |z_{ij}|=G(0;|z_{ij}|) 
\end{equation}
of the distance
\begin{align}
z_{ij}&=z_i-z_j
\end{align}
of two insertion points. The notation $G$ refers to the iterated integrals defined in \eqn{eqn:Gint} below.  In the configuration-space integrals, the Green's function appears in terms of the genus-zero Koba--Nielsen factor 
\begin{align}
	\label{eqn:KN}
	\KN^{\rm tree}_{12\ldots n}=\exp\Big(-\sum_{1\leq i<j\leq n} s_{ij} \CG^{\rm tree}_{ij}\Big) = \prod_{1\leq i<j\leq n} |z_{ij}|^{-s_{ij}}\,.
\end{align}
All configuration-space integrals for the calculation of open-string scattering amplitudes at tree level can be expressed as linear combinations of the integrals \cite{Mafra:2011nv, Broedel:2013tta, Azevedo:2018dgo}
\begin{equation}
	\label{eqn:Ztree}
	\Ztree_n(a_1,a_2,\ldots,a_n|1,2,\ldots,n)=\int\limits_{-\infty<z_{a_1}<\ldots<z_{a_n}<\infty} \frac{\dd z_1\cdots \dd z_n}{\text{vol}\,\text{SL}_2(\ZR)}\frac{\KN^{\rm tree}_{12\ldots n} }{z_{12}z_{23}\cdots z_{n-1\,n}z_{n1}}\, ,
\end{equation}
where the labels $a_1,\ldots ,a_n$ fix a certain succession of the insertion points on the disk boundary. An independent cyclic ordering selects the permutation of the inverse $z_{12}z_{23}\cdots z_{n-1\,n}z_{n1}$, the so-called Parke--Taylor factor.  For a given multiplicity and particular choice of the labels $a_i$ in the first slot, the collection of all integrals obtained for all permutations of the ordering $1,2,\ldots,n$ in the second slot is not independent: integrals over different Parke--Taylor factors are related by partial fraction and integration by parts \cite{Mafra:2011nv, Broedel:2013tta}. A convenient basis choice, which we are going to use throughout this article, consists of fixing the position of three of the labels in the Parke--Taylor factor, for example 
\begin{align}
	\label{eqn:Zbasisexample}
        \BZ^\tree_n=\begin{pmatrix}
        \Ztree_n(a_1,a_2,\ldots,a_n|1,\sigma,n,n{-}1)
        \end{pmatrix} \text{ for } \sigma \in \CP(2,3,\ldots,n{-}2)\, .
\end{align}
For the choice of fixing the $\SL_2(\ZR)$ redundancy via $(z_1,z_{n-1},z_n)=(0,1,\infty)$
and the ordering to $(a_1,a_2,\dots, a_n)=(1,2,\dots, n)$, the integrals are explicitly given by
\begin{align}\label{eqn:ZTreeExplicit}
\Ztree_n(1,2,\ldots,n|1,\sigma,n,n{-}1)&= - \! \! \! \! \! \! \! \! \int\limits_{0<z_{2}<\ldots<z_{n-2}<1}  \! \! \! \!\! \! \! \! \dd z_2  \cdots \dd z_{n-2}\frac{\prod_{1\leq i<j\leq n-1} |z_{ij}|^{-s_{ij}}}{z_{1\sigma(2)}z_{\sigma(2)\sigma(3)}\cdots z_{\sigma(n-3),\sigma(n-2)}}\,.
\end{align}
The dimension, that is, the length of the basis vector $\BZ^\tree_n$ for a fixed integration domain, is $(n{-}3)!$, which is precisely the number predicted by twisted cohomology and BCJ relations \cite{Bern:2008qj}. The basis dimension 
follows from results in twisted de Rham theory \cite{aomoto1987}, which have been interpreted in a string-theory context recently \cite{Mizera:2017cqs, Mizera:2019gea}. 

After taking its kinematic poles into account \cite{Mafra:2011nw, Broedel:2013tta, Brown:2019wna}, a $Z$-integral as defined in \eqn{eqn:Ztree} above is calculated by expanding the Koba--Nielsen-factor in $\ap$ (cf.~\eqn{eqn:Mandelstam}) and then evaluating each iterated integral separately. In particular, each of the $Z$-integrals can be expressed in terms of iterated integrals (multiple polylogarithms)\footnote{Our conventions for multiple polylogarithms agree with \rcites{Goncharov:2001iea, Duhr:2011zq}.}
\begin{align}
	\label{eqn:Gint}
	G(a_1,a_2,\ldots,a_r;z)=\int_0^z \frac{\dd z_1}{z_1-a_1}G(a_2,\ldots,a_r;z_1)\,
\end{align}
with $a_i\in\lbrace0,1\rbrace$ as well as $G(;z)=1$ and $z \in \mathbb C \setminus \lbrace 0,1\rbrace$.  For tree-level open-string integrals, the outermost integration variable, e.g.~one of the insertion points, can always be chosen to equal one by fixing the volume of $\mathrm{SL}_2(\ZR)$ in \eqn{eqn:Ztree}. Thus we will have to evaluate integrals of type \eqref{eqn:Gint} at $z=1$. Fortunately, all integrals appearing can be related to well-known representations of multiple zeta values (MZVs) using the identity: 

\begin{align}
	\label{eqn:MZV}
	\zeta_{n_1,n_2,\ldots,n_r}= \hspace{-2mm}\sum_{0<k_1<\ldots<k_r}\hspace{-2mm} k_1^{-n_1}\ldots k_r^{-n_r}=(-1)^r\,G(\underbrace{0,0,\ldots,0,1}_{n_r},\underbrace{0,0,\ldots,0,1}_{n_{r-1}},\ldots,\underbrace{0,0,\ldots,0,1}_{n_1};1)\,.
\end{align}
The integrals defined in \eqn{eqn:Gint} exhibit endpoint divergences if $a_r=0$ or $a_1=z$. Therefore, they will have to be regularized, which implies corresponding regularizations for MZVs and may have an echo in the kinematic poles
of the $Z$-integrals defined in \eqn{eqn:Ztree}. Throughout this article, we will always assume to work with regularized iterated integrals. For instance, the multiple polylogarithms $G(1,\ldots;1)$ and $G(\ldots,0;1)$
in (\ref{eqn:Gint}) will be shuffle-regularized based on $G(0;1)=G(1;1)=0$ which assigns regularized 
values to divergent MZVs (\ref{eqn:MZV}) with $n_r=1$ such as $\zeta_1=0$ \cite{Panzer:2015ida}.

As an example, let us state the first couple of orders of the series expansion of a typical integral $\Ztree_n$:
\begin{align}
&\Ztree_5(1,2,3,4,5|
 2,1,4,3,5)\nonumber\\
&= \frac{1}{s_{12}s_{34}} + \zeta_2 \Big(1- \frac{ s_{45}}{s_{12}}- \frac{ s_{15}}{s_{34}} \Big) \nonumber \\
&\phantom{=}-  \zeta_3  \Big(   \frac{ s_{45}  (s_{34} +s_{45})}{s_{12}}  + 
	\frac{ s_{15}(s_{12} + s_{15}) }{s_{34}}  -  2   s_{23}  - s_{12}-s_{34} \Big) + {\cal O}(s_{ij}^2)\,. \label{eqn:Zexample}
\end{align}
The analogous expressions for arbitrary orders in the $\ap$-expansion of $n$-point disk integrals can for instance be generated from the Drinfeld associator \cite{Broedel:2013aza, Kaderli:2019dny} or Berends--Giele recursions \cite{Mafra:2016mcc}.\footnote{Explicit results at $n\leq 7$ points can be downloaded from \rcite{MZVWebsite}, and explicit all-multiplicity expressions
up to and including $\ap^7$ can be generated from the code available to download from \rcite{gitrep}. Earlier work on $\alpha'$-expansions at $n\leq 7$ points including \cite{Oprisa:2005wu, Stieberger:2006te, Stieberger:2009rr, Boels:2013jua, Puhlfuerst:2015gta} took advantage of connections with hypergeometric functions.} The Berends--Giele method in \rcite{Mafra:2016mcc} applies to $Z$-integrals with arbitrary pairs of permutations $\Ztree_n(a_1,\ldots,a_n|b_1,\ldots,b_n)$
whose decomposition in the $(n{-}3)!$ bases expanded 
in \cite{Broedel:2013aza, Kaderli:2019dny} can be generated 
from the techniques in \rcite{Mafra:2016ltu}.

\subsection{One-loop level: genus one}

\begin{figure}[h!]
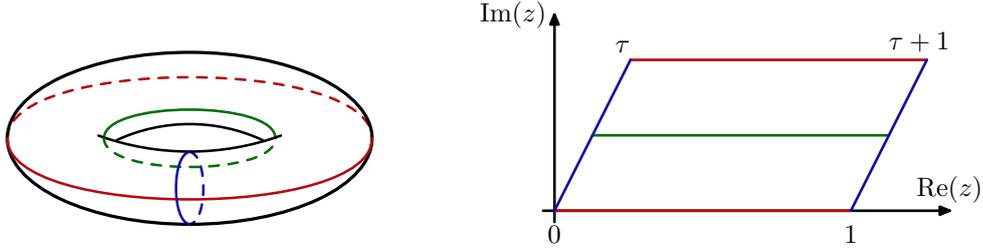

\begin{center}
	\mpostuse{fundamentaldomain}
\end{center}
\caption{
Torus with $A$- and $B$-cycle (red and blue) and their images as boundaries of the fundamental domain. The green line marks the additional cut necessary to obtain 
the cylinder and M\oe{}bius-strip worldsheets for the open string as the parallelogram below
if $\tau \in i \mathbb R$ and $\tau \in i\mathbb R+ \frac{1}{2}$, respectively.}
\label{toruspic}
\end{figure}

The calculation of one-loop open-string amplitudes requires consideration of configuration-space integrals on a genus-one surface with boundary. The latter can be constructed by starting from a genus-one Riemann surface (an elliptic curve or torus) whose geometry is usually parametrized by a modular parameter $\tau \in \mathbb C$ with $\Im \tau>0$.  The two homology cycles of the torus can be mapped to the boundaries of the fundamental domain of a lattice $\ZZ+\tau\ZZ$, where $\tau$ is the ratio of the respective lengths of the $B$- and $A$-cycle (see \figref{toruspic}).

Frequently, the modular parameter is used in an exponentiated version, 
\begin{align}
q&=e^{2 \pi i \tau}\,,
\end{align}
which appears in the Fourier expansions of the $\tau \rightarrow \tau + 1$ periodic functions to be used below. 

\begin{figure}[h!]
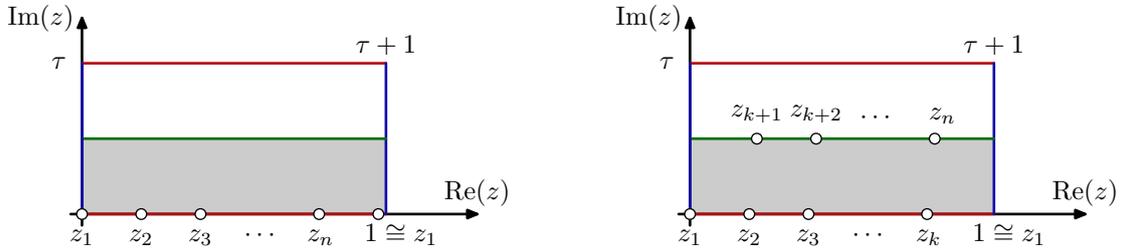

\begin{center}
	\mpostuse{fundamentaldomain2}
\end{center}
\caption{Insertion points for open-string configuration-space integrals on the cylinder can reside either on one boundary only (\textit{planar case}, left panel) or be located at both cylinder boundaries (\textit{non-planar case}, right panel). Following the parametrizations in \protect\cite{Green:1987mn, Polchinski:1998rq},
the cylinder is obtained from a torus with purely imaginary value for $\tau$.
}
\label{picnpl}
\end{figure}

One-loop open-string amplitudes receive contributions from worldsheets of cylinder and M\oe{}bius-strip topology which can be obtained from a torus through involutions described for instance in \rcite{Polchinski:1998rq}. The cylinder worldsheet with two boundaries at $\Im z= 0$ and $\Im z = \frac{1}{2} \Im \tau$ then arises from cutting the torus in two parts. When all insertion points $z_i$ are located at one boundary only, the resulting situation is called planar, while insertion points on two boundaries lead to non-planar integrals \cite{Green:1987mn} (see \figref{picnpl}).

The frameworks of elliptic multiple zeta values (eMZVs) \cite{Enriquez:Emzv} and elliptic polylogarithms \cite{Levin, BrownLev} allow to systematically perform the integrals over open-string punctures order by order in $\alpha'$ \cite{Broedel:2014vla}. For this purpose, the genus-one Green's function for planar open-string integrals is written as
\begin{align}
\CG^\tau_{ij}&=\Gamma\big(\begin{smallmatrix}1\\0\end{smallmatrix};|z_{ij}|\,\big|\tau\big)-\omega(1,0|\tau) 
\, ,
	\label{GFloop} 
\end{align}
see \rcite{Broedel:2017jdo} for their non-planar counterpart\footnote{For integration cycles with insertion points on two boundaries, one can as well define a suitable version of eMZVs, which is called twisted eMZVs \cite{Broedel:2017jdo} and described and explored in \rcite{CalaqueGonz}.}. The corresponding Koba--Nielsen factor is given by
\begin{align}
\KN^\tau_{12\ldots n}&=\exp\Big(-\sum\limits_{1\leq i<j\leq n} s_{ij}\CG^\tau_{ij}\Big)\, ,
\label{eqn:KNloop}
\end{align}
in direct analogy with (\ref{eqn:KN}) at genus zero.  The elliptic iterated integrals $\Gamma(\begin{smallmatrix}\ldots\\\ldots\end{smallmatrix};z|\tau)$ encoding the $z$-dependence of the genus-one Green's function (\ref{GFloop}) are generally defined by\footnote{Because of the non-holomorphic terms $\sim \frac{ \Im z}{\Im \tau}$ appearing in $f^{(k)}(z|\tau)$, the iterated integrals (\ref{eqn:Gammaint}) by themselves are not homotopy-invariant but can be lifted to homotopy-invariant iterated integrals by the methods of \cite{BrownLev} (also see section 3.1 of \rcite{Broedel:2014vla}).}
\begin{equation}
\label{eqn:Gammaint}
\Gamma(\smat{k_1 & k_2 & \cdots & k_r\\ a_1 & a_2 & \cdots & a_r};z|\tau)
=\int_0^z\,\dd z_1f^{(k_1)}(z_1-a_1|\tau)
  \Gamma(\smat{k_2 & \cdots & k_r\\a_2 & \cdots & a_r};z_1|\tau)
\end{equation}
with $\Gamma(;z|\tau)=1$ and $z\in \ZR$. In the same way as MZVs can be obtained as special values of iterated integrals on a genus-zero Riemann surface, see (\ref{eqn:MZV}), one can relate Enriquez' $A$-cycle eMZVs\footnote{Changing the integration path in \eqn{eqn:eMZV} to $(0,\tau)$ in the place of $(0,1)$ gives rise to $B$-cycle eMZVs \cite{Enriquez:Emzv} whose properties have for instance been discussed in \rcites{Broedel:2018izr, Zerbini:2018sox, Zerbini:2018hgs}.} as special values of the elliptic iterated integrals defined in \eqn{eqn:Gammaint}:
\begin{align}
	\label{eqn:eMZV}
	\omega(k_1,k_2,\ldots,k_r|\tau)=\Gamma(\smat{k_r & k_{r-1} & \cdots & k_1 \\ 0 & 0 & \cdots & 0};1|\tau)\,.
\end{align}

The integration kernels $f^{(k)}(z|\tau)$ in (\ref{eqn:Gammaint}) are generated
by a doubly-periodic version of a Kronecker--Eisenstein series \cite{Kronecker,BrownLev},
\begin{equation}
\Omega(z,\eta|\tau)=\exp\Big(2\pi i \eta \frac{\Im z}{\Im \tau}\Big)\frac{\theta'(0|\tau)\theta(z+\eta|\tau)}{\theta(z|\tau)\theta(\eta|\tau)}=\sum_{k=0}^\infty \eta^{k-1}f^{(k)}(z|\tau)\,,
\label{kroneis}
\end{equation}
where $\theta$ is the odd Jacobi theta function and $\theta'(0|\tau)$ its derivative in the first argument.  The double-periodicities of the series and the integration kernels are
\begin{equation}
	\label{eqn:fs}
	\Omega(z,\eta|\tau)=\Omega(z{+}1,\eta|\tau)= \Omega(z{+}\tau,\eta|\tau) \, ,\ \ \ \
	f^{(k)}(z|\tau)=f^{(k)}(z{+}1|\tau)=f^{(k)}(z{+}\tau|\tau)\, .
\end{equation}
Given the simple pole of $f^{(1)}(z|\tau)= \partial_z \log \theta(z|\tau) + 2\pi i \frac{ \Im z }{\Im \tau}$ at $z \in \mathbb Z + \tau \mathbb Z$, the integrals in \eqn{eqn:Gammaint} and thus eMZVs exhibit endpoint divergences
analogous to those in the tree-level scenario. Throughout this work, we will employ shuffle-regularization based on the prescription in section 2.2.1 of \cite{Broedel:2014vla} which assigns the following $q$-expansion to the constituents of the Green's function (\ref{GFloop}), 
\begin{align}
\omega(1,0|\tau) &= - \frac{i \pi}{2} +2\sum_{ k,l=1}^\infty
\frac{q^{kl} }{k}\,,  \\
\Gamma(\begin{smallmatrix}1\\0\end{smallmatrix};z|\tau) &= \log(1-e^{2\pi i z}) - i\pi z+ 2\sum_{ k,l=1}^\infty
\frac{1-\cos(2\pi k z) q^{kl} }{k} \,, \ \ \ \ z \in \mathbb R \, .
\label{qgammaexp}
\end{align}
From this $q$-expansion, the asymptotic behaviour for $0<z<1$ can be read off: the limit $z\to 0$ yields the logarithmic divergence

\begin{align}\label{eqn:asymptoticG0}
\Gamma(\begin{smallmatrix}1\\0\end{smallmatrix};z|\tau)&=\log(-2\pi i z)+\mathcal{O}(z)
\end{align}
while for $z\to 1$
\begin{align}\label{eqn:asymptoticG1}
\Gamma(\begin{smallmatrix}1\\0\end{smallmatrix};z|\tau)&=\log(-2\pi i (1{-}z))+\mathcal{O}(1-z)\,.
\end{align}
Apart from the constant $f^{(0)}(z|\tau)=1$ and $f^{(1)}(z|\tau)$ with a simple pole, the Kronecker--Eisenstein series (\ref{kroneis}) generates an infinity of kernels $f^{(k\geq 2)}(z|\tau)$ that do not have any poles in $z$.  Hence, the genus-one case involves an infinite number of differentials instead of the differential $\frac{\dd z_i}{z_{ij}}$ referring to finitely many $z_j$ in the genus-zero scenario.  Partial fraction, omnipresent for manipulating products of $\frac{1}{z_{ij}}$ in genus-zero integrands, is now replaced by the so-called Fay identity \cite{mumford1984tata}
\begin{equation}
	\label{eqn:Fay}
\Omega(z_{ki},\eta_a|\tau)\Omega(z_{kj},\eta_b|\tau)=\Omega(z_{ki},\eta_{a}{+}\eta_{b}|\tau)\Omega(z_{ij},\eta_b|\tau)+\Omega(z_{kj},\eta_{a}{+}\eta_{b}|\tau)\Omega(z_{ji},\eta_a|\tau)\, .
\end{equation}
The three partial derivatives of the Kronecker--Eisenstein series (\ref{kroneis}) are 
related through the mixed heat equation
\begin{equation}
	\label{eqn:mixedHeat}
2\pi i \partial_{\tau} \Omega(z,\eta|\tau) = \partial_z \partial_\eta \Omega(z,\eta|\tau) \, , \ \ \ \ \ \ z \in \mathbb R
\,.
\end{equation}

\subsubsection{\texorpdfstring{$\Zt$}{Ztau}-integrals at genus one}

In analogy to the genus-zero integrals $\Ztree_n$ defined in \eqn{eqn:Ztree}, let us now define a suitable class of genus-one integrals for open-string amplitudes of the bosonic and type I theories \cite{Mafra:2019xms}, 
\begin{align}
  \label{eqn:Zloop}
\Ztn{n}( 1,a_2,\ldots,a_n|1,2,\ldots,n)&=\int\limits_{0=z_1<z_{a_2}<\ldots<z_{a_n}<1}\dd z_2\cdots \dd z_n\,\KNt_{12\ldots n}\nnl
&\phantom{=}\times \Omega_{12}(\eta_{23\ldots n})\Omega_{23}(\eta_{3\ldots n})\ldots \Omega_{n\!-\!1\,n}(\eta_n)\,,
\end{align}
where we will always fix translation invariance by setting $z_1=0$. When writing a Kronecker--Eisenstein series where the first argument is of the form $z_{ij}$, we use the shorthand notation 
\begin{equation}
	\label{eqn:Omegashort}
	\Omega(z_{ij},\eta|\tau)=\Omega_{ij}(\eta)
\end{equation}
as well as
\begin{align}\label{eqn:AccumulatedEtaVariables}
\eta_{ij \ldots k} = \eta_i+\eta_j+\ldots+\eta_k
\end{align}
both of which will prove very handy below.
Similar to the genus-zero case, the labels $1,a_2,\ldots,a_n$ in the first slot refer to an integration domain. We have adapted (\ref{eqn:Zloop}) to planar genus-one integrals (cf.~\eqn{picnpl}), where $(1,a_2,\dots,a_n)$ specifies a cyclic ordering of insertion points on a single cylinder boundary\footnote{The integration domain in the non-planar situation is encoded by one cyclic ordering for both cylinder boundaries which can for instance be addressed by two-line labels $\Zt_{n}(\begin{smallmatrix}b_1,b_2,\ldots,b_r
\\ c_1,c_2,\ldots,c_s \end{smallmatrix} |1,2,\ldots,n)$ as in \cite{Mafra:2019xms}.}.

As a genus-one analogue of the so-called Parke--Taylor factor $(z_{12}\cdots z_{n-1,n}z_{n1})^{-1}$ in \eqn{eqn:Ztree}, the labels in the second slot of \eqn{eqn:Zloop} indicate products of the form
\begin{equation}
\label{eqn:PTloop}
f^{(k_1)}_{12} f^{(k_2)}_{23} \cdots f^{(k_{n-1})}_{n-1,n}
 \, , \ \ \ \ \ \ 
f^{(k)}_{ij} = f^{(k)}(z_{ij}|\tau)
\,.
\end{equation}
The absence of a factor $f^{(k_{n})}_{n,1}$ to close the cycle is reminiscent of Parke--Taylor factors in an 
${\rm SL}_2$-frame with $z_n \rightarrow \infty$, where they reduce to open chains like $(z_{12}z_{23}\cdots 
z_{n-2,n-1})^{-1}$ as in \eqn{eqn:ZTreeExplicit}. Instead of individual products (\ref{eqn:PTloop}), the integrands in (\ref{eqn:Zloop}) involve their generating series (\ref{kroneis}) where the combinations $\sum_{j=i}^n \eta_j$ of expansion variables are chosen for later convenience.

As a major advantage of the generating-series approach, the relations between different permutations of (\ref{eqn:Zloop}) take a simple form: by analogy with the genus-zero case, a basis of integrands can be found by taking the genus-one analogue of partial fraction\footnote{The genus-one analogue of integration-by-parts relations among Parke--Taylor factors in (\ref{eqn:Ztree}) does not relate permutations of the products (\ref{eqn:PTloop}) for generic choices of $k_i$. Instead, integration by parts at genus one will play an important role in later sections to find differential equations for various Koba--Nielsen integrals.} into account, the Fay identity \eqref{eqn:Fay}.

While the Fay identities among the products in (\ref{eqn:PTloop}) shift the overall weight $k_1+k_2$ between the two factors,
\begin{align}
	\label{eqn:Faycomp}
f^{(k_1)}(t-x) f^{(k_2)}(t)
 &=  - (-1)^{k_1} f^{(k_1+k_2)}(x) + \sum_{j=0}^{k_2} \binom{ k_1 - 1 + j}{j} f^{(k_2-j)}(x) f^{(k_1+j)}(t-x) \notag  \\
 & \phantom{=} + \sum_{j=0}^{k_1} \binom{k_2-1+j}{j} (-1)^{k_1+j} f^{(k_1-j)}(x) f^{(k_2+j)}(t)\,,
\end{align}
their series in (\ref{eqn:Zloop}) are simply related via \eqn{eqn:Fay}. 
After performing a simultaneous expansion of (\ref{eqn:Zloop}) in $\ap$ and $\eta_j$,
specific string integrals corresponding to particular integrands in \eqn{eqn:PTloop} 
can be retrieved by isolating suitable coefficients.

\subsubsection{Graphical notation}
\label{ssec:graphicalnotation}
All configuration-space integrals for string amplitudes appearing in this article exhibit the following features: they have a Koba--Nielsen-factor and a collection of integration kernels, which are labeled by (at least) the difference of two vertex positions: $z_{ij}$. Furthermore, there are vertex positions $z_i$, which are integrated over, and others, which remain unintegrated. For the discussion to follow, it is useful to define a graphical representation for the corresponding integrands, extending the graphical notation of \rcite{Kaderli:2019dny} to genus one: we are going to represent each occurring label as a vertex and each integration kernel as a directed edge 
\begin{equation}
	\frac{1}{z_{ij}}=\mpostuse{EKSeriesz},\quad \ \ \ \Omega_{ij}(\eta)=\mpostuse{EKSeries}\,,
\end{equation}
respectively.

In this graphical notation, both ${\rm SL}_2$-fixed Parke--Taylor factors (cf.~\eqn{eqn:Ztree}) and the integrals $\Zt_n$ with fixed cyclic symmetry at genus one (cf.~\eqn{eqn:Zloop}) exhibit a chain structure. As will be elaborated on below, partial-fraction relations and their one-loop analogue, the Fay relation (\ref{eqn:Fay}), allow to reduce tree-structures to chain-structures. The Fay identity \eqref{eqn:Fay}, for example, takes the following graphical form
\begin{align}\label{eqn:FayGraphical}
\mpostuse[align=b,vshift=-7pt]{FayLHS}&=\mpostuse[align=b,vshift=-7pt]{FayRHS1}+\mpostuse[align=b,vshift=-7pt]{PFRHS2}
\end{align}
which is -- not surprisingly -- equivalent to the graphical representation of partial fraction (here: $(z_{ki}z_{kj})^{-1}=(z_{ji}z_{kj})^{-1}+(z_{ij}z_{ki})^{-1}$): 
\begin{align}\label{eqn:PFGraphical}
\mpostuse[align=b,vshift=-7pt]{PFLHS}&=\mpostuse[align=b,vshift=-7pt]{PFRHS1}+\mpostuse[align=b,vshift=-7pt]{FayRHS2}.
\end{align}
The graphical representation of Kronecker--Eisenstein integrands described above will play a major role in the calculations of \secref{sec:derivatives}.


\subsubsection{eMZVs versus iterated Eisenstein integrals}

For the $A$-cycle eMZVs (\ref{eqn:eMZV}), another representation as iterated 
integrals of holomorphic Eisenstein series is available which exposes their relations 
over $\mathbb Q[{\rm MZV},(2\pi i)^{-1}]$. 
While eMZVs have been defined in \eqn{eqn:eMZV} in terms of special values of iterated integrals, which featured repeated integration in insertion points $z_i$, it is possible to write them in terms of $\tau$-iterated integrals. This is possible, because $\tau$-derivatives and $z$-derivatives of their integration kernels are related by the mixed-heat equation \eqref{eqn:mixedHeat}. Further details in converting the integrals into each other including integration constants at $\tau \rightarrow i\infty$ can be found in \rcite{Broedel:2015hia}. Here we would like to limit our attention to writing down the basic definitions and properties of two types of iterated Eisenstein integrals, which will be made use of below \cite{Broedel:2015hia}, 
\begin{align}
&\gm(k_1,k_2,\ldots,k_r|q) \nnl
&=
\frac{1}{4\pi^2} \int\limits_{0 < q'< q} \dlog q'\gm(k_1,\ldots,k_{r-1}|q') \GG{k_r}(q') 
\label{eqn:defEInew} \nnl
& = \frac{1}{(4\pi^2)^r}  \int\limits_{ 0 < q_1< q_2 <\ldots <q_{r} < q}
\dlog q_1\,\GG{k_1}(q_1) \ \dlog q_2 \ \GG{k_2}(q_2)\ \ldots \ \dlog q_r \GG{k_r}(q_r)  
\end{align}
and for $k_1\neq 0$
\begin{align}
&\gmz(k_1,k_2,\ldots,k_r|q) \nnl
&=
\frac{1}{4\pi^2} \int\limits_{0 < q' < q} \dlog q' \ \gmz(k_1,\ldots,k_{r-1}|q') \GGn{k_r}(q') 
\label{eqn:defEI} \nnl
&= \frac{1}{(4\pi^2)^r} \int\limits_{0 < q_1< q_2 <\ldots <q_{r} < q} 
\dlog q_1\  \GGn{k_1}(q_1) \ \dlog q_2 \ \GGn{k_2}(q_2) \ \ldots \ \dlog q_r \GGn{k_r}(q_r)  \, ,
\end{align}
where $\gamma(|q) = \gamma_0(|q)=1$ and the number $r$ of integrations will be referred to as the \textit{length} of either $\gm$ and $\gmz$. The integration kernels are holomorphic Eisenstein series\footnote{The case of $\GG{2}$ requires the Eisenstein summation prescription
\[
	\sum_{m,n \in \mathbb Z}a_{m,n}=
	\lim_{N \to \infty}\lim_{M \to \infty} \sum_{n=-N}^N\sum_{m=-M}^Ma_{m,n} \ .
\]
}
\begin{align}
  \label{eqn:eiss}
  \GG{0}(\tau)=-1,\quad \
  \GG{2k}(\tau)= \displaystyle \sum\limits_{m,n \in \mathbb Z \atop{(m,n) \neq (0,0)}} \frac{1}{(m+n\tau)^k},\quad  \
  \GG{2k+1}=0 \text{ for } k\in\mathbb N
\end{align}
or their modifications $\GGs^0$ with the constant term $2\zeta_{2k}$ removed for $k\neq0$,
\begin{align}
  \label{eqn:eiszero}
  \GGz{0}(\tau)=-1,\quad \
  \GGz{2k}(\tau)=\GG{2k}(\tau)-2\zm_{2k},\quad \ \GGz{2k+1}=0 \text{ for } k\in\mathbb N\,,
\end{align}
respectively. We will interchangeably refer to the arguments of $\GG{k}$, $\GGz{k}$ and related objects as $\tau$ or $q$. 

For both of the iterated Eisenstein integrals defined in \eqns{eqn:defEInew}{eqn:defEI} as well as for the eMZVs defined in \eqn{eqn:eMZV}, shuffle relations 
follow from the iterative definitions immediately:  
\begin{align}
\label{eqn:shuffle}
\om(n_1,n_2,\ldots,n_r|\tau) \om(m_1,m_2,\ldots,m_s|\tau) &= \om\big( (n_1,n_2,\ldots,n_r) \shuffle (m_1,m_2,\ldots,m_s) |\tau \big)\, ,
\nnl
\gm(n_1,n_2,\ldots,n_r|q) \gm(k_1,k_2,\ldots,k_s|q) &= \gm \!  \big( (n_1,n_2,\ldots,n_r) \shuffle (k_1,k_2,\ldots,k_s) |q
\big)\, ,
\nnl
\gmz(n_1,n_2,\ldots,n_r|q) \gmz(k_1,k_2,\ldots,k_s|q) &= \gmz \!  \big( (n_1,n_2,\ldots,n_r) \shuffle (k_1,k_2,\ldots,k_s) |q
\big)\,.
\end{align}
Regularized objects such as $\gamma(0|\tau) = \frac{\tau}{2\pi i}$ obtained by the tangential-base-point prescription \cite{Brown:mmv} preserve such shuffle relations.
Further identities implied by Fay relations as well as the precise relation between the spaces spanned by the respective iterated integrals have been investigated and spelt out in \rcites{Broedel:2015hia, Matthes:edzv, Matthes:Thesis}.

\subsubsection{Two-point example}

In order to wrap up this section, let us provide an example of a genus-one $Z$-integral (\ref{eqn:Zloop}) 
and express the leading orders of its expansion in $\ap$ and $\eta=\eta_2$ in two of the languages above: \small
\begin{align}
\label{eqn:Zloopexample}
\Zt_2(1,2|1,2)&=\int_{0}^1 \dd z_2 \, \KNt_{12} \Omega_{12}(\eta)  \nonumber \\
&=  \frac{1}{\eta} \Big[ 1 + s_{12}^2 \Big( \frac{ \om(0, 0, 2|\tau) }{2} + \frac{5 \zm_2}{12}  \Big)
+ s_{12}^3 \Big(
\frac{ \om(0, 0, 3, 0|\tau) }{18} - \frac{ 4 \zm_2}{3} \om(0, 0, 1, 0|\tau) +   \frac{ \zm_3}{12}
\Big) + {\cal O}(s_{12}^4) \Big]  \nonumber \\
&\phantom{=} +   \eta \Big[ {-}2 \zm_2
+   s_{12} \om(0, 3|\tau)
+  s_{12}^2 \Big( 3 \zm_2  \om(0, 0, 2|\tau) - \frac{ \om(0, 0, 4|\tau) }{2}+ 
    \frac{ 13 \zm_4 }{12 } \Big) + {\cal O}(s_{12}^3) \Big]
 \nonumber \\
&\phantom{=} + \eta^3 \Big[ {-}2 \zm_4 + s_{12} \Big( 
\om(0, 5|\tau) - 2  \zm_2 \om(0, 3|\tau) 
 \Big) 
 + {\cal O}(s_{12}^2) \Big] + {\cal O}(\eta^5)
 \nonumber \\
&= \frac{1}{\eta} \Big[ 1 + s_{12}^2 \Big( \frac{ \zm_2}{4} - 3 \gamma_0(4, 0|q) \Big)
+ s_{12}^3 \Big(
 \frac{  \zm_3}{4} + 24 \zm_2 \gamma_0(4, 0, 0|q) - 10 \gamma_0(6, 0, 0|q) 
\Big) + {\cal O}(s_{12}^4) \Big]  \nonumber \\
&\phantom{=} +   \eta \Big[ {-}2 \zm_2
+  3 s_{12} \gamma_0(4|q)
+  s_{12}^2 \Big({-} \frac{ 5 \zm_4}{4 }
- 18   \zm_2 \gamma_0(4, 0|q)
+ 10   \gamma_0(6, 0|q)  \Big) + {\cal O}(s_{12}^3) \Big]
 \nonumber \\
&\phantom{=} + \eta^3 \Big[ {-}2 \zm_4 + s_{12} \Big(  {-} 6  \zm_2 \gamma_0(4|q) +  5   \gamma_0(6|q) \Big) 
 + {\cal O}(s_{12}^2) \Big] + {\cal O}(\eta^5)    
\end{align} \normalsize
it does contain MZVs as well as eMZVs, which are still a function of the modular parameter $\tau$. This will be crucial for the constructions to be reviewed and discussed below. 


\section{Differential equations for one-loop open-string integrals}
\label{sec:DGLs}

In the last section, $Z$-integrals for tree-level and one-loop open-string amplitudes have been introduced. Most importantly, these integrals can be expressed in terms of iterated integrals $G$ and $\GL$ over punctures $z_i$ (cf.~\eqns{eqn:Gint}{eqn:Gammaint}), which -- if evaluated at special points -- lead to MZVs and eMZVs, respectively (cf.~eqs.\ \eqref{eqn:MZV}, \eqref{eqn:eMZV} and \eqref{eqn:defEI}). 

For iterated integrals with a particular class of differential forms, it is straightforward to infer differential equations - for example does \eqn{eqn:defEInew} immediately imply
\begin{align}
\label{eqn:diffgam}
2\pi i \partial_\tau \gm(k_1,k_2,\ldots,k_r|q) &= - \GG{k_r}(q)  \gm(k_1,k_2,\ldots,k_{r-1}|q)
\end{align}
while \eqn{eqn:Gammaint} leads to 
\begin{align}
\label{eqn:diffGamma}
\partial_z \Gamma(\smat{k_1 & k_2 & \cdots & k_r\\ a_1 & a_2 & \cdots & a_r};z|\tau)=f^{(k_1)}(z-a_1|\tau)
        \Gamma(\smat{k_2 & \cdots & k_r\\a_2 & \cdots & a_r};z|\tau)\,.
\end{align}
Starting from those simple equations, one can consider differential equations for complete $Z$-integrals.  In particular, we will study augmented variants of $Z$-integrals where an additional unintegrated puncture $z_0$ serves as a differentiation variable. This will require the evaluation of the action of derivatives on the integrands and in particular on the Koba--Nielsen factor. Suitable manipulations, partial fraction and integration by parts for $\Ztree_n$ integrals as well as Fay identities and integration by parts for the one-loop integrals $\Zt_n$, allow to frame differential equations as matrix equations, acting on a vector $\mathbf{Z}_{\text{basis}}$ whose elements form a (sometimes conjectural) basis of $Z$-type integrals and their augmented versions to be defined below: 
\begin{align}\label{eqn:modelDEQ}
	\dd\, \mathbf{Z}_{\text{basis}}&=\sum_i \nu_i\, r(D_i)\, \mathbf{Z}_{\text{basis}}\,.
\end{align}
Here $\nu_i$ are suitable differential forms in the alphabet for the iterated integrals that occur in the $\ap$-expansion of the respective $Z$-integral, whereas $r(D_i)$ denotes a particular square matrix representation of the coefficients of $\nu_i$, tailored to the basis choice.  The most crucial point of the game is the following: for all $Z$-type integrals we are going to consider, the representations $r(D_i)$ turn out to be \textit{linear} in the parameters $s_{ij}$, and thus in $\alpha'$, entering the Koba--Nielsen factors in \eqns{eqn:KN}{eqn:KNloop}. This will allow to solve the differential equation of the above form order by order in $\alpha'$, leading to the $\alpha'$-expansion of the $Z$-integrals. Note
that the linear appearance of $\alpha'$ in the above $r(D_i)$ is analogous to the $\epsilon$-form of differential equations for Feynman integrals, see e.g.\ \cite{Henn:2013pwa, Adams:2018yfj}, with $\alpha'$ taking the role of the dimensional-regularization parameter~$\epsilon$.

Considering the integrals $\Ztree_n$ defined in \eqn{eqn:Ztree}, the final result, i.e.\ the $\alpha'$-expansion, will contain numbers exclusively. In turn, a differential equation with respect to a variable which disappears during the evaluation of the iterated integral, is not very useful. The solution to this problem has been spelt out in both mathematics \cite{Aomoto, Terasoma} and physics \cite{Broedel:2013aza, Kaderli:2019dny} literature: one can introduce an additional auxiliary insertion point $z_0$ and establish a differential equation with respect to $z_0$ for a basis vector of augmented integrals $\BZ^\tree_{0,n}$. For the integrals $\Zt_n$ at genus one, a similar augmentation can be introduced leading to augmented one-loop integrals $\BZ^\tau_{0,n}$ whose constituents will be reviewed in \subsecref{ssec:z0lang} and whose differential equations in \secref{sec:derivatives} are a central result of this work. However, since the result in \eqn{eqn:Zloopexample} does still depend on the modular parameter $\tau$, one can readily use $\tau$ as a variable for differentiation when considering a vector $\BZ^\tau_n$ of one-loop integrals \eqn{eqn:Zloop} without $z_0$.  

By the choice of differential forms $\nu_i$ on the right-hand side of \eqn{eqn:modelDEQ}, the resulting system of differential equations is of Fuchsian type. Even more, on closer inspection one will find the equations to be of Knizhnik--Zamolodchikov(KZ) or Knizhnik--Zamolodchikov--Bernard type for $\BZ^{\tree}_{0,n}$ and $\BZ^{\tau}_{0,n}$, respectively, whose solution theory is well known \cite{Drinfeld:1989st,Aomoto, Drinfeld2, Terasoma, KZB, EnriquezEllAss, Hain}. By solving these differential equations along with suitable boundary conditions, one can then evaluate $Z$-integrals $Z^{\tree}_n$ and $\Zt_n$ at tree level and one loop order by order in $\ap$. 

Moreover, the matrix representations $r(D_i)$ we will encounter are linear in the Mandelstam variables $s_{ij}$ each of which comes with a parameter $\ap$ (cf.~\eqn{eqn:Mandelstam}). Hence, one can obtain solutions to all $Z$-type integrals in \eqn{eqn:modelDEQ} in terms of regularized iterated integrals, where the number of integrations is correlated with the power of $\ap$.\footnote{This property is often referred to as {\it uniform transcendentality} and a common theme of the $\ap$-expansion of configuration-space integrals in string amplitudes, see e.g.~\rcites{Terasoma,Broedel:2013aza,Puhlfuerst:2015gta, DHoker:2019blr, Mafra:2019xms, Broedel:2019gba, Gerken:2020yii}, and the $\epsilon$-expansion of dimensionally regularized Feynman integrals \cite{Kotikov:1990kg, ArkaniHamed:2010gh, Henn:2013pwa, Adams:2018yfj, Broedel:2018qkq}.} A major advantage of this concept is that the series expansion in $\ap$ follows from simple matrix algebra for the $r(D_i)$. Once the initial value for some limit of the differentiation variable is known, no integral has to be solved and the recursive nature of the solution algorithms allows to infer all higher-multiplicity $Z$-integrals at tree level and one loop from the knowledge of a single trivial tree-level three-point $Z$-integral.

In the following subsections, we are going to review the main structural points of three languages and corresponding algorithms: the $z_0$-language at genus zero in \subsecref{ssec:z0tree} and $\tau$- and $z_0$-languages at genus one in subsections \ref{ssec:taulang} and \ref{ssec:z0lang}. For each one, there is a basis of (augmented) $Z$-type integrals, a differential equation of type \eqref{eqn:modelDEQ} with suitable matrix representations and boundary values, which together allow to solve the differential equation recursively.


\subsection{\texorpdfstring{$z_0$}{z0}-language at genus zero}
\label{ssec:z0tree}
The simplest instance of the algorithm described above is the recursive
formalism for the evaluation of tree-level configuration-space integrals $\Ztree_n$. It has been put forward in \rcites{Broedel:2013aza, Kaderli:2019dny} and is based on \rcites{Aomoto, Terasoma}. At $n=4,5$ points, the augmented versions of
tree-level integrals \eqn{eqn:Ztree} with an extra marked point $z_0$ are given by
\begin{align}
\mathbf{Z}^{\rm tree}_{0,4} &=
\int_0^{z_0} \dd z_2 \, |z_2|^{-s_{12}}| 1{-}z_2 |^{-s_{23}} |z_{02}|^{-s_{02}} \vecb z_{12}^{-1}  \\ z_{32}^{-1} \vece  \, , \ \ \ \ 
(z_1,z_3)=(0,1) 
\label{eqn:Z0} \\
\mathbf{Z}^{\rm tree}_{0,5} &=
\! \! \! \int \limits_{0<z_2<z_3<z_0} \! \! \! \Big( \prod_{j=2}^3 \dd z_j |z_j|^{-s_{1j}}| z_{j4} |^{-s_{j4}} |z_{0j}|^{-s_{0j}} \Big) 
|z_{23}|^{-s_{23}} \vecb (z_{12} z_{23})^{-1} \\  (z_{13} z_{32})^{-1} \\
(z_{12} z_{43})^{-1} \\ (z_{13} z_{42})^{-1} \\
(z_{43} z_{32})^{-1} \\ (z_{42} z_{23})^{-1} \vece  \, , \ \ \ \ 
(z_1,z_4)=(0,1) \, , \notag
\end{align}
see the references for higher-multiplicity generalizations.
One can write down a differential equation with two types of poles: 
\begin{equation}
	\label{eqn:Z0diff}
	\partial_{z_0}\mathbf{Z}^{\rm tree}_{0,n}=\bigg(\frac{r^\tree_{0,n}(e_0)}{z_0}+\frac{r^\tree_{0,n}(e_1)}{z_0-1}\bigg)\mathbf{Z}^{\rm tree}_{0,n}\,.
\end{equation}
The above differential equation is of KZ-type and can be solved by considering
regularized boundary values
\begin{align}
	\mathbf{C}_{1,n}^\tree&=\lim_{z_0\to 1}(1{-}z_0)^{-r^\tree_{0,n}(e_1)} \mathbf{Z}^\tree_{0,n}\,,\nnl
	\mathbf{C}_{0,n}^\tree&=\lim_{z_0\to 0}(z_0)^{-r^\tree_{0,n}(e_0)}\mathbf{Z}^\tree_{0,n}
\end{align}
which are connected by the Drinfeld associator \cite{Drinfeld:1989st,Drinfeld2}
\begin{equation}\label{eqn:genus0AssocEq}
	\mathbf{C}_{1,n}^\tree=r^\tree_{0,n}(\Phi(e_0,e_1))\mathbf{C}_{0,n}^\tree\, .
\end{equation}
The Drinfeld associator can be expanded in terms of shuffle-regularized MZVs 
with $\zeta_1=0$~\cite{LeMura},
\begin{align}
\Phi(e_0,e_1)&=\sum_{w\geq 0}\sum_{k_1,\dots,k_w \geq 1}e_0^{k_w-1}e_1\dots e_0^{k_2-1}e_1e_0^{k_1-1}e_1\zeta_{k_1,k_2,\dots,k_w}\nnl
&=1-\zeta_2 [e_0,e_1]-\zeta_3\left([e_0,[e_0,e_1]]-[[e_0,e_1],e_1]\right)+\dots\,,
\label{eqn:leadingap}
\end{align}
where\footnote{Below, we will consider further representations $r_{n}$, $r_{0,n}$ and $r_{0,n}^{\E}$ of certain Lie algebra generators. Any such representation $r$ of generators $x_i$ will be assumed to form an algebra homomorphism such that $r(x_i x_j)=r(x_i)r(x_j)$ and $r(x_i+x_j)=r(x_i)+r(x_j)$.} $r^\tree_{0,n}(e_ie_j)=r^\tree_{0,n}(e_i)r^\tree_{0,n}(e_j)$ and $r^\tree_{0,n}(e_i+e_j)=r^\tree_{0,n}(e_i)+r^\tree_{0,n}(e_j)$. As discussed in \rcite{Broedel:2013aza}, the vector $\mathbf{C}_{0,n}^\tree$ can be shown to contain the integrals $\Ztree_{n{-}1}$ defined in \eqn{eqn:Ztree} for the $(n{-}1)$-point amplitude, while $\mathbf{C}_{1,n}^\tree$ contains integrals $\Ztree_n$ for the $n$-point amplitude.  In this formalism, the size of the matrix representations $r^\tree_{0,n}(e_i)$ is $(n{-}2)!$. 

In order to calculate for example the four-point disk integral, one would use the differential equation 
\begin{equation}
\partial_{z_0} \mathbf{Z}^{\rm tree}_{0,4}  = - \Bigg(\frac{(\smat{s_{12} +s_{02}& s_{23}\\0 & 0})}{z_0}+\frac{(\smat{0 & 0\\s_{12} & s_{23} + s_{02}})}{z_0-1}\Bigg) \mathbf{Z}^{\rm tree}_{0,4}\, . \label{kz4pt}
\end{equation}
In the kinematic limit $s_{0i} \rightarrow 0$ employed for the recursions of \rcite{Broedel:2013aza}, 
one can read off
\beq
r^\tree_{0,4}(e_0)= - \begin{pmatrix} s_{12} & s_{23} \\ 0 &0 \end{pmatrix}\, , \ \ \ \ \ \ 
r^\tree_{0,4}(e_1)= - \begin{pmatrix} 0 &0 \\ s_{12} & s_{23}  \end{pmatrix}
\label{2x2rep}
\eeq
by matching \eqn{kz4pt} with \eqn{eqn:Z0diff}, and the regularized boundary values turn out to be 
\begin{align}
\mathbf{C}_{1,4}^\tree&= \vecb \int_0^1   \dd z_2 \, |z_2|^{-s_{12}}| 1{-}z_2 |^{-s_{23}} z_{12}^{-1} \\ \ast \vece \,,\nnl
\mathbf{C}_{0,4}^\tree&= \vecb s_{12}^{-1} \\ 0 \vece \, .
\end{align}
The first entry of $\mathbf{C}_{1,4}^\tree$  may be recognized as the ${\rm SL}_2$-fixed disk
integral $-\Ztree_4(1,2,3,4|1,2,4,3)$ via \eqn{eqn:ZTreeExplicit}, and the more subtle computation of the 
second entry $\ast$ will not be needed here. The first entry of $\mathbf{C}_{0,4}^\tree$ combines the
kinematic poles $s_{12}^{-1}$ from the integration region $z_2 \rightarrow 0$ in \eqn{eqn:Z0}
with the three-point integral $\Ztree_3(1,2,3|1,2,3)=1$ at its residue.
With the leading orders of the Drinfeld associator in
\eqn{eqn:leadingap} and the matrix representations in \eqn{2x2rep},
one is indeed led to the known four-point $\alpha'$-expansion at tree level 
in the first component of the vector $\mathbf{C}_{1,4}^\tree$:
\begin{align}
\int_0^1   \dd z_2 \, |z_2|^{-s_{12}}| 1{-}z_2 |^{-s_{23}} z_{12}^{-1}
&= ( \mathbf{C}_{1,4}^\tree)_1|_{s_{0i}=0} \nonumber \\
&= \frac{1}{s_{12}}\big[ r^\tree_{0,n}(\Phi(e^0,e^1)) \big]_{1,1}|_{s_{0i}=0} \nonumber \\
&= \frac{1}{s_{12}} - \zeta_2 s_{23} - \zeta_3 s_{23}(s_{12}{+}s_{23}) + {\cal O}(s_{ij}^3) \,. 
\end{align}
Similarly, the $(n{-}3)!$ bases (\ref{eqn:Zbasisexample}) of $n$-point disk integrals can be retrieved
from the first $(n{-}3)!$ entries of the $(n{-}2)!$-component vectors $\mathbf{C}_{1,n}^\tree$.

\subsection{\texorpdfstring{$\tau$}{tau}-language at genus one}
\label{ssec:taulang}

In this subsection, we are going to review the differential equation of a vector
$\mathbf{Z}_n^\tau$ of genus-one $Z$-integrals \eqn{eqn:Zloop} without
augmentation through an extra puncture. The basis of these Kronecker--Eisenstein-type
integrals w.r.t.\ Fay relations and integration by parts is
$(n{-}1)!$ dimensional and spanned by \cite{Mafra:2019xms}
\begin{equation}
\mathbf{Z}_n^\tau = Z^\tau_{n}\big( \mathbb I_n | 1,\rho(2,3,\ldots,n)\big)\, .
\label{taulang.1}
\end{equation}
The permutations $\rho \in S_{n-1}$ of $\{2,3,\ldots,n\}$ act on both the punctures $z_j$ and
the expansion variables $\eta_j$ in the factors of $\Omega_{i-1,i}(\eta_{i\ldots n}|\tau)$ in \eqn{eqn:Zloop}.
The ordering $\mathbb I_n=1,2,\ldots,n$ in the first slot refers to a fixed planar integration domain $0=z_1<z_2<\ldots<z_n<1$ on the $A$-cycle of the torus (see \figref{toruspic}). 
We will usually sort the permutations in lexicographic order, e.g.
\begin{align}
\mathbf{Z}_2^\tau &= Z^\tau_{2}(1,2 | 1,2) \, , \ \ \ \ \ \
\mathbf{Z}_3^\tau =\vecb Z^\tau_{3}( \mathbb I_3 | 1,2,3) \\  Z^\tau_{3}( \mathbb I_3 | 1,3,2) \vece \, , \ \ \ \ \ \
\mathbf{Z}_4^\tau =
\vecb
Z^\tau_{4}( \mathbb I_4 | 1,2,3,4)\\
Z^\tau_{4}( \mathbb I_4 | 1,2,4,3)\\
Z^\tau_{4}( \mathbb I_4 | 1,3,2,4)\\
Z^\tau_{4}( \mathbb I_4 | 1,3,4,2)\\
Z^\tau_{4}( \mathbb I_4 | 1,4,2,3)\\
Z^\tau_{4}( \mathbb I_4 | 1,4,3,2)\\
\vece \, .
\label{taulang.2}
\end{align}
The $(n{-}1)!$-component vector \eqn{taulang.1} was conjectured to generate an
integral basis for arbitrary massless one-loop open-string amplitudes \cite{Mafra:2019xms} as
supported by its closure under $\tau$-derivatives to be reviewed below.

\subsubsection{Differential equation}\label{sec:tauGenusOneDEQ}

Based on the mixed heat equation (\ref{eqn:mixedHeat})
and the differential equation of the Koba--Nielsen factor
\begin{align}
    \label{eqn:kntau}
    2 \pi i \pd_\tau \KN^\tau_{12\ldots n}&= -\sum\limits_{1\leq i<j\leq n}s_{ij}(f_{ij}^{(2)}+2\zeta_2)\KN^\tau_{12\ldots n}
\end{align}
one can perform the $\tau$-derivative at the level of the integrand in \eqn{eqn:Zloop}, see
\eqn{eqn:PTloop} for the shorthand $f^{(k)}_{ij}$. The $z_i$-derivatives
of the $\Omega_{i-1,i}(\eta_{i \ldots n})$ generated by the mixed heat equation can be
integrated by parts to act on
\begin{align}
    \label{eqn:knz}
     \pd_{z_i} \KN^\tau_{12\ldots n}&= -\sum\limits_{\begin{smallmatrix}j=1\\j\neq i
     	\end{smallmatrix}}^{n}s_{ij} f_{ij}^{(1)} \KN^\tau_{12\ldots n} \, .
\end{align}
Finally, the combination of $f_{ij}^{(2)}$ and $f_{ij}^{(1)}$ due to the Koba--Nielsen derivatives can be
rewritten~via
\begin{equation}
(\partial_\eta + \partial_\xi) \Omega_{12}(\eta)\Omega_{21}(\xi)
= \big( \wp(\eta) - \wp(\xi) \big) \Omega_{12}(\eta-\xi) 
\label{taulang.3}
\end{equation}
such that the integrand of the $\tau$-derivative only depends on $z_i$ via Kronecker--Eisenstein series. Moreover, repeated use of the Fay identity (\ref{eqn:Zloop}) allows to rearrange their first arguments such that the $\tau$-derivatives are expressed in terms of the $(n{-}1)!$ components in \eqn{taulang.1}. 

As a result, the vector of genus-one $Z$-integrals in \eqn{taulang.1} obeys the linear and
homogeneous differential equation \cite{Mafra:2019xms}  
\begin{equation}
2\pi i \partial_\tau \mathbf{Z}_n^\tau  = D_n^\tau \mathbf{Z}_n^\tau 
\label{taulang.4}
\end{equation}
with an $(n{-}1)!{\times} (n{-}1)!$ matrix $D_n^\tau$. The entries of the latter are linear in $\ap$ by the Mandelstam invariants in the Koba--Nielsen derivatives in eqs.\ (\ref{eqn:kntau}), (\ref{eqn:knz}) and may comprise second derivatives in $\eta_j$ from an expansion of \eqn{taulang.3} around $\xi=0$.  Most importantly, the matrix $D_n^\tau$ solely depends on $\tau$ via holomorphic Eisenstein series \eqn{eqn:eiss} and can therefore be uniquely decomposed into
\begin{equation}
D_n^\tau = \sum_{k=0}^\infty (1{-}k) \GG{k}(\tau) \, r_n(\ep_k)\, ,
\label{taulang.5}
\end{equation}
where $\GG{0} = -1$ has been introduced for the $\tau$-independent piece. The appearance
of the $\GG{k}$ can be traced back to the expansion of the Weierstra\ss{} $\wp$-function in \eqn{taulang.3},
\begin{equation}
\wp(\eta|\tau) = \frac{1}{\eta^2} + \sum_{k=4}^\infty (k{-}1) \eta^{k-2} \GG{k}(\tau)\, .
\label{taulang.6}
\end{equation}
The $(n{-}1)!{\times} (n{-}1)!$ matrices $r_n(\ep_k)$ are still linear in $\ap$, comprise 
second derivatives in $\eta_j$ at $k=0$ and are conjectured to furnish matrix representations
of Tsunogai's derivation algebra \cite{Tsunogai}. By the appearance of these derivations in
the $\tau$-derivative of the KZB associator \cite{KZB, EnriquezEllAss, Hain}, their commutator relations 
encode the combinations of iterated Eisenstein integrals \eqn{eqn:defEInew} that occur among 
eMZVs \cite{Broedel:2015hia}. The relations in the derivation algebra have been studied in
\cite{LNT, Pollack, Broedel:2015hia} and were checked to be preserved by the $r_n(\ep_k)$
in \eqn{taulang.5} for a wide range of $n$ and $k$. An all-multiplicity proposal for $r_n(\ep_k)$
(with a detailed derivation in \cite{Gerken:2019cxz}) can be found in section 4 of \cite{Mafra:2019xms},
and its explicit form is encoded in the later eq.\ \eqref{dtaunpt}.

Note that the matrix $D^\tau_n$ does not depend on the choice of 
planar or non-planar integration cycle, i.e.\ it takes a universal form
for any $\mathbb I_n \rightarrow a_1,a_2,\ldots ,a_n$ in \eqn{taulang.1}.


\subsubsection{Solution via Picard iteration}

Given that the matrix $D_n^\tau$ in \eqn{taulang.4} is linear in $\ap$, the
$\ap$-expansion of the entire $\mathbf{Z}_n^\tau$ can be conveniently organized
by iterative use of
\begin{equation}
 \mathbf{Z}_n^\tau  =  \mathbf{Z}_n^{i\infty} + \frac{1}{2\pi i} \int^\tau_{i\infty} \dd \tau' \,  D_n^{\tau'} \, \mathbf{Z}_n^{\tau'} \, .
 \label{taulang.7}
 \end{equation}
Picard iteration of \eqn{taulang.7} leads to a perturbative solution to \eqn{taulang.4} with matrix products of order $(\ap)^\ell$ in the $\ell$'th term of 
\begin{equation}
 \mathbf{Z}_n^\tau  = \sum_{\ell=0}^\infty \Big( \frac{1}{2\pi i} \Big)^\ell \int^\tau_{i\infty} \dd \tau_1 
  \int^{\tau_1}_{i\infty} \dd \tau_2 \ldots  \int^{\tau_{\ell-1}}_{i\infty} \dd \tau_\ell\,
  D_n^{\tau_\ell}\ldots   D_n^{\tau_2}   D_n^{\tau_1}  \mathbf{Z}_n^{i\infty} 
 \, .
 \label{taulang.8}
 \end{equation}
By the decomposition \eqref{taulang.5} of the $D_n^{\tau_j} $ into holomorphic Eisenstein series,
all the integrals in \eqn{taulang.8} line up with the definition of $\gamma(k_1,\ldots,k_\ell|q)$ in \eqn{eqn:defEInew}. Hence, the entire $\tau$-dependence in the $\ap$-expansion of 
$ \mathbf{Z}_n^\tau$ enters via iterated Eisenstein integrals, and their coefficients are governed by matrix 
products $r_n(\ep_{k_1} \ep_{k_2}) = r_n(\ep_{k_1})r_n(\ep_{k_2})$ \cite{Mafra:2019xms},
\begin{align}
 \mathbf{Z}_n^\tau  = \sum_{\ell=0}^\infty \sum_{k_1,k_2,\ldots, k_\ell \atop{ =0,4,6,8,\ldots}}
 \Big( \prod_{j=1}^{\ell}(k_{j}{-}1)  \Big) \gamma(k_1,k_2,\ldots,k_\ell|q) r_n(\ep_{k_\ell} \ldots \ep_{k_2} \ep_{k_1})
 \mathbf{Z}_n^{i\infty} \, .
  \label{taulang.9}
\end{align}
The choice of summation range for the $k_j$ already incorporates the vanishing of $r_n(\ep_k)$ at odd $k$ and $k=2$ for any $n\geq 2$. Generating functions of torus integrals in closed-string one-loop amplitudes obey differential equations similar to \eqn{taulang.5} \cite{Gerken:2019cxz} which can be solved via combinations of derivations similar to \eqn{taulang.9} \cite{Gerken:2020yii}.


\subsubsection{Initial value at the cusp}
In order to extract the complete $\ap$-expansion of $\mathbf{Z}_n^\tau$ from \eqn{taulang.9}, it remains to analyze the initial values $ \mathbf{Z}_n^{i\infty} $ at the cusp. Given that the torus worldsheet degenerates to a nodal Riemann sphere as $\tau \rightarrow i\infty$, the initial data at $n$ points is expressible in terms of genus-zero integrals with two extra points, i.e.\ combinations of $(n{+}2)$-point disk integrals \eqn{eqn:Ztree} 
in the open-string case. At $n=2$ points, for instance, the genus-one $Z$-integral \eqref{eqn:Zloopexample}
degenerates to
\begin{align}
 Z^{i\infty}_{2}(1,2|1,2)  = \pi \cot(\pi \eta) \frac{ \Gamma(1{-} s_{12}) }{ \big[ \Gamma(1{-}\tfrac{s_{12}}{2}) \big]^2}\, ,
 \label{2ptin}
\end{align}
where the $\Gamma$-functions stem from a kinematic limit $s_{23} \rightarrow - \frac{ s_{12}}{2}$ 
of the Veneziano amplitude
\begin{align}
\Ztree_4(1,2,3,4|1,2,4,3)  =
 - \frac{ 1}{s_{12}} \frac{ \Gamma(1{-}s_{12}) \Gamma(1{-}s_{23} ) }{\Gamma(1{-}s_{12}{-}s_{23}) } 
 \label{more2ptin}
\end{align}
and yield the following $\ap$-expansion
\begin{equation}
\frac{ \Gamma(1{-}s_{12}) }{ \big[ \Gamma(1{-}\tfrac{s_{12}}{2}) \big]^2}= 1 + \frac{1}{4} s_{12}^2 \zeta_2 + \frac{1}{4} s_{12}^3 \zeta_3 + \frac{19}{160} s_{12}^4 \zeta_2^2
+ \frac{ 1}{16} s_{12}^5 \zeta_2 \zeta_3  + \frac{ 3}{16} s_{12}^5 \zeta_5+ {\cal O}(s_{12}^6) \, .
 \label{KNdrft2}
\end{equation}	
The $\eta$-dependent factor $ \pi \cot( \pi \eta)$ in \eqn{2ptin} stems from the $\tau \rightarrow i\infty$
limit of $\Omega_{12}(\eta)$. The detailed relation between \eqns{2ptin}{more2ptin} involving a
trigonometric factor $ \sin(\frac{\pi }{2}s_{12})$ from contour deformations can be found in 
sections 3.4 and 5 of the first reference in \rcite{Mafra:2019xms}.


\subsubsection{Two-point example}

In the two-point instance of the above setup, the target vector has a single component
$Z^\tau_{2}(1,2|1,2)$ and the operator $D^\tau_{2}$ in the differential equation (\ref{taulang.4})
is the following scalar instead of a matrix
\begin{equation}
2\pi i \partial_\tau Z^\tau_{2}(1,2|1,2) = D^\tau_{2} Z^\tau_{2}(1,2|1,2)\, , \ \ \ \ \ \
D^\tau_{2} = s_{12}  \Big( \frac{1}{2} \partial_{\eta}^2 - \wp(\eta|\tau)  - 2 \zeta_2 \Big)\, .
\label{2ptA}
\end{equation}
The decomposition \eqref{taulang.5} into holomorphic Eisenstein series allows to read off
conjectural scalar representations of the derivation algebra
\begin{align}
 r_{2}(\ep_{0})  &= s_{12} \Big( \frac{ 1}{\eta^{2}} +  2 \zeta_2 - \frac{1}{2} \partial_{\eta}^2\Big) \, , \ \ \ \ \ \
   r_{2}(\ep_{k})  = s_{12} \eta^{k-2} \, , \ \ \ \ \ \ k\geq 4 \ {\rm even}\, ,
 \label{2ptC}
\end{align}
which obey multiplicity-specific relations such as $[ r_{2}(\ep_{4}),r_{2}(\ep_{6}) ]=0$ that
no longer hold at $n\geq 3$ points. With the initial value \eqref{2ptin} involving a four-point
disk integral, the two-point instance of the Picard iteration \eqref{taulang.9} is given by
\begin{align}
Z^\tau_{2}(1,2|1,2)  = \frac{ \Gamma(1{-}s_{12}) }{ \big[ \Gamma(1{-}\tfrac{s_{12}}{2}) \big]^2}
 \sum_{\ell=0}^\infty \sum_{k_1,k_2,\ldots, k_\ell \atop{ =0,4,6,8,\ldots}}
\! \! \Big( \prod_{j=1}^{\ell}(k_{j}{-}1)  \Big) \gamma(k_1,k_2,\ldots,k_\ell|q) r_2(\ep_{k_\ell} \ldots \ep_{k_2} \ep_{k_1})
\pi \cot(\pi \eta) \, .
  \label{taulang.2pt}
\end{align}
Based on the expansions \eqref{KNdrft2} and $\pi \cot(\pi \eta) = \frac{1}{\eta} - 2 \sum_{k=1}^\infty \zm_{2k} \eta^{2k-1}$, one arrives at the leading orders in $\ap$ and $\eta$ spelt out in \eqn{eqn:Zloopexample}.


\subsection{\texorpdfstring{$z_0$}{z0}-language at genus one}
\label{ssec:z0lang}

In this subsection, we are going to sketch the formalism introduced in \rcite{Broedel:2019gba}. The formalism is the genus-one generalization of the tree-level recursion discussed in \subsecref{ssec:z0tree}: it utilizes an auxiliary point $z_0$ such that
\begin{align}
0&=z_1<z_2<\dots<z_n<z_0<1\,.
\end{align}
In the reference, the differential equation and thus the recursion was formulated for an infinitely long vector of genus-one $n$-point Selberg integrals, which are defined in terms of a Selberg seed
\begin{equation}
\label{eqn:SelbergSeed}
\SelEn(\tau)=\SIE{}{}(z_2,z_3,\dots,z_n,z_0,\tau)=
\prod_{0\leq i<j \leq n}
\exp\left(-s_{ij}\Gamma_{ij}\right) \, ,
\quad\quad \Gamma_{ij} = \Gamma\big(\begin{smallmatrix}1\\0\end{smallmatrix};|z_{ij}|\,\big|\tau\big) 
\end{equation}
with $z_1=0$ which agrees with an $(n{+}1)$-point Koba--Nielsen factor (\ref{eqn:KNloop}) upon multiplication by
$e^{s_{012\ldots n} \omega(1,0|\tau)}$.
The Selberg seed (\ref{eqn:SelbergSeed}) serves as starting point for the recursive definition\footnote{On the locus of the purely real integration paths considered here and in \rcite{Broedel:2019gba}, the integration kernels $g^{(k)}$ in the reference equal $f^{(k)}$ here. Furthermore, in the reference the auxiliary point is called $z_2$ and the insertion points are labeled differently, close to the notation and the notion of \textit{admissible} integrals in \rcite{aomoto1987}.}
\begin{align}
	\label{eqn:Selberg}
\SIE{k_{\ell},\dots,k_2}{i_{\ell},\dots,i_2}(z_{\ell+1},\dots,z_n,z_0)&=\int_0^{z_{\ell+1}}\dd z_{\ell}\, f^{(k_{\ell})}_{\ell,i_{\ell}}\SIE{k_{\ell-1},\dots,k_2}{i_{\ell-1},\dots,i_2}(z_{\ell},z_{\ell+1},\dots,z_{n},z_0)
\end{align}
with $\ell=2,3,\ldots,n$ as well as $z_{n+1}=z_0$ and
\begin{align}\label{eqn:admissibility}
i_\ell\in \{0,1,\ell{+}1,\ell{+}2,\dots,n\}\,,
\end{align}
where the shorthand notation for the integration kernels $f^{(k)}_{ij}$ was defined in \eqn{eqn:PTloop}. 


\subsubsection{Differential equation and boundary values}

The differential equation of the (infinitely long) vector of Selberg integrals 
\begin{align}
\SelbldEn(z_0|\tau)&=\begin{pmatrix}
\SIE{k_{n},\dots,k_2}{i_{n},\dots,i_2}(z_0)
\end{pmatrix} \text{ for } k_\ell\geq 0,\, i_\ell\in\{0,1,\ell{+}1,\ell{+}2,\dots,n\}
\label{selbdefn}
\end{align}
%
can formally be brought into the KZB form
\begin{equation}\label{eqn:KZBSelE}
	\partial_{z_0}\SelbldEn(z_0|\tau)=\sum_{k=0}^\infty f^{(k)}_{01}r^{\E}_{0,n}(x_k)\SelbldEn(z_0|\tau)\,,
\end{equation}
where the representations $r^{\E}_{0,n}$ of $x_k$ are block-(off-)diagonal
and proportional to $\alpha'$. Similar to the tree-level scenario described in
\subsecref{ssec:z0tree}, one now considers the two regularized boundary values
\begin{align}
	\label{eqn:z0LanguageBoundaryValues}
	\mathbf{C}_{1,n}^{\E}(\tau)&=\lim_{z_0\to 1}(-2\pi i(1{-}z_0))^{-r^{\E}_{0,n}(x_1)}\SelbldEn(z_0|\tau)\,, \nnl
	\mathbf{C}_{0,n}^{\E}(\tau)&=\lim_{z_0\to 0}(-2 \pi i z_0)^{-r^{\E}_{0,n}(x_1)}\SelbldEn(z_0|\tau)
\end{align}
which turn out to contain $n$-point one-loop integrals $\Ztn{n}$ in $\mathbf{C}_{1,n}^{\E}$ and $(n{+}2)$-point tree-level integrals $\Ztree_{n{+}2}$ in $\mathbf{C}_{0,n}^{\E}$. As shown in the next subsection, these two boundary values can be related to each other, such that knowing $\mathbf{C}_{0,n}^{\E}$ from the tree-level recursion described above, allows to infer the $\ap$-expansion of all one-loop Selberg integrals and thus -- as will be elaborated on below -- all integrals $\Zt_n$. 

\subsubsection{The elliptic KZB associator}\label{subsec:eKZB}
As argued in sec.\ 3.3 of \cite{Broedel:2019gba} for a general solution of the elliptic KZB equation of the form \eqref{eqn:KZBSelE}, the regularized boundary values $\mathbf{C}_{0,n}^{\E}$ and $\mathbf{C}_{1,n}^{\E}$ are related according to
\begin{align}\label{eqn:assocEqSE}
\mathbf{C}_{1,n}^{\E}&=r^{\E}_{0,n}(\Phi^{\tau}(x_k))\mathbf{C}_{0,n}^{\E}
\end{align}
by the elliptic KZB associator \cite{Enriquez:EllAss} 
\begin{align}\label{eqn:ellKZBAssocSE}
\Phi^{\tau}(x_k)&=\sum_{w\geq 0}\sum_{k_1,k_2,\dots,k_w\geq 0}x_{k_1} x_{k_2} \dots x_{k_w}\omega(k_w,\dots,k_2,k_1|\tau)\nonumber\\
&=1 + x_0 - 2 \zeta_2 x_2 + \frac{1}{2} x_0x_0 - [x_0,x_1]\omega(0, 1|\tau)  - \zeta_2 \{x_0,x_2\} \nnl
&\phantom{=}+ [x_1,x_2]\big(\omega(0,3|\tau)-2\zeta_2\omega(0,1|\tau)\big)
-[x_0,x_3]\,\omega(0,3|\tau) \nnl
&\phantom{=}+ \zeta_4 (-\{x_0,x_4\}+5 x_2 x_2 -2 x_4)+ \cdots\,.
\end{align}
Since the matrices $r_{0,n}^{\E}(x_{k})$ are proportional to $\alpha'$, the associator \eqn{eqn:assocEqSE} yields the $\alpha'$-expansion of $\mathbf{C}_{1,n}^{\E}$. Furthermore, by the lower triangular block structure of the matrices $r_{0,n}^{\E}(x_{k})$ with $k\geq 2$, only finitely many terms of $r^{\E}_{0,n}(\Phi^{\tau})$ contribute to each order of $\alpha'$. 

Instead of reviewing the formalism in detail here, we will rephrase and discuss it in the next section: therein we are going to replace the infinite-dimensional vector $\SelbldE(z_0)$ by a finite-dimensional vector $\BZ^{\tau}_{0,n}$ of integrals $Z^{\tau}_{0,n}$ that serve as a generating series of Selberg integrals (see appendix \ref{app:recoverSelbergInt} for details).  These vectors $\BZ^{\tau}_{0,n}$ augment the vectors $\BZ^{\tau}_{n}$ of section \ref{ssec:taulang} by an auxiliary point $z_0$. The detailed discussion of the differential equation and boundary values will be performed for vectors $\BZ^{\tau}_{0,n}$. While being already closer to the formalism to be spelt out in \subsecref{sec:z0Deriv} below, rewriting in terms of the generating series $Z^{\tau}_{0,n}$ renders the representation of the matrices $x_{k}$ finite-dimensional.  

\subsubsection{Two-point example}
The two-point example elaborated in \cite{Broedel:2019gba} can be summarized as follows: the genus-one Selberg integrals (\ref{eqn:Selberg}) for $n=2$ are
\begin{equation}
\SIEzwei{k_2}{i_2}(z_0)=\int_0^{z_0} \dd z_2 \Sel_2^\El f^{(k_2)}_{2, i_2}\,,\quad  \Sel_2^\El=\exp\left({-}s_{12} \Gamma_{12}-s_{01}\Gamma_{01}-s_{02}\Gamma_{02}\right)\,,
\end{equation}
where $k_2\geq 0$ and $i_2\in \{0,1\}$. Note that not all of these integrals are independent: due to the triviality of $f^{(0)}=1$ for $k_2=0$ and integration by parts for $k_2=1$, there are the relations
\begin{align}
\SIEzwei{0}{0}(z_0)&=\SIEzwei{0}{1}(z_0)\,,\qquad \ \ \SIEzwei{1}{0}(z_0)=-\frac{s_{12}}{s_{02}}\SIEzwei{1}{1}(z_0)\,.
\end{align}
The corresponding vector of independent integrals 
\begin{equation}\label{eqn:2PtExSelVec}
\Selbld^\El_2(z_0)=\begin{pmatrix}
\SIEzwei{0}{1}(z_0)\\\SIEzwei{1}{1}(z_0)\\ \SIEzwei{2}{1}(z_0)\\ \SIEzwei{2}{0}(z_0)\\\vdots
\end{pmatrix}\,
\end{equation}
satisfies the differential equation \eqref{eqn:KZBSelE} with the block-off-diagonal matrices 
\begin{equation}\label{eqn:2ptExamplex01}
r_{0,2}^{\E}(x_{0})=\begin{pmatrix}
0&-s_{12}&0&0&\dots\\
0&0&s_{02}&s_{02}&\dots\\
0&0&0&0&\dots\\
0&0&0&0&\dots\\
\vdots&\vdots&\vdots&\vdots&\ddots
\end{pmatrix}\,,\qquad \! \! \! \! \! 
r_{0,2}^{\E}(x_2)=\begin{pmatrix}
0&0&0&0&\dots\\
s_{02}&0&0&0&\dots\\
0&-s_{12}&0&0&\dots
\\
0&-s_{12}&0&0&\dots\\
\vdots&\vdots&\vdots&\vdots&\ddots
\end{pmatrix}\,,\qquad \! \! \! \! \!  \dots
\end{equation}
and the block-diagonal one
\begin{align}\label{eqn:2ptExamplex2}
r_{0,2}^{\E}(x_{1})=\begin{pmatrix}
-s_{01}&0&0&0&\dots\\
0&\! -s_{012} \!&0&0&\dots\\
0&0&\! -s_{01}{-}s_{02} \!&s_{02}&\dots\\
0&0&s_{12}&\!-s_{01}{-}s_{12} \!&\dots\\
\vdots&\vdots&\vdots&\vdots&\ddots
\end{pmatrix}\, .
\end{align}
On the one hand, the only non-vanishing entry of the boundary value $\mathbf{C}_{0,2}^{\E}$ is proportional to the tree-level Veneziano amplitude (\ref{more2ptin})
\begin{align}
\mathbf{C}_{0,2}^{\E}&=\begin{pmatrix}0\\
-\frac{1}{s_{12}}\frac{\Gamma(1-s_{12})\Gamma(1-s_{02})}{\Gamma(1-s_{12}-s_{02})}\\
0\\\vdots
\end{pmatrix}
\label{exC0.2}
\end{align}
with the two independent four-point, tree-level Mandelstam variables $s_{12}$ and $s_{02}$ associated to the punctures $0,z_2,1$ and the well-known gamma function $\Gamma$. On the other hand, the first entry of $\mathbf{C}_{1,2}^{\E}$ is proportional to the simplest two-point, one-loop configuration-space integral with Mandelstam variable $\tilde s_{12}=s_{12}+s_{02}$
\begin{align}
\mathbf{C}_{1,2}^{\E}&=\begin{pmatrix}
\int_0^1 \dd z_2 \exp(-\tilde s_{12} \Gamma_{21})\\\vdots
\end{pmatrix} 
\label{exC1.2}
\end{align}
that occurs at the $\eta^{-1}$-order of \eqn{eqn:Zloopexample}. 
Therefore, due to the block-off-diagonality of $r_{0,2}^{\E}(x_{k})$ for $k\geq 2$, the three $4\times 4$ submatrices shown in \eqns{eqn:2ptExamplex01}{eqn:2ptExamplex2} of $r_{0,2}^{\E}(x_0)$, $r_{0,2}^{\E}(x_1)$ and $r_{0,2}^{\E}(x_2)$ are sufficient to calculate the first entry $\int_0^1 \dd z_2 \exp(-\tilde s_{12} \Gamma_{21})$ of $\mathbf{C}_{1,2}^{\E}$ up to the second order in $\alpha'$ using the associator equation \eqref{eqn:assocEqSE}, which results in the expansion
\begin{align}
\int_0^1 \dd z_2 \exp(-\tilde s_{12} \Gamma_{21}) &= - \frac{1}{s_{12}}  \frac{\Gamma(1-s_{12})\Gamma(1-s_{02})}{\Gamma(1-s_{12}-s_{02})}\big[ r^{\E}_{0,2}(\Phi^{\tau}) \big]_{1,2}\nonumber
\\
&=1-(s_{12}{+}s_{02})\omega(1,0|\tau)+(s_{12}{+}s_{02})^2\omega(1,1,0|\tau)+\CO(s_{ij}^3)\,.
\end{align}
Upon multiplication by $e^{(s_{12}{+}s_{02})\omega(1,0|\tau)}$, this agrees with the leading orders of  \eqn{eqn:Zloopexample} since $\omega(1,1,0|\tau)-\frac{1}{2} \omega(1,0|\tau)^2 = \frac{1}{2} \omega(0,0,2|\tau)
+ \frac{ 5 \zeta_2 }{12}$ \cite{Broedel:2015hia}. The two-point associator only contributes through its entry 
$\big[ r^{\E}_{0,2}(\Phi^{\tau}) \big]_{1,2}$ because the only non-zero entry of $\mathbf{C}_{0,2}^{\E}$ occurs
in the second line of \eqn{exC0.2} and the desired one-loop integral occurs in the first line of $\mathbf{C}_{1,2}^{\E}$
in  \eqn{exC1.2}.


\section{Differential equations for the integrals \texorpdfstring{$\BZtzn{n}$}{Ztau(0,n)}}
\label{sec:derivatives}

In order to link the two formalisms described in subsections \ref{ssec:taulang} and \ref{ssec:z0lang} above, we will now introduce genus-one integrals $\Ztzn{n}$ augmented by an auxiliary insertion position $z_0$. In particular, we will evaluate their derivatives with respect to both $z_0$ and the modular parameter $\tau$ in closed form.  While the integrals $\Ztzn{n}$ contain all genus-one Selberg integrals\footnote{Strictly speaking, the $\Ztzn{n}$ in this section and the $Z^\tau_n$ in \eqn{eqn:Zloop} are genus-one Selberg integrals as well since their definition is not tied to the admissibility condition \cite{aomoto1987}. Still, we will only refer to the integrals $\SelbldEn$ in section \ref{ssec:z0lang} subject to the admissibility condition as Selberg integrals.}
from \subsecref{ssec:z0lang} in their expansion with respect to the variables $\eta_j$, they differ from the integrals $\Ztn{n}$ used in \subsecref{ssec:taulang} only by the inclusion of an auxiliary point $z_0$: they are ideal to bridge the gap between the different languages, in particular between the KZB-type differential equations in $z_0$ and $\tau$. 

\subsection{Integrals with auxiliary point}
\label{graphsec}

In order to augment the space spanned by the integrals $\Ztn{n}$ defined in \eqn{eqn:Zloop}, we introduce an auxiliary point $z_0$, which is, however, not integrated over. Thus, we will consider integrals associated to the configuration space of the twice-punctured torus, with punctures $z_1=0$ and $z_0$. In order to compactly write down a conjectural\footnote{Also in presence of the augmentation by $z_0$, Koba--Nielsen-type integrals over cycles of 
$f^{(k)}_{ij}$-functions such as $f^{(k_1)}_{12}f^{(k_2)}_{23}f^{(k_3)}_{31}$ are expected to be 
expressible in terms of chain integrands as in \eqn{eqn:PTloop} and the $\eta_j$-expansion of~\eqn{eqn:Ztzbasis}.
Since we do not present a proof of this claim here, the $\BZ_{0,n}^{\tau}$ below are a {\it conjectural} basis of augmented Koba--Nielsen integrals (with supporting evidence from their closure under $z_0$- and $\tau$-derivatives). We will not repeat the word ,,conjectural'' when referring to the $\BZ_{0,n}^{\tau}$ as a basis later on.} basis $\Ztzn{n}$ of such integrals, let us define a \textit{chain} of Kronecker--Eisenstein series by
\begin{align}
  \phiChain(1,2,\ldots,p)&=\Omega_{12}(\eta_{23\ldots p})\Omega_{23}(\eta_{3\ldots p})\dots \Omega_{p-1,p}(\eta_{p})\,, \label{eqn:defChain}\\
\phiChain(a_1)&=1\, , \label{eqn:defTrivialChain}
\end{align}
and $\phiChain(a_1,a_2,\dots,a_p)$ is obtained from simultaneous permutations of $z_i \rightarrow z_{a_i}$ and $\eta_i \rightarrow \eta_{a_i}$.  The individual factors of Kronecker--Eisenstein series in such a chain $\phiChain(a_1,a_2,\dots,a_p)$ accumulate their $\eta$-variables from right to left and, thus, the chain is said to begin at $a_p$ and end at $a_1$. Chains of Kronecker--Eisenstein series will turn out to be a very versatile tool in the description of integrals $\Ztzn{n}$, and already the $Z^\tau$-integrals \eqref{eqn:Zloop} without augmentation feature $\phiChain(1,2,\ldots,n)$ in the integrand. 

Using the notation in \eqn{eqn:defChain}, the vector of the $n!$ basis integrals is given by 
\begin{align}
\label{eqn:Ztzbasis}
\BZ_{0,n}^{\tau} &= \int_{\gamma_{12\ldots n0}} \dd z_2\, \dd z_3 \, \ldots \, \dd z_n\, {\rm KN}^\tau_{0123\ldots n}   \\
&\ \ \ \ \times
\vecb
\rho[\phiChain(1,2,\dots,n)]
\\
\rho[\phiChain(1,2,\dots,n{-}1)\phiChain(0,n)]
\\
\rho[\phiChain(1,2,\dots,n{-}2)\phiChain(0,n,n{-}1)]
\\
\vdots
\\
\rho[\phiChain(1,2)\phiChain(0,n,n{-}1,\dots,3)]
\\
\rho[\phiChain(0,n,n{-}1,\dots,2)]
\vece \,,
\notag
\end{align}
where we keep on setting $z_1=0$, and the permutations $\rho \in {\cal P}(2,3,\ldots,n)$
acting on the integrand are again lexicographically ordered. The original Koba--Nielsen-factor 
\eqn{eqn:KNloop} is extended by additional variables $s_{0j}$ with $j=1,2,\ldots,n$ as in \eqn{eqn:SelbergSeed}, 
\begin{equation}
\label{eqn:KNt0}
{\rm KN}_{012\ldots n}^\tau =\exp \Big( 
	-\sum_{0\leq i<j \leq n} s_{ij} {\cal G}^\tau_{ij}
\Big)\, ,
\end{equation}
and we use the following shorthand notation for the integration domain,
\begin{equation}
\gamma_{12\ldots n0} = \{z_2,z_3,\ldots,z_n \in \mathbb R\, |\, 0=z_1<z_2<z_3<\ldots <z_n<z_0 \} \, .
\label{shortdom}
\end{equation}
The basis integrals in the components can be denoted by
\begin{align}\label{eqn:notationChain}
	Z^{\tau}_{0,n}( (1,A),(0,B))&=\int_{\gamma_{12\ldots n0}}
	\dd z_2\, \dd z_3 \, \ldots \, \dd z_n\, \KN^\tau_{0123\ldots n} \phiChain(1,A)\phiChain(0,B)\,,
\end{align}
where $A=(a_1,a_2,\dots,a_p)$ and $B=(b_1,b_2,\dots,b_q)$ are disjoint sequences without repetitions such that $A\cup B = \{2,3,\dots,n\}$. For simplicity, we denote $ \Ztzn{n}((1,A),(0)) =  \Ztzn{n}(1,A)$ for an empty sequence $B$ and similarly $ \Ztzn{n}((1),(0,B)) =  \Ztzn{n}(0,B)$ for $A=\emptyset$. This notation directly exhibits the chain structure of the products of Kronecker--Eisenstein series: the dictionary of section \ref{ssec:graphicalnotation} assigns
two rooted chains -- trees without branching points and root vertices $0$ and $1$ -- to the integrals (\ref{eqn:notationChain}) with non-empty $A,B$. 

Similar to the $Z^\tau$-integrals \eqref{taulang.1} without augmentation, we are studying the fixed planar integration domain \eqref{shortdom} for any choice of $A,B$ throughout this section.  Accordingly, the two elements $(1,A),(0,B)$ of notation in \eqn{eqn:notationChain} both refer to the integrand, and we suppress a separate slot $Z^{\tau}_{0,n}( \cdot | (1,A),(0,B))$ specifying the integration domain to avoid cluttering. Nevertheless, the differential equations to be derived below are universal to all integration domains $0=z_1<z_{\sigma(2)}<\ldots <z_{\sigma(n)}<z_0<1$ with any permutation $\sigma \in \CP (2,3,\ldots,n)$. 

Not surprisingly, the augmented integrals \eqref{eqn:Ztzbasis} can also be obtained from the $(n{+}1)$-point basis $\mathbf{Z}_{n+1}^\tau$ in \eqn{taulang.1} by the following formal operations: dropping the integration over the last puncture, identifying $z_{n+1} \rightarrow z_0$ according to the integration domain $\mathbb I_{n+1}$ and peeling off a factor of $\Omega_{1,0}$. The latter can be enforced to appear in each component of $\mathbf{Z}_{n+1}^\tau  |^{\eta_{n+1} \rightarrow \eta_0}_{z_{n+1} \rightarrow z_0}$ by expanding the permutations $\rho[\phiChain(1,2,\dots,n,0)]$ in the integrand in a basis of $\phiChain(\ldots,0,1,\ldots)$ via Fay relations.\footnote{The augmented integrals $\BZ_{0,2}^{\tau}$ at two points are for instance obtained by starting from the integrands $\Omega_{12}(\eta_{20}) \Omega_{20}(\eta_0)$ and $\Omega_{10}(\eta_{20}) \Omega_{02}(\eta_2)$ of $\mathbf{Z}_{3}^\tau |^{\eta_3 \rightarrow \eta_0}_{z_3\rightarrow z_0}$ and rewriting the former as $\Omega_{10}(\eta_{0})  \Omega_{12}(\eta_{2}) -\Omega_{10}(\eta_{20}) \Omega_{02}(\eta_2)$. After peeling off the factors of $\Omega_{10}(\eta_{0})$ and $\Omega_{10}(\eta_{20})$, one is left with the integrands $ \Omega_{12}(\eta_{2})$ and $\Omega_{02}(\eta_{2})$ in the first and second component of \eqn{eqn:Ztzbasis} at $n=2$, respectively.} These operations resemble the construction of fibration bases for configuration-space integrals at genus zero \cite{Mizera:2019gea}.

\subsubsection{Further integrals from Fay identities}\label{sec:411}

The main ingredients to the integrals in the basis \eqref{eqn:notationChain} are chains, which can conveniently be associated with a chain graph using the dictionary from \secref{ssec:graphicalnotation}, for example\footnote{For notational simplicity, in graphs we use the convention $\eta_{a_2a_3\dots a_p}=\eta_{a_{23\dots p}}$ and similarly for other sums of $\eta$-variables associated to sequences of the form $(a_2,a_3,\dots a_p)$.} of the form
\begin{equation}
\phiChain(1,2,3,4)=\Omega_{12}(\eta_{234})\Omega_{23}(\eta_{34})\Omega_{34}(\eta_{4})\quad=\quad \mpostuse[align=b,vshift=-42pt]{example4Pt}\,.
\end{equation}
Multiplying chains corresponds to combining chain graphs. Whenever the same label appears in two different chains, the Fay identity \eqref{eqn:Fay} can be used to link the chains and produce tree graphs. In fact, repeated use of Fay identities implies various identities shown and proven in \appref{app:chainIdentities} satisfied by products of chains $\phiChain$. Such identities among chains are not only useful for the translation to basis integrals, but also for expressing differential equations satisfied by these basis integrals in closed form in subsections \ref{sec:z0Deriv} and \ref{sec:dtau} below. 
For each disconnected pair of tree graphs, the associated product of Kronecker--Eisenstein series can be represented as a linear combination of the integrals in \eqn{eqn:notationChain}, resulting in integrals of the form 
\begin{align}
\label{eqn:Ztaug}
\int_{\gamma_{12\ldots n0}}\dd z_2\cdots \dd z_n\,\KNt_{012\ldots n}\,\Omega_{i_2 2}(\xi_{2}) 
\Omega_{i_3 3}(\xi_{3})  \ldots   \Omega_{i_n n}(\xi_{n})\, .
\end{align}
Since the indices $i_k\in\{0,\ldots,n\}$, $i_k\neq k$ of the $\Omega_{i_k k}(\xi_{k})$ in the integrand are related to the basis \eqref{eqn:Ztzbasis} via Fay identities with graphical form \eqn{eqn:FayGraphical}, the associated edges between vertices $i_k$ and $k$ form tree graphs. The counting of vertices and edges only admits one or two connected components, and the vertices $0,1$ cannot be in the same connected component. Moreover, each vertex $k\neq 0,1$ has only one outgoing edge, while the vertices $k=0,1$ have none. 

The precise form of the variables $\xi_k$ in \eqn{eqn:Ztaug} can be deduced from the graph: for the edge pointing away from the vertex~$k$, the associated $\xi_k$ is a combination of $\eta_2,\eta_3,\ldots,\eta_n$ obtained by accumulating (adding) all $\eta_j$-labels from edges higher in the tree pointing towards the vertex~$k$.
In other words, $\xi_k$ for a given edge is the sum of all the $\eta_j$ of those vertices $j$ which become disconnected from $0$ or $1$ through deletion of the edge under consideration.

As detailed in appendix \ref{app:recoverSelbergInt} the integrals \eqn{eqn:Ztaug} with suitable restrictions on the $i_k$ generate the Selberg integrals \eqn{eqn:Selberg} upon expansion in the $\xi_k$. From this observation 
as well as the aforementioned formal relation between the $n!$ bases $\BZ_{0,n}^{\tau}$ and $\mathbf{Z}_{n+1}^\tau$, one can already anticipate the potential of the augmented $Z^\tau$-integrals \eqn{eqn:Ztzbasis} to relate
the two approaches of \rcites{Mafra:2019xms, Broedel:2019gba} to genus-one $\ap$-expansions.

\subsubsection{Five-point example}\label{sec:412}

A typical example for (\ref{eqn:Ztaug}) at $n=5$ is the integral
\begin{align}
	\label{eqn:Ztaugexample}
	\int_{\gamma_{123450}}\dd z_2\cdots \dd z_5\,\KNt_{01\ldots 5}\,\Omega_{12}(\eta_{234})\,\Omega_{23}(\eta_{3})\,\Omega_{24}(\eta_{4})\,\Omega_{05}(\eta_{5})\,,
\end{align}
which is represented by the following diagram:
\begin{equation}\label{eqn:example1}
	\Omega_{12}(\eta_{234})\,\Omega_{23}(\eta_{3})\,\Omega_{24}(\eta_{4})\,\Omega_{05}(\eta_{5})=	\mpostuse[align=b,vshift=-14pt]{example1}\,,
\end{equation}
where the $\eta$-variables with multiple indices have been defined in \eqn{eqn:AccumulatedEtaVariables}\,.
In parallel to the Fay identities for genus-one $Z$-integrals, one can apply Fay identities to rewrite \eqn{eqn:Ztaugexample} as  
\begin{align}
	\label{eqn:Ztaugexampletwo}
	&\int_{\gamma_{123450}}\dd z_2\cdots \dd z_5\,\KNt_{01\ldots 5}\,\Omega_{12}(\eta_{234})\,\Omega_{23}(\eta_{3})\,\Omega_{24}(\eta_{4})\,\Omega_{05}(\eta_{5})\nonumber\\
	&=\int_{\gamma_{123450}}\dd z_2\cdots \dd z_5\,\KNt_{01\ldots 5}\,\Omega_{12}(\eta_{234})\Omega_{23}(\eta_{34})\Omega_{34}(\eta_{4})\Omega_{05}(\eta_{5})\nnl
	&\phantom{=} +\int_{\gamma_{123450}}\dd z_2\cdots \dd z_5\,\KNt_{01\ldots 5}\,\Omega_{12}(\eta_{234})\Omega_{24}(\eta_{34})\Omega_{43}(\eta_{3})\Omega_{05}(\eta_{5}) \nonumber \\
	&= Z^{\tau}_{0,5}( (1,2,3,4),(0,5)) + Z^{\tau}_{0,5}( (1,2,4,3),(0,5))
	\,,
\end{align}
which is based on the following application of the graphical Fay identity \eqref{eqn:FayGraphical}:
\begin{equation}
	\label{eqn:example2}
	\mpostuse[align=b,vshift=-14pt]{example1}\qquad=\qquad\mpostuse[align=b,vshift=-14pt]{example1RHS1}\qquad+\qquad\mpostuse[align=b,vshift=-14pt]{example1RHS2}\,.
\end{equation}


\subsection{\texorpdfstring{$z_0$}{z0}-derivative of \texorpdfstring{$\BZtzn{n}$}{Ztau(0,n)}}\label{sec:z0Deriv}
Here, we will rewrite the differential equation of \subsecref{ssec:z0lang} in the language of integrals $\Ztzn{n}$, which will be the main players in \secref{sec:translation}. The genus-one Selberg integrals (\ref{eqn:Selberg}) with an auxiliary point used in \rcite{Broedel:2019gba} can be obtained by the methods described in \appref{app:ZtzTogenSelberg}. 

Starting from the basis choice for the $n$-point integrals $\Ztzn{n}$ in \eqn{eqn:Ztzbasis},  we will now demonstrate that the $z_0$-derivatives $\partial_0 \Ztzn{n}((1,A),(0,B))$, where we denote $\partial_0=\partial_{z_0}$, are expressible in terms of the basis integrals $\Ztzn{n}$. In the $n!$-component vector notation of \eqn{eqn:Ztzbasis}, we will derive a differential equation of the form 
\begin{align}\label{eqn:z0DerivE} 
	\partial_{0} \BZtzn{n}&= \XXX \BZtzn{n}\,,
\end{align}
where the entries of the $n!{\times} n!$ matrix $\XXX$ are linear in $s_{ij}$, comprise first derivatives in the $\eta_j$ and will be explicitly determined at any $n$. Moreover, the sole $z_0$- and $\tau$-dependence of $X^\tau_{0,n}$ occurs via $f^{(k)}_{01}=f^{(k)}(z_{01}|\tau)$, i.e.\ one can uniquely identify $z_0$- and $\tau$-independent $n!{\times} n!$ matrices $r_{0,n}(x_{k})$ that cast (\ref{eqn:z0DerivE}) into the form
\begin{align}\label{eqn:KZBforZ}
\partial_0 \BZtzn{n}&=\sum_{k=0}^{\infty} f^{(k)}_{01} r_{0,n}(x_{k}) \BZtzn{n}\,.
\end{align}
Hence, the main result of this section to be derived below is that the $n!$-component vector $\BZtzn{n}$ of augmented $Z^\tau$-integrals satisfies an elliptic KZB equation in the auxiliary puncture $z_0$.

\subsubsection{Deriving the \texorpdfstring{$n$}{n}-point formula}
The first step in calculating $\partial_{0} \BZtzn{n}$ is to use $\partial_{z_i} f^{(1)}_{ij} = -\partial_{z_j} f^{(1)}_{ij}$ and integration by parts such that all the partial derivatives only act on the Koba--Nielsen factor, followed by an application of eq.\ \eqref{eqn:knz} with an extra point $z_0$
\begin{align}\label{eqn:zDerivKN}
\pd_{z_i} \KN^\tau_{012\ldots n}&= -\sum\limits_{\begin{smallmatrix}
	j=0\\j\neq i
	\end{smallmatrix}}^{n}s_{ij} f_{ij}^{(1)} \KN^\tau_{012\ldots n} \, .
\end{align}
Using the notation $A=(a_1,a_2,\dots,a_p)$ and $B=(b_1,b_2,\dots,b_q)$ for the disjoint sequences with $A\cup B=\{2,3,\dots,n\}$ and additionally denoting $a_0=1$, $b_0=0$, this amounts to 
\begin{align}
\partial_0 \Ztzn{n}((1,A),(0,B))
&=\int_{\gamma_{12\ldots n0}} \prod_{i=2}^n\dd z_i\,\left(\sum_{j=0}^q \partial_{b_j}\KN^{\tau}_{01\dots n}\right)\phiChain(1,A)\phiChain(0,B)\label{eqn:partial0Z}\nonumber\\
&=\int_{\gamma_{12\ldots n0}} \prod_{i=2}^n\dd z_i\,  \KN^{\tau}_{01\dots n} \sum_{k=0}^p\sum_{j=0}^q \Big(s_{a_k,b_j}f^{(1)}_{a_k,b_j}\phiChain(1,A)\phiChain(0,B)\Big)\,.
\end{align}
The second step of the calculation consists of rewriting the term in parenthesis using the Fay identity \eqref{eqn:Fay} and the chain identities \eqref{eqn:chainConcatenation} to \eqref{eqn:shuffleIdentityChainWith1And0AlternatingSum}. The rewriting process is cumbersome, but can be cast into an elegant form using a couple of additional notations and tools. The complete derivation can be found in \appref{app:z0Deriv} and the result is the following: for a sequence $C=(c_1,c_2,\dots,c_m)$, a sum of $\eta$-variables is denoted by
\begin{align}
\eta_{C}&=\sum_{i=1}^m \eta_{c_i}\,,
\end{align}
then, the $\eta$-variables
\begin{align}\label{eqn:eta01}
\eta_0&=-\eta_{B} \, , \ \ \ \ \ \ \eta_1=-\eta_{A}\
\end{align}
are assigned to the unintegrated punctures $z_0$ and $z_1=1$. Thus, defining the decomposition of a sequence $C$ into subsequences $C_{ij}$
\begin{align}\label{eqn:subsequences}
C=(c_1,\dots,c_{i-1}, \underbrace{c_i,c_{i+1}\dots,c_{j-1}}_{C_{i,j}=C_{ij}},c_j,c_{j+1}\dots,c_m)\,,
\end{align}
with
\begin{align}
C_{ji}=\emptyset\,\text{ for }j\geq i\,, \ \ \ \ \ \
C_{1,m+1} =C\,,\ \ \ \ \ \ 
\tilde{C}_{ij}=(c_{j-1},c_{j-2},\dots, c_i)\,,
\end{align}
where a tilde denotes the reversal of a sequence, the following closed formula can be derived as shown in \appref{app:z0DerivClosed}: 
\begin{align}\label{eqn:derZ0Closed}
&\partial_0 \Ztzn{n}((1,A),(0,B))\nonumber\\
&=- \left(s_{(1,A),(0,B)}f^{(1)}_{01}+\sum_{k=1}^p s_{a_k,(0,B)}\partial_{\eta_{a_k}}-\sum_{j=1}^qs_{(1,A),b_j}\partial_{\eta_{b_j}}\right)\Ztzn{n}((1,A),(0,B))\nonumber\\
&\phantom{=}+\sum_{k=1}^p\sum_{j=1}^q s_{a_k,b_j}\sum_{i=1 }^{k}\sum_{l=1 }^{j}(-1)^{k+j-i-l}\Omega_{01}(\eta_{B_{l,q+1}})\nonumber\\
&\phantom{=\sum_{i=1 }^{k}\sum_{l=1 }^{j}}\Ztzn{n}\left(\left(1,A_{1i}\shuffle(a_k,(\tilde{A}_{i,k}\shuffle A_{k+1,p+1})\shuffle (b_j,\tilde{B}_{l,j}\shuffle B_{j+1,q+1}))\right),\left(0,B_{1l}\right)\right)\nonumber\\
&\phantom{=}+\sum_{k=1}^p\sum_{j=1}^q s_{a_k,b_j}\sum_{i=1 }^{k}\sum_{l=1 }^{j}(-1)^{k+j-i-l}\Omega_{01}(-\eta_{A_{i,p+1}})\nonumber\\
&\phantom{=\sum_{i=1 }^{k}\sum_{l=1 }^{j}}\Ztzn{n}\left(\left(1,A_{1i}\right),\left(0,B_{1l}\shuffle(b_j,(\tilde{B}_{l,j}\shuffle B_{j+1,q+1})\shuffle (a_k,\tilde{A}_{i,k}\shuffle A_{k+1,p+1}))\right)\right)\nonumber\\
&\phantom{=}+\sum_{j=1}^q s_{1,b_j}\sum_{l=1 }^{j}(-1)^{j-l}\Omega_{01}(\eta_{B_{l,q+1}})\Ztzn{n}\left(\left(1,A\shuffle (b_j,\tilde{B}_{l,j}\shuffle B_{j+1,q+1})\right),\left(0,B_{1l}\right)\right)\nonumber\\
&\phantom{=}+\sum_{k=1}^p s_{a_k,0}\sum_{i=1 }^{k}(-1)^{k-i}\Omega_{01}(-\eta_{A_i,p+1})\Ztzn{n}\left(\left(1,A_{1i}\right), \left(0,B\shuffle (a_k,\tilde{A}_{i,k}\shuffle A_{k+1,p+1})\right)\right)\,.
\end{align}
Moreover, $s_{(1,A),(0,B)},s_{a_k,(0,B)}$ and $s_{(1,A),b_j}$ denote 
sums of Mandelstam invariants according to the following general definition
for sequences $P=(p_1,p_2,\dots,p_l)$ and $Q=(q_1,q_2,\dots,q_m)$ 
\begin{equation}
s_{P,Q} = \sum_{i=1}^l \sum_{j=1}^m s_{p_i q_j} \, .
\label{shortmand}
\end{equation}
Upon writing the partial differential equation (\ref{eqn:derZ0Closed}) for the vector of 
integrals $\BZtzn{n}$ in matrix form, we arrive at the central result (\ref{eqn:z0DerivE}) previewed above.
The entries of the matrix $\XXX$ are determined by the linear combinations in  eq.\ \eqref{eqn:derZ0Closed}, and expanding the Kronecker--Eisenstein series therein in terms of the functions $f^{(k)}_{01}$, we arrive at the elliptic KZB equation (\ref{eqn:KZBforZ}) satisfied by the $n!$-component vector $\BZtzn{n}$ of augmented $Z^\tau$-integrals. This generalizes the KZB-type \eqn{eqn:KZBSelE} for the genus-one Selberg integrals to their generating series $\BZtzn{n}$.

The matrices $r_{0,n}(x_{k})$ in \eqn{eqn:KZBforZ} are independent of $z_0$ and $\tau$, and one can see from \eqn{eqn:derZ0Closed} that they are linear in the Mandelstam variables $s_{ij}$ and of homogeneity degree $k{-}1$ in the variables $\eta_j$. The matrix $r_{0,n}(x_{0})$ additionally involves first derivatives in $\eta_j$ that are counted as homogeneity degree $-1$.
 
\subsubsection{Alternative form in terms of the \texorpdfstring{$S$}{S}-map}
\label{smapz}

The $z_0$-derivative in \eqn{eqn:derZ0Closed} can be compactly rewritten using the so-called $S$-map defined by 
\begin{align}
\phiChain&( S[a_1a_2\ldots a_p, b_1 b_2\ldots b_q]) = \sum_{i=1}^p \sum_{j=1}^q (-1)^{i-j+p-1} s_{a_i b_j}   \label{eq3.6}\nonumber \\
&\times \phiChain\big( (a_1 a_2\ldots a_{i-1} \shuffle a_p a_{p-1}\ldots a_{i+1}),a_i,b_j, (b_{j-1} b_{j-2} \ldots b_{1} \shuffle b_{j+1} b_{j+2}\ldots b_q) \big) \, .
\end{align}
The $S$-map \eqref{eq3.6} has been firstly studied in \rcite{Mafra:2014oia} to rewrite BCJ 
relations \cite{Bern:2008qj} and can
therefore be used to bring integration-by-parts relations among disk integrals \eqref{eqn:Ztree} into the form
\beq
\Ztree_n(a_1,\ldots,a_n| S[P,Q],n) =0 
\label{smaptree}
\eeq
with arbitrary disjoint sequences $P,Q$ such that $P \cup Q=\{1,2,\ldots,n{-}1\}$.
In a genus-one context, the $S$-map featured in the $n$-point proposal for
the $\tau$-derivatives of $Z^\tau$-integrals \eqref{eqn:Zloop}
in \rcite{Mafra:2019xms}, 
\begin{align}
2\pi i \partial_\tau Z^\tau_n(\mathbb I_n|1,2,\ldots,n) &= 
\Big( \frac{1}{2} \sum_{i=2}^n s_{1i} \partial_{\eta_i}^2 + \frac{1}{2} \sum_{2\leq i<j}^n s_{ij} \big( \partial_{\eta_i}{-} \partial_{\eta_j} \big)^2 - 2 \zm_2 s_{12\ldots n} \Big) Z^\tau_n(\mathbb I_n|1,2,\ldots,n) 
\notag \\
& \ \ \ \ - \sum_{i=2}^n \wp(\eta_i{+}\eta_{i+1}{+}\ldots{+}\eta_n|\tau) Z^\tau_n(\mathbb I_n| S[12\ldots i{-}1, i(i{+}1)\ldots n])  \label{dtaunpt}
\end{align}
which was rigorously derived in \rcite{Gerken:2019cxz}.

As will be derived in \appref{app:z0DerivSMap}, an alternative form of \eqn{eqn:derZ0Closed} 
is given by the following formula,
\begin{align}\label{eqn:derZ0SMap}
&\partial_0 \Ztzn{n}((1,A),(0,B))\nonumber\\
&=- \left(s_{(1,A),(0,B)}f^{(1)}_{01}+\sum_{k=1}^ps_{a_k,(0,B)}\partial_{\eta_{a_k}}-\sum_{j=1}^qs_{(1,A),b_j}\partial_{\eta_{b_j}}\right)\Ztzn{n}((1,A),(0,B))\nonumber\\
&\phantom{=}+\sum_{l=1 }^{q}\Omega_{01}(\eta_{B_{l,q+1}})
\Ztzn{n}\big( (0,B_{1l}) ,(S[(1,A),B_{l,q+1}])  \big)  \nonumber\\
&\phantom{=}+\sum_{l=1}^p\Omega_{01}(-\eta_{A_{l,p+1}})
\Ztzn{n}\big(
(1,A_{1l}), (S[(0,B),A_{l,p+1}]) \big)\,,
\end{align}
where e.g.\ (recall that $b_0=0$)
\begin{align}\label{eqn:SMapTerm}
\phiChain(S[(0,B),A_{l,p+1}])&=\sum_{k=l}^p\sum_{j=0}^q(-1)^{q-j+k-l} s_{a_k,b_j}\phiChain( B_{0j}\shuffle \tilde B_{j+1,q+1},b_j,a_k,\tilde{A}_{l k}\shuffle A_{k+1,p+1})\,.
\end{align}
The individual terms in the shuffle $\phiChain( B_{0j}\shuffle\ldots)$
do not necessarily have the label $0$ in the first entry of the chain and
thereby involve integrands outside the basis of $\Ztzn{n}(
(1,A), (0,B)) $ in \eqn{eqn:notationChain}. Hence, it remains to apply 
combinations of Fay identities in the Kleiss--Kuijf form \cite{Kleiss:1988ne, Schocker}
\begin{align}
\Ztzn{n}\big(
(P,1,Q) ,(0,B) \big) &= (-1)^{|P|} \Ztzn{n}\big(
(1,\tilde P \shuffle Q),(0,B)  \big) \,,\notag
\\
\Ztzn{n}\big(
(1,A), (P,0,Q) \big) &= (-1)^{|P|} \Ztzn{n}\big(
(1,A) ,(0,\tilde P \shuffle Q) \big)
\label{kleissk}
\end{align}
in order to manifest the $n!$ entries of $\BZtzn{n}$ on the right-hand side of \eqn{eqn:derZ0SMap}, where $|P|$ denotes the number of labels in $P$. The combination of \eqns{kleissk}{eqn:derZ0SMap} encodes all-multiplicity expressions for the matrix $\XXX$ in \eqn{eqn:z0DerivE}.

\subsubsection{Two-point example}
The simplest example can be found at two points, where the basis vector \eqref{eqn:Ztzbasis} of augmented $Z^\tau$-integrals is given by
\begin{align}\label{eqn:2PointExampleIntegral}
\BZtzn{2}&=\int_{0}^{z_0}\dd z_2\,\KN^{\tau}_{012}\begin{pmatrix}
\Omega_{12}(\eta)\\\Omega_{02}(\eta)
\end{pmatrix}=\begin{pmatrix}
\Ztzn{2}(1,2)\\
\Ztzn{2}(0,2)\\
\end{pmatrix}\,.
\end{align}
The partial differential equation \eqref{eqn:z0DerivE} follows from the closed formula 
\eqref{eqn:derZ0Closed} or its reformulation in section \ref{smapz}. Both approaches yield
\begin{align}\label{eqn:2PtZ0}
\partial_0 \BZtzn{2}&=\begin{pmatrix}
-(s_{01}+s_{02})f^{(1)}_{01}-s_{02}\partial_{\eta}&s_{02}\Omega_{01}(-\eta)\\
s_{12}\Omega_{01}(\eta)&-(s_{01}+s_{12})f^{(1)}_{01}-s_{12}\partial_{\eta}
\end{pmatrix}\BZtzn{2}
\end{align}
and expose the matrix $\XXXn{2}$ in the notation of \eqn{eqn:z0DerivE}.
The expansion  $\XXXn{2}=\sum_{k=0}^{\infty}f^{(k)}_{01} r_{0,2}(x_k)$
leads to the elliptic KZB equation (\ref{eqn:KZBforZ}), where the
matrices $r_{0,2}(x_k)$ are given by
\begin{align}\label{eqn:2PointEMatrices}
r_{0,2}(x_0)&=\begin{pmatrix}
-s_{02}\partial_{\eta}&-s_{02}/\eta\\s_{12}/\eta&s_{12}\partial_{\eta}
\end{pmatrix}\,, \notag \\
r_{0,2}(x_1)&=\begin{pmatrix}
-(s_{01}+s_{02})&s_{02}\\s_{12}&-(s_{01}+s_{12}) 
\end{pmatrix}\,, \nonumber\\
r_{0,2}(x_k)&=\eta^{k-1}\begin{pmatrix}
0&(-1)^{k-1}s_{02}\\s_{12}&0
\end{pmatrix} \, , \ \ \ \ \ \ k\geq 2\, . 
\end{align}


\subsubsection{Three-point example}
\label{sec:z03pt}

The three-point basis vector \eqref{eqn:Ztzbasis}
\begin{align}\label{eqn:3PointExampleIntegral}
\BZtzn{3}&=\int_{0}^{z_0}\dd z_3 \int^{z_3}_0 \dd z_2 \,\KN^{\tau}_{0123}\begin{pmatrix}
\Omega_{12}(\eta_{23}) \Omega_{23}(\eta_{3}) \\ 
\Omega_{13}(\eta_{23}) \Omega_{32}(\eta_{2}) \\ 
 \Omega_{12}(\eta_{2})  \Omega_{03}(\eta_{3}) \\
  \Omega_{13}(\eta_{3})  \Omega_{02}(\eta_{2}) \\
\Omega_{03}(\eta_{23}) \Omega_{32}(\eta_{2}) \\ 
\Omega_{02}(\eta_{23}) \Omega_{23}(\eta_{3})
\end{pmatrix}=\begin{pmatrix}
\Ztzn{3}(1,2,3)\\
\Ztzn{3}(1,3,2)\\
\Ztzn{3}((1,2),(0,3))\\
\Ztzn{3}((1,3),(0,2))\\
\Ztzn{3}(0,3,2)\\
\Ztzn{3}(0,2,3)
\end{pmatrix}
\end{align}
obeys the following differential equation according to \eqns{eqn:derZ0Closed}{eqn:derZ0SMap},
\begin{small}
\begin{align}
&\partial_0 \BZtzn{3}= \left[
\te{diag} \vecb {-}s_{02} \partial_{\eta_2}{-}s_{03} \partial_{\eta_3}{-}s_{0,123} f^{(1)}_{01} 
\\
{-}s_{02} \partial_{\eta_2}{-}s_{03} \partial_{\eta_3}{-} s_{0,123}  f^{(1)}_{01} 
\\
{-}(s_{02}{+}s_{23}) \partial_{\eta_2}{+}(s_{13}{+}s_{23}) \partial_{\eta_3}{-} s_{03,12} f^{(1)}_{01} 
 \\
 {-}(s_{03}{+}s_{23}) \partial_{\eta_3}{+}(s_{12}{+}s_{23}) \partial_{\eta_2}{-}s_{02,13} f^{(1)}_{01} 
\\
s_{12} \partial_{\eta_2}{+}s_{13} \partial_{\eta_3}{-}s_{1,023} f^{(1)}_{01} 
\\
s_{12} \partial_{\eta_2}{+}s_{13} \partial_{\eta_3} {-} s_{1,023}   f^{(1)}_{01} \vece
\right.  \\[5pt]
&+\left.\left( 
\setlength{\arraycolsep}{-2.4pt}
	\begin{array}{cccccc}
0&0&-s_{03}\Omega_{10}(\eta_{3}) &0&s_{03}\Omega_{10}(\eta_{23})&-s_{02}\Omega_{10}(\eta_{23})\\
0&0&0 &-s_{02}\Omega_{10}(\eta_{2}) &-s_{03}\Omega_{10}(\eta_{23})&s_{02}\Omega_{10}(\eta_{23})\\
(s_{13}{+}s_{23})\Omega_{01}(\eta_{3}) & s_{31}\Omega_{01}(\eta_{3})& 0 & 0& -\left(s_{02}{+}s_{23}\right)\Omega_{10}(\eta_2) & -s_{02}\Omega_{10}(\eta_{2})\\
s_{21}\Omega_{01}(\eta_{2})&(s_{12}{+}s_{23})\Omega_{01}(\eta_{2})&0 &0&-s_{03}\Omega_{10}(\eta_{3}) &-\left(s_{03}{+}s_{23}\right)\Omega_{10}(\eta_3)\\
-s_{12}\Omega_{01}(\eta_{23})&s_{13}\Omega_{01}(\eta_{23})& s_{12}\Omega_{01}(\eta_{2}) &0&0&0 \\
s_{12}\Omega_{01}(\eta_{23}) &-s_{13}\Omega_{01}(\eta_{23}) &0 &s_{13}\Omega_{01}(\eta_{3})  &0&0
\end{array} \right)\right]  \BZtzn{3}\nonumber
\end{align}
\end{small}
see \eqn{shortmand} for the $s_{P,Q}$ notation in the first line.
By matching with the general form \eqref{eqn:KZBforZ} of the elliptic KZB equation,
one can read off the following $6\times 6$ matrices $r_{0,3}(x_k)$:
\begin{small}
\begin{align}
r_{0,3}(x_0) &=\te{diag} \vecb
-s_{02} \partial_{\eta_2}{-}s_{03} \partial_{\eta_3} \\
-s_{02} \partial_{\eta_2}{-}s_{03} \partial_{\eta_3}  \\
\! \! -(s_{02}{+}s_{23}) \partial_{\eta_2}{+}(s_{13}{+}s_{23}) \partial_{\eta_3} \! \! \\
\! \! -(s_{03}{+}s_{23}) \partial_{\eta_3}{+}(s_{12}{+}s_{23}) \partial_{\eta_2} \! \! \\
s_{12} \partial_{\eta_2}{+} s_{13} \partial_{\eta_3} \\
s_{12} \partial_{\eta_2}{+} s_{13} \partial_{\eta_3}
\vece
+  \left( 
\setlength{\arraycolsep}{-1pt}
\begin{array}{cccccc}
0
&0 &-\frac{s_{03}}{\eta_3}  &0 &\frac{s_{03}}{\eta_{23}} &- \frac{s_{02}}{\eta_{23}} 
\\[8pt]
0 &0 &0 
&- \frac{s_{02}}{\eta_{2}} &-\frac{s_{03}}{\eta_{23}}   &\frac{s_{02}}{\eta_{23}} 
\\[8pt]
\frac{s_{13}{+}s_{23}}{\eta_{3}}  &\frac{s_{13}}{\eta_{3}} & 
0
&0 &- \frac{s_{02}{+}s_{23}}{\eta_{2}} & - \frac{s_{02}}{\eta_{2}} 
\\[8pt]
\frac{s_{12}}{\eta_{2}} &\frac{s_{12}{+}s_{23}}{\eta_{2}} &0 
&0
&- \frac{s_{03}}{\eta_{3}} &- \frac{s_{03}{+}s_{23}}{\eta_{3}} 
\\[8pt]
- \frac{s_{12}}{\eta_{23}} &\frac{s_{13}}{\eta_{23}}  &\frac{s_{12}}{\eta_{2}}  
&0 & 0
&0 
\\[8pt]
\frac{s_{12}}{\eta_{23}}  &-\frac{s_{13}}{\eta_{23}} &0 &\frac{s_{13}}{\eta_{3}}  &0 &
0
\end{array} \right) \! \! \,,  \label{x03pt} 
\\
r_{0,3}(x_1) &= 
\ccccccb
- s_{0,123} 
&0
&s_{03}
&0
&-s_{03}
&s_{02}
\\
0
&-s_{0,123}
&0
&s_{02}
&s_{03}
&-s_{02}
\\
s_{13}{+}s_{23}
&s_{13}
&-s_{03,12}
&0
&s_{02}{+}s_{23}
&s_{02}
\\
s_{12}
&s_{12}{+}s_{23}
&0
&-s_{02,13}
&s_{03}
&s_{03}{+}s_{23}
\\
-s_{12}
&s_{13}
&s_{12}
&0
&-s_{023,1}
&0
\\
s_{12}
&-s_{13}
&0
&s_{13}
&0
&-s_{023,1}
\cccccce \,,
\label{alteq2.25}
\\
r_{0,3}(x_k) &= 
\left( 
\setlength{\arraycolsep}{-2pt}
\begin{array}{cccccc}
0 
&0
&s_{03}({-}\eta_{3})^{k-1}
&0
&-s_{03}({-}\eta_{23})^{k-1}
&s_{02}({-}\eta_{23})^{k-1}
\\
0
&0
&0
&s_{02}({-}\eta_{2})^{k-1}
&s_{03} ({-}\eta_{23})^{k-1}
&-s_{02}({-}\eta_{23})^{k-1}
\\
(s_{13}{+}s_{23}) \eta_{3}^{k-1}
&s_{13}\eta_{3}^{k-1}
&0
&0
&(s_{02}{+}s_{23}) ({-}\eta_{2})^{k-1}
&s_{02} ({-}\eta_{2})^{k-1}
\\
s_{12} \eta_{2}^{k-1}
&(s_{12}{+}s_{23}) \eta_{2}^{k-1}
&0
&0
&s_{03}({-}\eta_{3})^{k-1}
&(s_{03}{+}s_{23}) ({-}\eta_{3})^{k-1}
\\
-s_{12}\eta_{23}^{k-1}
&s_{13}\eta_{23}^{k-1}
&s_{12} \eta_{2}^{k-1}
&0
&0
&0
\\
s_{12}\eta_{23}^{k-1}
&-s_{13}\eta_{23}^{k-1}
&0 
&s_{13}\eta_{3}^{k-1}
&0
&0
\end{array} \right) \, ,\notag \\ 
\label{alteq2.26}
\end{align}
\end{small}
where $k\geq 2$.


\subsection{\texorpdfstring{$\tau$}{tau}-derivative of \texorpdfstring{$\BZtzn{n}$}{Ztau(0,n)}}
\label{sec:dtau}
As we will show in this section, the $n!$ basis $\BZtzn{n}$ in \eqn{eqn:Ztzbasis} also closes under $\tau$-derivatives. Similar to the homogeneous first-order equation (\ref{eqn:z0DerivE}) in $z_0$, the $\tau$-derivative of $ \BZtzn{n}$ will be cast into the form
\begin{align}\label{eqn:tauDerivE}
	2\pi i \partial_{\tau} \BZtzn{n}&=
	(\DDD+\BBB)
	\BZtzn{n}\, ,
\end{align}
where the explicit form of the $n!{\times} n!$ matrices $\DDD,\BBB$ will be determined below.
We have grouped the matrices according to the $\tau$- and $z_0$-dependence which is solely carried
by $f_{01}^{(k)}$ and Eisenstein series $\GG{k}$,
\begin{align}
\DDD&=-r_{0,n}(\ep_0)+\sum_{k=4}^{\infty}(1{-}k)\GG{k}r_{0,n}(\ep_k)\,,\nonumber
\\
\BBB&=\sum_{k=2}^{\infty}(k{-}1)f_{01}^{(k)}r_{0,n}(b_k)\, ,
\label{expanddtau}
\end{align}
where the entries of $\DDD, \BBB$ and therefore all the $r_{0,n}(b_k),r_{0,n}(\ep_k)$ are again linear in $s_{ij}$. By construction, the $n!{\times} n!$ matrices $r_{0,n}(b_k),r_{0,n}(\ep_k)$ no longer depend on $z_0$ and $\tau$. The appearance of $f_{01}^{(k)}$ in the differential equation in $\tau$ is shared by the operator \eqn{eqn:KZBforZ} in the $z_0$ derivative, and we will see from two perspectives that the accompanying $n!{\times} n!$ matrices are related by
\beq
r_{0,n}(b_k)=r_{0,n}(x_{k-1}) \, , \ \ \ \ \ \ k \geq 2 \, .
\label{bequalx}
\eeq
Note that the additional zero in the subscript distinguishes the $n!{\times} n!$ matrices
$r_{0,n}(\ep_k)$ in \eqn{expanddtau} from the $(n{-}1)!{\times} (n{-}1)!$ matrices
$r_{n}(\ep_k)$ in the differential equation (\ref{taulang.5}) of the $Z^\tau$-integrals without augmentation.

\subsubsection{Deriving the \texorpdfstring{$n$}{n}-point formula}
The evaluation of the $\tau$-derivative $2\pi i \partial_{\tau} \Ztzn{n}((1,A),(0,B))$ with $A=(a_1,a_2,\dots,a_p)$ and $B=(b_1,b_2,\dots,b_q)$ follows the same steps as the $z_0$-derivative in \secref{sec:z0Deriv}. While the details are shown in \appref{appTauDerivative}, we give an overview in the following paragraphs. 

First, the mixed heat equation (\ref{eqn:mixedHeat}) and integration by parts can be used to find an expression where all the derivatives only act on the Koba--Nielsen factor. Then, for the $z_j$-derivatives \eqn{eqn:zDerivKN} and for the $\tau$-derivative the equation
\begin{align}
2 \pi i \pd_\tau \KN^\tau_{012\ldots n}&= -\sum\limits_{0\leq i<j\leq n}s_{ij}(f_{ij}^{(2)}+2\zeta_2)\KN^\tau_{012\ldots n}
\label{dtauofKN}
\end{align}
can be applied, which leads to the expression 
\begin{align}
&2\pi i \partial_{\tau} \Ztzn{n}((1,A),(0,B))
\nonumber \\
&=-\int_{\gamma_{12\ldots n0}} \dd z_2\, \ldots \, \dd z_n \, \KN^{\tau}_{01\dots n} \Bigg(
2\zm_2 s_{01\dots n} \phiChain(1,A) \phiChain(0,B)\nonumber\\
&\phantom{=}+ \sum_{k=1}^p\sum_{j=0}^{k-1} s_{a_k,a_j}\left(f^{(1)}_{a_k,a_j}\left(\partial_{\eta_{a_k}}-
\theta_{j\geq 1}\partial_{\eta_{a_{j}}}\right) +f^{(2)}_{a_k,a_j}\right)\phiChain(1,A) \phiChain(0,B)\label{eqn:SMapConjectureTau}\nonumber\\
&\phantom{=}+ \sum_{k=1}^q\sum_{j=0}^{k-1}s_{b_k,b_j} \left(f^{(1)}_{b_k,b_j}\left(\partial_{\eta_{b_k}}-
\theta_{j\geq 1} \partial_{\eta_{b_{j}}}\right)+f^{(2)}_{b_k,b_j}\right)\phiChain(1,A)\phiChain(0,B) \nonumber\\
&\phantom{=}+ \sum_{k=0}^p\sum_{j=0}^{q}s_{a_k,b_j} \left(f^{(1)}_{a_k,b_j}\left(
\theta_{k\geq 1}
\partial_{\eta_{a_k}}-
\theta_{j\geq 1} \partial_{\eta_{b_{j}}}\right)-f^{(2)}_{b_j,a_k} \right)\phiChain(1,A) \phiChain(0,B) \Bigg)\, ,
\end{align}
where the step function $\theta_{j\geq 1} $ is taken to be $1$ for $j\geq 1$ and zero for $j=0$.
From this equation similar identities as for the $z_0$-derivative, mainly based on the Fay identity of Kronecker--Eisenstein series, lead to the following expression involving the $S$-map in \eqns{eqn:SMapTerm}{kleissk}
\begin{align}
&2\pi i \partial_{\tau}\Ztzn{n}((1,A),(0,B))\nonumber\\
&= \bigg(\frac{1}{2}\sum_{j=2}^n(s_{0j}{+}s_{1j}) \partial^2_{\eta_{j}} +\frac{1}{2}\sum_{2\leq i<j\leq n} \! \! \! s_{ij} (\partial_{\eta_{i}}{-}\partial_{\eta_{j}})^2- 2\zeta_2s_{01\dots n}-s_{(1,A),(0,B)}f^{(2)}_{01}\bigg)\Ztzn{n}((1,A),(0,B))\nonumber\\
&\phantom{=}+\sum_{l=1 }^{q}\Bigg(\Omega^+_{01}(\eta_{B_{l,q+1}})
\Ztzn{n} \big( (0,B_{1l}),(S[(1,A),B_{l,q+1}]) \big) \label{eqn:tauDerivSMap}\nonumber\\
&\phantom{=-\sum_{l=1 }^{j}\Big(}-\wp(\eta_{B_{l,p+1}})
\Ztzn{n}\big( (1,A),(S[(0,B_{1,l}),B_{l,q+1}])\big) \Bigg)\nonumber\\
&\phantom{=}+\sum_{l=1}^p\Bigg(\Omega^-_{01}(-\eta_{A_l,p+1})
\Ztzn{n}\big((1,A_{1l}),(S[(0,B),A_{l,p+1}]) \big)\nonumber\\
&\phantom{=-\sum_{l=1 }^{j}\Big(}-\wp(\eta_{A_{l,p+1}})
\Ztzn{n}\big((0,B),(S[(1,A_{1,l}),A_{l,p+1}]) \big)\Bigg)\, , 
\end{align}
where
\begin{align}
\Omega^{\pm}_{01}(\pm\xi)&=\pm \partial_{\xi}\Omega_{01}(\pm\xi)\,.
\label{delKE}
\end{align}

When all the $S$-maps in \eqn{eqn:tauDerivSMap}
are expanded in terms of $\Ztzn{n}((1,P),(0,Q))$, the
result takes the closed form
\begin{align}
&2\pi i \partial_{\tau} \Ztzn{n}((1,A),(0,B))\nonumber\\
&= \bigg(\frac{1}{2}\sum_{j=2}^n(s_{0j}{+}s_{1j}) \partial^2_{\eta_{j}} +\frac{1}{2}\sum_{2\leq i<j\leq n} \! \! \! s_{ij} (\partial_{\eta_{i}}{-}\partial_{\eta_{j}})^2- 2\zeta_2s_{01\dots n}-s_{(1,A),(0,B)}f^{(2)}_{01}\bigg)\Ztzn{n}((1,A),(0,B))
\nonumber\\
&\phantom{=}-\sum_{k=1}^q\sum_{j=0}^{k-1} s_{b_k,b_j}\sum_{l=j+1}^k\wp(\eta_{B_{l,q+1}})(-1)^{k-l}\Ztzn{n}\left((1,A),\left(0,B_{1,j},b_j, B_{j,l}\shuffle (b_k,\tilde B_{l,k}\shuffle B_{k+1,q+1})\right)\right)\nonumber\\
&\phantom{=}-\sum_{k=1}^p\sum_{j=0}^{k-1} s_{a_k,a_j}\sum_{l=j+1}^k\wp(\eta_{A_{l,p+1}})(-1)^{k-l}\Ztzn{n}\left(\left(1,A_{1,j},a_j, A_{j,l}\shuffle (a_k,\tilde A_{l,k}\shuffle A_{k+1,p+1})\right),(0,B)\right)\nonumber\\
&\phantom{=}+\sum_{k=1}^p\sum_{j=1}^q s_{a_k,b_j}\sum_{i=1 }^{k}\sum_{l=1 }^{j}(-1)^{k+j-i-l}\Omega^+_{01}(\eta_{B_{l,q+1}})  \label{eqn:tauDerivClosed}\nonumber \\
&\phantom{=\sum_{i=1 }^{k}\sum_{l=1 }^{j}}\Ztzn{n}\left(\left(1,A_{1i}\shuffle(a_k,(\tilde{A}_{i,k}\shuffle A_{k+1,p+1})\shuffle (b_j,\tilde{B}_{l,j}\shuffle B_{j+1,q+1}))\right),\left(0,B_{1l}\right)\right)\nonumber\\
&\phantom{=}+\sum_{k=1}^p\sum_{j=1}^q s_{a_k,b_j}\sum_{i=1 }^{k}\sum_{l=1 }^{j}(-1)^{k+j-i-l}\Omega^-_{01}(-\eta_{A_{i,p+1}})\nonumber\\
&\phantom{=\sum_{i=1 }^{k}\sum_{l=1 }^{j}}\Ztzn{n}\left(\left(1,A_{1i}\right),\left(0,B_{1l}\shuffle(b_j,(\tilde{B}_{l,j}\shuffle B_{j+1,q+1})\shuffle (a_k,\tilde{A}_{i,k}\shuffle A_{k+1,p+1}))\right)\right)\nonumber\\
&\phantom{=}+\sum_{j=1}^q s_{1,b_j}\sum_{l=1 }^{j}(-1)^{j-l}\Omega^+_{01}(\eta_{B_{l,q+1}})\Ztzn{n}\left(\left(1,A\shuffle (b_j,\tilde{B}_{l,j}\shuffle B_{j+1,q+1})\right),\left(0,B_{1l}\right)\right)\nonumber\\
&\phantom{=}+\sum_{k=1}^p s_{a_k,0}\sum_{i=1 }^{k}(-1)^{k-i}\Omega^-_{01}(-\eta_{A_i,p+1})\Ztzn{n}\left(\left(1,A_{1i}\right), \left(0,B\shuffle (a_k,\tilde{A}_{i,k}\shuffle A_{k+1,p+1})\right)\right)\,. 
\end{align}
Note that the all-multiplicity formul\ae{} \eqref{eqn:tauDerivSMap} and \eqref{eqn:tauDerivClosed} for $\tau$-derivatives strongly resemble \eqns{eqn:derZ0SMap}{eqn:derZ0Closed} for $z_0$-derivatives, respectively. The only differences concern the diagonal term (the coefficients of $\Ztzn{n}((1,A),(0,B))$ on the right-hand sides), the $\eta_j$-derivatives $\Omega^+_{01}(\eta_{B_{l,q+1}})$, $\Omega^-_{01}(-\eta_{A_l,p+1})$ instead of $\Omega_{01}(\eta_{B_{l,q+1}})$, $\Omega_{01}(-\eta_{A_l,p+1})$, respectively, and the additional appearance of Weierstra\ss{} $\wp$-functions in the $\tau$-derivatives.

The closed formula \eqref{eqn:tauDerivClosed} or (\ref{eqn:tauDerivSMap}) along with \eqn{kleissk} lead to the matrix equation (\ref{eqn:tauDerivE}) for the basis vector \eqn{eqn:Ztzbasis}, i.e.\ determine the entries of the $n!{\times} n!$ matrices $\DDD,\BBB$. With the expansion of the Kronecker--Eisenstein series and $\wp$-functions in the $\tau$-derivatives in terms of $f^{(k)}_{01}$ and ${\rm G}_k$, we can read off the explicit form of the matrices $r_{0,n}(b_k)$ and $r_{0,n}(\ep_k)$ defined by \eqn{expanddtau},
\begin{align}\label{eqn:tauDerivF}
2\pi i \partial_{\tau}  \BZtzn{n}&=\left(-r_{0,n}(\eps_0)+\sum_{k=4}^{\infty}(1{-}k)\GG{k}r_{0,n}(\eps_k)+\sum_{k= 2}^{\infty}(k{-}1)f_{01}^{(k)}r_{0,n}(b_k)\right) \BZtzn{n}\,.
\end{align}
The $r_{0,n}(b_k)$ and $r_{0,n}(\ep_k)$ are all linear in $s_{ij}$ and independent of $z_0$ and $\tau$.  Moreover, their instances at $k\geq 2$ are both homogeneous polynomials of degree $k{-}2$ in the $\eta_j$, whereas $r_{0,n}(\eps_0)$ is a combination of $2 \zeta_2 s_{012\ldots n}, \eta_{j}^{-2}$ and $\partial_{\eta_j}^2$. As previewed in \eqn{bequalx}, one can confirm from \eqn{eqn:tauDerivClosed} or \eqn{eqn:tauDerivSMap} that all the $r_{0,n}(b_k)$ with $k\geq 2$ agree with the operators $r_{0,n}(x_{k-1})$ in the $z_0$-derivative \eqn{eqn:KZBforZ}. An alternative derivation of \eqn{bequalx} will be given in section~\ref{commderiv}.


\subsubsection{Two-point example}

According to \eqns{eqn:tauDerivSMap}{eqn:tauDerivClosed}, the $\tau$-derivative of the
$n=2$ example \eqn{eqn:2PointExampleIntegral} is given by 
\begin{align}
2\pi i \partial_{\tau}\BZtzn{2}&= \left(\frac{1}{2}(s_{02}+s_{12})\partial_{\eta}^2-2\zeta_2 s_{012}\right)\BZtzn{2} \label{eqn:2PtTau}\nonumber\\
&\phantom{=}+\begin{pmatrix}
-s_{12}\wp(\eta)-(s_{01}+s_{02})f^{(2)}_{01}&-s_{02}\partial_{\eta}\Omega_{01}(-\eta)\\
s_{12}\partial_{\eta}\Omega_{01}(\eta)&-s_{02}\wp(\eta)-(s_{01}+s_{12})f^{(2)}_{01}
\end{pmatrix}\BZtzn{2} \, .
\end{align}
Upon comparison with (\ref{eqn:tauDerivF}), one can read off the $r_{0,2}(\eps_k)$ matrices
\begin{align}\label{eqn:2PointTauDeriv}
r_{0,2}(\eps_0)&=\frac{1}{\eta^2}\begin{pmatrix}
s_{12}&s_{02}\\s_{12}&s_{02}
\end{pmatrix}+ \Big( 2\zeta_2 s_{012}-\frac{1}{2}(s_{02}+s_{12})\partial_{\eta}^2 \Big) \mathds{1}_{2{\times} 2}\,,\nonumber \\
r_{0,2}(\eps_k)&=\eta^{k-2}\begin{pmatrix}
s_{12}&0\\0&s_{02} 
\end{pmatrix} \, , \ \ \ \ \ \  k\geq 4  \,,
\end{align}
and $r_{0,2}(b_k)$ matrices
\begin{align}
r_{0,2}(b_2)&= \begin{pmatrix}
 -(s_{01}+s_{02} )&s_{02} \\ s_{12} &-(s_{01}+s_{12}) 
\end{pmatrix} = r_{0,2}(x_{1})
\label{eqn:2ptb} \,,\nonumber\\
r_{0,2}(b_k)&=\eta^{k-2}\begin{pmatrix}
0&(-1)^{k-2}s_{02}\\s_{12}&0
\end{pmatrix}=r_{0,2}(x_{k-1}) \, , \ \ \ \ \ \ k \geq 3\, ,
\end{align}
in agreement with the $r_{0,2}(x_{k-1})$ in \eqn{eqn:2PointEMatrices}.


\subsubsection{Three-point example}
\label{sec:tau3}

For the six-component vector $\BZtzn{3}$ in \eqn{eqn:3PointExampleIntegral}, the $\tau$-derivative
resulting from \eqns{eqn:tauDerivSMap}{eqn:tauDerivClosed} is
spelt out in appendix \ref{app:tau3}.
The $r_{0,3}(\cdot)$ matrices obtained by matching \eqn{bigeq}
with \eqn{eqn:tauDerivF} are given by
\begin{align}
r_{0,3}(\eps_{0}) &= \Big( 2\zeta_2 s_{0123} - \frac{1}{2}(s_{02}{+}s_{12}) \partial_{\eta_2}^2 
- \frac{1}{2} (s_{03}{+}s_{13}) \partial_{\eta_3}^2
- \frac{1}{2}  s_{23}(\partial_{\eta_2}{-}\partial_{\eta_3})^2 \Big) \mathds{1}_{6\times 6} \label{eq2.23} \nonumber\\
&\phantom{=}+ \ccccccb
\frac{s_{12}}{\eta_{23}^2}+ \frac{s_{13}{+}s_{23}}{\eta_{3}^2} 
&\frac{s_{13}}{\eta_{3}^2} - \frac{s_{13}}{\eta_{23}^2} 
&\frac{s_{03}}{\eta_{3}^2}
&0
&-\frac{s_{03}}{\eta_{23}^2}
& \frac{s_{02}}{\eta_{23}^2}
\\
\frac{s_{12}}{\eta_{2}^2}-\frac{s_{12}}{\eta_{23}^2}
&\! \frac{s_{13}}{\eta_{23}^2}+\frac{s_{12}{+}s_{23}}{\eta_{2}^2} \! 
&0
&\frac{s_{02}}{\eta_{2}^2}
&\frac{s_{03}}{\eta_{23}^2}
&-\frac{s_{02}}{\eta_{23}^2}
\\
\frac{(s_{13}{+}s_{23})}{\eta_{3}^2}
&\frac{s_{13}}{\eta_{3}^2}
&\! \! \! \frac{s_{12}}{\eta_{2}^2} + \frac{s_{03}}{\eta_{3}^2}  \! \! \!
&0
&\frac{(s_{02}{+}s_{23})}{\eta_{2}^2}
&\frac{s_{02}}{\eta_{2}^2}
\\
\frac{s_{12}}{\eta_{2}^2}
&\frac{(s_{12}{+}s_{23})}{\eta_{2}^2}
&0
&\! \! \! \frac{s_{13}}{\eta_{3}^2} + \frac{s_{02}}{\eta_{2}^2} \! \! \!
&\frac{s_{03}}{\eta_{3}^2}
&\frac{(s_{03}{+}s_{23})}{\eta_{3}^2}
\\
-\frac{s_{12}}{\eta_{23}^2}
&\frac{s_{13}}{\eta_{23}^2}
&\frac{s_{12}}{\eta_{2}^2}
&0
& \! \frac{s_{03}}{\eta_{23}^2}+\frac{(s_{02}{+}s_{23})}{\eta_{2}^2} \!
&\frac{s_{02}}{\eta_{2}^2} - \frac{s_{02}}{\eta_{23}^2}
\\
\frac{s_{12}}{\eta_{23}^2}
&- \frac{s_{13}}{\eta_{23}^2}
&0
&\frac{s_{13}}{\eta_{3}^2}
&\frac{s_{03}}{\eta_{3}^2} - \frac{s_{03}}{\eta_{23}^2}
&\frac{s_{02}}{\eta_{23}^2} + \frac{(s_{03}{+}s_{23})}{\eta_{3}^2}
\cccccce\,,
\\
r_{0,3}(\eps_{k}) &= {\rm diag} \bigg\{
\ccb
s_{12}\eta_{23}^{k-2}+ (s_{13}{+}s_{23})\eta_{3}^{k-2} 
& s_{13}\eta_{3}^{k-2} - s_{13}\eta_{23}^{k-2} 
\\
s_{12}\eta_{2}^{k-2}-s_{12}\eta_{23}^{k-2}
&s_{13}\eta_{23}^{k-2}+(s_{12}{+}s_{23})\eta_{2}^{k-2}
\cce , \  \notag \\
&\ \ \ \ \ \
\qquad\ccb
s_{12}\eta_{2}^{k-2} + s_{03}\eta_{3}^{k-2} 
&0
\\
0
&s_{13}\eta_{3}^{k-2} + s_{02}\eta_{2}^{k-2}
\cce , \ \notag \\
&\ \ \ \ \ \
\qquad\ccb
s_{03}\eta_{23}^{k-2}+(s_{02}{+}s_{23})\eta_{2}^{k-2}
&s_{02}\eta_{2}^{k-2} - s_{02}\eta_{23}^{k-2}
\\
s_{03}\eta_{3}^{k-2} - s_{03}\eta_{23}^{k-2}
&s_{02}\eta_{23}^{k-2} + (s_{03}{+}s_{23})\eta_{3}^{k-2}
\cce
\bigg\}\, , \ \ \ \ k \geq 4\, ,
\label{eq2.24}
\end{align}
as well as the $r_{0,3}(x_{k-1})$ in \eqns{alteq2.25}{alteq2.26},
\begin{align}
r_{0,3}(b_k) = r_{0,3}(x_{k-1})
\, , \ \ \ \ \ \ k\geq 2\, .
\label{eq2.26new}
\end{align}

\subsection{Elliptic KZB system on the twice-punctured torus}
\label{commderiv}

In the previous subsections, the two partial differential equations \eqref{eqn:KZBforZ} and \eqref{eqn:tauDerivF} satisfied by the vector $\BZ_{0,n}^{\tau}$ defined in \eqn{eqn:Ztzbasis} have been identified. Together, they form the system of differential equations
\begin{align}
\partial_{z_0}\BZ_{0,n}^{\tau}&=\sum_{k=0}^{\infty} f^{(k)}_{01} r_{0,n}(x_{k}) \BZ_{0,n}^{\tau}\,,
\label{eqn:ZSystemDEQ} \nonumber \\
2\pi i \partial_{\tau} \BZ_{0,n}^{\tau}&=\left(-r_{0,n}(\eps_0)+\sum_{k=4}^{\infty}(1{-}k) 
\GG{k}
r_{0,n}(\eps_k)+\sum_{k=2}^{\infty}(k{-}1)f_{01}^{(k)}r_{0,n}(b_k)\right) \BZ_{0,n}^{\tau}\,,
\end{align}
which is an elliptic KZB system on the moduli space of the twice-punctured torus, with (fixed) puncture $z_1=0$ and (variable) puncture $z_0$ \cite{KZB}. While we have considered the two differential equations separately so far, in this section properties of the corresponding matrices $r_{0,n}(x_k)$, $r_{0,n}(b_k)$ and $r_{0,n}(\eps_k)$ are determined employing the interplay of both differential equations. In order to investigate these commutation relations, an unspecified system is considered, which has the same structure
\begin{align}
\partial_{z_0} {\cal I}^{\tau}_{0}&= \sum_{k=0}^{\infty} f^{(k)}_{01}  x_{k} {\cal I}^{\tau}_{0}
\label{eq1.28}\,, \notag \\
2\pi i\partial_{\tau} {\cal I}^{\tau}_{0} &= \Big\{{-}  \eps_{0} + \sum_{k=4}^{\infty} (1{-}k){\rm G}_{k}  \eps_{k}
+  \sum_{k=2}^{\infty}(k{-}1) f^{(k)}_{01} b_{k}
\Big\} {\cal I}^{\tau}_{0}\,,
\end{align}
where unspecified representations of braid matrices $x_{k}$, derivations $ \eps_{k}$ and further generators $b_{k}$ act on an abstract solution ${\cal I}^{\tau}_{0}$. The commutativity of the mixed second derivatives (Schwarz integrability condition)
\begin{align}
\partial_{z_0} \big( 2\pi i \partial_\tau {\cal I}^{\tau}_{0} \big) &= \sum_{k=2}^\infty \partial_{z_0} f^{(k)}_{01}(k{-}1)  b_{k} {\cal I}^{\tau}_{0} 
\label{eq1.29}\notag  \\
&\phantom{=}+ \Big\{  
{-}\ep_0 + \sum_{k=4}^\infty (1{-}k) {\rm G}_k \ep_k + \sum_{k=2}^{\infty} f^{(k)}_{01} (k{-}1) b_{k} 
\Big\} \sum_{\ell = 0}^\infty f^{(\ell)}_{01} x_\ell {\cal I}^{\tau}_{0}\,,\notag  \\
2\pi i \partial_\tau \big(  \partial_{z_0} {\cal I}^{\tau}_{0} \big) &= \sum_{k=0}^\infty  2 \pi i \partial_\tau f^{(k)}_{01}x_{k} {\cal I}^{\tau}_{0} 
\notag \\
&\phantom{=}+ \sum_{\ell = 0}^\infty f^{(\ell)}_{01} x_\ell  \Big\{  
{-}\ep_0 + \sum_{k=4}^\infty (1{-}k) {\rm G}_k \ep_k + \sum_{k=2}^{\infty} f^{(k)}_{01} (k{-}1) b_{k} 
\Big\} {\cal I}^{\tau}_{0} 
\end{align}
imposes various constraints on the $x_{k},\ep_{k},b_{k}$ which will serve as cross-checks for the $n!{\times} n!$ representations $r_{0,n}(\cdot)$ encoded in \eqns{eqn:derZ0SMap}{eqn:tauDerivSMap}.  By the components $2\pi i \partial_{\tau} f^{(k)}_{01} = k \partial_{z_0}  f^{(k+1)}_{01} $ of the mixed heat equation (\ref{eqn:mixedHeat}) for real $z_{01}$, the commutator
\begin{align}
[2\pi i \partial_\tau , \partial_{z_0} ]  {\cal I}^{\tau}_{0} &= \sum_{k=2}^{\infty} (k{-}1) \partial_{z_0} f_{01}^{(k)} (x_{k-1} - b_k) {\cal I}^{\tau}_{0}
\label{eq1.30} \notag\\
&\phantom{=}+ \sum_{\ell = 0}^\infty f^{(\ell)}_{01} \Big\{
{-}[x_\ell , \ep_0] 
+ \sum_{k=4}^{\infty} (1{-}k) {\rm G}_k [x_\ell , \ep_k]
+ \sum_{k=2}^{\infty} (k{-}1) f^{(k)}_{01} [x_\ell , b_k]
\Big\}  {\cal I}^{\tau}_{0} 
\end{align}
has to vanish. In particular, the first line of \eqn{eq1.30} has to vanish separately,
\begin{equation}
b_k = x_{k-1} \ , \ \ \ \ \ \ k\geq 2\, ,
\label{eq1.31} 
\end{equation}
since the derivatives $ \partial_{z_0}  f^{(k+1)}_{01} $ can be rewritten 
as \cite{Gerken:2019cxz}\footnote{The simplest examples of \eqn{eq1.33} are
\[
\partial_{z_0} f^{(1)}_{01} = - f^{(1)}_{01} f^{(1)}_{01} + 2 f^{(2)}_{01} - \widehat {\rm G}_2 \, , \ \ \ \ \ \
\partial_{z_0} f^{(2)}_{01} = - f^{(1)}_{01} f^{(2)}_{01} + 3 f^{(3)}_{01} - \widehat {\rm G}_2 f^{(1)}_{01}
 \, ,
\]
while the simplest examples of \eqn{gen1.36} below are
\[
2 f^{(1)}_{01} f^{(3)}_{01} - f^{(2)}_{01} f^{(2)}_{01} = 2 f^{(4)}_{01} - 3 {\rm G}_4
\, , \ \ \ \ \ \ 
3 f^{(1)}_{01} f^{(4)}_{01} - f^{(2)}_{01} f^{(3)}_{01} = 5 f^{(5)}_{01} - 3 f^{(1)}_{01} {\rm G}_4\, .
\]}
\begin{equation}
\partial_{z_0} f^{(n)}_{01} = - f^{(1)}_{01} f^{(n)}_{01} + (n{+}1) f^{(n+1)}_{01} 
- \widehat {\rm G}_2 f^{(n-1)}_{01} - \sum_{k=4}^{n+1} {\rm G}_k f^{(n+1-k)}_{01} \, , \ \ \ \ \ \ n \geq 1
\label{eq1.33}
\end{equation}
%
%
and generate terms $\sim \widehat {\rm G}_2= \GG{2}- \frac{ \pi }{\Im \tau}$ that do not occur in the second line of \eqn{eq1.30}. By repeating these arguments for the KZB system (\ref{eqn:ZSystemDEQ}) obeyed by the $n!$-component vector $\BZ_{0,n}^{\tau}$ in \eqn{eqn:Ztzbasis}, we arrive at \eqn{bequalx} independent of the explicit form of $r_{0,n}(b_k)$ and $r_{0,n}(x_k)$ obtained in sections \ref{sec:z0Deriv} and \ref{sec:dtau}. 

Hence, the leftover integrability constraints after imposing \eqn{eq1.31} are
\begin{align}
0&=  \sum_{k=2}^{\infty} (k{-}1) f^{(k)}_{01} [x_0,x_{k-1}] 
-  \sum_{\ell = 0}^\infty f^{(\ell)}_{01} [x_\ell , \ep_0] 
+  \sum_{\ell = 0}^\infty f^{(\ell)}_{01}  \sum_{k=4}^{\infty} (1{-}k) \GG{k} [x_\ell , \ep_k]
\notag \\
&\phantom{=}+ \sum_{1\leq a<b}^\infty  (b f_{01}^{(a)} f_{01}^{(b+1)} - a f_{01}^{(a+1)} f_{01}^{(b)}  ) [x_a, x_b]
\, .
\label{eq1.34} 
\end{align}
In order to infer relations among the commutators, the products $f_{01}^{(a)} f_{01}^{(b+1)} $ and $ f_{01}^{(a+1)} f_{01}^{(b)} $ in the second line have to be rewritten in terms of the combinations ${\rm G}_k f^{(\ell)}_{01}$ or $f^{(k+\ell)}_{01}$ in the first line. The required identities valid for $a+b\geq 2$ are generated by the $\xi$- and $\eta$-expansion of \eqn{taulang.3},
\begin{align}
&b f_{01}^{(a)} f_{01}^{(b+1)} - a f_{01}^{(a+1)} f_{01}^{(b)} \nonumber\\
&= \frac{ (b{-}a) (a{+}b{+}1)! }{(a{+}1)!(b{+}1)!} f_{01}^{(a+b+1)}  - (-1)^b (a{+}b) \GG{a+b+1} 
+ a \GG{a+1} \theta_{a\geq 3} f^{(b)}_{01}
- b \GG{b+1} \theta_{b\geq 3} f^{(a)}_{01}  \label{gen1.36}  \nonumber\\
&\phantom{=}
+ \sum_{k=4}^a { a{+}b{-}k \choose b{-}1 } (k{-}1) \GG{k} f_{01}^{(a+b+1-k)}
- \sum_{k=4}^b { a{+}b{-}k \choose a{-}1 } (k{-}1) \GG{k} f_{01}^{(a+b+1-k)}\, .
\end{align}
%
Since the step functions $\theta_{c\geq 3}$ in the second line are taken to be $1$ for $c\geq 3$ and zero for $c\leq 2$, the aforementioned $\widehat {\rm G}_2$ does not occur in \eqn{eq1.34} or \eqn{gen1.36}. After rewriting the second line of \eqn{eq1.34} via \eqn{gen1.36}, the coefficients of ${\rm G}_k f^{(\ell)}_{01}$ and $f^{(k+\ell)}_{01}$ have to vanish separately, and one can read off identities like
\begin{align}
[x_0,\ep_0 ] &= 0 \, , & [x_0,\ep_4 ] &= - [x_1, x_2]\,, \notag  \\
[x_1,\ep_0 ] &= 0 \, , & [x_1,\ep_4 ] &= - [x_1, x_3] \,, \notag \\
[x_2,\ep_0 ] &= [x_0, x_1] \, ,  & [x_2,\ep_4 ] &= - [x_1, x_4]- [x_2, x_3] \,,  \notag \\
[x_3,\ep_0 ] &= 2 [x_0, x_2] \, , &[x_0,\ep_6 ] &= - [x_1, x_4]+ [x_2, x_3] \,,  \label{eq1.37} \\
[x_4,\ep_0 ] &= 3 [x_0, x_3]+ 2  [x_1, x_2] \, , &[x_1, \ep_6] &= - [x_1, x_5]\,,  \notag \\
[x_5,\ep_0 ] &= 4 [x_0, x_4]  + 5  [x_1, x_3] \, , &[x_2, \ep_6] &= - [x_1, x_6] - [x_2, x_5] \,, \notag \\
[x_6,\ep_0 ] &= 5 [x_0, x_5]+ 9  [x_1, x_4]  + 5 [x_2, x_3] \, , \ \ \ &[x_0, \ep_8] &= - [x_1,x_6] + [x_2, x_5] - [x_3, x_4]  \, , \notag
\end{align}
and more generally (with $\ell \geq 1$ and $k\geq 4$)  
\begin{align}
[x_\ell,\ep_0] &=  \sum_{j=0}^{\lfloor \ell/2 \rfloor -1} {\ell \choose j}
\frac{ (\ell{-}1{-}2j) }{(j{+}1)  } [x_j, x_{\ell-1-j}] 
\notag \\
[x_0, \ep_k] &= \sum_{j=1}^{k/2-1} (-1)^j [x_j , x_{k-1-j}]
\label{gen1.37} \\
[x_\ell, \ep_k]&= - \sum_{j=0}^{\ell-1} {\ell{-}1 \choose j} [x_{j+1} , x_{k+\ell - j-2} ]\, .
\notag
\end{align}
By iterating the first relation, 
the adjoint action ${\rm ad}_{\ep_0}(\cdot) = [\ep_0,\cdot]$ 
turns out to be nilpotent when acting on $x_\ell$,
\begin{equation}
{\rm ad}^\ell_{\ep_0}(x_\ell) = 0 \, , \ \ \ \ \ \ \ell \geq 1 \, ,
\label{eq1.38} 
\end{equation}
for instance $[\ep_0,[\ep_0,x_2]]=0$. This follows from the fact that \eqn{gen1.37} relates $[x_\ell,\ep_0]$ to $[x_a,x_b]$ with $a{+}b=\ell{-}1$, so the $k^{\rm th}$ adjoint action ${\rm ad}_{\ep_0}^k(x_\ell)$ yields nested commutators of $x_{a_1},x_{a_2},\ldots,x_{a_{k+1}}$ with $\sum_{j=1}^{k+1} a_j = \ell{-}k$. After $k=\ell{-}1$ steps, only $x_0$ and $x_1$ are left in the nested commutators which are both annihilated by the $\ell^{\rm th}$ application of ${\rm ad}_{\ep_0}$, see \eqn{eq1.37}. 

By repeating the above arguments for the KZB system (\ref{eqn:ZSystemDEQ}) obeyed by the $n!$-component vector $\BZ_{0,n}^{\tau}$ in \eqn{eqn:Ztzbasis}, the commutator relations (\ref{gen1.37}) are found to be preserved under $x_k  \rightarrow r_{0,n}(x_k) $ and $\ep_k \rightarrow r_{0,n}(\eps_k)$, for instance $[r_{0,n}(x_0) , r_{0,n}(\ep_0)]=0$ and ($ \ell \geq 1 , \ k\geq 4$)
\begin{equation}
[r_{0,n}(x_\ell), r_{0,n}(\ep_k)]= - \sum_{j=0}^{\ell-1} {\ell{-}1 \choose j} [r_{0,n}(x_{j+1} ), r_{0,n}( x_{k+\ell - j-2} ) ]  \, .\label{r1.38}
\end{equation}
As a consistency check of the results for $\partial_{z_0}\BZ_{0,n}^{\tau}$ and $\partial_{\tau}\BZ_{0,n}^{\tau}$, we have confirmed validity of the above relations for $[r_{0,n}(x_\ell), r_{0,n}(\ep_k)]$ for numerous configurations of $(n,\ell,k)$.

The nilpotency property (\ref{eq1.38}) is known to also hold for Tsunogai's derivations \cite{Tsunogai},
\begin{equation}
{\rm ad}^{k-1}_{\ep_0}(\ep_k) = 0 \, .
\label{eq1.39}
\end{equation}
The $Z^\tau$-integrals \eqref{taulang.1} without augmentation introduce conjectural $(n{-}1)! \times (n{-}1)!$ matrix representations $r_n(\ep_k)$ through their differential equations (\ref{taulang.5}) which have been tested to preserve \eqn{eq1.39} for a wide range of $k$ and $n$ \cite{Mafra:2019xms}. For the analogous $n!{\times} n!$ matrices $r_{0,n}(\ep_k)$ seen in the KZB system (\ref{eqn:ZSystemDEQ}) of the augmented $\BZ_{0,n}^{\tau}$, we conjecture that they furnish another representation of the derivation algebra. This conjecture not only applies to \eqn{eq1.39},
\begin{equation}
{\rm conjecture:} \ \ \ \ \ \ {\rm ad}^{k-1}_{r_{0,n}(\ep_0)}\big(r_{0,n}(\ep_k) \big) = 0 \, ,
\label{eq1.39again}
\end{equation}
but also to the additional depth-two relations in the derivation algebra besides \eqn{eq1.39}
\cite{LNT, Pollack, Broedel:2015hia} such as
\begin{align}
0 &=[\ep_{10},\ep_4]-3[\ep_{8},\ep_6] \,, \notag \\
0&=2 [\ep_{14},\ep_4] - 7[\ep_{12},\ep_6] + 11 [\ep_{10},\ep_8]  \,.\label{extras}
\end{align}
%
Given that the Schwarz integrability condition (\ref{eq1.30}) does not allow to derive constraints for commutators $[\ep_k,\ep_\ell]$, we have supported our conjecture through numerous successful case-by-case tests using the representations $r_{0,n}(\ep_k)$ only. However, the depth-three relations in the derivation algebra starting from
\begin{align}
0&=80[\ep_{12},[\ep_4,\ep_{0}]] + 16 [\ep_4,[\ep_{12},\ep_0]] - 250 [\ep_{10},[\ep_6,\ep_0]] \notag \\
&\phantom{=} - 125 [\ep_6,[\ep_{10},\ep_0]] + 280 [\ep_8,[\ep_8,\ep_0]]- 462 [\ep_4,[\ep_4,\ep_8]] - 1725 [\ep_6,[\ep_6,\ep_4]] \label{extra3}
\end{align}
do not seem to carry over to the $r_{0,n}(\epsilon_k)$, e.g.\ \eqn{extra3} with $\ep_k  \rightarrow r_{0,n}(\ep_k)$ is already violated at $n=2$. Instead, \eqn{extra3} and some of its higher-weight 
analogues \cite{Pollack, Broedel:2015hia} have been checked to hold upon assigning
\beq
\ep_k  \rightarrow \left\{ 
\begin{array}{cl}
r_{0,n}(\ep_0)&: \ k=0\,,\\
  r_{0,n}(\ep_k)+r_{0,n}(x_{k-1}) &: \ k \geq 4\, .\\
 \end{array} \right.
\label{shiftep}
\eeq
The deformation in the second line resonates with the additional appearance of $b_k=x_{k-1}$ in the augmented differential \eqn{eqn:ZSystemDEQ} with respect to $\tau$ compared to the non-augmented one in \eqn{taulang.4}.
As we will see in \eqn{eqn:UpperBVEigenvalueEqsProjected}, the combination 
$r_{0,n}(\ep_k)+r_{0,n}(x_{k-1}) $ in \eqn{shiftep} appears naturally when relating the representations 
$r_{0,n}(\ep_k)$ and $r_{n}(\ep_k)$ of the derivations. Note that depth-two relations such as
\eqn{extras} were tested to hold for both assignments \eqn{shiftep} and $\ep_k \rightarrow r_{0,n}(\ep_k)$.


\subsection{Total differential of \texorpdfstring{$\Ztzn{n}$}{Ztau(0,n)} integrals}
In summary of the differential equations in sections \ref{sec:z0Deriv}, \ref{sec:dtau} and as a way of manifesting the Schwarz integrability conditions, we will now spell out the total differential of the $\Ztzn{n}$-integrals \eqref{eqn:notationChain}, to bring the differential of $\BZ_{0,n}^{\tau}$ into the form \eqref{eqn:modelDEQ}. After eliminating the $r_{0,n}(b_k)$, the total differential $\dd = \dd z_0 \partial_{z_0} + \dd \tau \partial_{\tau}$ following from the KZB system (\ref{eqn:ZSystemDEQ}) takes the form
%
\begin{align}
\dd\!\BZ_{0,n}^{\tau}&= \Big\{ \dd z_0 \, r_{0,n}(x_{0})  - \frac{\dd \tau}{2\pi i}  \, r_{0,n}(\ep_{0}) 
- \frac{\dd \tau}{2\pi i}  \sum_{k=4}^{\infty} (k{-}1) \GG{k} r_{0,n}(\ep_{k}) + 
 \sum_{k=1}^{\infty} \psi_{01}^{(k)} \, r_{0,n}(x_{k})
\Big\} \BZ_{0,n}^{\tau}\, ,
 \label{eq5.14} 
\end{align}
where the characteristic combinations
\begin{equation}
\psi_{01}^{(k)} = 
\dd z_0 \, f^{(k)}_{01} + \frac{ k \, \dd \tau}{2\pi i} \, f^{(k+1)}_{01} 
 \label{eq5.15} 
\end{equation}
in the last term of \eqn{eq5.14} agree with the closed
one-forms $\omega^{(k)}_{ij}$ in eqn.~(3.15) of \cite{Broedel:2018iwv}
since $z_0,z_1\in \mathbb R$.
The same one-forms $\psi^{(k)}_{01}$ in \eqn{eq5.15} appear when the $z_0$-
and $\tau$-derivatives \eqref{eqn:derZ0SMap} and \eqref{eqn:tauDerivSMap}
are combined to the following explicit form of the total differential:
\begin{align}
\dd &\Ztzn{n}((1,2,3,\ldots,p),(0,n,n{-}1,\ldots,p{+}1) )
= \Big\{
\dd z_0 \Big[  \sum_{j=p+1}^n s_{j,123\ldots p} \partial_{\eta_j} - \sum_{j=2}^p s_{j,0 n(n-1)\ldots p+1} \partial_{\eta_j} \Big]
\notag \\
& \ \ \ \ \ \ \ \ + \frac{ \dd \tau }{2\pi i} \Big[
 \frac{1}{2} \sum_{j=2}^n (s_{0j}{+}s_{1j}) \partial_{\eta_j}^2 
 + \frac{1}{2} \sum_{2\leq i<j}^n s_{ij}( \partial_{\eta_i}{-}\partial_{\eta_j})^2  - 2 \zeta_2 s_{012\ldots n} \Big] \notag \\
 &\ \ \ \ \ \ \ \  - s_{12\ldots p,0 n(n-1)\ldots (p+1)} \psi^{(1)}_{01}  \Big\}
\Ztzn{n}((1,2,3,\ldots,p),(0,n,n{-}1,\ldots,p{+}1))\notag \\
 &\ \ \ \ +  \sum_{j=2}^p \Big\{
\Big( {-} \frac{ \dd z_0}{\eta_{j(j+1)\ldots  p}} - \frac{ \dd \tau}{2\pi i \eta_{j(j+1)\ldots  p}^2}   
+ \sum_{k=1}^{\infty} (-\eta_{j(j+1)\ldots  p})^{k-1} \psi_{01}^{(k)} \Big)
\notag \\
& \ \ \ \ \ \ \ \ \ \ \ \ \ \ \ \ \times
\Ztzn{n}((1,2,\ldots,j{-}1) ,(S[0 n(n{-}1) \ldots (p{+}1) , j(j{+}1)\ldots  p])) \notag \\
&\ \ \ \ \ \ \ \
 -  \frac{ \dd \tau}{2\pi i} \wp(\eta_{j(j+1)\ldots  p})  \Ztzn{n} ((0, n,n{-}1, \ldots ,p{+}1) ,
(S[12\ldots(j{-}1) , j(j{+}1)\ldots  p]) )
 \Big\} \notag \\
 &\ \ \ \ +  \sum_{j=p+1}^n \Big\{
 \Big(  \frac{ \dd z_0}{\eta_{j(j-1)\ldots p+1}} - \frac{ \dd \tau}{2\pi i \eta_{j(j-1)\ldots p+1}^2}   
+ \sum_{k=1}^{\infty} (\eta_{j(j-1)\ldots p+1})^{k-1} \psi_{01}^{(k)} \Big)
 \notag \\
& \ \ \ \ \ \ \ \ \ \ \ \ \ \ \ \ \times
   \Ztzn{n}((0,n,\ldots,j{+}1),  (S[12 \ldots p , j(j{-}1)\ldots  p{+}1])) \notag \\
&\ \ \ \ \ \ \ \
 - \frac{ \dd \tau}{2\pi i}  \wp(\eta_{j(j-1)\ldots p+1}) \Ztzn{n}((1,2,\ldots,p) ,
(S[ 0 n \ldots j{+}1 , j(j{-}1)\ldots  p{+}1]) )\Big\}    \, .
 \label{eq5.16}
\end{align}
In the third and sixth line from below, we have used that
\begin{equation}
 \Omega_{01}^{\pm}(\eta) \frac{ \dd \tau }{2\pi i} + \Omega_{01}(\pm \eta) \dd z_0
= \pm \frac{ \dd z_0 }{\eta} - \frac{ \dd \tau }{2\pi i \eta^2} + \sum_{k=1}^\infty (\pm \eta)^{k-1} \psi_{01}^{(k)}\, .
 \label{eq5.17}
\end{equation}
The fact that the $f^{(k \neq 0)}_{01}$ on the right-hand side of \eqn{eq5.16} combine to $\psi^{(k)}_{01}$ manifests the equality of the operators $ r_{0,n}(b_{k})$ and $r_{0,n}(x_{k-1})$ in the KZB system (\ref{eqn:ZSystemDEQ}). Based on the total differential \eqn{eq5.14}, the formalism of \cite{Broedel:2018iwv} can be used to obtain the coaction of the augmented $Z^\tau$-integrals, also see section 7.2 of \cite{Mafra:2019xms} for the coaction of plain $Z^\tau$-integrals~\eqref{eqn:Zloop}.


\section{Identification and translation}
\label{sec:translation}
While $z_0$- and $\tau$-derivatives of the integrals $\Ztzn{n}$ have been discussed in full generality in the previous section, let us now compare the two resulting approaches by investigating their boundary conditions, limits and solutions, respectively. As a guiding principle, we will explore how the representations $r_n$ and $r_{0,n}^{\E}$ of the algebra generators\footnote{Sometimes these generators are referred to as \textit{letters} and their respective entirety as \textit{alphabets}.} $\ep_k$ and $x_{k}$ in \eqns{taulang.5}{eqn:KZBSelE}, respectively, are related to each other and to the representations $r_{0,n}(\ep_k)$ in \eqns{eqn:KZBforZ}{eqn:tauDerivE} above.


\subsection{Overview}
\label{sec:over}
Boundary values $\BC^{\tau}_{0,n}$ and $\BC^{\tau}_{1,n}$ of $\BZ^{\tau}_{0,n}$ for the limits $z_0\to 0$ and $z_0\to 1$, respectively, are particularly important since they allow for an explicit expansion of the integrals $\BZ^{\tau}_{n}$ in $\alpha'$ using the elliptic KZB associator. Simultaneously, it is those boundary values, which finally allow to find the link between the matrices $r_{n}(\ep_k)$, $r_{0,n}^{\E}(x_k)$, $r_{0,n}(\ep_k)$ and $r_{0,n}(x_k)$. 
Due to the poles of $f^{(1)}_{01}$ at $z_0=0$ and $z_0=1$ in the differential equation \eqref{eqn:KZBforZ}, these limits are singular. The regularization leading to the corresponding non-vanishing finite values will be derived and related to the genus-zero and genus-one $Z$-integrals $\Ztree_n$ and $\Ztn{n}$, respectively, in this section.

The boundary values $\BC^\tau_{0,n}$ and $\BC^\tau_{1,n}$ to be considered here are the finite-length cousins of the infinitely long boundary vectors $\mathbf{C}_{0,n}^{\E}$ and $\mathbf{C}_{1,n}^{\E}$ in \eqn{eqn:z0LanguageBoundaryValues} for the genus-one Selberg integrals. The main results to be derived below are the expressions 
\begin{align}
\BC_{0,n}^{\tau} &= - e^{s_{012\dots n}\omega(1,0|\tau)}  U^{\text{BCJ}}_{n} \BZ^{\tree}_{n+2}\,,
\label{spoiler.1}
\\
P^{\boldshuffle}_n\BC^{\tau}_{1,n} &= e^{s_{01}\omega(1,0|\tau)} \mathbf{Z}_n^\tau
\label{spoiler.2}
\end{align}
in terms of bases of genus-zero integrals $\BZ^{\tree}_{n+2}$ in \eqn{eqn:ZTreeExplicit} and genus-one integrals $\mathbf{Z}_n^\tau$ in \eqn{eqn:Zloop}. The entries of the $n! \times (n{-}1)!$
matrices $U^{\text{BCJ}}_{n}$ and $(n{-}1)! \times n!$ matrices $P^{\boldshuffle}_n$ are rational functions of
the $s_{ij}$ with $0{\leq} i{<}j{\leq} n$ and will be defined in the discussions around \eqns{bcjdec}{eqn:defProjectionPn}, respectively.

Based on the associator relation
\begin{align}\label{eqn:assocEqAugmentedZInt}
	\BC_{1,n}^{\tau}&=r_{0,n}\left(\Phi^{\tau}(x_k)\right)\BC_{0,n}^{\tau}
\end{align}
adapted to the $n!{\times} n!$ matrices $r_{0,n}(x_k)$ constructed in section \ref{sec:z0Deriv}, eqs.\ (\ref{spoiler.1}) and (\ref{spoiler.2}) connect the genus-one integrals $\mathbf{Z}_n^\tau$ with their genus-zero counterparts $\BZ^{\tree}_{n+2}$,
\beq
 \mathbf{Z}_n^\tau =   - e^{(s_{012\dots n}-s_{01})\omega(1,0|\tau)} P^{\boldshuffle}_n \, r_{0,n}\left(\Phi^{\tau}(x_k)\right) U^{\text{BCJ}}_{n} \BZ^{\tree}_{n+2}\, .
 \label{spoiler.3}
\eeq
Using the expansion of the elliptic KZB associator \cite{Enriquez:EllAss},
\begin{align}\label{eqn:ellKZBAssocPre}
r_{0,n}\left(\Phi^{\tau}(x_k)\right)&=\sum_{w\geq 0}\sum_{k_1,\dots,k_w\geq 0}r_{0,n}(x_{k_1}\dots x_{k_w})\omega(k_w,\dots,k_1|\tau)\,,
\end{align}
each term in the expansion of the genus-one integrals in $\alpha'$ and $\eta_j$ can be obtained via elementary operations. Eq.\ (\ref{spoiler.3}) is the generating-function reformulation of the method in \cite{Broedel:2019gba}, where the matrices $r_{0,n}(x_k)$ are now finite dimensional. 

The application of the formalism to integrals $Z^\tau_n$ with kinematic poles (i.e.\ factors of $f^{(1)}_{ij}$ in the integrand) have not been investigated in \rcite{Broedel:2019gba}. However, when doing so using the method of \rcite{Broedel:2019gba}, the matrices $P^{\boldshuffle}_n$ allow to project on the desired configuration-space integrals in the same way as they do in the language using generating functions in the current paper. 


\subsection{Lower boundary value \texorpdfstring{$\BC^{\tau}_{0,n}$}{Ctau(0,n)} in the \texorpdfstring{$z_0$}{z0}-language}

In this subsection, we will derive the expression \eqref{spoiler.1} for the
lower boundary value $\BC^{\tau}_{0,n}$ which is defined as the regularized limit 
\begin{align}
	\label{eqn:C0Limes}
	\BC_{0,n}^{\tau}&=\lim_{z_0\to 0}(-2\pi i z_0)^{-r_{0,n}(x_{1})}\BZ^{\tau}_{0,n} \, .
\end{align}
The vector $\BZ^{\tau}_{0,n}$ has been introduced in \eqn{eqn:Ztzbasis}, and $r_{0,n}(x_1)$ is the representation of the letter $x_1$ appearing in the KZB equation \eqref{eqn:KZBforZ} along with the singular $f^{(1)}_{01}$.  Here we are going to derive the main mechanism necessary for evaluating \eqn{eqn:C0Limes}. 

\subsubsection{Recovering Parke--Taylor integrals}

Let us start by examining the limiting behaviour of the Koba--Nielsen factor: following the reasoning of \rcite{Broedel:2019gba} and using \eqn{eqn:asymptoticG0}, we find by the change of variables $z_i=z_0x_i$ and recalling that $0<z_i<z_0$ for $1\leq i\leq n$, that the Koba--Nielsen factor degenerates as follows in the limit $z_0\to 0$:
\begin{align}
	\label{eqn:KNlowerlimit}
\KN^{\tau}_{012\dots n}&= (-2\pi i z_0)^{-s_{012\dots n}}e^{s_{012\dots n}\omega(1,0|\tau)}\prod_{1\leq i<j\leq n} x_{ji}^{-s_{ij}} \prod_{k=2}^n (1{-}x_k)^{-s_{0k}} +\mathcal{O}(z_0^{-s_{012\ldots n}+1})\nonumber\\
&=(-2\pi i z_0)^{-s_{012\dots n}}e^{s_{012\dots n}\omega(1,0|\tau)}\KN^{\tree}_{12\dots n+2}+\mathcal{O}(z_0^{-s_{012\ldots n}+1})\,.
\end{align}
We have identified
\begin{align}
x_{n+1}&=x_{0}=1 \, , \ \ \ \ \ \ s_{j,n+1}= s_{0j}
\label{idmands}
\end{align}
and $\KN^{\tree}_{12\dots n{+}2}$ is the $(n{+}2)$-point tree-level Koba--Nielsen factor \eqref{eqn:KN} with the fixed variables
\begin{equation}
	\label{eqn:treefixing}
(x_1,x_{n+1},x_{n+2})=(0,1,\infty)\,.
\end{equation}
The asymptotics of the functions $f^{(k)}(z|\tau)$ with $z$ approaching zero from the positive real line
\begin{align}\label{eqn:asymptBehaviourf1}
f^{(k)}(z|\tau)&=\begin{cases}
\frac{1}{z}+\mathcal{O}(z)&\text{if } k=1\,,\\
\mathcal{O}(1)&\text{otherwise}\, ,\\
\end{cases}
\end{align}
determine the asymptotic behaviour of the Kronecker--Eisenstein series for $z\to 0$
\begin{align}
  \label{eqn:asymptBehaviourOmega}
  \Omega(z,\eta|\tau)&=\frac{1}{z}+\mathcal{O}(1)\,.
\end{align}
Using this equation, the integrand in the augmented genus-one $Z$-integrals in \eqn{eqn:notationChain} without the Koba--Nielsen factor degenerates to the tree-level integrand
\begin{align}\label{eqn:varphiAsymptBehaLowerBound}
\lim_{z_0\to 0}\prod_{k=2}^n \dd z_k\,\phiChain(1,A)\phiChain(0,B)&=\prod_{k=2}^n \dd x_k \bigg[ \frac{1}{(-x_{a_1})} \prod_{t=2}^p\frac{1}{x_{a_{t-1},a_t}} \bigg] \, \bigg[ \frac{1}{1{-}x_{b_1}}\prod_{r=2}^q\frac{1}{x_{b_{r-1},b_r}} \bigg]\,,
\end{align}
where the limit is performed at fixed $x_i$ in $z_i = z_0 x_i$, and
$A=(a_1,a_2,\dots,a_p), \ B=(b_1,b_2,\dots,b_q)$ are again disjoint sequences without repetitions such that $A\cup B=\{2,3,\dots, n\}$. Note that only the contribution from the simple poles of the Kronecker--Eisenstein series in \eqn{eqn:asymptBehaviourOmega} survives the change of variables $\dd z_i=z_0\,\dd x_i$ followed by the limit $z_0 \rightarrow 0$ in the product \eqn{eqn:varphiAsymptBehaLowerBound}. 

Combining \eqns{eqn:KNlowerlimit}{eqn:varphiAsymptBehaLowerBound}, we find that the asymptotic behaviour of the augmented $Z^\tau$-integrals \eqn{eqn:notationChain} yields genus-zero integrals \eqref{eqn:Ztree} of Parke--Taylor type,
\begin{align}
	\label{eqn:Zdegeneration1}
	\lim_{z_0\to 0}(-2\pi i z_0)^{s_{012\dots n}}\Ztzn{n}((1,A),(0,B))&=-(-1)^{|B|}e^{s_{012\dots n}\omega(1,0|\tau)}Z^{\tree}_{n{+}2}( \mathbb I_{n+2} |1,A,n{+}2,\tilde B,n{+}1)\,,
\end{align}
with the identification \eqn{idmands} of Mandelstam invariants and $|B|=q$ denoting the number of labels in $B$.  The integrals $Z^{\tree}_{n{+}2}$ have been identified by matching \eqn{eqn:varphiAsymptBehaLowerBound} with their expression \eqref{eqn:ZTreeExplicit} in the ${\rm SL}_2$-frame \eqn{eqn:treefixing}. In particular, \eqn{eqn:Zdegeneration1} at $B= \emptyset$ specializes to
\begin{align}
\label{eqn:510}
\lim_{z_0\to 0}(-2\pi i z_0)^{s_{012\dots n}}\Ztzn{n}(1,A)&= - e^{s_{012\dots n}\omega(1,0|\tau)}Z^{\tree}_{n{+}2}( \mathbb I_{n+2} |1,A,n{+}2,n{+}1)\,,
\end{align}
for $A=(a_2,a_3,\dots,a_n)$ a permutation of $(2,3,\dots, n)$. Therefore, the first $(n{-}1)!$ entries of the vector $\BZtzn{n}$ degenerate to the BCJ basis $ \BZ_{n+2}^{\tree} = Z^{\tree}_{n{+}2}( \mathbb I_{n+2} |1,\sigma,n{+}2,n{+}1), \ \sigma \in {\cal P}(2,3,\ldots,n)$ of $(n{+}2)$-point genus-zero integrals at fixed integration domain $\mathbb I_{n+2}$, cf.\ \eqn{eqn:Zbasisexample}. The remaining Parke--Taylor orderings in \eqn{eqn:Zdegeneration1} descending from $\Ztzn{n}((1,A),(0,B))$ at $B\neq \emptyset$ can be reduced to the $(n{-}1)!$ basis integrals $Z^{\tree}_{n{+}2}( \mathbb I_{n+2} |1,A,n{+}2,n{+}1)$ in \eqn{eqn:510} via BCJ relations~\cite{Bern:2008qj}. Their unique decomposition into an $(n{-}1)!$ BCJ basis defines the entries of the following $n!{\times} (n{-}1)!$ matrix $U^{\text{BCJ}}_{n}$ indexed by permutations $\rho \in S_{n-1}$:
\begin{equation}
\vecb Z^{\tree}_{n{+}2}( \mathbb I_{n+2} |1,\rho(2,3,\ldots,n),n{+}2,n{+}1) \\
- Z^{\tree}_{n{+}2}( \mathbb I_{n+2} |1,\rho(2,3,\ldots,n{-}1),n{+}2,\rho(n),n{+}1) \\
Z^{\tree}_{n{+}2}( \mathbb I_{n+2} |1,\rho(2,3,\ldots,n{-}2),n{+}2,\rho(n{-}1,n),n{+}1) \\
\hdots \\
(-1)^nZ^{\tree}_{n{+}2}( \mathbb I_{n+2} |1,\rho(2),n{+}2,\rho(3,\ldots,n),n{+}1) \\
-(-1)^n Z^{\tree}_{n{+}2}( \mathbb I_{n+2} |1,n{+}2,\rho(2,3,\ldots,n), n{+}1) \vece
= U^{\text{BCJ}}_{n} \BZ_{n+2}^{\tree} \, .
\label{bcjdec}
\end{equation}
Given the order of the $\Ztzn{n}((1,A),(0,B))$ integrals in the $\BZtzn{n}$-vector \eqn{eqn:Ztzbasis}, 
the degeneration \eqref{eqn:Zdegeneration1} and the BCJ basis decomposition \eqref{bcjdec} yield
\begin{align}\label{eqn:C0}
\lim_{z_0\to 0} (-2\pi i z_0)^{s_{012\dots n}}\BZtzn{n}&= - e^{s_{012\dots n}\omega(1,0|\tau)}\begin{pmatrix}
\BZ_{n+2}^{\tree}\\
\ast
\end{pmatrix} = - e^{s_{012\dots n}\omega(1,0|\tau)}  U^{\text{BCJ}}_{n} \BZ_{n+2}^{\tree}  \,. 
\end{align}
In the second step, we have used that the upper block of size $(n{-}1)!{\times} (n{-}1)!$ in $U^{\text{BCJ}}_{n}$ is the identity matrix. 

\subsubsection{The maximal \texorpdfstring{$r_{0,n}(x_1)$}{r(0,n)(x1)} eigenvalue}

In order to derive our earlier claim \eqn{spoiler.1} for the lower boundary value, we will show how the finite value $\lim_{z_0\to 0} (-2\pi i z_0)^{s_{01\dots n}}\BZtzn{n}$ in \eqn{eqn:C0} emerges from the regularized boundary value $\BC_{0,n}^{\tau}$ in \eqn{eqn:C0Limes}. The key observation to be demonstrated below is that the matrix $r_{0,n}(x_1)$ in the exponent of \eqn{eqn:C0Limes} can be set to its eigenvalue $- s_{012\ldots n}$ when acting on the $z_0 \rightarrow 0$ asymptotics of $\BZ^{\tau}_{0,n}$. With the asymptotic result \eqref{eqn:C0}, the task is therefore to show that
\begin{align}\label{eqn:C0TauProperRegularization}
(-2\pi i z_0)^{-r_{0,n}(x_1)} U^{\text{BCJ}}_{n}
&= (-2\pi i z_0)^{s_{012\ldots n}} U^{\text{BCJ}}_{n} \, ,
\end{align}
which will turn out to be independent on the BCJ basis vector $\BZ_{n+2}^{\tree}$ that
$U^{\text{BCJ}}_{n}$ acts on.

Our proof of \eqn{eqn:C0TauProperRegularization} is based on the continuity of $(2\pi i z_0)^{s_{012\ldots n}} 2\pi i\partial_{\tau} \BZ^{\tau}_{0,n}$ at $z_0=0$ due to the absence of singular terms in \eqn{eqn:tauDerivF}. We can therefore equate the two orders of performing the limit $z_0 \rightarrow 0$ and the $\tau$-derivative.
\begin{itemize}
\item On the one hand, \eqn{eqn:C0} can be differentiated with respect to $\tau$ {\it after} taking
the $z_0\rightarrow0$ limit, which only acts via $2\pi i\partial_{\tau}\omega(1,0|\tau)=\GG{2}-2\zeta_2$ and yields
\begin{align}\label{eqn:pdTauz00}
  2\pi i\partial_{\tau} \lim_{z_0\to 0}(-2\pi i z_0)^{s_{012\dots n}} \BZ^{\tau}_{0,n}
  =s_{012\dots n} \left(\GG{2}-2\zeta_2\right)e^{s_{012\dots n}\omega(1,0|\tau)} U^{\text{BCJ}}_{n} \BZ_{n+2}^{\tree}\,.
\end{align}
\item On the other hand, exchanging the limit and the partial derivative in \eqn{eqn:pdTauz00} leads -- according to \eqn{eqn:tauDerivF} -- to the identity 
\begin{align}\label{eqn:z00pdTau}
&\lim_{z_0\to 0}(-2\pi i z_0)^{s_{012\ldots n}} 2\pi i\partial_{\tau} \BZ^{\tau}_{0,n}\nonumber\\
&=\left(-r_{0,n}(\ep_0)-\GG{2}r_{0,n}(x_1) +\sum_{k=4}^{\infty}(1{-}k)\GG{k}\left(r_{0,n}(\ep_k)+r_{0,n}(x_{k-1})\right)\right)\nonumber\\
&\phantom{=}\times e^{s_{012\dots n}\omega(1,0|\tau)} U^{\text{BCJ}}_{n} \BZ_{n+2}^{\tree}\,,
\end{align}
where we have used that $r_{0,n}(b_k) =r_{0,n}(x_{k-1})$ (cf.~\eqn{eq1.31}) and that for $k\geq 2$
(and $z_0 \in \mathbb R$ in case of $k=2$)
\begin{align}
\lim_{z_0\to 0}f^{(k)}_{01}&= {-} \GG{k}\, .
\end{align}
\end{itemize}
Since the derivative $(2\pi i z_0)^{s_{012\ldots n}} 2\pi i\partial_{\tau} \BZ^{\tau}_{0,n}$ is continuous at $z_0=0$, \eqns{eqn:pdTauz00}{eqn:z00pdTau} have to agree. Comparing the coefficients of ${\rm G}_k$ in the two equations leads to the eigenvalue equations:
\begin{align}
r_{0,n}(x_1)U^{\text{BCJ}}_{n}&=-s_{01\dots n}U^{\text{BCJ}}_{n}\,,\nonumber\\
r_{0,n}(\ep_0)U^{\text{BCJ}}_{n}&=2\zeta_2s_{01\dots n}  U^{\text{BCJ}}_{n}\,, \label{eqn:lowerBVMatrixEqs} \\
\left(r_{0,n}(\ep_k)+r_{0,n}(x_{k-1})\right)U^{\text{BCJ}}_{n}&=0\,,\quad k\geq 4\,. \nonumber
\end{align}
These equations imply that the columns of $U^{\text{BCJ}}_{n}$ are eigenvectors of $r_{0,n}(x_1)$ and $r_{0,n}(\ep_0)$ for the eigenvalues $-s_{012\dots n}$ and $2\zeta_2s_{012\dots n}$, respectively. This proves the lemma
\eqref{eqn:C0TauProperRegularization} and ultimately the main claim \eqref{spoiler.1} of this subsection. Moreover, we see that the representations $r_{0,n}(\ep_0)$ and $r_{0,n}(x_1)$ as well as $r_{0,n}(\ep_k)$ and $r_{0,n}(x_{k-1})$ for $k=4,6,\dots$ acting on $U^{\text{BCJ}}_{n}$ and thus, on the lower boundary value, are equivalent up to constant factors. 

\subsubsection{Two-point example}

Let us approve the above findings on the two-point example $\BZtzn{2}$ from \eqn{eqn:2PointExampleIntegral}. The finite part at $z_0=0$ according to \eqn{eqn:Zdegeneration1} and $s_{23}=s_{02}$ is given by
\begin{align}\label{eqn:2ptExampleC0}
\lim_{z_0\to 0}(-2\pi i z_0)^{s_{012}} \BZtzn{2}&= e^{s_{012}\omega(1,0|\tau)} \vecb - \Ztree_4(\mathbb I_4 |1,2,4,3) \\
 \Ztree_4(\mathbb I_4 |1,4,2,3) \vece \\
&= e^{s_{012}\omega(1,0|\tau)} \vecb \frac{1}{s_{12}} \\
 -\frac{1}{s_{02}} \vece \frac{ \Gamma(1{-}s_{12}) \Gamma(1{-}s_{02}) }{\Gamma(1{-}s_{12}{-}s_{02}) }
 \notag
\end{align}
which can be rewritten using the following BCJ matrix in \eqn{eqn:C0}
\begin{align}\label{eqn:twoPointExampleC0}
\lim_{z_0\to 0}(-2\pi i z_0)^{s_{012}} \BZtzn{2}&=-e^{s_{012}\omega(1,0|\tau)}U^{\text{BCJ}}_{2}\Ztree_4(\mathbb I_4 |1,2,4,3)\,,\qquad U^{\text{BCJ}}_{2}=\begin{pmatrix}
1\\-\frac{s_{12}}{s_{02}}
\end{pmatrix}\,.
\end{align}
The eigenvalue equations \eqref{eqn:lowerBVMatrixEqs} can immediately be checked using the explicit form of the $r_{0,2}(\cdot)$ in \eqns{eqn:2PointEMatrices}{eqn:2PointTauDeriv}. In particular,
\begin{align}\label{eqn:2PointEVal}
r_{0,2}(x_1)U^{\text{BCJ}}_{2}&=
-s_{012}U^{\text{BCJ}}_{2}
\end{align}
together with \eqn{eqn:twoPointExampleC0} lead to
\begin{align}
\BC_{0,2}^{\tau}&=\lim_{z_0\to 0}(-2\pi i z_0)^{-r_{0,2}(x_1)}\BZtzn{2}
\nonumber\\
&= - e^{s_{012}\omega(1,0|\tau)}U^{\text{BCJ}}_{2}\Ztree_4(\mathbb I_4 |1,2,4,3) \notag  \\
&= e^{s_{012}\omega(1,0|\tau)} \vecb \frac{1}{s_{12}} \\
 -\frac{1}{s_{02}} \vece \frac{ \Gamma(1{-}s_{12}) \Gamma(1{-}s_{02}) }{\Gamma(1{-}s_{12}{-}s_{02}) }\,.
\end{align}
Note that the other two eigenvalue equations of \eqref{eqn:lowerBVMatrixEqs} 
are also straightforwardly checked via \eqns{eqn:2PointEMatrices}{eqn:2PointTauDeriv}.

\subsubsection{Three-point example}

At three points, the explicit form of $\BZtzn{3}$ can be found in \eqn{eqn:3PointExampleIntegral},
and its finite part at $z_0=0$ is determined as follows by \eqn{eqn:Zdegeneration1}:
\begin{align}\label{eqn:3ptExampleC0}
\lim_{z_0\to 0}(-2\pi i z_0)^{s_{0123}} \BZtzn{3}&= e^{s_{0123}\omega(1,0|\tau)} \vecb 
- \Ztree_5(\mathbb I_5 |1,2,3,5,4) \\
- \Ztree_5(\mathbb I_5 |1,3,2,5,4) \\
 \Ztree_5(\mathbb I_5 |1,2,5,3,4) \\
\Ztree_5(\mathbb I_5 |1,3,5,2,4) \\
- \Ztree_5(\mathbb I_5 |1,5,2,3,4) \\
- \Ztree_5(\mathbb I_5 |1,5,3,2,4)  \vece  \, .
\end{align}
BCJ relations among five-point disk integrals give rise to the following matrix $U^{\text{BCJ}}_{3}$ in
\eqn{bcjdec},
\begin{align}\label{eqn:threePointExampleC0}
U^{\text{BCJ}}_{3}=\begin{pmatrix}
1 &0 
\\
0 &1
\\
-\frac{(s_{13} {+} s_{23})}{s_{03}}  & -\frac{s_{13}}{s_{03}}
\\
-\frac{s_{12}}{s_{02}} &-\frac{(s_{12} {+} s_{23})}{ s_{02}}
\\
\frac{s_{12} (s_{03} {+} s_{13} {+} s_{23})}{s_{03} s_{023} } &  \frac{( s_{12} {-}s_{03} ) s_{13}}{s_{03} s_{023}}
\\
 \frac{s_{12} (s_{13} {-} s_{02} )}{ s_{02} s_{023}} & \frac{s_{13} (s_{02} {+} s_{12} {+} s_{23})}{s_{02} s_{023}}
\end{pmatrix}\,,
\end{align}
which can be checked to obey the eigenvalue equations \eqref{eqn:lowerBVMatrixEqs} 
using the expressions for $r_{0,3}(\cdot)$ from sections \ref{sec:z03pt} and \ref{sec:tau3}.
By \eqns{bcjdec}{eqn:3ptExampleC0}, we arrive at
\begin{align}
\BC_{0,3}^{\tau}&=\lim_{z_0\to 0}(-2\pi i z_0)^{-r_{0,3}(x_1)}\BZtzn{3}
\nonumber\\
&= - e^{s_{0123}\omega(1,0|\tau)}U^{\text{BCJ}}_{3}
\vecb 
\Ztree_5(\mathbb I_5 |1,2,3,5,4) \\
\Ztree_5(\mathbb I_5 |1,3,2,5,4)\vece
\,.
\end{align}
%


\subsection{Upper boundary value \texorpdfstring{$\BC^{\tau}_{1,n}$}{Ctau(1,n)} in the \texorpdfstring{$z_0$}{z0}-language}
\label{sec:regC1}

In order to derive the claim \eqref{spoiler.2} for the upper boundary value $\BC^{\tau}_{1,n}$, one will have to evaluate
\begin{align}
	\label{eqn:C1Limes}
\BC_{1,n}^{\tau}&=\lim_{z_0\to 1}(-2\pi i (1{-}z_0))^{-r_{0,n}(x_1)}\BZtzn{n}. 
\end{align}
Similar to the procedure in the last subsection, we will evaluate the limit $z_0\to 1$ separately for the Koba-Nielsen part and the remainder of the integrand, before commenting on the action of the matrix representation $r_{0,n}$ in the exponent of the regulating factor. 

\subsubsection{Recovering genus-one integrands}

Following the same steps as in \rcite{Broedel:2019gba}, the Koba--Nielsen factor degenerates for $z_0\to 1$ along the unit interval as follows: 
\begin{align}
\KN_{012\dots n}
&=(-2\pi i (1{-}z_0))^{-s_{01}}e^{s_{01}\omega(1,0|\tau)
} \KN^{\tau}_{12\dots n}|_{\tilde s_{ij}}+\mathcal{O}((1{-}z_0)^{-s_{01}+1})\,,
\label{eqn:KNdegeneration1}
\end{align}
where
\begin{align}
\KN^{\tau}_{12\dots n}|_{\tilde s_{ij}}&= \prod_{1\leq i<j\leq n}e^{-\tilde s_{ij}\mathcal{G}^{\tau}_{ij}}
\end{align}
is the $n$-point Koba--Nielsen factor with shifted Mandelstam invariants 
$\tilde s_{ij}$ with $i<j$ given by
\begin{equation}
\tilde s_{ij}= \begin{cases}
s_{1j}+s_{0j}& \text{if } i=1\,,\\
s_{ij}& \text{otherwise} \, .
\end{cases}
\label{shiftmands}
\end{equation}
The Kronecker--Eisenstein chains in the integrand of $\Ztzn{n}((1,A),(0,B))$ are regular for $z_0\to 1$ since $0\not \in A$ and $1\not \in B$, and, using the periodicity \eqref{eqn:fs} of the Kronecker--Eisenstein series and the shuffle identity \eqref{eqn:shuffleProductChains}, given by
\begin{align}\label{eqn:shuffleLimit}
\lim_{z_0\to 1}\phiChain(1,A)\phiChain(0,B)&=\phiChain(1,A\shuffle B)\,.
\end{align}
In addition to the higher-order terms in \eqn{eqn:KNdegeneration1}, there is another subtlety in taking the limit $\lim_{z_0\to 1}\Ztzn{n}((1,A),(0,B))$: as discussed in \appref{restTerms}, further subleading terms will appear at certain orders in $(1{-}z_0)^{-s_{01}+\Delta}$ for some $\Delta$ composed of further $s_{ij}$ with $\Re(\Delta)>0$. They originate from the difference between the integration domain $\int_0^{z_0}$ entering the definition of $\Ztzn{n}((1,A),(0,B))$ and the larger integration domain $\int^1_0$ of the integrals $Z^{\tau}_{n}$. On the one hand, the interval $(z_0,1)$ distinguishing the integration domains of $\Ztzn{n}((1,A),(0,B))$ and $Z^{\tau}_{n}$ becomes arbitrarily small as $z_0 \rightarrow 1$. On the other hand, the poles of $\phiChain(1,A)\phiChain(0,B) $ lead to $n{-}1$ subleading contributions to the integral over $(z_0,1)$ due to the merging of $k{+}2$ punctures $z_0< z_{n-k+1}<\dots < z_{n}<1\cong z_1$ for $0<k<n$ as $z_0\to 1$. Still, these subleading contributions scale as $(-2\pi i (1{-}z_0))^{-s_{01\,(n-k+1)\dots n}}$, where according to \eqn{assumpGenusOne} the exponent has a real part larger than $-\Re(s_{01})$. For the boundary value $\BC_{1,n}^{\tau}$ in (\ref{eqn:C1Limes}), these finite terms only pose a problem in an eigenspace of $-r_{0,n}(x_1)$ where the associated eigenvalue has a real part larger than $-\Re(s_{01})$. By multiplication with $(-2\pi i(1{-}z_0))^{s_{01}}$ and taking the limit $z_0\to 1$, the lowest order contribution can be isolated. The corresponding non-vanishing value is determined by combining \eqns{eqn:KNdegeneration1}{eqn:shuffleLimit}:
\begin{align}\label{eqn:finiteEntriesC1}
\lim_{z_0\to 1}(-2\pi i(1{-}z_0))^{s_{01}}\Ztzn{n}((1,A),(0,B))
&=e^{s_{01}\omega(1,0|\tau)}
Z^{\tau}_{n}(\mathbb I_n | 1,(A \shuffle B))|_{\tilde{s}_{ij}}\,.
\end{align}
The right-hand side involves the genus-one integrals \eqref{eqn:Zloop} with the shifted Mandelstam variables $\tilde s_{ij}$ from \eqn{shiftmands} in the Koba--Nielsen factor. The special cases of \eqn{eqn:finiteEntriesC1} with $B= \emptyset$ yield
a particularly simple form for the first $(n{-}1)!$ components of,
\begin{align}\label{eqn:C1}
\lim_{z_0\to 1} (-2\pi i (1{-}z_0))^{s_{01}}\BZtzn{n}&=e^{s_{01}\omega(1,0|\tau)}\begin{pmatrix}
	\BZ^{\tau}_{n}|_{\tilde{s}_{ij}}\\
\ast
\end{pmatrix}\,,
\end{align}
where $\BZ_n^{\tau}$ defined in \eqn{taulang.1} comprises $(n{-}1)!$ basis integrals $Z^{\tau}_{n}(\mathbb I_n | 1,A)$.  Also the remaining components of \eqn{eqn:finiteEntriesC1} fall into this basis: evaluating the shuffles on the right-hand side of \eqn{eqn:finiteEntriesC1} defines a $n! \times (n{-}1)!$ matrix $U^{\boldshuffle}_n$ with entries in $ \{0,1\}$
\beq
\vecb Z^{\tau}_{n}\big(\mathbb I_n | 1,\rho(2,3,\ldots,n)) \\
Z^{\tau}_{n}\big(\mathbb I_n | 1,[\rho(2,3,\ldots,n{-}1) \shuffle \rho(n) ] \big) \\
Z^{\tau}_{n}\big(\mathbb I_n | 1,[\rho(2,3,\ldots,n{-}2) \shuffle \rho(n{-}1,n) ] \big) \\
\hdots \\
Z^{\tau}_{n}\big(\mathbb I_n | 1,[\rho(2) \shuffle \rho(n,\ldots,3) ] \big) \\
Z^{\tau}_{n}\big(\mathbb I_n | 1,\rho(n,\ldots ,3,2) \big) \vece
= U^{\boldshuffle}_n \BZ_n^{\tau} 
\label{shuffdec}
\eeq
such that
\begin{align}\label{eqn:genusOneBCJ}
\lim_{z_0\to 1}(-2\pi i(1{-}z_0))^{s_{01}}\BZtzn{n}&= e^{s_{01}\omega(1,0|\tau)}
U^{\boldshuffle}_n \BZ^{\tau}_{n}|_{\tilde{s}_{ij}}\,.
\end{align}
Note that the shuffle decomposition \eqref{shuffdec} of genus-one integrals in the context of the boundary value $\BC_{1,n}^{\tau}$ is the analogue of the BCJ decomposition \eqref{bcjdec} relevant for the genus-zero integrals in $\BC_{0,n}^{\tau}$.  In particular, both $n!{\times} (n{-}1)!$ matrices $ U^{\text{BCJ}}_{n}$ and $U^{\boldshuffle}_n$ feature a unit matrix within their upper $(n{-}1)! \times (n{-}1)!$ block.

\subsubsection{The minimal \texorpdfstring{$r_{0,n}(x_1)$}{r(o,n)(x1)} eigenvalue}

What remains to be discussed here is the regulating factor: similar to the previous subsection, it turns out that the finite value in \eqn{eqn:genusOneBCJ} emerges from the regularized boundary value $\BC^{\tau}_{1,n}$. However, recovering the genus-one $Z^\tau$-integrals is more subtle than the tree-level integrals from $\BC^{\tau}_{0,n}$, since the validity of \eqn{eqn:finiteEntriesC1}, i.e.\ the absence of rest terms, relies on the exponent $s_{01}$ of the regulating factor $(-2\pi i (1{-}z_0))^{s_{01}}$ and would generally fail if there were further contributions $s_{ij}$. Therefore, one has to make sure that the dominating eigenvalue of $r_{0,n}(x_1)$ in $\BC_{1,n}^{\tau}$, c.f.\ \eqn{eqn:C1Limes}, is $-s_{01}$, i.e.\ the eigenvalue with the minimal real part as opposed to the eigenvalue $-s_{012\dots n}$ with maximal real part in the calculation \eqref{eqn:C0TauProperRegularization}. This can be achieved by employing a projection to the eigenspace of $-s_{01}$ as follows: interchanging limit and $\tau$-derivative of $(-2\pi i (1{-}z_0))^{s_{01}}\BZtzn{n}$ using the continuity of the latter leads again on the one hand according to \eqns{taulang.4}{eqn:C1} to
\begin{align}\label{eqn:PartialTauLimReg}
&2\pi i \partial_{\tau}\lim_{z_0\to 1}(-2\pi i (1{-}z_0))^{s_{01}}\BZtzn{n}\nonumber\\
&= 2\pi i \partial_{\tau}e^{s_{01}\omega(1,0|\tau)}U^{\boldshuffle}_n 
\BZ^{\tau}_{n}|_{\tilde{s}_{ij}}
\\
&= e^{s_{01}\omega(1,0|\tau)} U^{\boldshuffle}_n\left(-\left(r_n(\ep_0)|_{\tilde{s}_{ij}}+2\zeta_2s_{01}\right)+s_{01} \GG{2}+\sum_{k=4}^\infty (1{-}k) \GG{k} \, r_n(\ep_k)|_{\tilde{s}_{ij}}\right) \BZ^{\tau}_{n}|_{\tilde{s}_{ij}} \, .\nonumber
\end{align}
On the other hand, we find similar to the calculation in \eqn{eqn:z00pdTau}
\begin{align}
&\lim_{z_0\to 1}(-2\pi i (1{-}z_0))^{s_{01}}2\pi i \partial_{\tau}\BZtzn{n} \label{eqn:tauDerivativeLimit}\\
&= e^{s_{01}\omega(1,0|\tau)}\left(-r_{0,n}(\ep_0)-\GG{2}r_{0,n}(x_1) +\sum_{k=4}^{\infty}(1{-}k)\GG{k}\left(r_{0,n}(\ep_k)+r_{0,n}(x_{k-1})\right)\right)  U^{\boldshuffle}_n \BZ^{\tau}_{n}|_{\tilde{s}_{ij}}\,,
\nonumber
\end{align}
such that comparing the coefficients of ${\rm G}_k$ leads to the matrix equations 
\begin{align}
r_{0,n}(x_1)U^{\boldshuffle}_n&=-s_{01}U^{\boldshuffle}_n\,,\nonumber\\
r_{0,n}(\ep_0)U^{\boldshuffle}_n&=U^{\boldshuffle}_n\left(r_n(\ep_0)|_{\tilde{s}_{ij}}+2\zeta_2s_{01}\right)\,,\label{eqn:UpperBVEigenvalueEqs}\\
\left(r_{0,n}(\ep_k)+r_{0,n}(x_{k-1})\right)U^{\boldshuffle}_n&=U^{\boldshuffle}_nr_{n}(\ep_k)|_{\tilde{s}_{ij}}\,,\quad k\geq 4\,. \nonumber
\end{align}
While the second and third equations show the action of the representations $r_{0,n}(\ep_k)$, $r_{0,n}(x_k)$ and $r_{n}(\ep_k)$ on the vector space spanned by the columns of $U^{\boldshuffle}_n$, which will be discussed in more detail in \subsecref{subsec:PicardAugmentedIntegrals}, we shall next elaborate on the first one.

\subsubsection{Projecting to the \texorpdfstring{$r_{0,n}(x_1)$}{r(0,n)(x1)} eigenspace of the minimal eigenvalue}

The first eigenvalue equation (\ref{eqn:UpperBVEigenvalueEqs}) determines 
the first $(n{-}1)!$ column vectors of the $n!\times n!$ basis transformation $U_n$ diagonalizing the matrix $r_{0,n}(x_1)$:
\begin{align}\label{eqn:eValDec}
r_{0,n}(x_1)&=U_n\, \text{diag}(\underbrace{-s_{01},\dots,-s_{01}}_{(n-1)!},\dots,\underbrace{-s_{01\dots n},\dots,-s_{01\dots n}}_{(n-1)!} ) U_n^{-1}\,,
\end{align}
with
\begin{align}\label{eqn:defProjectionPn}
U_n&=\begin{pmatrix}
U^{\boldshuffle}_n&\dots&U^{\BCJ}_n
\end{pmatrix}\,,\qquad U_n^{-1}=\begin{pmatrix}
P^{\boldshuffle}_n\\\vdots
\end{pmatrix}\,,
\end{align}
where the last $(n{-}1)!$ column vectors of $U_n$ are determined by the first eigenvalue equation in \eqn{eqn:lowerBVMatrixEqs}. The $(n{-}1)! \times n!$ matrix $P^{\boldshuffle}_n$ defined by \eqn{eqn:defProjectionPn} is composed of the dual vectors of $U^{\boldshuffle}_n$, such that
\begin{align}\label{eqn:PU}
P^{\boldshuffle}_n\,U^{\boldshuffle}_n&=\mathds{1}_{(n-1)!}\,.
\end{align}
This equation and the eigenvalue decomposition \eqref{eqn:eValDec} imply that the matrix $P^{\boldshuffle}_n$ acts on the regularizing factor $(-2\pi i(1{-}z_0))^{-r_{0,n}(x_1)}$ in $\BC^{\tau}_{1,n}$ by projecting to the eigenspace associated with the eigenvalue $-s_{01}$,
\begin{align}
P^{\boldshuffle}_n(-2\pi i(1{-}z_0))^{-r_{0,n}(x_1)}&=(-2\pi i(1{-}z_0))^{s_{01}}P^{\boldshuffle}_n\,.
\end{align}
Combining this with \eqn{eqn:C1} shows that the boundary value $\BC^{\tau}_{1,n}$ can be projected to
the $(n{-}1)!$-vector $\BZ^{\tau}_{n}$ of genus-one $Z$-integrals by left-multiplication with $P^{\boldshuffle}_n$:
\begin{align}\label{eqn:C1TauProperRegularization}
P^{\boldshuffle}_n\BC^{\tau}_{1,n}&=\lim_{z_0\to 1} (-2\pi i (1{-}z_0))^{s_{01}}P^{\boldshuffle}_n\BZtzn{n}\nonumber\\
&=e^{s_{01}\omega(1,0|\tau)}
\BZ^{\tau}_{n}|_{\tilde{s}_{ij}}\,,
\end{align}
which finishes the proof of \eqn{spoiler.2}.

\subsubsection{Two-point example}

Let us explicitly determine the upper boundary value $\BC^\tau_{1,2}$ and the associated
matrices $U^{\boldshuffle}_2, P^{\boldshuffle}_2$ for the two-point example. The finite part 
in \eqn{eqn:C1} for $\BZtzn{2}$ is given by 
\begin{align}\label{eqn:2ptC1}
\lim_{z_0\to 1}(-2\pi i (1{-}z_0))^{s_{01}}\BZtzn{2}
&=\lim_{z_0\to 1}e^{s_{01}\omega(1,0|\tau)}\int_0^{z_0}\dd z_2\, e^{-s_{02}\CG_{02}^{\tau}-s_{12}\CG_{12}^{\tau}}\begin{pmatrix}
\Omega_{12}(\eta)\\\Omega_{02}(\eta)
\end{pmatrix}(1+\mathcal{O}(1{-}z_0))\nonumber\\
&=e^{s_{01}\omega(1,0|\tau)}\int_0^{1}\dd z_2\, e^{-(s_{02}+s_{12})\CG_{12}^{\tau}}\begin{pmatrix}
\Omega_{12}(\eta)\\\Omega_{12}(\eta)
\end{pmatrix}\nonumber\\
&=e^{s_{01}\omega(1,0|\tau)}U^{\boldshuffle}_{2} \BZ^{\tau}_2|_{\tilde s_{ij}}
\end{align}
where $\tilde s_{12}=s_{02}+s_{12}$, and the symmetry $\Omega(1{-}z_2,\eta|\tau)=\Omega(-z_2,\eta|\tau)$ determines 
\begin{align}
U^{\boldshuffle}_{2}&=\begin{pmatrix}
1\\1
\end{pmatrix}\,.
\end{align}
The integral $\BZ^{\tau}_2=Z^{\tau}_2(\mathbb I_2|1,2)$ is given by the two-point, genus-one integral
\begin{align}
\BZ^{\tau}_2&=\int_{0}^{1}\dd z_2\, e^{-s_{12}\CG^{\tau}_{12}} 
\Omega_{12}(\eta)\,.
\label{twoptint}
\end{align}
The three eigenvalue equations (\ref{eqn:UpperBVEigenvalueEqs}) can be easily verified using the expressions
\eqref{2ptC} for the scalar derivations $r_{2}(\ep_{k})|_{\tilde{s}_{ij}} $ as well as \eqns{eqn:2PointEMatrices}{eqn:2PointTauDeriv} for $r_{0,2}(x_k)$ and $r_{0,2}(\epsilon_k)$, respectively. In particular, we have
\begin{align}
r_{0,2}(x_1)U^{\boldshuffle}_{2}&= -s_{01}U^{\boldshuffle}_{2}
\end{align}
and can diagonalize $r_{0,2}(x_1)$ via
\begin{align}
r_{0,2}(x_1)&=\begin{pmatrix}
-(s_{01}+s_{02})&s_{02}\\s_{12}&-(s_{01}+s_{12}) 
\end{pmatrix}\nonumber\\
&= \underbrace{ \begin{pmatrix}
1&1\\1&-\frac{s_{12}}{s_{02}}
\end{pmatrix} }_{U_2}\begin{pmatrix}
-s_{01}&0\\0&-s_{012}
\end{pmatrix} \underbrace{ \frac{1}{s_{02}+s_{12}}\begin{pmatrix}s_{12}&s_{02}\\s_{02}&-s_{02}
\end{pmatrix} }_{U_2^{-1}}\,,
\end{align}
By isolating the first row of $U_2^{-1}$ as prescribed by \eqn{eqn:defProjectionPn}, one can read off
\begin{align}\label{eqn:2PointProjection}
P^{\boldshuffle}_2&=\frac{1}{s_{02}+s_{12}}\begin{pmatrix}s_{12}&s_{02}
\end{pmatrix}
\end{align}
and, hence,
\begin{align}\label{eqn:C1TauProperRegularization2Point}
P^{\boldshuffle}_2\BC^{\tau}_{1,2}&=\lim_{z_0\to 1} (-2\pi i (1{-}z_0))^{s_{01}}P^{\boldshuffle}_2\BZtzn{2}\nonumber\\
&=e^{s_{01}\omega(1,0|\tau)}P^{\boldshuffle}_2 U^{\boldshuffle}_{2} \BZ^{\tau}_2|_{\tilde s_{ij}}\nonumber\\
&=e^{s_{01}\omega(1,0|\tau)} \BZ^{\tau}_2|_{\tilde s_{ij}}\,.
\end{align}

\subsubsection{Three-point example}
In the three-point case $n=3$, the eigenvalue decomposition \eqref{eqn:eValDec} leads to the matrix
\beq
U_3= \begin{pmatrix}
1& 0& -\frac{s_{02}}{s_{12}}& 0& 1& 0 \\
0& 1& 0& -\frac{s_{03}}{s_{13}}& 0& 1 \\
 1& 1& -\frac{s_{02}}{s_{12}}& 1& -\frac{(s_{13} {+} s_{23})}{s_{03}}& -\frac{s_{13}}{s_{03}} \\
 1& 1& 1& -\frac{s_{03}}{s_{13}}& -\frac{s_{12}}{s_{02}}& - \frac{(s_{12} {+} s_{23})}{s_{02}} \\
0& 1& 0& 1& \frac{s_{12} (s_{03} {+} s_{13} {+} s_{23})}{ s_{03} s_{023}}& \frac{( s_{12}{-}s_{03}) s_{13}}{   s_{03} s_{023}} \\
1& 0& 1& 0& \frac{s_{12} (s_{13}{-}s_{02} )}{s_{02} s_{023}}& \frac{  s_{13} (s_{02} {+} s_{12} {+} s_{23})}{s_{02} s_{023}}
\end{pmatrix}
\eeq
of eigenvectors of $r_{0,3}(x_1)$, written out in \eqn{alteq2.25}. The first two columns of $U_3$  correspond to the eigenvalue $-s_{01}$ and furnish the matrix $U_3^{\shuffle}$ such that 
\begin{align}\label{eqn:3ptC1}
\lim_{z_0\to 1}(-2\pi i (1{-}z_0))^{s_{01}}\BZtzn{3}
&=e^{s_{01}\omega(1,0|\tau)}U^{\boldshuffle}_{3} \BZ^{\tau}_3|_{\tilde s_{ij}}\,,
\end{align}
while the last two columns reproduce the matrix $U^{\BCJ}_3$ in \eqn{eqn:threePointExampleC0} and are the eigenvectors to $-s_{0123}$. The projection $P^{\boldshuffle}_3$ given by the first two rows of $U_3^{-1}$ takes the form
\begin{align}
P^{\boldshuffle}_3 &=
\frac{1}{(s_{02} {+} s_{03} {+} s_{12} {+} s_{13} {+} s_{23})}
\left( \begin{array}{ccc}
\frac{ s_{12} (s_{02} {+} s_{12} {+} s_{13} {+} s_{23})}{ (s_{02} {+} s_{12})}& 
-\frac{s_{02} s_{13}}{(s_{02} {+} s_{12})}  &\ldots \\
 - \frac{s_{03} s_{12}}{(s_{03} {+} s_{13})}& 
 \frac{s_{13} (s_{03} {+} s_{12} {+} s_{13} {+} s_{23})}{(s_{03} {+} s_{13})}&\ldots
\end{array} \right.
\\
& \ \ \ \ \ \ \ \ \ \ \ \ \left. \begin{array}{ccccc}
\ldots &
 \frac{s_{03} s_{12}}{(s_{02} {+} s_{12})} &
\frac{ s_{02} s_{13}}{(s_{02} {+} s_{12})} &
 -\frac{s_{03} s_{12}}{(s_{02} {+} s_{12})} & 
 \frac{ s_{02} (s_{02} {+} s_{03} {+} s_{12} {+} s_{23})}{ (s_{02} {+} s_{12})} \\
\ldots &
 \frac{s_{03} s_{12}}{(s_{03} {+} s_{13})}& 
\frac{s_{02} s_{13}}{(s_{03} {+} s_{13})}& 
\frac{s_{03} (s_{02} {+} s_{03} {+} s_{13} {+} s_{23})}{( s_{03} {+} s_{13})}& 
-\frac{s_{02} s_{13}}{(s_{03} {+} s_{13})}
\end{array} \right) \notag\,.
\end{align}

\subsection{Applying the languages: two solution strategies for \texorpdfstring{$\BC^{\tau}_{1,n}$}{Ctau(1,n)}}\label{subsec:application}
The system of differential equations \eqref{eqn:ZSystemDEQ} for $Z^\tau_{0,n}$ contains the fundamental equation \eqref{taulang.4} for $Z^{\tau}_n$ as well as the elliptic KZB equation \eqref{eqn:KZBSelE} for $\SelbldEn(z_0|\tau)$. As a consequence, both corresponding solution strategies can be applied to calculate the integrals $Z^{\tau}_n$ appearing in the regularized boundary value $\BC^{\tau}_{1,n}$ given in \eqn{eqn:C1TauProperRegularization}. The corresponding calculations presented in the following two subsections are applications of the methods developed in \rcites{Mafra:2019xms, Broedel:2019gba} and reviewed in \subsecref{ssec:taulang} and \subsecref{ssec:z0lang} which builds upon the analysis of $\BC^{\tau}_{0,n},\BC^{\tau}_{1,n}$ in the previous subsections.

\subsubsection{The elliptic KZB associator}\label{ssec:eKZB}
Having the elliptic KZB equation \eqref{eqn:KZBforZ} at hand, analogously to the discussion in \subsecref{ssec:z0lang} the regularized boundary values identified in \eqns{eqn:C0TauProperRegularization}{eqn:C1TauProperRegularization} are related according to the associator equation \eqref{eqn:assocEqAugmentedZInt} by the elliptic KZB associator $\Phi^{\tau}$ defined in \eqn{eqn:ellKZBAssocSE}. Applying the projection $P^{\boldshuffle}_n$ from \eqn{eqn:defProjectionPn} and rearranging factors leads to the associator equation for the genus-one $Z$-integrals 
\begin{align}\label{eqn:assocEqAugmentedZIntegrals}
\BZ^{\tau}_{n}|_{\tilde{s}_{ij}}&=- e^{\left(s_{012\dots n}-s_{01}\right)  \omega(1,0|\tau)}P^{\boldshuffle}_n \,r_{0,n}\left(\Phi^{\tau}(x_k)\right) U^{\text{BCJ}}_{n}\BZ^{\tree}_{n+2}\,.
\end{align}
The associator equation \eqref{eqn:assocEqAugmentedZIntegrals} is the backbone in
calculating the $\alpha'$-expansions of $\BZ^{\tau}_{n}$ from differential equations in $z_0$.
It relates the $n$-point, genus-one integrals $\BZ^{\tau}_{n}$ containing the planar, one-loop 
configuration-space integrals with (arbitrary) Mandelstam variables $\tilde s_{ij}$ for $1{\leq} i{<}j{\leq} n$ 
to the $(n{+}2)$-point, tree-level $Z$-integrals $\BZ^{\tree}_{n+2}$:
the elliptic KZB associator can be represented by the generating series of eMZVs with the letters being the matrices $r_{0,n}^{\tau}(x_k)$ appearing in the elliptic KZB equation of $\BZtzn{n}$ \cite{Enriquez:EllAss}
\begin{align}\label{eqn:ellKZBAssoc}
r_{0,n}\left(\Phi^{\tau}(x_k)\right)&=\sum_{w\geq 0}\sum_{k_1,\dots,k_w\geq 0}r_{0,n}(x_{k_1}\dots x_{k_w})\omega(k_w,\dots,k_1|\tau)\,.
\end{align}
The matrices $r_{0,n}^{\tau}(x_k)$ are proportional to $s_{ij}$ and therefore to $\alpha'$ (cf.~\eqn{eqn:Mandelstam}), such that \eqn{eqn:ellKZBAssoc} is simply the $\alpha'$-expansion of $\Phi^{\tau}$.
When plugged into the associator equation \eqref{eqn:assocEqAugmentedZIntegrals}, it yields the $\alpha'$-expansion of the genus-one integrals $\BZ^{\tau}_{n}$ from the $\ap$-expansion of the genus-zero integrals $\BZ^{\tree}_{n+2}$. To obtain the $\alpha'$-expansion of $\BZ^{\tau}_{n}$ up to the order $o_{\alpha,\max}^{\tau}$, words $r_{0,n}^{\tau}(x_{k_1}\dots x_{k_w})$ up to the maximal word length $0\leq w\leq w_{\max}$ with
\begin{align}
w_{\max}&=o_{\alpha,\max}^{\tau}-o_{\alpha,\min}^{\tree}
\end{align}
have to be included, where
\begin{align}
o_{\alpha,\min}^{\tree}&=1-n
\end{align}
is the minimal order in $\alpha'$ of $\BZ^{\tree}_{n+2}$.
However, the actual genus-one configuration-space integrals appearing in one-loop open-string amplitudes are the coefficients of the $\eta$-variables from the integrals $\BZ^{\tau}_{n}$. Since the matrices $r_{0,n}^{\tau}(x_{k_i})$ are homogeneous in these variables of degree $k_i{-}1\geq -1$, a configuration-space integral which is given by an $\eta$-degree $k\geq 1{-}n$ coefficient of $\BZ^{\tau}_{n}$ receives at most non-trivial contributions from words $r_{0,n}^{\tau}(x_{k_1}\dots x_{k_w})$ with word length $0\leq w\leq w_{\max}$ satisfying
\begin{align}
w&=\left(k_1+\dots +k_w\right)-k\,.
\end{align}
To summarize, in order to calculate the $\alpha'$-expansion of an $n$-point, genus-one configuration-space integral appearing as an $\eta$-coefficient\footnote{See \rcite{Mafra:2019xms} for details on the extraction of the appropriate $\eta$-coefficient from a $Z$-integral $\Ztn{n}$.} of degree $k$ of $\BZ^{\tau}_{n}$ up to the order $o_{\alpha,\max}^{\tau}$, only the finitely many words $r_{0,n}^{\tau}(x_{k_1}\dots x_{k_w})$ with
\begin{align}\label{eqn:condCalculation}
0\leq w\leq o_{\alpha,\max}^{\tau}+n-1\,,\qquad w=\left(k_1+\dots +k_w\right)-k
\end{align}
contribute to the corresponding $\eta$-coefficients of the elliptic KZB associator \eqref{eqn:ellKZBAssoc} and, thus, have to be included in the associator \eqn{eqn:assocEqAugmentedZIntegrals}. Upon rewriting the eMZVs in \eqn{eqn:ellKZBAssoc} in terms of iterated Eisenstein integrals \cite{Broedel:2015hia}, we have checked \eqn{eqn:assocEqAugmentedZIntegrals} to reproduce the $\ap$-expansion generated by \eqn{taulang.9} for a wide range of orders in $\alpha'$ and $\eta_j$. 

Note that the results of \eqn{eqn:assocEqAugmentedZIntegrals} for the integrals $\BZ^{\tau}_{n}|_{\tilde{s}_{ij}}$
relevant to one-loop open-string amplitudes no longer depend on $s_{01}$, 
which is why the conditions (\ref{assumpGenusOne}) on its real part do not pose any restrictions
on the physical applications.

\subsubsection{Two-point example}\label{ssec:eKZB2pt}

Let us investigate the two-point working example. From \eqns{more2ptin}{eqn:twoPointExampleC0}, the tree-level integrals are known to be given by 
\begin{align}\label{eqn:Ztree2}
U^{\text{BCJ}}_{2}\BZ^{\tree}_{4} & =
\begin{pmatrix}
-\frac{ 1}{s_{12}} \\\frac{1}{s_{02}}
\end{pmatrix} \frac{ \Gamma(1{-}s_{12}) \Gamma(1{-}s_{02} ) }{\Gamma(1{-}s_{12}{-}s_{02}) }\,,
\end{align}
while the matrices $r_{0,2}(x_k)$ are spelled out in \eqn{eqn:2PointEMatrices}. The projection $P^{\boldshuffle}_2$ is given in \eqn{eqn:2PointProjection}, such that upon combining these quantities the associator \eqn{eqn:assocEqAugmentedZIntegrals} for the two-point integral $\BZ^{\tau}_2$ in \eqn{twoptint} takes the form 
\begin{align}\label{eqn:assocEqAugmentedZIntegrals2Point} 
\BZ^{\tau}_2 \big|_{\tilde s_{12}}&=e^{\left(s_{02}+s_{12}\right)  \omega(1,0|\tau)}\frac{1}{s_{02}{+}s_{12}}\begin{pmatrix}s_{12}&s_{02}
\end{pmatrix} r_{0,2}\left(\Phi^{\tau}(x_k)\right) \begin{pmatrix}
\frac{ 1}{s_{12}} \\-\frac{1}{s_{02}}
\end{pmatrix} \frac{ \Gamma(1{-}s_{12}) \Gamma(1{-}s_{02} ) }{\Gamma(1{-}s_{12}{-}s_{02}) }\,.
\end{align}
Let us extract the configuration-space integral for the two-point, one-loop integral
\begin{align}
I_2^{\tau}&=\int_{0}^{1}\dd z_2  \, e^{-(s_{02}+s_{12})\CG^{\tau}_{12}} = \BZ^{\tau}_2 \big|_{\tilde s_{12}} \, \big|_{\eta^{-1}}
\end{align}
on the left-hand side of \eqn{eqn:assocEqAugmentedZIntegrals2Point} and calculate its $\alpha'$-expansion up to order $o_{\alpha,\max}^{\tau}=2$. Since according to \eqn{eqn:Ztree2} the minimal order of the tree-level integral is $o_{\alpha,\min}^{\tau}=-1$, words $r_{0,n}^{\tau}(x_{k_1}\dots x_{k_w})$ with word length $0\leq w\leq 3$ have to be considered. However, the condition \eqref{eqn:condCalculation} only selects the words which are explicitly written down in the following to contribute non-trivially:
\begin{align}\label{eqn:r02Phi}
r_{0,2}\left(\Phi^{\tau}(x_k)\right)
&=r_{0,2}(x_0)\,\omega(0|\tau)+r_{0,2}([x_1,x_0])\omega(0,1|\tau)\nonumber\\
&\phantom{=}+r_{0,2}\left(x_2 x_0 x_0+x_0x_0x_2\right)\omega(0,0,2|\tau)+r_{0,2}(x_0x_2 x_0)\omega(0,2,0|\tau)\nonumber\\
&\phantom{=}+r_{0,2}\left(x_1 x_1 x_0+x_0x_1x_1\right)\omega(0,1,1|\tau)+r_{0,2}(x_1x_0 x_1)\omega(1,0,1|\tau)+\mathcal{O}(\eta^0,s_{ij}^4)\nonumber\\ &=r_{0,2}(x_0)+r_{0,2}([x_1,x_0])\omega(0,1|\tau)\nonumber\\
&\phantom{=}+r_{0,2}\left([x_0,[x_0,x_2]]\right)\omega(0,0,2|\tau)-r_{0,2}(x_0x_2 x_0)\zeta_2\nonumber\\
&\phantom{=}+r_{0,2}\left([x_1,[x_1,x_0]]\right)\Big(\frac{5}{12}\zeta_2+\frac{1}{2}\omega(0,1|\tau)^2 +\frac{1}{2}\omega(0,0,2|\tau) \Big)\nonumber\\
&\phantom{=}+\mathcal{O}(\eta^0,s_{ij}^4)\,.
\end{align}
Therefore, denoting the words written down above and the corresponding higher-order terms $\mathcal{O}(s_{ij}^4)$ which give the order $\eta^{-1}$ of the elliptic KZB associator by $r_{0,2}\left(\Phi^{\tau}(x_k)\right)|_{\eta^{-1}}$, we obtain the equation for the configuration-space integral
\begin{align}\label{eqn:assocEqAugmentedZIntegrals2PointComponent}
I_2^{\tau}&=e^{\left(s_{02}+s_{12}\right)  \omega(1,0|\tau)}\frac{1}{s_{02}+s_{12}}\begin{pmatrix}s_{12}&s_{02}
\end{pmatrix} r_{0,2}\left(\Phi^{\tau}(x_k)\right)|_{\eta^{-1}} \begin{pmatrix}
\frac{ 1}{s_{12}} \\ -\frac{1}{s_{02}}
\end{pmatrix} \frac{ \Gamma(1{-}s_{12}) \Gamma(1{-}s_{02} ) }{\Gamma(1{-}s_{12}{-}s_{02}) }\nonumber\\
&= 1+\frac{1}{4} (s_{02}{+}s_{12})^2\big(\zeta_2-12\gamma(4,0|q)\big)+\mathcal{O}(s_{ij}^3) 
\end{align}
in agreement with \eqn{eqn:Zloopexample} and \cite{Mafra:2019xms}.

\subsubsection{Two organization schemes for \texorpdfstring{$\ap$}{alphaprime}-expansions}\label{subsec:PicardAugmentedIntegrals}

The other expansion method for the genus-one $Z$-integrals $\BZ^{\tau}_n$ put forward in \rcite{Mafra:2019xms} is reviewed in \subsecref{ssec:taulang} and consists of solving the differential \eqn{taulang.4} in $\tau$ by Picard iteration. The resulting $\ap$-expansion (\ref{taulang.9}) is organized in terms of iterated Eisenstein integrals and $(n{-}1)!{\times}(n{-}1)!$ matrix representations $r_n(\epsilon_k)$ and furnishes an alternative to the expanded form of \eqn{eqn:assocEqAugmentedZIntegrals} in terms of eMZVs and $n! \times n!$ matrix representations $r_{0,n}(x_k)$. The equivalence of \eqns{taulang.9}{eqn:assocEqAugmentedZIntegrals}, i.e.\
\begin{align}
&\sum_{\ell=0}^\infty \sum_{k_1,k_2,\ldots, k_\ell \atop{ =0,4,6,8,\ldots}}
\Big( \prod_{j=1}^{\ell}(k_{j}{-}1)  \Big) \gamma(k_1,k_2,\ldots,k_\ell|q) r_n(\ep_{k_\ell} \ldots \ep_{k_2} \ep_{k_1})
\mathbf{Z}_n^{i\infty} \, \big|_{\tilde s_{ij}} \notag \\
&\ \ = - e^{\left(s_{012\dots n}-s_{01}\right)  \omega(1,0|\tau)} \sum_{w=0}^\infty \sum_{k_1,\ldots,k_w\geq0}  \omega(k_1,k_2,\ldots,k_w|\tau) \label{allequal} \\
& \ \ \ \ \ \ \ \ \ \ \ \ \ \ \ \  \ \ \ \ \ \ \ \ \ \ \ \ \times P^{\boldshuffle}_n \,r_{0,n}(x_{k_w}\ldots x_{k_2}x_{k_1}) U^{\text{BCJ}}_{n}\BZ^{\tree}_{n+2} \, ,\notag
\end{align}
is not obvious from the first glance at these types of series but guaranteed by the arguments in \rcites{Mafra:2019xms, Broedel:2019gba} and the previous sections. The initial value $ \mathbf{Z}_n^{i\infty}$ on the left-hand side is related to $\BZ^{\tree}_{n+2}$ on the right-hand side by an $s_{ij}$- and $\eta_j$ dependent $(n{-}1)!{\times}(n{-}1)!$ matrix described in \rcite{Mafra:2019xms}, see \eqn{2ptin} for the two-point example.

The discussion of the previous subsections yields a streamlined way of showing {\it directly} that both sides of \eqn{allequal} obey the same differential equation in $\tau$. The differential \eqn{taulang.4} (with $\tilde s_{ij}$ in the place of $s_{ij}$) holds for the left-hand side by construction, and the analogous equation for $P^{\boldshuffle}_n\BC^{\tau}_{1,n}=e^{s_{01}\omega(1,0|\tau)}\BZ^{\tau}_{n}|_{\tilde{s}_{ij}}$ on the right-hand side can be inferred from properties of the augmented $Z$-integrals: according to the calculation \eqref{eqn:tauDerivativeLimit}, this differential equation for $P^{\boldshuffle}_n\BC^{\tau}_{1,n}$ can be written as
\begin{align}\label{eqn:deq2}
&2\pi i \partial_{\tau}P^{\boldshuffle}_n\BC^{\tau}_{1,n} =\lim_{z_0\to 1}(-2\pi i (1{-}z_0))^{s_{01}}P^{\boldshuffle}_n2\pi i \partial_{\tau}\BZtzn{n}\nonumber\\
&=P^{\boldshuffle}_n\left(-r_{0,n}(\ep_0)-\GG{2}r_{0,n}(x_1) +\sum_{k=4}^{\infty}(1{-}k)\GG{k}\left(r_{0,n}(\ep_k)+r_{0,n}(x_{k-1})\right)\right)U^{\boldshuffle}_n\nonumber\\
&\phantom{=}\times e^{-s_{01}\omega(1,0|\tau)} \BZ^{\tau}_{n}|_{\tilde{s}_{ij}}\,,
\end{align}
where the identities \eqref{eqn:UpperBVEigenvalueEqs} together with \eqn{eqn:PU}
can be used to relate the matrix representations of different sizes
\begin{align}
P^{\boldshuffle}_n\,r_{0,n}(x_1)U^{\boldshuffle}_n&=-s_{01} \mathds{1}_{(n-1)!{\times} (n-1)!}\,, \nonumber\\
P^{\boldshuffle}_n\,r_{0,n}(\ep_0)U^{\boldshuffle}_n&=\left(r_n(\ep_0)|_{\tilde{s}_{ij}}+2\zeta_2s_{01}\right)\,, \label{eqn:UpperBVEigenvalueEqsProjected}\\
P^{\boldshuffle}_n\left(r_{0,n}(\ep_k)+r_{0,n}(x_{k-1})\right)U^{\boldshuffle}_n&=r_{n}(\ep_k)|_{\tilde{s}_{ij}}\,,\quad k\geq 4\,. \nonumber  
\end{align}
As a consequence, \eqn{eqn:C1TauProperRegularization} can be simplified to
\begin{align}
2\pi i &\partial_{\tau} \big( e^{s_{01}\omega(1,0|\tau)}\BZ^{\tau}_{n}|_{\tilde{s}_{ij}} \big) =
2\pi i \partial_{\tau}P^{\boldshuffle}_n\BC^{\tau}_{1,n} \label{equivdtau} \\
&= \left(-\left(r_n(\ep_0)|_{\tilde{s}_{ij}}+2\zm_2s_{01}\right)+s_{01} \GG{2}+\sum_{k=4}^\infty (1{-}k) \GG{k} \, r_n(\ep_k)|_{\tilde{s}_{ij}}\right)e^{s_{01}\omega(1,0|\tau)} \BZ^{\tau}_{n}|_{\tilde{s}_{ij}}\,, \notag
\end{align}
which is equivalent to \eqn{taulang.4} for the Mandelstam variables $\tilde s_{ij}$ defined in \eqn{shiftmands} after employing 
\begin{equation}
 2\pi i\partial_{\tau}\omega(1,0|\tau)=\GG{2}-2\zeta_2\,.
\end{equation}
This concludes the direct proof that both sides of \eqn{allequal} obey the same differential equation in $\tau$. A direct comparison of the respective initial values as $\tau \rightarrow i\infty$ may be challenging, but the consistency in this limit is guaranteed since both sides have been derived in \rcites{Mafra:2019xms, Broedel:2019gba} and the previous sections. Note that the combination $r_{0,n}(\ep_k)+r_{0,n}(x_{k-1})$ in the third line of \eqn{eqn:UpperBVEigenvalueEqsProjected} also arises when adapting depth-three
relations \eqn{extra3} in the derivation algebra to the twice punctured torus.


\section{Conclusion}\label{sec:conclusion}

In \rcites{Mafra:2019xms,Broedel:2019gba} two different constructions for the $\alpha'$-expansion of configuration-space integrals in one-loop open-string amplitudes have been put forward.  In both references the (elliptic) multiple zeta values in the $\alpha'$-expansions are derived from different types of differential equations.  Here we have connected these two approaches within a more general framework and, in particular, shown that
\begin{itemize}
	\item both approaches and the definitions therein can be traced back to one class of iterated integrals, called augmented (genus-one) $Z$-integrals, a vector of $n$-point basis integrals $\BZtzn{n}$ is defined in \eqn{eqn:Ztzbasis}. Besides the usual fixed puncture $z_1=0$ on the torus $\ZC/(\ZZ{+}\tau \ZZ)$ appearing in the definition of the genus-one $Z$-integrals of \rcite{Mafra:2019xms}, the integrals $\BZtzn{n}$ are augmented by a second unintegrated puncture $z_0$ with $z_1=0<z_0<1$ as in \rcite{Broedel:2019gba}. Thus -- apart from Mandelstam invariants $s_{ij}$ -- the augmented integrals depend on two parameters: the modular parameter $\tau$ of the torus and the additional puncture $z_0$.  
	\item differentiation of $\BZtzn{n}$ with respect to the two parameters $\tau$ and $z_0$ leads to a homogeneous linear system of two partial differential equations -- an elliptic KZB system \eqref{eqn:ZSystemDEQ} on the twice-punctured torus 
	\item the genus-one Selberg integrals from \rcite{Broedel:2019gba} are linear combinations of the components of the augmented $Z$-integrals $\BZtzn{n}$; they can be recovered according to the discussion in \appref{app:recoverSelbergInt}.  
	\item the genus-one $Z$-integrals from \rcite{Mafra:2019xms} -- and hence the configuration-space integrals in $n$-point, planar, one-loop open-string amplitudes -- are recovered as regularized boundary values of the augmented integrals in $\BZtzn{n}$ as $z_0\to 1$.  Correspondingly, the two differential equations in the system \eqref{eqn:ZSystemDEQ} can be solved independently for the string integrals in the limit of $z_0 \rightarrow 1$ via integration w.r.t.\ $\tau$ or $z_0$. The respective initial values at $\tau \rightarrow i \infty$ and $z_0 \rightarrow 0$ are reduced to $(n{+}2)$-point genus-zero integrals whose $\alpha'$-expansion in terms of multiple zeta values is known from several all-multiplicity methods, see e.g.\ \cite{Broedel:2013tta, Broedel:2013aza, Mafra:2016mcc, Kaderli:2019dny}. As summarized in \subsecref{subsec:application}, this yields the two approaches in \rcites{Mafra:2019xms,Broedel:2019gba} to calculate the $\ap$-expansion of one-loop open-string amplitudes in terms of (elliptic) multiple zeta values and iterated Eisenstein integrals.
	\item calculating the $\alpha'$-expansion using the integrals $\BZtzn{n}$ involves elementary operations only: differentiation in formal expansion variables $\eta_j$ and matrix algebra, with the matrices being determined by the elliptic KZB system \eqref{eqn:ZSystemDEQ}. The entries of the corresponding $n!{\times} n!$ matrix representation determine the coefficients in the $\alpha'$-expansion and are explicitly given for an arbitrary number of points $n$ in \eqns{eqn:derZ0Closed}{eqn:tauDerivClosed}.
	\item the $(n{-}1)!{\times}(n{-}1)!$ matrix representation in the differential equation in \rcite{Mafra:2019xms} is reproduced from the matrices appearing in the elliptic KZB system \eqref{eqn:ZSystemDEQ} according to \eqn{eqn:UpperBVEigenvalueEqsProjected}.  
\item the operators appearing in an elliptic KZB system of the form \eqref{eqn:ZSystemDEQ} satisfy the commutation relations \eqref{gen1.37} which serve as consistency checks for our explicit matrix representations.  
\end{itemize}
Our construction of particular integral representations $\BZtzn{n}$ for the solution of an elliptic KZB system on the twice-punctured curve leads to the question as how it may be embedded into the existing Mathematics literature about similar systems. In particular, its connection to \rcite{KZB} and the representation of the algebra generators therein  appearing in the differential equations as nested commutators, should be clarified.

\section*{Acknowledgments}

We are grateful to Federico Zerbini for discussions and to Carlos Mafra, Nils Matthes, Carlos Rodriguez and Federico Zerbini for comments on the manuscript.  AK is supported by the International Max Planck Research School for Mathematical and Physical Aspects of Gravitation, Cosmology and Quantum Field Theory. OS is supported by the European Research Council under ERC-STG-804286 UNISCAMP. 

\appendix

\section{Kronecker--Eisenstein chain identities}\label{app:chainIdentities}
In this appendix, we prove various identities used for chains \eqref{eqn:defChain} of Kronecker--Eisenstein series in the proofs of the $n$-point formul\ae{} for the $z_0$- and $\tau$-derivatives of the augmented $Z$-integrals in \appref{app:z0Deriv} and \appref{appTauDerivative}, respectively. Some of these identities involve chains with shifts or replacements in certain $\eta$-variables. For a replacement of $\eta_{a_k}$ by $\xi$ (or a shift by $\xi-\eta_{a_k}$, respectively), this will be denoted as follows,
\begin{align}\label{eqn:defChainShifted}
\phiChain(a_1,a_2,\dots,a_p)|_{\eta_{a_k}\to \xi}&=\prod_{i=2}^k\Omega_{a_{i-1},a_i}(\eta_{a_i a_{i+1}\dots a_p}+\xi-\eta_{a_k})\prod_{i=k+1}^p\Omega_{a_{i-1},a_i}(\eta_{a_i a_{i+1}\dots a_p})\,,
\end{align}
where we recall the shorthand $\Omega_{ij}(\eta)=\Omega(z_{ij},\eta|\tau)$.
Moreover, we generally assign an $\eta$-variable to the first index of a chain $\phiChain(a_1,a_2,\dots,a_p)$ 
such that the overall sum vanishes, i.e.\
\begin{align}\label{eqn:etaConvention}
\eta_{a_1}&=-\eta_{a_2}-\dots -\eta_{a_p}\,,
\end{align}
which is in agreement with \eqn{eqn:eta01} for the chains  $\phiChain(1,A)$ and $ \phiChain(0,B)$.

Throughout this and the following section, we accompany the crucial identities by the graphical notation for chains of Kronecker--Eisenstein series in order to facilitate the readability of the proofs for the $n$-point $z_0$- and $\tau$-derivatives. Let us briefly recall the corresponding definitions and conventions from \subsecref{ssec:graphicalnotation}:
\begin{itemize}
	\item A Kronecker--Eisenstein series $\Omega_{ij}(\eta)$ is represented by a directed edge with weight $\eta$ from vertex $j$ to vertex $i$
	\begin{equation}
	\Omega_{ij}(\eta)=\mpostuse{EKSeries}\,.
	\end{equation}
	\item A chain of Kronecker--Eisenstein series $\phiChain(A)$ labeled by the sequence $A=(a_1,a_2,\dots,a_p)$ is represented by a chain of directed edges connecting the corresponding vertices $a_i$ and~$a_{i+1}$. If the $\eta$-variable $\eta_{a_{i}\dots a_{p}}$ of the factor $\Omega_{a_{i-1},a_i}(\eta_{a_{i}\dots a_{p} })$ is not explicitly depicted as a weight of the edge and unless stated otherwise, it is determined by the vertices pointing to the vertex $i$ through a chain of arrows: each vertex $j$ which has an edge pointing in the direction of $i$ (possibly via further directed edges pointing towards $i$) contributes a term $\eta_j$, such that the edge pointing away from $i$ is given by the corresponding sum
	\begin{equation}
		\phiChain(A)=\quad\mpostuse[align=b,vshift=-14pt]{phiChainA}\quad=\quad\prod_{i=2}^p\quad\mpostuse[align=b,vshift=-14pt]{phiChainAFactors}=\quad\prod_{i=2}^p\Omega_{a_{i-1},a_i}(\eta_{a_i a_{i+1}\dots a_p})\,,
	\end{equation}
where we use the convention $\eta_{a_i a_{i+1}\dots a_p}=\eta_{a_{i,i+1 \dots p}}$ for sums of $\eta$-variables associated to a sequence $(a_i, a_{i+1},\dots, a_p)$ in graphs. It should always be clear from the context whether edges without weights refer to the genus-zero notation or the genus-one notation where the weights are only implicit. In particular, in this and the next section, we exclusively discuss the genus-one case.
	\item The same accumulation of the $\eta$-variables is used for directed tree graphs, if it is not denoted explicitly: the weight of the edge pointing away from the vertex $i$ is the sum of all the $\eta$-variables associated to the edges pointing towards $i$ via a chain of Kronecker--Eisenstein series. For example
	\begin{equation}\label{eqn:exampleTree}
	\Omega_{12}(\eta_{234})\Omega_{23}(\eta_{3})\Omega_{24}(\eta_{4})=\quad\mpostuse[align=b,vshift=-14pt]{ExampleTreeWeights}\quad=\quad\mpostuse[align=b,vshift=-14pt]{ExampleTree}\,.
	\end{equation}
\end{itemize}

The following identities, which are proven in \appref{app:shuffleIdentities}, are particularly useful: for finite, disjoint sequences $A=(a_1,\dots, a_p)$, $B=(b_1,\dots,b_q)$, $C=(c_1,\dots,c_m)$ and $D=(d_1,\dots, d_l)$ as well as distinct labels $r,r_0,r_1$ not contained in any of the sequences $A,B,C,D$, we find
\begin{itemize}
	\item the concatenation of two chains with shifted $\eta$-variables of the first chain 
	\begin{align}\label{eqn:chainConcatenation}
	&\phiChain(A,r)|_{\eta_r\to\eta_r+\eta_B}\phiChain(r,B)=\phiChain(A,r,B)\,,
	\end{align}
	i.e.\ 
	\begin{equation}
	\mpostuse[align=b,vshift=-40pt]{concatenationLHS} \quad= \quad\mpostuse[align=b,vshift=-40pt]{concatenationRHS}
	\end{equation}
	\item the shuffle product of two chains ending at the same point $r$
	\begin{align}\label{eqn:shuffleProductChains}
	\phiChain(r,A)\phiChain(r,B)&=\phiChain(r,A\shuffle B)\,,
	\end{align}
	i.e.
	\begin{align}
	\mpostuse[align=b,vshift=-40pt]{shuffleSameEndingLHS}\quad=\quad\mpostuse[align=b,vshift=-40pt]{shuffleSameEndingRHS}
	\end{align}
	\item the shuffle product of two chains beginning at the same point $r$ 
	\begin{align}\label{eqn:shuffleChainsSameBeginning}
	\phiChain(C,r)|_{\eta_r\to \eta_A}\phiChain(D,r)|_{\eta_r\to \eta_r+\eta_B}
	&=\phiChain(C\shuffle D,r)|_{\eta_r\to\eta_r+\eta_A+\eta_B}\,,
	\end{align}
	where according to \eqn{eqn:etaConvention} $\eta_{c_1}= -\eta_A-\eta_{c_2,\dots,c_m}$ and $\eta_{d_1}= -\eta_r-\eta_B-\eta_{d_2,\dots,d_l}$, i.e.\
	\begin{align}
	\mpostuse{shuffleSameBeginningLHS}&=\quad\mpostuse{shuffleSameBeginningRHS}
	\end{align}
	\item the reflection property
	\begin{align}\label{eqn:reflectionChain}
	\phiChain(r,a_1,\dots,a_p)&=(-1)^{p}\phiChain(a_p,\dots,a_1,r)\,,
	\end{align}
	i.e.\
	\begin{align}
	\mpostuse[align=b,vshift=-40pt]{reflectionLHS}\quad&=\quad(-1)^p\quad\mpostuse[align=b,vshift=-40pt]{reflectionRHS}
	\end{align}
	where again by our convention $\eta_r=-\eta_A=-\eta_{a_1\dots a_p}$.
	\item and the shifting of two labels $r_0$ and $r_1$ next to each other
	\begin{align}\label{eqn:shuffleIdentityChainWith1And0AlternatingSum}
	&\phiChain(r_0,A,r_1,B)\nonumber\\
	&=\sum_{i=1 }^{p+1}(-1)^{p+1-i}\phiChain(r_0,a_1,a_2,\dots,a_{i-1})\phiChain(r_0,r_1,\left(a_p,a_{p-1},\dots, a_i\right)\shuffle B)\,,
	\end{align}
	i.e.\
	\begin{equation}
	\mpostuse[align=b,vshift=-40pt]{phir0Ar1BLHS}\quad=\quad\sum_{i=1}^{p+1}(-1)^{p+1-i}\quad\mpostuse[align=b,vshift=-40pt]{phir0Ar1BRHS}\,.
	\end{equation}
	%
\end{itemize}
\subsection{Shuffle and concatenation identities}\label{app:shuffleIdentities}
The first identity is the concatenation \eqref{eqn:chainConcatenation} of two chains, where 
one has a shifted $\eta$-variable
\begin{align}
\phiChain(A,r)|_{\eta_r\to\eta_r+\eta_B}\phiChain(r,B)&=\phiChain(A,r,B)\,.
\end{align}
For $A=\emptyset$ or $B=\emptyset$, this relation is trivial due to \eqn{eqn:defTrivialChain}, while for $A=(a_1,\dots,a_p),B=(b_1,\dots b_q)\not=\emptyset$, it follows from the definitions \eqref{eqn:defChain} and \eqref{eqn:chainConcatenation}
\begin{align}
&\phiChain(A,r)|_{\eta_r\to\eta_r+\eta_B}\phiChain(r,B)\nonumber\\
&=\Omega_{a_1,a_2}(\eta_{A}+\eta_{r}+\eta_B,\tau)\dots \Omega_{a_{p-1},a_p}(\eta_{a_p}+\eta_{r}+\eta_B,\tau)\Omega_{a_{p},r}(\eta_{r}+\eta_B,\tau)\nonumber\\
&\phantom{=}\times \Omega_{r,b_1}(\eta_{B},\tau)\Omega_{b_1,b_2}(\eta_{b_2,\dots,b_q},\tau)\dots \Omega_{b_{q-1},b_q}(\eta_{b_q},\tau)\nonumber\\
&=\phiChain(A,r,B)\,.
\end{align} 

The second identity is the shuffle relation \eqref{eqn:shuffleProductChains}, which can be proven by induction in the length of the sequence $A$ (and by the symmetry in $A$ and $B$). Thus, let us assume that $A=\emptyset$ or $B=\emptyset$, then it is trivially satisfied according to the definition \eqref{eqn:defTrivialChain}, i.e.\ $\phiChain(r)=1$. For $A=(a_1)$ and $B=(b_1)$, we simply find the Fay identity \eqref{eqn:Fay} for the Kronecker--Eisenstein series
\begin{align}
\phiChain(r,a_1)\phiChain(r,b_1)&=\Omega_{r,a_1}(\eta_{a_1})\Omega_{r,b_1}(\eta_{b_1})\nonumber\\
&=\Omega_{r,a_1}(\eta_{a_1,b_1})\Omega_{a_1,b_1}(\eta_{b_1})+\Omega_{r,b_1}(\eta_{a_1,b_1})\Omega_{b_1,a_1}(\eta_{a_1})\nonumber\\
&=\phiChain(r,a_1,b_1)+\phiChain(r,b_1,a_1)\nonumber\\
&=\phiChain(r,a_1\shuffle b_1)\,.
\end{align}
Now, let us assume that it holds for $(a_2,\dots,a_p)$ and $B=(b_1,\dots,b_q)$, as well as for $A=(a_1,\dots,a_p)$ and $(b_2,\dots,b_q)$ and use the Fay identity for the induction step to show the identity for $A$ and $B$
\begin{align}
\phiChain(r,A)\phiChain(r,B)&=\phiChain(A)\phiChain(B)\phiChain(r,a_1)|_{\eta_{a_1}\to \eta_A}\phiChain(r,b_1)|_{\eta_{b_1}\to \eta_B}\nonumber\\
&=\phiChain(A)\phiChain(B)\left(\phiChain(r,a_1,b_1)+\phiChain(r,b_1,a_1)\right)|_{\eta_{a_1},\eta_{b_1}\to \eta_A,\eta_B}\nonumber\\
&=\phiChain(a_1,a_2,\dots,a_p)\phiChain(r,a_1,B)|_{\eta_{a_1}\to \eta_A}+\phiChain(r,b_1,A)|_{\eta_{b_1}\to \eta_B}\phiChain(b_1,b_2,\dots,b_q)\nonumber\\
&=\phiChain(r,a_1)|_{\eta_{a_1}\to \eta_A+\eta_B}\phiChain(a_1,a_2,\dots,a_p)\phiChain(a_1,B)\nonumber\\
&\phantom{=}+\phiChain(r,b_1)|_{\eta_{b_1}\to \eta_B+\eta_A}\phiChain(b_1,b_2,\dots,b_q)\phiChain(b_1,A)\nonumber\\
&=\phiChain(r,a_1,(a_2,\dots,a_p)\shuffle B)+\phiChain(r,b_1,A\shuffle (b_2,\dots,b_q))\nonumber\\
&=\phiChain(r,A\shuffle B)\,,
\end{align}
where we have used the concatenation property in the intermediate step.

The next identity \eqref{eqn:shuffleChainsSameBeginning},
\begin{align}
&\phiChain(C,r)|_{\eta_r\to \eta_A}\phiChain(D,r)|_{\eta_r\to \eta_r+\eta_B}\nonumber\\
&=\phiChain(C\shuffle D,r)|_{\eta_r\to\eta_r+\eta_A+\eta_B, \eta_{c_1}\to -\eta_A-\eta_{c_2,\dots,c_m},\eta_{d_1}\to -\eta_r-\eta_B-\eta_{d_2,\dots,d_l} }\,,
\end{align}
is similar to the one before, but we have to be more careful with the shifts in the $\eta$-variables. For $C=\emptyset$ or $D=\emptyset$, it is trivial. Thus, let $C=(c_1,\dots,c_m)$ and $D=(d_1,\dots,d_l)$ with $m,l\neq 0$. For $m=l=1$, it is simply the Fay identity. For $m,l\geq 2$, we can iteratively apply the Fay identity:
\begin{align}
&\phiChain(C,r)|_{\eta_r\to \eta_A}\phiChain(D,r)|_{\eta_r\to \eta_r+\eta_B}\nonumber\\
&=\phiChain(c_1,\dots,c_{m-1},c_m,r)|_{\eta_r\to \eta_A}\phiChain(d_1,\dots,d_{l-1},d_l,r)|_{\eta_r\to \eta_r+\eta_B}\nonumber\\
&=\phiChain(c_1,\dots,c_{m-1},c_m)|_{\eta_{c_m}\to \eta_{c_m}+ \eta_A}\phiChain(d_1,\dots,d_{l},c_m)|_{\eta_{c_m}\to \eta_r+\eta_B}\phiChain(c_m,r)|_{\eta_{r}\to \eta_r+\eta_A+\eta_B}\nonumber\\
&\phantom{=}+\phiChain(c_1,\dots,c_{m},d_l)|_{\eta_{d_l}\to \eta_A}\phiChain(d_1,\dots,d_{l-1},d_l)|_{\eta_{d_l}\to \eta_{d_l}+\eta_r+\eta_B}\phiChain(d_l,r)|_{\eta_{r}\to \eta_r+\eta_A+\eta_B}\nonumber\\
&=\phiChain((c_1,\dots,c_{m-1})\shuffle D,c_m,r)|_{\eta_{r}\to \eta_r+\eta_A+\eta_B}\nonumber\\
&\phantom{=}+\phiChain(C\shuffle (d_1,\dots,d_{l-1}),d_l,r)|_{\eta_{r}\to \eta_r+\eta_A+\eta_B}\nonumber\\
&=\phiChain(C\shuffle D,r)|_{\eta_{r}\to \eta_r+\eta_A+\eta_B}\,,
\end{align}
where according to \eqn{eqn:etaConvention}, $\eta_{c_1}=-\eta_A-\eta_{c_2,\dots, c_m}$ and $\eta_{d_1}= -\eta_{r}-\eta_B-\eta_{d_2,\dots, d_l}$

In a similar inductive proof of 
the reflection property \eqref{eqn:reflectionChain},
\begin{align}
\phiChain(r,a_1,\dots,a_p)&=(-1)^{p}\phiChain(a_p,\dots,a_1,r)\,,
\end{align}
with $\eta_r=-\eta_A$, the $p=1$ case simply reduces to
the antisymmetry property of the Kronecker--Eisenstein series,
\begin{align}
\phiChain(r,a_1)&=\Omega_{r,a_1}(\eta_{a_1},\tau)=-\Omega_{a_1,r}(-\eta_{a_1},\tau)=-\phiChain(a_1,r)\,.
\end{align}
The inductive step is done by concatenation
\begin{align}
\phiChain(r,a_1,\dots,a_{p-1},a_p)&=\phiChain(r,a_1,\dots,a_{p-1})|_{\eta_{a_{p-1}}\to\eta_{a_{p-1},a_p}}\phiChain(a_{p-1},a_p)\nonumber\\
&=(-1)^{p-1}\phiChain(a_{p-1},\dots,a_1,r)|_{\eta_{r}=-\eta_{a_{1},\dots,a_{p}}}\phiChain(a_{p-1},a_p)\nonumber\\
&=(-1)^{p}\phiChain(a_p,a_{p-1})|_{\eta_{a_{p-1}}\to-\eta_{a_p}}\phiChain(a_{p-1},\dots,a_1,r)|_{\eta_{r}=-\eta_{a_{1},\dots,a_{p}}}\nonumber\\
&=(-1)^{p}\phiChain(a_p,a_{p-1},\dots,a_1,r)|_{\eta_{r}=-\eta_{a_{1},\dots,a_p}}\,.
\end{align}

To finish, let us proof the identity \eqref{eqn:shuffleIdentityChainWith1And0AlternatingSum},
\begin{align}\label{eqn:shuffleIdentityChainWith1And0AlternatingSumApp}
\phiChain(r_0,A,r_1,B)&=\sum_{i=1 }^{p+1}(-1)^{p+1-i}\phiChain(r_0,A_{1i})\phiChain(r_0,r_1,\tilde{A}_{i,p+1}\shuffle B)\,,
\end{align}
for $A=(a_1,a_2,\dots,a_p)$ and $B=(b_1,b_2,\dots,b_q)$, where we already make use of the notation in \eqn{eqn:subsequences} for subsequences. It is trivially satisfied for $A=\emptyset$, thus, let us assume that it holds for $A=(a_1,\dots,a_p)$ and show it for $A^0=(a_0,A)=(a_0,a_1,\dots,a_p)$. We do this in two steps. First, the following combinatorial identity is proven 
\begin{align}\label{eqn:combinatorialStep}
\sum_{i=1}^{p+1}(-1)^i (a_0,A_{1,i})\shuffle \tilde A_{i,p+1}&=-(\tilde A,a_0)\,.
\end{align}
It is trivially satisfied for $A=\emptyset$ and for $A=(a_1)$, it takes the form
\begin{align}
(-1)(a_0)\shuffle (a_1)+(a_0,a_1)&=-(a_1,a_0)\,.
\end{align}
The induction step can be obtained using the recursive definition of the shuffle product
\begin{align}
\sum_{i=1}^{p+1}(-1)^i (a_0,A_{1,i})\shuffle \tilde A_{i,p+1}&=\sum_{i=1}^{p+1}(-1)^i (a_0,(A_{1,i}\shuffle \tilde A_{i,p+1}))+\sum_{i=1}^{p}(-1)^i (a_p,(a_0,A_{1,i}\shuffle \tilde A_{i,p}))\nonumber\\
&=- (a_0, \tilde A)+\sum_{i=2}^{p+1}(-1)^i (a_0,(A_{1,i}\shuffle \tilde A_{i,p+1})) -(\tilde A,a_0)\nonumber\\
&=- (a_0, \tilde A)-\sum_{i=2}^{p+1}(-1)^{i-1} (a_0,(a_1,A_{2,i}\shuffle \tilde A_{i,p+1})) -(\tilde A,a_0)\nonumber\\
&=- (a_0, \tilde A)+(a_0,\tilde A)-(\tilde A,a_0)\nonumber\\
&=-(\tilde A,a_0)\,.
\end{align}
%
Second, we use concatenation and the induction step to write 
\begin{align}
&\phiChain(r_0,a_0,A,r_1,B)\nonumber\\
&=\Omega_{r_0,a_0}(\eta_{a_0}+\eta_A+\eta_{r_1}+\eta_{B})\phiChain(a_0,A,r_1,B)\nonumber\\
&=\Omega_{r_0,a_0}(\eta_{a_0}+\eta_A+\eta_{r_1}+\eta_{B})\sum_{i=1 }^{p+1}(-1)^{p+1-i}\phiChain(a_0,A_{1i})\phiChain(a_0,r_1,\tilde{A}_{i,p+1}\shuffle B)
\end{align}
and apply the Fay identity 
\begin{align}
&\Omega_{r_0,a_0}(\eta_{a_0}+\eta_A+\eta_{r_1}+\eta_{B})\Omega_{a_0,r_1}(\eta_{a_i,\dots,a_p}+\eta_{r_1}+\eta_{B})\nonumber\\
&=\Omega_{r_0,a_0}(\eta_{a_0,\dots,a_{i-1}})\Omega_{r_0,r_1}(\eta_{a_i,\dots,a_p}+\eta_{r_1}+\eta_{B})\nonumber\\
&\phantom{=}-\Omega_{r_0,r_1}(\eta_{a_0}+\eta_A+\eta_{r_1}+\eta_{B})\Omega_{r_1,a_0}(\eta_{a_i,\dots,a_p}+\eta_{r_1}+\eta_{B})
\end{align}
to obtain 
\begin{align}
&\phiChain(r_0,a_0,A,r_1,B)\nonumber\\
&=\Omega_{r_0,a_0}(\eta_{a_0}+\eta_A+\eta_{r_1}+\eta_{B})\sum_{i=1 }^{p+1}(-1)^{p+1-i}\phiChain(a_0,A_{1i})\phiChain(a_0,r_1,\tilde{A}_{i,p+1}\shuffle B)\nonumber\\
&=\sum_{i=1 }^{p+1}(-1)^{p+1-i}\phiChain(r_0,a_0,A_{1i})\phiChain(r_0,r_1,\tilde{A}_{i,p+1}\shuffle B)\nonumber\\
&\phantom{=}-\Omega_{r_0,r_1}(\eta_{a_0}+\eta_A+\eta_{r_1}+\eta_{B})\sum_{i=1 }^{p+1}(-1)^{p+1-i}\phiChain(r_1,a_0,A_{1i})\phiChain(r_1,\tilde{A}_{i,p+1}\shuffle B)\nonumber\\
&=\sum_{i=1 }^{p+1}(-1)^{p+1-i}\phiChain(r_0,a_0,A_{1i})\phiChain(r_0,r_1,\tilde{A}_{i,p+1}\shuffle B)\nonumber\\
&\phantom{=}-\Omega_{r_0,r_1}(\eta_{a_0}+\eta_A+\eta_{r_1}+\eta_{B})\sum_{i=1 }^{p+1}(-1)^{p+1-i}\phiChain(r_1,(a_0,A_{1i})\shuffle \tilde{A}_{i,p+1}\shuffle B)\nonumber\\
&=\sum_{i=2 }^{p+2}(-1)^{p+2-i}\phiChain(r_0,A^0_{1i})\phiChain(r_0,r_1,\tilde{A}^0_{i,p+1}\shuffle B)\nonumber\\
&\phantom{=}+(-1)^{p+1}\Omega_{r_0,r_1}(\eta_{a_0}+\eta_A+\eta_{r_1}+\eta_{B})\phiChain(r_1,(\tilde{A},a_0)\shuffle B)\nonumber\\
&=\sum_{i=1 }^{p+2}(-1)^{p+2-i}\phiChain(r_0,A^0_{1i})\phiChain(r_0,r_1,\tilde{A}^0_{i,p+2}\shuffle B)\,,
\end{align}
where we have used \eqn{eqn:combinatorialStep} for the second equality from below. Thus, if the identity \eqref{eqn:shuffleIdentityChainWith1And0AlternatingSum} holds for $A$, it also holds for the longer sequence $A^0=(a_0,A)$, as shown by the calculation above, which proves its general validity by induction.

\section{Derivation of the \texorpdfstring{$n$}{n}-point \texorpdfstring{$z_0$}{z0}-derivative}\label{app:z0Deriv}
In this section, we derive the $n$-point formula for the $z_0$-derivative. The starting point is \eqn{eqn:partial0Z}, 
\begin{align}\label{eqn:z0DerivativeStart}
&\partial_0 \Ztzn{n}((1,A),(0,B))=\sum_{k=0}^p\sum_{j=0}^q\int_{ \gamma} \prod_{i=2}^n\dd z_i\,  \,\te{KN}^{\tau}_{01\dots n} \Big(s_{a_k,b_j}f^{(1)}_{a_k,b_j}\phiChain(1,A)\phiChain(0,B)\Big)\,,
\end{align}
where $A=(a_1,\dots, a_p)$ and $B=(b_1,\dots,b_q)$ are disjoint sequences without repetitions such that $A\cup B=\{2,3,\dots, n\}$, and the proof is split into three parts. Moreover,
we will write $\gamma$ in the place of $\gamma_{12\ldots n0}$ for the integration domain \eqref{shortdom}
throughout the appendices. First, we derive some preliminary identities which will be useful to rewrite the term
\begin{align}\label{eqn:z0DerivCrucialTerm}
f^{(1)}_{a_k,b_j}\phiChain(1,A)\phiChain(0,B)&=\left(\Omega_{a_k,b_j}(\xi)\phiChain(1,A)\phiChain(0,B)\right)|_{\xi^0}\,.
\end{align}
Using these identities, we then give the proof of the closed formula \eqref{eqn:derZ0Closed}.

Throughout this section, we accompany the crucial identities by the graphical notation for chains of Kronecker--Eisenstein series in order to facilitate the readability of the proofs for the $n$-point $z_0$- and $\tau$-derivatives.
\subsection{Preliminary identities}
Instead of only investigating the term \eqref{eqn:z0DerivCrucialTerm} with the factor $f^{(1)}_{a_k,b_j}$, we consider the corresponding generating series and, thus, the product of chains
\begin{equation}
\Omega_{a_k,b_j}(\xi)\phiChain(1,A)\phiChain(0,B)\quad=\quad\label{eqn:startzzero}
\mpostuse[align=b,vshift=-40pt]{startzzero}\,.
\end{equation}
In particular, we will use the following identities: for $k,j\neq 0$, i.e.\ $a_k\neq 1=a_0$ and $b_j\neq 0=b_0$,
\begin{align}\label{eqn:3OmegaIdentity}
&\Omega_{1,a_k}(\eta_a)\Omega_{a_k,b_j}(\xi)\Omega_{0,b_j}(\eta_b)\nonumber\\
&=\left(\Omega_{1,a_k}(\eta_a-\xi)\Omega_{1,b_j}(\xi)-\Omega_{b_j,a_k}(\eta_a-\xi)\Omega_{1,b_j}(\eta_a)\right)\Omega_{0,b_j}(\eta_b)\nonumber\\
&=\Omega_{1,a_k}(\eta_a-\xi)\left(\Omega_{1,b_j}(\eta_b+\xi)\Omega_{01}(\eta_b)+\Omega_{1,0}(\xi)\Omega_{0,b_j}(\eta_b+\xi)\right)\nonumber\\
&\phantom{=}-\Omega_{b_j,a_k}(\eta_a-\xi)\left(\Omega_{1,0}(\eta_a)\Omega_{0,b_j}(\eta_a+\eta_b)+\Omega_{1,b_j}(\eta_a+\eta_b)\Omega_{01}(\eta_b)\right)\nonumber\\
&=\left(\Omega_{1,a_k}(\eta_a-\xi)\Omega_{1,b_j}(\eta_b+\xi)-\Omega_{b_j,a_k}(\eta_a-\xi)\Omega_{1,b_j}(\eta_a+\eta_b)\right)\Omega_{01}(\eta_b)\nonumber\\
&\phantom{=}+\Omega_{1,a_k}(\eta_a-\xi)\Omega_{1,0}(\xi)\Omega_{0,b_j}(\eta_b+\xi)-\Omega_{b_j,a_k}(\eta_a-\xi)\Omega_{1,0}(\eta_a)\Omega_{0,b_j}(\eta_a+\eta_b)\nonumber\\
&=\Omega_{1,a_k}(\eta_a+\eta_b)\Omega_{a_k,b_j}(\eta_b+\xi)\Omega_{01}(\eta_b)\nonumber\\
&\phantom{=}+\Omega_{1,a_k}(\eta_a-\xi)\Omega_{1,0}(\xi)\Omega_{0,b_j}(\eta_b+\xi)-\Omega_{b_j,a_k}(\eta_a-\xi)\Omega_{1,0}(\eta_a)\Omega_{0,b_j}(\eta_a+\eta_b)\nonumber\\
&\phantom{=}+\Omega_{1,a_k}(\eta_a-\xi)\Omega_{1,0}(\xi)\Omega_{0,b_j}(\eta_b+\xi)-\Omega_{b_j,a_k}(\eta_a-\xi)\Omega_{1,0}(\eta_a)\Omega_{0,b_j}(\eta_a+\eta_b)\nonumber\\
&=\Omega_{01}(\eta_b)\Omega_{1,a_k}(\eta_a+\eta_b)\Omega_{a_k,b_j}(\eta_b+\xi)+\Omega_{01}(-\eta_a)\Omega_{0,b_j}(\eta_a+\eta_b)\Omega_{b_j,a_k}(\eta_a-\xi)\nonumber\\
&\phantom{=}+\Omega_{1,0}(\xi)\Omega_{1,a_k}(\eta_a-\xi)\Omega_{0,b_j}(\eta_b+\xi)\,,
\end{align}
which can be depicted as
\begin{equation}
\label{eqn:fundamentalIdentity}
\mpostuse[align=b,vshift=-14pt]{fundamentalIdentityLHS}\quad=\quad\mpostuse[align=b,vshift=-14pt]{fundamentalIdentityRHS1}\quad+\quad\mpostuse[align=b,vshift=-14pt]{fundamentalIdentityRHS2}\quad+\quad\mpostuse[align=b,vshift=-14pt]{fundamentalIdentityRHS3}\,,
\end{equation}
while for $k\neq 0$ and $j=0$
\begin{align}\label{eqn:3OmegaIdentityJ}
\Omega_{1,a_k}(\eta_a)\Omega_{a_k,0}(\xi)&=\Omega_{01}(-\eta_a)\Omega_{0,a_k}(\eta_a-\xi)+\Omega_{1,0}(\xi)\Omega_{1,a_k}(\eta_a- \xi)\,,
\end{align}
depicted by
\begin{equation}
\label{eqn:fundamentalIdentityJ}
\mpostuse[align=b,vshift=-7pt]{fundamentalIdentityJLHS}\quad=\quad\mpostuse[align=b,vshift=-14pt]{fundamentalIdentityJRHS1}\quad+\quad\mpostuse[align=b,vshift=-14pt]{fundamentalIdentityJRHS2}\,,
\end{equation}
and for $k= 0$ and $j\neq 0$
\begin{align}\label{eqn:3OmegaIdentityK}
\Omega_{1,b_j}(\xi)\Omega_{0,b_j}(\eta_b)&=\Omega_{01}(\eta_b)\Omega_{1,b_j}(\eta_b+\xi)+\Omega_{1,0}(\xi)\Omega_{0,b_j}(\eta_b+\xi)\,,
\end{align}
i.e.\
\begin{equation}
\label{eqn:fundamentalIdentityK}
\mpostuse[align=b,vshift=-7pt]{fundamentalIdentityKLHS}\quad=\quad\mpostuse[align=b,vshift=-14pt]{fundamentalIdentityKRHS1}\quad+\quad\mpostuse[align=b,vshift=-14pt]{fundamentalIdentityKRHS2}\,.
\end{equation}
The procedure to rewrite the $\Omega_{a_k,b_j}$ in \eqn{eqn:z0DerivCrucialTerm} is the following: first, we move the indices $a_k$ and $b_j$ in $\phiChain(1,A)\phiChain(0,B)$ next to 1 and 0, respectively, by means of eq.\ \eqref{eqn:shuffleIdentityChainWith1And0AlternatingSum}, which yields for $k\neq 0$
\begin{align}\label{eqn:ASum}
\phiChain(1,A)&=\phiChain(1,A_{1,k},a_k,A_{k+1,p+1})\nonumber\\
&=\sum_{i=1 }^{k}(-1)^{k-i}\phiChain(1,A_{1i})\phiChain(1,a_k,\tilde{A}_{i,k}\shuffle A_{k+1,p+1})\nonumber\\
&=\sum_{i=1 }^{k}(-1)^{k-i}\phiChain(1,A_{1i})\phiChain(a_k,\tilde{A}_{i,k}\shuffle A_{k+1,p+1})\Omega_{1,a_k}(\eta_{A_i,p+1})\,,
\end{align}
i.e.\ 
\begin{equation}
\label{eqn:phiA1ak}
\mpostuse[align=b,vshift=-40pt]{phiA}\quad=\quad\sum_{i=1}^{k}(-1)^{k-i}\quad\mpostuse[align=b,vshift=-40pt]{phi1ak}\,,
\end{equation}
and for $j\neq 0$
\begin{align}\label{eqn:BSum}
\phiChain(0,B)&=\phiChain(0,B_{1,j},b_j,B_{j+1,q+1})\nonumber\\
&=\sum_{l=1 }^{j}(-1)^{j-l}\phiChain(0,B_{1l})\phiChain(0,b_j,\tilde{B}_{l,j}\shuffle B_{j+1,q+1})\nonumber\\
&=\sum_{l=1 }^{j}(-1)^{j-l}\phiChain(0,B_{1l})\phiChain(b_j,\tilde{B}_{l,j}\shuffle B_{j+1,q+1})\Omega_{0,b_j}(\eta_{B_{l,q+1}})\,,
\end{align}
i.e.\ 
\begin{equation}
\label{eqn:phiB1ak}
\mpostuse[align=b,vshift=-40pt]{phiB}\quad=\quad\sum_{l=1}^{j}(-1)^{j-l}\quad\mpostuse[align=b,vshift=-40pt]{phi0bj}\,.
\end{equation}
As a consequence, for $k,j\neq 0$
\begin{align}\label{eqn:prelimEq}
&\Omega_{a_k,b_j}(\xi)\phiChain(1,A)\phiChain(0,B)\nonumber\\
&=\sum_{i=1 }^{k}\sum_{l=1 }^{j}(-1)^{k+j-i-l}\phiChain(1,A_{1i})\phiChain(a_k,\tilde{A}_{i,k}\shuffle A_{k+1,p+1})\phiChain(0,B_{1l})\phiChain(b_j,\tilde{B}_{l,j}\shuffle B_{j+1,q+1})\nonumber\\
&\phantom{=\sum_{i=1 }^{k}\sum_{l=1 }^{j}}\Omega_{1,a_k}(\eta_{A_i,p+1})\Omega_{a_k,b_j}(\xi)\Omega_{0,b_j}(\eta_{B_{l,q+1}})\,,
\end{align}
i.e.\
\begin{equation}
\mpostuse[align=b,vshift=-40pt]{startzzero}\quad=\quad\sum_{i=1 }^{k}\sum_{l=1 }^{j}(-1)^{k+j-i-l}\quad\mpostuse[align=b,vshift=-40pt]{phi1ak0bj}\,,
\end{equation}
for $k\neq 0,\ j=0$
\begin{align}\label{eqn:prelimEqJ}
&\Omega_{a_k,0}(\xi)\phiChain(1,A)\phiChain(0,B)\nonumber\\
&=\sum_{i=1 }^{k}(-1)^{k-i}\phiChain(1,A_{1i})\phiChain(a_k,\tilde{A}_{i,k}\shuffle A_{k+1,p+1})\phiChain(0,B)\Omega_{1,a_k}(\eta_{A_i,p+1})\Omega_{a_k,0}(\xi)
\end{align}
i.e.\
\begin{equation}
\label{eqn:phiA1akJ0}
\mpostuse[align=b,vshift=-40pt]{phiAphiBxiJ0}\quad=\quad\sum_{i=1}^{k}(-1)^{k-i}\quad\mpostuse[align=b,vshift=-40pt]{phi1ak0}\,,
\end{equation}
and for $k=0,\ j\neq 0$
\begin{align}\label{eqn:prelimEqK}
&\Omega_{a_k,b_j}(\xi)\phiChain(1,A)\phiChain(0,B)\nonumber\\
&=\sum_{l=1 }^{j}(-1)^{j-l}\phiChain(1,A)\phiChain(0,B_{1l})\phiChain(b_j,\tilde{B}_{l,j}\shuffle B_{j+1,q+1})\Omega_{1,b_j}(\xi)\Omega_{0,b_j}(\eta_{B_{l,q+1}})\,,
\end{align}
i.e.\
\begin{equation}
\label{eqn:phiB1k0}
\mpostuse[align=b,vshift=-40pt]{phiAphiBk0}\quad=\quad\sum_{l=1}^{j}(-1)^{j-l}\quad\mpostuse[align=b,vshift=-47pt]{phik0bj}\,.
\end{equation}
At this point, we can apply the identity \eqref{eqn:3OmegaIdentity} in \eqref{eqn:prelimEq}, \eqref{eqn:3OmegaIdentityJ} in \eqref{eqn:prelimEqJ} and \eqref{eqn:3OmegaIdentityK} in \eqref{eqn:prelimEqK} to the triangles or squares formed by the vertices $1$, $0$, $a_k$ and/or $b_j$ in the corresponding graphs, which leads to two (for $k=0$ or $j=0$) or three distinct sums. For $k,j\neq 0$, we obtain
\begin{align}\label{eqn:OmegaABCchains}
&\Omega_{a_k,b_j}(\xi)\phiChain(1,A)\phiChain(0,B)\nonumber\\
&=\sum_{i=1 }^{k}\sum_{l=1 }^{j}(-1)^{k+j-i-l}\phiChain(1,A_{1i})\phiChain(a_k,\tilde{A}_{i,k}\shuffle A_{k+1,p+1})\phiChain(0,B_{1l})\phiChain(b_j,\tilde{B}_{l,j}\shuffle B_{j+1,q+1})\nonumber\\
&\phantom{=\sum_{i=1 }^{k}\sum_{l=1 }^{j}}\Big(\Omega_{01}(\eta_{B_{l,q+1}})\Omega_{1,a_k}(\eta_{A_i,p+1}+\eta_{B_{l,q+1}})\Omega_{a_k,b_j}(\eta_{B_{l,q+1}}+\xi)\nonumber\\
&\phantom{=\sum_{i=1 }^{k}\sum_{l=1 }^{j}}+\Omega_{01}(-\eta_{A_i,p+1})\Omega_{0,b_j}(\eta_{A_i,p+1}+\eta_{B_{l,q+1}})\Omega_{b_j,a_k}(\eta_{A_i,p+1}-\xi)\nonumber\\
&\phantom{=\sum_{i=1 }^{k}\sum_{l=1 }^{j}}+\Omega_{1,0}(\xi)\Omega_{1,a_k}(\eta_{A_i,p+1}-\xi)\Omega_{0,b_j}(\eta_{B_{l,q+1}}+\xi)\Big)\,,
\end{align}
i.e.\
\begin{align}
&\mpostuse[align=b,vshift=-40pt]{startzzero}\quad=\quad\sum_{i=1 }^{k}\sum_{l=1 }^{j}(-1)^{k+j-i-l} \nonumber\\[8pt]
&\times \mpostuse[align=b,vshift=-40pt]{phi1akFullFactor}\left(
\mpostuse[align=b,vshift=-26pt]{fundamentalIdentityRHS1FullFactor}+
\mpostuse[align=b,vshift=-26pt]{fundamentalIdentityRHS2FullFactor}+
\mpostuse[align=b,vshift=-26pt]{fundamentalIdentityRHS3FullFactor}\right)
\mpostuse[align=b,vshift=-40pt]{phi0bjFullFactor}
\end{align}
The last sum, with the factor $\Omega_{1,0}(\xi)\Omega_{1,a_k}(\eta_{A_i,p+1}-\xi)\Omega_{0,b_j}(\eta_{B_{l,q+1}}+\xi) $, can be rewritten in terms of the original chains, using eqs.\ \eqref{eqn:ASum} and \eqref{eqn:BSum} in the reverse direction, leaving a shift of $\mp\xi$ in the variables $\eta_{a_k}$ and $\eta_{b_j}$, respectively,
\begin{align}
&\sum_{i=1 }^{k}\sum_{l=1 }^{j}(-1)^{k+j-i-l}\phiChain(1,A_{1i})\phiChain(a_k,\tilde{A}_{i,k}\shuffle A_{k+1,p+1})\phiChain(0,B_{1l})\phiChain(b_j,\tilde{B}_{l,j}\shuffle B_{j+1,q+1})\nonumber\\
&\phantom{=\sum_{i=1 }^{k}\sum_{l=1 }^{j}}\Omega_{1,0}(\xi)\Omega_{1,a_k}(\eta_{A_i,p+1}-\xi)\Omega_{0,b_j}(\eta_{B_{l,q+1}}+\xi)\nonumber\\
&=\Omega_{1,0}(\xi)\phiChain(1,A)|_{\eta_{a_k}\to \eta_{a_k}-\xi}\phiChain(0,B)|_{\eta_{b_j}\to \eta_{b_j}+\xi}\,.
\end{align}
The remaining two sums in eq.\ \eqref{eqn:OmegaABCchains} can either be written in terms of products of two chains, the first starting at 1 and the second at 0. This will yield the closed formula for the derivatives of $Z_{0,n}^{\tau}$. Or, they can be expressed in terms of the S-map. Both formul\ae{} are derived in the next two subsections.

To summarize, we have so far for $k,j\neq 0$ the identity 
\begin{align}\label{eqn:OmegaABCchainsTrivial}
&\Omega_{a_k,b_j}(\xi)\phiChain(1,A)\phiChain(0,B) =\Omega_{1,0}(\xi)\phiChain(1,A)|_{\eta_{a_k}\to \eta_{a_k}-\xi}\phiChain(0,B)|_{\eta_{b_j}\to \eta_{b_j}+\xi}\nonumber\\
&\phantom{=}+\sum_{i=1 }^{k}\sum_{l=1 }^{j}(-1)^{k+j-i-l}\phiChain(1,A_{1i})\phiChain(a_k,\tilde{A}_{i,k}\shuffle A_{k+1,p+1})\phiChain(0,B_{1l})\phiChain(b_j,\tilde{B}_{l,j}\shuffle B_{j+1,q+1})\nonumber\\
&\phantom{=\sum_{i=1 }^{k}\sum_{l=1 }^{j}}\Big(\Omega_{01}(\eta_{B_{l,q+1}})\Omega_{1,a_k}(\eta_{A_i,p+1}+\eta_{B_{l,q+1}})\Omega_{a_k,b_j}(\eta_{B_{l,q+1}}+\xi)\nonumber\\
&\phantom{=\sum_{i=1 }^{k}\sum_{l=1 }^{j}}+\Omega_{01}(-\eta_{A_i,p+1})\Omega_{0,b_j}(\eta_{A_i,p+1}+\eta_{B_{l,q+1}})\Omega_{b_j,a_k}(\eta_{A_i,p+1}-\xi)\Big)\,,
\end{align}
i.e.\ 
\begin{align}\label{eqn:graphicalMainEq}
&\mpostuse[align=b,vshift=-40pt]{startzzero}\quad=\quad\mpostuse[align=b,vshift=-49.6pt]{startzzeroXiDown}+\sum_{i=1 }^{k}\sum_{l=1 }^{j}(-1)^{k+j-i-l} \nonumber\\
&\times \left(\mpostuse[align=b,vshift=-40pt]{phi1ak0bj1}+\mpostuse[align=b,vshift=-40pt]{phi1ak0bj2} \right)\,,
\end{align}
for $k\neq 0, \ j=0$
\begin{align}\label{eqn:OmegaABCchainsTrivialJ}
&\Omega_{a_k,0}(\xi)\phiChain(1,A)\phiChain(0,B)= 
\Omega_{1,0}(\xi)\phiChain(1,A)|_{\eta_{a_k}\to \eta_{a_k}-\xi}\phiChain(0,B)\nonumber\\
&\phantom{=}+\sum_{i=1 }^{k}(-1)^{k-i}\phiChain(1,A_{1i})\phiChain(a_k,\tilde{A}_{i,k}\shuffle A_{k+1,p+1})\phiChain(0,B)\Omega_{01}(-\eta_{A_i,p+1})\Omega_{0,a_k}(\eta_{A_i,p+1}-\xi)\,,
\end{align}
i.e.\ 
\begin{equation}\label{eqn:graphicalMainEqj}
\mpostuse[align=b,vshift=-40pt]{phiAphiBxiJ0}\quad=\quad \mpostuse[align=b,vshift=-49.6pt]{startzzeroXiDownj0}\quad+\quad\sum_{i=1}^{k}(-1)^{k-i}\quad\mpostuse[align=b,vshift=-49.6pt]{phiak01}\,,
\end{equation}
and for $k=0,\ j\neq 0$
\begin{align}\label{eqn:OmegaABCchainsTrivialK}
&\Omega_{1,b_j}(\xi)\phiChain(1,A)\phiChain(0,B) =\Omega_{1,0}(\xi)\phiChain(1,A)\phiChain(0,B)|_{\eta_{b_j}\to \eta_{b_j}+\xi}\nonumber\\
&\phantom{=}+\sum_{l=1 }^{j}(-1)^{j-l}\phiChain(1,A)\phiChain(0,B_{1l})\phiChain(b_j,\tilde{B}_{l,j}\shuffle B_{j+1,q+1})\Omega_{01}(\eta_{B_{l,q+1}})\Omega_{1,b_j}(\eta_{B_{l,q+1}}+\xi)\,.
\end{align}
i.e.\
\begin{equation}\label{eqn:graphicalMainEqk}
\mpostuse[align=b,vshift=-40pt]{phiAphiBk0}\quad=\quad \mpostuse[align=b,vshift=-49.6pt]{startzzeroXiDownk0}\quad+\quad\sum_{l=1}^{j}(-1)^{j-l}\quad\mpostuse[align=b,vshift=-49.6pt]{phik0bj}\,.
\end{equation}

\subsection{Closed formula}\label{app:z0DerivClosed}
The closed-form expression (\ref{eqn:derZ0Closed}) is obtained from eqs.\ \eqref{eqn:OmegaABCchainsTrivial} to \eqref{eqn:OmegaABCchainsTrivialK} by multiple applications of the shuffle identity \eqref{eqn:shuffleProductChains} and the concatenation \eqref{eqn:chainConcatenation}. 

For $k,j\neq 0$, the first sum in eq.\ \eqref{eqn:OmegaABCchainsTrivial} is given by 
\begin{align}\label{eqn:OmegaABCchainsTrivialFirstSum}
&\sum_{i=1 }^{k}\sum_{l=1 }^{j}(-1)^{k+j-i-l}\phiChain(1,A_{1i})\phiChain(a_k,\tilde{A}_{i,k}\shuffle A_{k+1,p+1})\phiChain(0,B_{1l})\phiChain(b_j,\tilde{B}_{l,j}\shuffle B_{j+1,q+1})\nonumber\\
&\phantom{=\sum_{i=1 }^{k}\sum_{l=1 }^{j}}\Omega_{01}(\eta_{B_{l,q+1}})\Omega_{1,a_k}(\eta_{A_i,p+1}+\eta_{B_{l,q+1}})\Omega_{a_k,b_j}(\eta_{B_{l,q+1}}+\xi)\nonumber\\
&=\sum_{i=1 }^{k}\sum_{l=1 }^{j}(-1)^{k+j-i-l}\Omega_{01}(\eta_{B_{l,q+1}})\phiChain(0,B_{1l})\phiChain(1,A_{1i})\nonumber\\
&\phantom{=\sum_{i=1 }^{k}\sum_{l=1 }^{j}}\Omega_{1,a_k}(\eta_{A_i,p+1}+\eta_{B_{l,q+1}})\phiChain(a_k,\tilde{A}_{i,k}\shuffle A_{k+1,p+1})\phiChain(a_k,b_j,\tilde{B}_{l,j}\shuffle B_{j+1,q+1})|_{\eta_{b_j}\to \eta_{b_j}+\xi}\nonumber\\
&=\sum_{i=1 }^{k}\sum_{l=1 }^{j}(-1)^{k+j-i-l}\Omega_{01}(\eta_{B_{l,q+1}})\phiChain(0,B_{1l})\phiChain(1,A_{1i})\nonumber\\
&\phantom{=\sum_{i=1 }^{k}\sum_{l=1 }^{j}}\phiChain(1,a_k,(\tilde{A}_{i,k}\shuffle A_{k+1,p+1})\shuffle (b_j,\tilde{B}_{l,j}\shuffle B_{j+1,q+1}))|_{\eta_{a_k}\to \eta_{a_k}-\xi,\eta_{b_j}\to \eta_{b_j}+\xi}\nonumber\\
&=\sum_{i=1 }^{k}\sum_{l=1 }^{j}(-1)^{k+j-i-l}\Omega_{01}(\eta_{B_{l,q+1}})\phiChain(0,B_{1l})\nonumber\\
&\phantom{=\sum_{i=1 }^{k}\sum_{l=1 }^{j}}\phiChain(1,A_{1i}\shuffle(a_k,(\tilde{A}_{i,k}\shuffle A_{k+1,p+1})\shuffle (b_j,\tilde{B}_{l,j}\shuffle B_{j+1,q+1})))|_{\eta_{a_k}\to \eta_{a_k}-\xi,\eta_{b_j}\to \eta_{b_j}+\xi}
\end{align}
and similarly for the second sum 
\begin{align}\label{eqn:OmegaABCchainsTrivialSecondSum}
&\sum_{i=1 }^{k}\sum_{l=1 }^{j}(-1)^{k+j-i-l}\phiChain(1,A_{1i})\phiChain(a_k,\tilde{A}_{i,k}\shuffle A_{k+1,p+1})\phiChain(0,B_{1l})\phiChain(b_j,\tilde{B}_{l,j}\shuffle B_{j+1,q+1})\nonumber\\
&\phantom{=\sum_{i=1 }^{k}\sum_{l=1 }^{j}}+\Omega_{01}(-\eta_{A_i,p+1})\Omega_{0,b_j}(\eta_{A_i,p+1}+\eta_{B_{l,q+1}})\Omega_{b_j,a_k}(\eta_{A_i,p+1}-\xi)\nonumber\\
&=\sum_{i=1 }^{k}\sum_{l=1 }^{j}(-1)^{k+j-i-l}\Omega_{01}(-\eta_{A_{i,p+1}})\phiChain(1,A_{1i})\nonumber\\
&\phantom{=\sum_{i=1 }^{k}\sum_{l=1 }^{j}}\phiChain(0,B_{1l}\shuffle(b_j,(\tilde{B}_{l,j}\shuffle B_{j+1,q+1})\shuffle (a_k,\tilde{A}_{i,k}\shuffle A_{k+1,p+1})))|_{\eta_{a_k}\to \eta_{a_k}-\xi,\eta_{b_j}\to \eta_{b_j}+\xi}\,.
\end{align}
The sum in in eq.\ \eqref{eqn:OmegaABCchainsTrivialJ} for $k\neq 0, \ j= 0$ can be rewritten as 
\begin{align}\label{eqn:OmegaABCchainsTrivialJSum}
&\sum_{i=1 }^{k}(-1)^{k-i}\phiChain(1,A_{1i})\phiChain(a_k,\tilde{A}_{i,k}\shuffle A_{k+1,p+1})\phiChain(0,B)\Omega_{01}(-\eta_{A_i,p+1})\Omega_{0,a_k}(\eta_{A_i,p+1}-\xi)\nonumber\\
&=\sum_{i=1 }^{k}(-1)^{k-i}\Omega_{01}(-\eta_{A_i,p+1})\phiChain(1,A_{1i})\phiChain(0,a_k,\tilde{A}_{i,k}\shuffle A_{k+1,p+1})|_{\eta_{a_k}\to \eta_{a_k}-\xi}\phiChain(0,B)\nonumber\\
&=\sum_{i=1 }^{k}(-1)^{k-i}\Omega_{01}(-\eta_{A_i,p+1})\phiChain(1,A_{1i})\phiChain(0,B\shuffle (a_k,\tilde{A}_{i,k}\shuffle A_{k+1,p+1}))|_{\eta_{a_k}\to \eta_{a_k}-\xi}
\end{align}
and, similarly, the one in eq.\ \eqref{eqn:OmegaABCchainsTrivialK} for $k= 0, \ j\neq 0$ as follows
\begin{align}\label{eqn:OmegaABCchainsTrivialKSum}
&\sum_{l=1 }^{j}(-1)^{j-l}\phiChain(1,A)\phiChain(0,B_{1l})\phiChain(b_j,\tilde{B}_{l,j}\shuffle B_{j+1,q+1})\Omega_{01}(\eta_{B_{l,q+1}})\Omega_{1,b_j}(\eta_{B_{l,q+1}}+\xi)\nonumber\\
&=\sum_{l=1 }^{j}(-1)^{j-l}\Omega_{01}(\eta_{B_{l,q+1}})\phiChain(0,B_{1l})\phiChain(1,A\shuffle (b_j,\tilde{B}_{l,j}\shuffle B_{j+1,q+1}))|_{\eta_{b_j}\to \eta_{b_j}+\xi}\, .
\end{align}
Plugging eqs.\ \eqref{eqn:OmegaABCchainsTrivialFirstSum} and \eqref{eqn:OmegaABCchainsTrivialSecondSum} into \eqn{eqn:OmegaABCchainsTrivial}, we can conclude that for $k,j\neq 0$
\begin{align}\label{eqn:OmegaABCchainsClosedFormula}
&\Omega_{a_k,b_j}(\xi)\phiChain(1,A)\phiChain(0,B) =\Omega_{1,0}(\xi)\phiChain(1,A)|_{\eta_{a_k}\to \eta_{a_k}-\xi}\phiChain(0,B)|_{\eta_{b_j}\to \eta_{b_j}+\xi}\nonumber\\
&\phantom{=}+\sum_{i=1 }^{k}\sum_{l=1 }^{j}(-1)^{k+j-i-l}\Omega_{01}(\eta_{B_{l,q+1}})\phiChain(0,B_{1l})\nonumber\\
&\phantom{=\sum_{i=1 }^{k}\sum_{l=1 }^{j}}\phiChain(1,A_{1i}\shuffle(a_k,(\tilde{A}_{i,k}\shuffle A_{k+1,p+1})\shuffle (b_j,\tilde{B}_{l,j}\shuffle B_{j+1,q+1})))|_{\eta_{a_k}\to \eta_{a_k}-\xi,\eta_{b_j}\to \eta_{b_j}+\xi}\nonumber\\
&\phantom{=}+\sum_{i=1 }^{k}\sum_{l=1 }^{j}(-1)^{k+j-i-l}\Omega_{01}(-\eta_{A_{i,p+1}})\phiChain(1,A_{1i})\nonumber\\
&\phantom{=\sum_{i=1 }^{k}\sum_{l=1 }^{j}}\phiChain(0,B_{1l}\shuffle(b_j,(\tilde{B}_{l,j}\shuffle B_{j+1,q+1})\shuffle (a_k,\tilde{A}_{i,k}\shuffle A_{k+1,p+1})))|_{\eta_{a_k}\to \eta_{a_k}-\xi,\eta_{b_j}\to \eta_{b_j}+\xi}\,,
\end{align}
the corresponding graphical equation is obtained from \eqn{eqn:graphicalMainEq} by simply folding back any branch fork to a sum of single chains using the shuffle product, while all the $\eta$-variables and the corresponding shifts $\pm \xi$ stay the same (and will not be depicted in the following equation for notational simplicity)
\begin{align}\label{eqn:graphMainResult}
&\mpostuse[align=b,vshift=-40pt]{startzzero}\quad=\quad\mpostuse[align=b,vshift=-49.6pt]{startzzeroXiDown}+\sum_{i=1 }^{k}\sum_{l=1 }^{j}(-1)^{k+j-i-l} \nonumber\\[4pt]
&\times \left(\mpostuse[align=b,vshift=-40pt]{phi1ak0bj1Chain}+\mpostuse[align=b,vshift=-40pt]{phi1ak0bj2Chain}\right)\,,
\end{align}
and similarly using \eqn{eqn:OmegaABCchainsTrivialJSum} in \eqn{eqn:OmegaABCchainsTrivialJ} yields for $k\neq 0,\ j=0$
\begin{align}\label{eqn:OmegaABCchainsClosedFormulaJ}
&\Omega_{a_k,0}(\xi)\phiChain(1,A)\phiChain(0,B) =\Omega_{1,0}(\xi)\phiChain(1,A)|_{\eta_{a_k}\to \eta_{a_k}-\xi}\phiChain(0,B)\nonumber\\
&\phantom{=}+\sum_{i=1 }^{k}(-1)^{k-i}\Omega_{01}(-\eta_{A_i,p+1})\phiChain(1,A_{1i})\phiChain(0,B\shuffle (a_k,\tilde{A}_{i,k}\shuffle A_{k+1,p+1}))|_{\eta_{a_k}\to \eta_{a_k}-\xi}\,,
\end{align}
which is also obtained from the graphs in \eqn{eqn:graphicalMainEqj} by folding back any branches to a single chain using the shuffle product
\begin{equation}
\mpostuse[align=b,vshift=-40pt]{phiAphiBxiJ0}\quad =\quad \mpostuse[align=b,vshift=-49.6pt]{startzzeroXiDownj0}\quad+\quad\sum_{i=1}^{k}(-1)^{k-i}\quad\mpostuse[align=b,vshift=-49.6pt]{phiak01Chain}\,,
\end{equation}
and eq.\ \eqref{eqn:OmegaABCchainsTrivialK} in \eqn{eqn:OmegaABCchainsTrivialKSum} for $k=0,\ j\neq 0$
\begin{align}\label{eqn:OmegaABCchainsClosedFormulaK}
&\Omega_{1,b_j}(\xi)\phiChain(1,A)\phiChain(0,B) =\Omega_{1,0}(\xi)\phiChain(1,A)\phiChain(0,B)|_{\eta_{b_j}\to \eta_{b_j}+\xi}\nonumber\\
&\phantom{=}+\sum_{l=1 }^{j}(-1)^{j-l}\Omega_{01}(\eta_{B_{l,q+1}})\phiChain(0,B_{1l})\phiChain(1,A\shuffle (b_j,\tilde{B}_{l,j}\shuffle B_{j+1,q+1}))|_{\eta_{b_j}\to \eta_{b_j}+\xi}\,,
\end{align}
i.e.\ \eqn{eqn:graphicalMainEqk} folded back
\begin{equation}
\mpostuse[align=b,vshift=-40pt]{phiAphiBk0}\quad=\quad \mpostuse[align=b,vshift=-49.6pt]{startzzeroXiDownk0}+\quad\sum_{l=1}^{j}(-1)^{j-l}\quad\mpostuse[align=b,vshift=-49.6pt]{phik0bjChain}\,.
\end{equation}

The above formul\ae{} \eqref{eqn:OmegaABCchainsClosedFormula} to \eqref{eqn:OmegaABCchainsClosedFormulaK} yield the closed expressions for the partial derivatives of $\Ztzn{n}((1,A),(0,B))$, since they contain two chains beginning at zero and one. For the $z_0$-derivative, we can continue from \eqn{eqn:z0DerivativeStart}
using the fact that $f^{(1)}_{a_k,b_j}$ is the order zero $\xi^0$-term of $\Omega_{a_k,b_j}(\xi)$, 
see \eqn{eqn:z0DerivCrucialTerm}.
Thus, eqs.\ \eqref{eqn:OmegaABCchainsClosedFormula} to \eqref{eqn:OmegaABCchainsClosedFormulaK} can be applied to extract the $\xi^0$-term in
\begin{align}
&\partial_0 \Ztzn{n}((1,A),(0,B))=\sum_{k=0}^p\sum_{j=0}^q s_{a_k,b_j}\int_{\gamma} \prod_{i=2}^n\dd z_i\,  \,\te{KN}^{\tau}_{01\dots n} \Big(\Omega_{a_k,b_j}(\xi)\phiChain(1,A)\phiChain(0,B)\Big)|_{\xi^0}\,.
\label{againd0}
\end{align}
The non-trivial extraction of the $\xi^0$-term occurs in the term proportional to $\Omega_{1,0}(\xi)$ in eqs.\ \eqref{eqn:OmegaABCchainsClosedFormula} to \eqref{eqn:OmegaABCchainsClosedFormulaK}, i.e.\ $\Omega_{1,0}(\xi)\phiChain(1,A)\phiChain(0,B)$ with some shifts in the $\eta$-variables, and can be treated using the expansion 
\begin{align}
\Omega_{1,0}(\xi)&=\frac{1}{\xi}+f^{(1)}_{1,0}+\mathcal{O}(\xi)
\end{align}
in $\xi$ of the single factors to project out differential operators from the chains. The extraction in the remaining terms can be done by (expanding factors of the form $\Omega_{ij}(\eta+\xi)$ around $\xi$ and) simply setting $\xi=0$. For a single factor of the Kronecker--Eisenstein series, we find for $k\neq 0$
\begin{align}\label{eqn:projectOutDerivative}
&\left(\Omega_{1,0}(\xi)\Omega_{a_{k-1},a_{k}}(\eta_{a_k}-\xi)\right)|_{\xi^0}\nonumber\\
&=\left(\left(\frac{1}{\xi}+f^{(1)}_{1,0}+\mathcal{O}(\xi)\right)\left(\Omega_{a_{k-1},a_{k}}(\eta_{a_k})-\xi \partial_{\eta_{a_k}}\Omega_{a_{k-1},a_{k}}(\eta_{a_k})+\mathcal{O}(\xi)\right)\right)|_{\xi^0}\nonumber\\
&=-\left(f^{(1)}_{01}+\partial_{\eta_{a_k}}\right)\Omega_{a_{k-1},a_{k}}(\eta_{a_k})\,.
\end{align}
For a whole product, i.e.\ a chain, this procedure exactly reproduces the product rule due to the cross-terms and leads to
\begin{align}
&\Big(\Omega_{1,0}(\xi)\phiChain(1,A)|_{\eta_{a_k}\to \eta_{a_k}-\xi}\phiChain(0,B)|_{\eta_{b_j}\to \eta_{b_j}+\xi}\Big)|_{\xi^0}\nonumber\\
&=-\left(f^{(1)}_{01}+\partial_{\eta_{a_k}}-\partial_{\eta_{b_j}}\right)\phiChain(1,A)\phiChain(0,B)
\end{align}
for $k,j\neq 0$. Putting everything together, one indeed arrives at the closed formula  \eqn{eqn:derZ0Closed} for 
$\partial_0 \Ztzn{n}((1,A),(0,B))$.

\subsection{\texorpdfstring{$S$}{S}-map formula}\label{app:z0DerivSMap}
Alternatively, the derivatives of  $ \Ztzn{n}((1,A),(0,B))$ can be expressed in terms of the $S$-map as in \eqn{eqn:derZ0SMap}. For this purpose, we proceed from eq.\ \eqref{eqn:OmegaABCchainsTrivial} for $k,j\neq 0$ again using the shuffle identity \eqref{eqn:shuffleProductChains} and the reflection property \eqref{eqn:reflectionChain}, but slightly differently than in the previous subsection: we want to keep a factor of $\Omega_{a_k,b_j}(\eta_{B_{l,q+1}}+\xi)$ or $\Omega_{b_j,a_k}(\eta_{A_i,p+1}-\xi)$ as a bridge between the chains of $A$ and $B$. For the first sum in eq.\ \eqref{eqn:OmegaABCchainsTrivial}, the calculation amounts to
\begin{align}
&\sum_{i=1 }^{k}\sum_{l=1 }^{j}(-1)^{k+j-i-l}\phiChain(1,A_{1i})\phiChain(a_k,\tilde{A}_{i,k}\shuffle A_{k+1,p+1})\phiChain(0,B_{1l})\phiChain(b_j,\tilde{B}_{l,j}\shuffle B_{j+1,q+1})\nonumber\\
&\phantom{=\sum_{i=1 }^{k}\sum_{l=1 }^{j}}\Omega_{01}(\eta_{B_{l,q+1}})\Omega_{1,a_k}(\eta_{A_i,p+1}+\eta_{B_{l,q+1}})\Omega_{a_k,b_j}(\eta_{B_{l,q+1}}+\xi)\nonumber\\
&=\sum_{l=1 }^{j}(-1)^{j-l}\Omega_{01}(\eta_{B_{l,q+1}})\phiChain(0,B_{1l})\phiChain(a_k,b_j,\tilde{B}_{l,j}\shuffle B_{j+1,q+1})|_{\eta_{b_j}\to \eta_{b_j}+\xi}\nonumber\\
&\phantom{=\sum_{i=1 }^{k}}\sum_{i=1 }^{k}(-1)^{k-i}\phiChain(1, A_{1,i})\Omega_{1,a_k}(\eta_{A_i,p+1}+\eta_{B_{l,q+1}})\phiChain(a_k,\tilde{A}_{i,k})\phiChain(a_k, A_{k+1,p+1})\nonumber\\
&=\sum_{l=1 }^{j}(-1)^{j-l}\Omega_{01}(\eta_{B_{l,q+1}})\phiChain(0,B_{1l})\phiChain(a_k,b_j,\tilde{B}_{l,j}\shuffle B_{j+1,q+1})|_{\eta_{b_j}\to \eta_{b_j}+\xi}\nonumber\\
&\phantom{=\sum_{i=1 }^{k}}\sum_{i=1 }^{k}(-1)^{k-i}\phiChain(1, A_{1,i})\phiChain(1,a_k,\tilde{A}_{i,k})|_{\eta_{a_k}\to \eta_{A_{k,p+1}}+\eta_{B_{l,q+1}}}\phiChain(a_k, A_{k+1,p+1})\nonumber\\
&=\sum_{l=1 }^{j}(-1)^{j-l}\Omega_{01}(\eta_{B_{l,q+1}})\phiChain(0,B_{1l})\phiChain(a_k,b_j,\tilde{B}_{l,j}\shuffle B_{j+1,q+1})|_{\eta_{b_j}\to \eta_{b_j}+\xi}\nonumber\\
&\phantom{=\sum_{i=1 }^{k}}(-1)^{p-k}\phiChain(1, A_{1,k},a_k)|_{\eta_{a_k}\to \eta_{A_{k,p+1}}+\eta_{B_{l,q+1}}}\phiChain(\tilde A_{k+1,p+1},a_k)|_{\eta_{a_k}\to -\eta_{A_{k+1,p+1}}}\nonumber\\
&=\sum_{l=1 }^{j}(-1)^{p-k+j-l}\Omega_{01}(\eta_{B_{l,q+1}})\phiChain(0,B_{1l})\phiChain(a_k,b_j,\tilde{B}_{l,j}\shuffle B_{j+1,q+1})|_{\eta_{b_j}\to \eta_{b_j}+\xi}\nonumber\\
&\phantom{=\sum_{i=1 }^{k}}\phiChain((1, A_{1,k})\shuffle \tilde A_{k+1,p+1},a_k)|_{\eta_{a_k}\to \eta_{a_{k}}+\eta_{B_{l,q+1}}}
\nonumber\\
&=\sum_{l=1 }^{j}(-1)^{p-k+j-l}\Omega_{01}(\eta_{B_{l,q+1}})\phiChain(0,B_{1l})\nonumber\\
&\phantom{=\sum_{i=1 }^{k}}\phiChain((1, A_{1,k})\shuffle \tilde A_{k+1,p+1},a_k,b_j,\tilde{B}_{l,j}\shuffle B_{j+1,q+1})|_{\eta_{a_k}\to \eta_{a_k}-\xi, \eta_{b_j}\to \eta_{b_j}+\xi}\,,
\end{align}
where we applied the shuffle identity \eqref{eqn:shuffleProductChains} and \eqn{eqn:shuffleIdentityChainWith1And0AlternatingSum} in the reverse direction in the first and third equality, respectively. The same calculation leads to a similar result for the second sum in eq.\ \eqref{eqn:OmegaABCchainsTrivial}
\begin{align}
&\sum_{i=1 }^{k}\sum_{l=1 }^{j}(-1)^{k+j-i-l}\phiChain(1,A_{1i})\phiChain(a_k,\tilde{A}_{i,k}\shuffle A_{k+1,p+1})\phiChain(0,B_{1l})\phiChain(b_j,\tilde{B}_{l,j}\shuffle B_{j+1,q+1})\nonumber\\
&\phantom{=\sum_{i=1 }^{k}\sum_{l=1 }^{j}}\Omega_{01}(-\eta_{A_i,p+1})\Omega_{0,b_j}(\eta_{A_i,p+1}+\eta_{B_{l,q+1}})\Omega_{b_j,a_k}(\eta_{A_i,p+1}-\xi)\nonumber\\
&=\sum_{i=1}^k(-1)^{q-j+k-i}\Omega_{01}(-\eta_{A_i,p+1})\phiChain(1,A_{1i})\nonumber\\
&\phantom{=\sum_{i=1 }^{k}}\phiChain((0, B_{1,j})\shuffle \tilde B_{j+1,q+1},b_j,a_k,\tilde{A}_{i,k}\shuffle A_{k+1,p+1})|_{\eta_{a_k}\to \eta_{a_k}-\xi, \eta_{b_j}\to \eta_{b_j}+\xi}\,.
\end{align}
Similarly, we find for the sum in \eqref{eqn:OmegaABCchainsTrivialJ} for $k\neq 0,\ j=0$
\begin{align}
&\sum_{i=1 }^{k}(-1)^{k-i}\phiChain(1,A_{1i})\phiChain(a_k,\tilde{A}_{i,k}\shuffle A_{k+1,p+1})\phiChain(0,B)\Omega_{01}(-\eta_{A_i,p+1})\Omega_{0,a_k}(\eta_{A_i,p+1}-\xi)\nonumber\\
&=\sum_{i=1 }^{k}(-1)^{q+k-i}\phiChain(\tilde B,0,1,A_{1i})|_{\eta_{0}\to \eta_{0}+A_{i,p+1}}\phiChain(0,a_k,\tilde{A}_{i,k}\shuffle A_{k+1,p+1})|_{\eta_{a_k}\to \eta_{a_k}-\xi}\nonumber\\
&=\sum_{i=1 }^{k}(-1)^{q+k-i}\phiChain(0,1,A_{1i})\phiChain(\tilde B,0,a_k,\tilde{A}_{i,k}\shuffle A_{k+1,p+1})|_{\eta_{a_k}\to \eta_{a_k}-\xi,\eta_0\to \eta_0+\xi}\nonumber\\
&=\sum_{i=1}^k(-1)^{q+k-i}\Omega_{01}(-\eta_{A_i,p+1})\phiChain(1,A_{1i})\nonumber\\
&\phantom{=\sum_{i=1 }^{k}}\phiChain(\tilde B,0,a_k,\tilde{A}_{i,k}\shuffle A_{k+1,p+1})|_{\eta_{a_k}\to \eta_{a_k}-\xi, \eta_{0}\to \eta_{0}+\xi}
\end{align}
and for the sum in eq.\ \eqref{eqn:OmegaABCchainsTrivialK} with $k=0,\ j\neq 0$
\begin{align}
&\sum_{l=1 }^{j}(-1)^{j-l}\phiChain(1,A)\phiChain(0,B_{1l})\phiChain(b_j,\tilde{B}_{l,j}\shuffle B_{j+1,q+1})\Omega_{01}(\eta_{B_{l,q+1}})\Omega_{1,b_j}(\eta_{B_{l,q+1}}+\xi)\nonumber\\
&=\sum_{l=1 }^{j}(-1)^{p+j-l}\Omega_{01}(\eta_{B_{l,q+1}})\phiChain(0,B_{1l})\phiChain( \tilde A,1,b_j,\tilde{B}_{l,j}\shuffle B_{j+1,q+1})|_{\eta_{1}\to \eta_{1}-\xi, \eta_{b_j}\to \eta_{b_j}+\xi}\,.
\end{align}
Thus, the identity \eqref{eqn:OmegaABCchainsTrivial}, valid for $k,j\neq 0$ can be rewritten as follows 
\begin{align}
&\Omega_{a_k,b_j}(\xi)\phiChain(1,A)\phiChain(0,B) =\Omega_{1,0}(\xi)\phiChain(1,A)|_{\eta_{a_k}\to \eta_{a_k}-\xi}\phiChain(0,B)|_{\eta_{b_j}\to \eta_{b_j}+\xi}\nonumber\\
&\phantom{=}+\sum_{l=1 }^{j}(-1)^{p+1-k+j-l}\Omega_{01}(\eta_{B_{l,q+1}})\phiChain(0,B_{1l})\nonumber\\
&\phantom{=\sum_{i=1 }^{k}}\phiChain((1, A_{1,k})\shuffle \tilde A_{k+1,p+1},a_k,b_j,\tilde{B}_{l,j}\shuffle B_{j+1,q+1})|_{\eta_{a_k}\to \eta_{a_k}-\xi, \eta_{b_j}\to \eta_{b_j}+\xi}\nonumber\\
&\phantom{=}+\sum_{i=1}^k(-1)^{q+1-j+k-i}\Omega_{01}(-\eta_{A_i,p+1})\phiChain(1,A_{1i})\nonumber\\
&\phantom{=\sum_{i=1 }^{k}}\phiChain((0, B_{1,j})\shuffle \tilde B_{j+1,q+1},b_j,a_k,\tilde{A}_{i,k}\shuffle A_{k+1,p+1})|_{\eta_{a_k}\to \eta_{a_k}-\xi, \eta_{b_j}\to \eta_{b_j}+\xi}\,.
\end{align}
Graphically this equation is obtained from \eqn{eqn:graphicalMainEq} by absorbing the sum over $i$ into the second term and the sum over $j$ into the third term, to connect the vertices $a_i$ with $a_{i-1}$ and $b_l$ with $b_{l-1}$, respectively, leading to
\begin{align}
\mpostuse[align=b,vshift=-40pt]{startzzero}&=\mpostuse[align=b,vshift=-49.6pt]{startzzeroXiDown} \nonumber\\
&\phantom{=}+ \sum_{l=1 }^{j}(-1)^{p+1-k+j-l}\mpostuse[align=b,vshift=-40pt]{phi1ak0bj1SMap}\nonumber\\
&\phantom{=}+ \sum_{i=1 }^{k}(-1)^{q+1-j+k-i}\mpostuse[align=b,vshift=-40pt]{phi1ak0bj2SMap}\,.
\end{align}
Similarly, the identity \eqref{eqn:OmegaABCchainsTrivialJ}
for $k\neq 0,\ j=0$ is given by 
\begin{align}
&\Omega_{a_k,0}(\xi)\phiChain(1,A)\phiChain(0,B)=\Omega_{1,0}(\xi)\phiChain(1,A)|_{\eta_{a_k}\to \eta_{a_k}-\xi}\phiChain(0,B)\nonumber\\
&\phantom{=}+\sum_{i=1}^k(-1)^{q+k-i}\Omega_{01}(-\eta_{A_i,p+1})\phiChain(1,A_{1i})\phiChain(\tilde B,0,a_k,\tilde{A}_{i,k}\shuffle A_{k+1,p+1})|_{\eta_{a_k}\to \eta_{a_k}-\xi, \eta_{0}\to \eta_{0}+\xi}\,,
\end{align}
i.e.\ obtained from \eqn{eqn:graphicalMainEqj} by reverting the chain $\phiChain(0,B)$
\begin{equation}
\mpostuse[align=b,vshift=-40pt]{phiAphiBxiJ0}\quad=\quad \mpostuse[align=b,vshift=-49.6pt]{startzzeroXiDownj0}+\sum_{i=1}^{k}(-1)^{q+k-i}\quad\mpostuse[align=b,vshift=-49.6pt]{phiak01SMap}\,,
\end{equation}
and eq.\ \eqref{eqn:OmegaABCchainsTrivialK} for $k=0,\ j\neq 0$ by
\begin{align}
&\Omega_{1,b_j}(\xi)\phiChain(1,A)\phiChain(0,B) =\Omega_{1,0}(\xi)\phiChain(1,A)\phiChain(0,B)|_{\eta_{b_j}\to \eta_{b_j}+\xi}\nonumber\\
&\phantom{=}+\sum_{l=1 }^{j}(-1)^{p+j-l}\Omega_{01}(\eta_{B_{l,q+1}})\phiChain(0,B_{1l})\phiChain( \tilde A,1,b_j,\tilde{B}_{l,j}\shuffle B_{j+1,q+1})|_{\eta_{1}\to \eta_{1}-\xi, \eta_{b_j}\to \eta_{b_j}+\xi}\,,
\end{align}
i.e.\ 
\begin{equation}
\mpostuse[align=b,vshift=-40pt]{phiAphiBk0}\quad=\quad \mpostuse[align=b,vshift=-49.6pt]{startzzeroXiDownk0}\quad+\quad\sum_{l=1}^{j}(-1)^{p+j-l}\quad\mpostuse[align=b,vshift=-49.6pt]{phik0bjSMap}\,.
\end{equation}
Note that these three identities indeed all have a factor of $\Omega_{a_k,b_j}$, which is the backbone in the formulation in terms of the $S$-map. This can be seen by summing these identities over all $0\leq k\leq p$ and $0\leq j\leq q$ 
\begin{align}\label{eqn:SumShiftedSMapRep}
&\sum_{k=0}^p\sum_{j=0}^q s_{a_k,b_j}\Omega_{a_k,b_j}(\xi)\phiChain(1,A)\phiChain(0,B)\nonumber\\
&=\sum_{k=0}^p\sum_{j=0}^q s_{a_k,b_j}\Omega_{1,0}(\xi)\phiChain(1,A)|_{\eta_{a_k}\to \eta_{a_k}-\xi}\phiChain(0,B)|_{\eta_{b_j}\to \eta_{b_j}+\xi}\nonumber\\
&\phantom{=}+\sum_{k=0}^p\sum_{j=1}^q s_{a_k,b_j}\sum_{l=1 }^{j}(-1)^{p-k+j-l}\Omega_{01}(\eta_{B_{l,q+1}})\phiChain(0,B_{1l})\nonumber\\
&\phantom{=\sum_{i=1 }^{k}}\phiChain( A_{0,k}\shuffle \tilde A_{k+1,p+1},a_k,b_j,\tilde{B}_{l,j}\shuffle B_{j+1,q+1})|_{\eta_{a_k}\to \eta_{a_k}-\xi, \eta_{b_j}\to \eta_{b_j}+\xi}\nonumber\\
&\phantom{=}+\sum_{k=1}^p\sum_{j=0}^q s_{a_k,b_j}\sum_{i=1}^k(-1)^{q-j+k-i}\Omega_{01}(-\eta_{A_i,p+1})\phiChain(1,A_{1i})\nonumber\\
&\phantom{=\sum_{i=1 }^{k}}\phiChain( B_{0,j}\shuffle \tilde B_{j+1,q+1},b_j,a_k,\tilde{A}_{i,k}\shuffle A_{k+1,p+1})|_{\eta_{a_k}\to \eta_{a_k}-\xi, \eta_{b_j}\to \eta_{b_j}+\xi}\nonumber\\
&=\left(\sum_{k=0}^p\sum_{j=0}^q s_{a_k,b_j}\Omega_{1,0}(\xi)\phiChain(1,A)\phiChain(0,B)\right)|_{\eta_{a_k}\to \eta_{a_k}-\xi,\eta_{b_j}\to \eta_{b_j}+\xi}\nonumber\\
&\phantom{=}+\sum_{l=1 }^{q}\Omega_{01}(\eta_{B_{l,q+1}})\phiChain(0,B_{1l})\sum_{k=0}^p\sum_{j=l}^q(-1)^{p-k+j-l} s_{a_k,b_j}\nonumber\\
&\phantom{=\sum_{i=1 }^{k}}\phiChain( A_{0,k}\shuffle \tilde A_{k+1,p+1},a_k,b_j,\tilde{B}_{l,j}\shuffle B_{j+1,q+1})|_{\eta_{a_k}\to \eta_{a_k}-\xi, \eta_{b_j}\to \eta_{b_j}+\xi}\nonumber\\
&\phantom{=}+\sum_{i=1}^p\Omega_{01}(-\eta_{A_i,p+1})\phiChain(1,A_{1i})\sum_{k=i}^p\sum_{j=0}^q(-1)^{q-j+k-i} s_{a_k,b_j}\nonumber\\
&\phantom{=\sum_{i=1 }^{k}}\phiChain( B_{0,j}\shuffle \tilde B_{j+1,q+1},b_j,a_k,\tilde{A}_{i,k}\shuffle A_{k+1,p+1})|_{\eta_{a_k}\to \eta_{a_k}-\xi, \eta_{b_j}\to \eta_{b_j}+\xi}\,.
\end{align}
Comparing with \eqn{eq3.6}, we see that the second and third sum explicitly involve the definition of the $S$-map, except for the shift $\mp \xi$ in the variables $\eta_{a_k}$ and $\eta_{b_j}$. 

In order to obtain the $S$-map representation of the $z_0$-derivative \eqn{againd0}
we simply need to extract the $\xi^0$ part of eq.\ \eqref{eqn:SumShiftedSMapRep}, where we immediately recover the $S$-map formula \eqref{eqn:derZ0SMap}. 

\section{Derivation of the \texorpdfstring{$n$-point $\tau$-derivative}{n-point tau-derivative}}\label{appTauDerivative}
In this section, we determine the action of $2\pi i \partial_{\tau}$ on the integrals $\Ztzn{n}$ in \eqn{eqn:notationChain}
to derive the corresponding formul\ae{} \eqref{eqn:tauDerivSMap} and \eqref{eqn:tauDerivClosed}. The techniques in
this appendix generalize those in \cite{Mafra:2019xms, Gerken:2019cxz}, where the $\tau$-derivatives of $Z^\tau$-integrals without augmentation were studied\footnote{See in particular section 4 and appendix A of the first
reference in \rcite{Mafra:2019xms} as well as
section 4 and appendix E of \rcite{Gerken:2019cxz}.}.
First, we recall that the $\tau$-derivative of the Koba--Nielsen factor is given by \eqn{dtauofKN}.
Second, the action (up to integration by parts) on a chain
\begin{align}
\phiChain(C)&=\prod_{i=2}^m\Omega_{c_{i-1,i}}(\eta_{c_i\dots c_m}) \, , \ \ \ \ \ \ 
\eta_{c_i\dots c_m}=\sum_{j=i}^m \eta_{c_j}\,,
\end{align}
with $C=(c_1,c_2,\dots, c_m)$ can be expressed using the mixed heat equation 
\begin{align}
2\pi i \partial_{ \tau } \Omega_{ij}(\eta)&=\partial_i\partial_{\eta}\Omega_{ij}(\eta)=-\partial_j\partial_{\eta}\Omega_{ij}(\eta)
\end{align}
for real $z_i,z_j$ as follows:  
\begin{align}
&2\pi i \partial_{\tau}\phiChain(C)\nonumber\\
&=\sum_{i=2}^m \prod_{\begin{smallmatrix}
	j=2\\
	j\neq i
	\end{smallmatrix}}^m \Omega_{c_{j-1},c_j}(\eta_{c_j\dots c_m})2\pi i \partial_{\tau}\Omega_{c_{i-1},c_i}(\eta_{c_i\dots c_m})\nonumber\\
&=-\sum_{i=2}^m \prod_{\begin{smallmatrix}
	j=2\\
	j\neq i
	\end{smallmatrix}}^m \Omega_{c_{j-1},c_j}(\eta_{c_j\dots c_m}) \partial_{c_i}\partial_{\eta_{c_i\dots c_m}}\Omega_{c_{i-1},c_i}(\eta_{c_i\dots c_m})\nonumber\\
&=-\sum_{i=2}^m \left(\partial_{\eta_{c_i}}-  \theta_{i\geq 3}  
\partial_{\eta_{c_{i-1}}}\right)\prod_{\begin{smallmatrix}
	j=2\\
	j\neq i
	\end{smallmatrix}}^m \Omega_{c_{j-1},c_j}(\eta_{c_j\dots c_m}) \partial_{c_i}\Omega_{c_{i-1},c_i}(\eta_{c_i\dots c_m})\nonumber\\
&=\sum_{i=2}^m \left(\partial_{\eta_{c_i}}- \theta_{i\geq 3}  
\partial_{\eta_{c_{i-1}}}\right)\prod_{\begin{smallmatrix}
	j=2\\
	j\neq i+1
	\end{smallmatrix}}^m \Omega_{c_{j-1},c_j}(\eta_{c_j\dots c_m}) \left(\partial_{c_i}\Omega_{c_{i},c_{i+1}}(\eta_{c_i\dots c_m})+\Omega_{c_{i},c_{i+1}}(\eta_{c_i\dots c_m})\partial_{c_i}\right)
\nonumber\\
&=\sum_{i=2}^m \left(\partial_{\eta_{c_i}}- \theta_{i\geq 3}  
\partial_{\eta_{c_{i-1}}}\right)\prod_{\begin{smallmatrix}
	j=2\\
	j\neq i+1
	\end{smallmatrix}}^m \Omega_{c_{j-1},c_j}(\eta_{c_j\dots c_m}) \left(-\partial_{c_{i+1}}\Omega_{c_{i},c_{i+1}}(\eta_{c_i\dots c_m})+\Omega_{c_{i},c_{i+1}}(\eta_{c_i\dots c_m})\partial_{c_i}\right)\nonumber\\
&=\sum_{i=2}^m \left(\partial_{\eta_{c_i}}- \theta_{i\geq 3}  
\partial_{\eta_{c_{i-1}}}\right)\prod_{
	j=2}^m \Omega_{c_{j-1},c_j}(\eta_{c_j\dots c_m}) \left(\sum_{k=i}^m\partial_{c_k}\right)
\nonumber\\
&=\sum_{i=2}^m \left(\left(\partial_{\eta_{c_i}}- \theta_{i\geq 3}  
 \partial_{\eta_{c_{i-1}}}\right)\phiChain(C) \right)\left(\sum_{k=i}^m\partial_{c_k}\right)\,.
\end{align}
We again use a step function $\theta_{j\geq k}$ which is taken to be 1 for $j \geq k$ and zero for $j<k$.
Therefore, denoting $(1,A)=(a_0,a_1,\dots,a_p)$ and $ (0,B)=(b_0,b_1,\dots,b_q)$ we find
\begin{align}\label{eqn:tauDerivApp}
&2\pi i \partial_{\tau}\left(\KN^{\tau}_{01\dots n} \phiChain(1,A)\phiChain(0,B)\right)\nonumber\\
&=-\sum_{0\leq i<j\leq n}s_{ij}\left(f_{ij}^{(2)}+2\zeta_2\right)\KN^{\tau}_{01\dots n}\phiChain(1,A) \phiChain(0,B)\nonumber\\
&\phantom{=}+\sum_{i=1}^p \left(\sum_{k=i}^p\partial_{a_k}\KN^{\tau}_{01\dots n} \right)\phiChain(0,B) \left(\partial_{\eta_{a_i}}-\theta_{i\geq 2}  
\partial_{\eta_{a_{i-1}}}\right)\phiChain(1,A) \nonumber\\
&\phantom{=}+\sum_{i=1}^q \left(\sum_{k=i}^q\partial_{b_k}\KN^{\tau}_{01\dots n} \right)\phiChain(1,A) \left(\partial_{\eta_{b_i}}- \theta_{i\geq 2}  
\partial_{\eta_{b_{i-1}}}\right)\phiChain(0,B)\nonumber\\
&=-\sum_{0\leq i<j\leq n}s_{ij}\left(f_{ij}^{(2)}+2\zeta_2\right)\KN^{\tau}_{01\dots n}\phiChain(1,A) \phiChain(0,B)\nonumber\\
&\phantom{=}-\KN^{\tau}_{01\dots n}\sum_{i=1}^p \left(\sum_{k=i}^p\sum_{j=0}^{i-1}s_{a_k,a_j}f^{(1)}_{a_k,a_j}+\sum_{k=i}^p\sum_{j=0}^{q}s_{a_k,b_j}f^{(1)}_{a_k,b_j} \right)\phiChain(0,B) \left(\partial_{\eta_{a_i}}- \theta_{i\geq 2}  
\partial_{\eta_{a_{i-1}}}\right)\phiChain(1,A) \nonumber\\
&\phantom{=}-\KN^{\tau}_{01\dots n}\sum_{i=1}^q \left(\sum_{k=i}^q\sum_{j=0}^{i-1}s_{b_k,b_j}f^{(1)}_{b_k,b_j}+\sum_{k=i}^q\sum_{j=0}^p s_{b_k,a_j}f^{(1)}_{b_k,a_j} \right)\phiChain(1,A) \left(\partial_{\eta_{b_i}}- \theta_{i\geq 2}  
\partial_{\eta_{b_{i-1}}}\right)\phiChain(0,B)\nonumber\\
&=-\sum_{0\leq i<j\leq n}s_{ij}\left(f_{ij}^{(2)}+2\zeta_2\right)\KN^{\tau}_{01\dots n}\phiChain(1,A) \phiChain(0,B)\nonumber\\
&\phantom{=}-\KN^{\tau}_{01\dots n} \left(\sum_{k=1}^p\sum_{j=0}^{k-1}s_{a_k,a_j}f^{(1)}_{a_k,a_j}\sum_{i=j+1}^k+\sum_{k=1}^p\sum_{j=0}^{q}s_{a_k,b_j}f^{(1)}_{a_k,b_j}\sum_{i=1}^k \right)\nonumber\\
&\phantom{=} \ \ \ \ \times\phiChain(0,B) \left(\partial_{\eta_{a_i}}- \theta_{i\geq 2}  
 \partial_{\eta_{a_{i-1}}}\right)\phiChain(1,A) \nonumber\\
&\phantom{=}-\KN^{\tau}_{01\dots n} \left(\sum_{k=1}^q\sum_{j=0}^{k-1}s_{b_k,b_j}f^{(1)}_{b_k,b_j}\sum_{i=j+1}^k+\sum_{k=1}^q\sum_{j=0}^p s_{b_k,a_j}f^{(1)}_{b_k,a_j}\sum_{i=1}^k \right)\nonumber\\
&\phantom{=}\ \ \ \ \times\phiChain(1,A) \left(\partial_{\eta_{b_i}}- \theta_{i\geq 2}  
 \partial_{\eta_{b_{i-1}}}\right)\phiChain(0,B)\nonumber\\
&=-\sum_{0\leq i<j\leq n}s_{ij}\left(f_{ij}^{(2)}+2\zeta_2\right)\KN^{\tau}_{01\dots n}\phiChain(1,A) \phiChain(0,B)\nonumber\\
&\phantom{=}-\KN^{\tau}_{01\dots n} \left(\sum_{k=1}^p\sum_{j=0}^{k-1}s_{a_k,a_j}f^{(1)}_{a_k,a_j}\left(\partial_{\eta_{a_k}}- \theta_{j\geq 1}   
\partial_{\eta_{a_{j}}}\right) \right)\phiChain(1,A) \phiChain(0,B)\nonumber\\
&\phantom{=}-\KN^{\tau}_{01\dots n} \left(\sum_{k=1}^q\sum_{j=0}^{k-1}s_{b_k,b_j}f^{(1)}_{b_k,b_j}\left(\partial_{\eta_{b_k}}- \theta_{j\geq 1}  
\partial_{\eta_{b_{j}}}\right)\right)\phiChain(1,A) \phiChain(0,B) \nonumber\\
&\phantom{=}-\KN^{\tau}_{01\dots n} \left(\sum_{k=0}^p\sum_{j=0}^{q}s_{a_k,b_j}f^{(1)}_{a_k,b_j}\left( \theta_{k\geq 1}  
 \partial_{\eta_{a_k}}- \theta_{j\geq 1}   
 \partial_{\eta_{b_{j}}}\right) \right)\phiChain(1,A) \phiChain(0,B)\nonumber\\
&=-s_{01\dots n}2\zeta_2\KN^{\tau}_{01\dots n}\phiChain(1,A) \phiChain(0,B)\nonumber\\
&\phantom{=}-\KN^{\tau}_{01\dots n}\sum_{k=1}^p\sum_{j=0}^{k-1} s_{a_k,a_j}\left(f^{(1)}_{a_k,a_j}\left(\partial_{\eta_{a_k}}- \theta_{j\geq 1}  
 \partial_{\eta_{a_{j}}}\right) +f^{(2)}_{a_k,a_j}\right)\phiChain(1,A)\phiChain(0,B) \nonumber\\
&\phantom{=}-\KN^{\tau}_{01\dots n}\sum_{k=1}^q\sum_{j=0}^{k-1}s_{b_k,b_j} \left(f^{(1)}_{b_k,b_j}\left(\partial_{\eta_{b_k}}-\theta_{j\geq 1}   
 \partial_{\eta_{b_{j}}}\right)+f^{(2)}_{b_k,b_j}\right)\phiChain(1,A)\phiChain(0,B) \nonumber\\
&\phantom{=}-\KN^{\tau}_{01\dots n}\sum_{k=0}^p\sum_{j=0}^{q}s_{a_k,b_j} \left(f^{(1)}_{a_k,b_j}\left( \theta_{k\geq 1}  
\partial_{\eta_{a_k}}- \theta_{j\geq 1}  
 \partial_{\eta_{b_{j}}}\right)-f^{(2)}_{a_k,b_j} \right)\phiChain(1,A) \phiChain(0,B)\,,
\end{align}
which implies \eqn{eqn:SMapConjectureTau}. As for \eqn{eqn:z0DerivativeStart} in the calculation of the $z_0$-derivative, this equation is the starting point to determine the $\tau$-derivative of $\Ztzn{n}((1,A),(0,B))$. In the following two subsections, we give the corresponding formula in terms of the $S$-map and a closed expression. 

\subsection{\texorpdfstring{$S$}{S}-map formula}
Let us start with deriving the $S$-map formula, continuing from \eqn{eqn:tauDerivApp} similar to the calculation of the $z_0$-derivative. In order to rewrite the last sum, the operator in front of the product $\phiChain(1,A) \phiChain(0,B)$ is expressed as follows
\begin{align}
&\left(f^{(1)}_{a_k,b_j}\left(  \theta_{k\geq 1}
\partial_{\eta_{a_k}}-  \theta_{j\geq 1}
 \partial_{\eta_{b_{j}}}\right)-f^{(2)}_{b_j,a_k} \right)\phiChain(1,A) \phiChain(0,B)\nonumber\\
&=\left(   \theta_{k\geq 1}
\partial_{\eta_{a_k}}-  \theta_{j\geq 1}
\partial_{\eta_{b_{j}}}-\partial_{\xi}\right) \Omega_{a_k,b_j}(\xi)\phiChain(1,A) \phiChain(0,B)|_{\xi^0}\,,
\end{align}
such that eq.\ \eqref{eqn:SumShiftedSMapRep} can be used again. Due to the shifts $\mp\xi$ in the variables $\eta_{a_k}-\xi$ and $\eta_{b_j}+\xi$, the additional differential operator $\partial_{\eta_{a_k}}-\partial_{\eta_{b_{j}}}-\partial_{\xi}$ acts only non-trivially on the factor $\Omega_{01}$ in eq.\ \eqref{eqn:SumShiftedSMapRep}. For the first sum in eq.\ \eqref{eqn:SumShiftedSMapRep}, this amounts to projecting out a second derivative in analogy to eq.\ \eqref{eqn:projectOutDerivative}, since for $k\neq 0$
\begin{align}
&\left(\partial_{\eta_{a_k}}-\partial_{\eta_{b_{j}}}-\partial_{\xi}\right)\left(\Omega_{1,0}(\xi)\Omega_{a_{k-1},a_{k}}(\eta_{a_k}-\xi)\right)|_{\xi^0}\nonumber\\
&=\left(-\frac{1}{\xi^2}+f^{(2)}_{1,0}+\mathcal{O}(\xi)\right)\left(\Omega_{a_{k-1},a_{k}}(\eta_{a_k})-\xi \partial_{\eta_{a_k}}\Omega_{a_{k-1},a_{k}}(\eta_{a_k})+\frac{\xi^2}{2} 
\partial_{\eta_{a_k}}^2\Omega_{a_{k-1},a_{k}}(\eta_{a_k})+\mathcal{O}(\xi)\right)|_{\xi^0}\nonumber\\
&=\left(f^{(2)}_{01}-\frac{1}{2}\partial^2_{\eta_{a_k}}\right)\Omega_{a_{k-1},a_{k}}(\eta_{a_k})\,,
\end{align}
and more generally
\begin{align}
&\left(  \theta_{k\geq 1}
 \partial_{\eta_{a_k}}-   \theta_{j\geq 1}
 \partial_{\eta_{b_{j}}}-\partial_{\xi}\right)\Big(\Omega_{1,0}(\xi)\phiChain(1,A)|_{\eta_{a_k}\to \eta_{a_k}-\xi}\phiChain(0,B)|_{\eta_{b_j}\to \eta_{b_j}+\xi}\Big)|_{\xi^0}\nonumber\\
&=\left(f^{(2)}_{01}-\frac{1}{2}\left(  \theta_{k\geq 1}
\partial_{\eta_{a_k}}-  \theta_{j\geq 1}
 \partial_{\eta_{b_{j}}}\right)^2\right)\phiChain(1,A)\phiChain(0,B)\,.
\end{align}
Thus, if we apply $ ( \theta_{k\geq 1}
 \partial_{\eta_{a_k}}-  \theta_{j\geq 1}
 \partial_{\eta_{b_{j}}}-\partial_{\xi} )$ on the first sum in eq.\ \eqref{eqn:SumShiftedSMapRep}, its $\xi^0$-part is given~by
\begin{align}
&\left(\sum_{k=0}^p\sum_{j=0}^q s_{a_k,b_j}\left( \theta_{k\geq 1}
 \partial_{\eta_{a_k}}- \theta_{j\geq 1}
 \partial_{\eta_{b_{j}}}-\partial_{\xi}\right)\Omega_{1,0}(\xi)\phiChain(1,A)\phiChain(0,B)|_{\eta_{a_k}\to \eta_{a_k}-\xi,\eta_{b_j}\to \eta_{b_j}+\xi}\right)|_{\xi^0}\nonumber\\
&=\left(s_{(1,A),(0,B)}f^{(2)}_{01}-\frac{1}{2}\sum_{k=0}^p\sum_{j=0}^q s_{a_k,b_j}\left( \theta_{k\geq 1}
\partial_{\eta_{a_k}}- \theta_{j\geq 1}
 \partial_{\eta_{b_{j}}}\right)^2\right)\phiChain(1,A)\phiChain(0,B)\,,
\end{align}
such that the identity \eqref{eqn:SumShiftedSMapRep} implies 
\begin{align}
&\sum_{k=0}^p\sum_{j=0}^{q}s_{a_k,b_j} \left(f^{(1)}_{a_k,b_j}\left( \theta_{k\geq 1} 
 \partial_{\eta_{a_k}}- \theta_{j\geq 1}
 \partial_{\eta_{b_{j}}}\right)-f^{(2)}_{b_j,a_k} \right)\phiChain(1,A) \phiChain(0,B)\nonumber\\
&=\left(\frac{1}{2}\sum_{k=0}^p\sum_{j=0}^q s_{a_k,b_j}\left( \theta_{k\geq 1}
 \partial_{\eta_{a_k}}- \theta_{j\geq 1} 
 \partial_{\eta_{b_{j}}}\right)^2-s_{(1,A),(0,B)}f^{(2)}_{01}\right)\phiChain(1,A)\phiChain(0,B)\nonumber\\
&\phantom{=}-\sum_{l=1 }^{q}\Omega^{+}_{01}(\eta_{B_{l,q+1}})\phiChain(0,B_{1l})\phiChain(S[(1,A),B_{l,q+1}])\nonumber\\
&\phantom{=}-\sum_{i=1}^p\Omega^{-}_{01}(-\eta_{A_i,p+1})\phiChain(1,A_{1i})\phiChain(S[(0,B),A_{l,q+1}])\,,
\end{align}
where
\begin{align}
\Omega^{\pm}_{01}(\pm\xi)&=\pm \partial_{\xi}\Omega_{01}(\pm\xi)\,.
\end{align}
The second and third sum in \eqn{eqn:tauDerivApp} can be calculated similarly. We find for example 
\begin{align}\label{eqn:tauSpecificSum}
&\sum_{k=1}^p\sum_{j=0}^{k-1} s_{a_k,a_j}\left(f^{(1)}_{a_k,a_j}\left(\partial_{\eta_{a_k}}- \theta_{j\geq 1}
 \partial_{\eta_{a_{j}}}\right) +f^{(2)}_{a_k,a_j}\right)\phiChain(1,A)\phiChain(0,B)\nonumber\\
&=\phiChain(0,B)\sum_{k=1}^p\sum_{j=0}^{k-1} s_{a_k,a_j}\Big(\left(\partial_{\eta_{a_k}}- \theta_{j\geq 1}
 \partial_{\eta_{a_j}}+\partial_{\xi}\right)\Omega_{a_k,a_j}(\xi)\phiChain(1,A)\Big)|_{\xi^0}\,,
\end{align}
where using eq.\ \eqref{eqn:shuffleIdentityChainWith1And0AlternatingSum} gives
\begin{align}\label{eqn:SecondSumTau}
&\Omega_{a_k,a_j}(\xi)\phiChain(1,A)\nonumber\\
&=\Omega_{a_k,a_j}(\xi)\phiChain(A_{0,j+1})|_{\eta_{j}\to \eta_{A_{j,p+1}}}\phiChain(a_j,A_{j+1,k},a_k,A_{k+1,p+1})\nonumber\\
&=\Omega_{a_k,a_j}(\xi)\phiChain(A_{0,j+1})|_{\eta_{j}\to \eta_{A_{j,p+1}}}\sum_{l=j+1}^k(-1)^{k-l}\phiChain(a_j,A_{j+1,l})\phiChain(a_j,a_k,\tilde A_{l,k}\shuffle A_{k+1,p+1})\nonumber\\
&=\phiChain(A_{0,j+1})|_{\eta_{j}\to \eta_{A_{j,p+1}}}\sum_{l=j+1}^k(-1)^{k-l}\phiChain(a_j,A_{j+1,l})\nonumber\\
&\phantom{=} \ \ \ \ \times \Omega_{a_k,a_j}(\xi)\Omega_{a_j,a_k}(\eta_{A_{l,p+1}})\phiChain(a_k,\tilde A_{l,k}\shuffle A_{k+1,p+1})\,.
\end{align}
With the identity \cite{BrownLev, Enriquez:Emzv}
\begin{align}
(\partial_{\eta}+\partial_{\xi})\Omega_{k,j}(\xi)\Omega_{j,k}(\eta)&=\left(\wp(\eta)-\wp(\xi)\right)\Omega_{j,k}(\eta-\xi)
\end{align}
and extracting its $\xi^0$ contribution
\begin{align}
\left((\partial_{\eta}+\partial_{\xi})\Omega_{k,j}(\xi)\Omega_{j,k}(\eta)\right)|_{\xi^0}&=\left(\wp(\eta)-\frac{1}{2}\partial_{\eta}^2\right)\Omega_{j,k}(\eta)
\end{align}
we can continue to find 
\begin{align}\label{eqn:SecondSumTauCalculation}
&\Big(\left(\partial_{\eta_{a_k}}- \theta_{j\geq 1} 
\partial_{\eta_{a_j}}+\partial_{\xi}\right)\Omega_{a_k,a_j}(\xi)\phiChain(1,A)\Big)|_{\xi^0}\nonumber\\
&=\phiChain(A_{0,j+1})|_{\eta_{j}\to \eta_{A_{j,p+1}}}\sum_{l=j+1}^k(-1)^{k-l}\phiChain(a_j,A_{j+1,l})\nonumber\\
&\phantom{=}\Big(\left(\partial_{\eta_{a_k}}+\partial_{\xi}\right)\Omega_{a_k,a_j}(\xi)\Omega_{a_j,a_k}(\eta_{A_{l,p+1}})\Big)|_{\xi^0}\phiChain(a_k,\tilde A_{l,k}\shuffle A_{k+1,p+1})\nonumber\\
&=\phiChain(A_{0,j+1})|_{\eta_{j}\to \eta_{A_{j,p+1}}}\sum_{l=j+1}^k(-1)^{k-l}\phiChain(a_j,A_{j+1,l})\nonumber\\
&\phantom{=}\left(\wp(\eta_{A_{l,p+1}})-\frac{1}{2}\partial_{\eta_{a_k}}^2\right)\Omega_{a_j,a_k}(\eta_{A_{l,p+1}})\phiChain(a_k,\tilde A_{l,k}\shuffle A_{k+1,p+1})
\nonumber\\
&=-\frac{1}{2}\left(\partial_{\eta_{a_k}}-  \theta_{j\geq 1} 
 \partial_{\eta_{a_j}}\right)^2\phiChain(1,A)\nonumber\\
&\phantom{=}+\sum_{l=j+1}^k\wp(\eta_{A_{l,p+1}})(-1)^{k-l}\phiChain(A_{0,j+1})|_{\eta_{j}\to \eta_{A_{j,p+1}}}(-1)^{l-j-1}\phiChain(\tilde A_{j+1,l},a_j)|_{\eta_{a_j}\to -\eta_{A_{j+1,l}}}\nonumber\\
&\phantom{=+}\phiChain(a_j,a_k,\tilde A_{l,k}\shuffle A_{k+1,p+1})\nonumber\\
&=-\frac{1}{2}\left(\partial_{\eta_{a_k}}-  \theta_{j\geq 1} 
 \partial_{\eta_{a_j}}\right)^2\phiChain(1,A)\nonumber\\
&\phantom{=}+\sum_{l=j+1}^k\wp(\eta_{A_{l,p+1}})(-1)^{k-j-1}\phiChain(A_{0,j}\shuffle \tilde A_{j+1,l},a_j,a_k,\tilde A_{l,k}\shuffle A_{k+1,p+1})\,,
\end{align}
where we have used the reflection property \eqref{eqn:reflectionChain} and again the identity \eqref{eqn:shuffleIdentityChainWith1And0AlternatingSum} in the reverse direction, to pull out the appropriate second derivative for the second last equality. 
Thus, we finally obtain
\begin{align}
&\Big(\left(\partial_{\eta_{a_k}}-  \theta_{j\geq 1} 
 \partial_{\eta_{a_j}}+\partial_{\xi}\right)\Omega_{a_k,a_j}(\xi)\phiChain(1,A)\Big)|_{\xi^0}
=-\frac{1}{2}\left(\partial_{\eta_{a_k}}-  \theta_{j\geq 1}  
 \partial_{\eta_{a_j}}\right)^2\phiChain(1,A)\nonumber\\
&\phantom{=}+\sum_{l=j+1}^k\wp(\eta_{A_{l,p+1}})(-1)^{k-j-1}\phiChain(A_{0,j}\shuffle \tilde A_{j+1,l},a_j,a_k,\tilde A_{l,k}\shuffle A_{k+1,p+1})\,.
\end{align}
Therefore, plugging the above identity into the sum \eqref{eqn:tauSpecificSum} yields
\begin{align}
&\sum_{k=1}^p\sum_{j=0}^{k-1} s_{a_k,a_j}\left(f^{(1)}_{a_k,a_j}\left(\partial_{\eta_{a_k}}-  \theta_{j\geq 1} 
 \partial_{\eta_{a_{j}}}\right) +f^{(2)}_{a_k,a_j}\right)\phiChain(1,A)\phiChain(0,B)\nonumber\\
&=-\sum_{k=1}^p\sum_{j=0}^{k-1} \frac{1}{2}\left(\partial_{\eta_{a_k}}-  \theta_{j\geq 1} 
 \partial_{\eta_{a_j}}\right)^2s_{a_k,a_j}\phiChain(1,A)\phiChain(0,B)\nonumber\\
&\phantom{=}+\sum_{k=1}^p\sum_{j=0}^{k-1} s_{a_k,a_j}\sum_{l=j+1}^k\wp(\eta_{A_{l,p+1}})(-1)^{k-j-1}\phiChain(A_{0,j}\shuffle \tilde A_{j+1,l},a_j,a_k,\tilde A_{l,k}\shuffle A_{k+1,p+1})\phiChain(0,B)\nonumber\\
&=-\sum_{k=1}^p\sum_{j=0}^{k-1} \frac{1}{2}\left(\partial_{\eta_{a_k}}-  \theta_{j\geq 1} 
 \partial_{\eta_{a_j}}\right)^2s_{a_k,a_j}\phiChain(1,A)\phiChain(0,B)\nonumber\\
&\phantom{=}+\sum_{l=1}^p\wp(\eta_{A_{l,p+1}})\phiChain(0,B)\phiChain(S[(1,A_{1,l}),A_{l,p+1}])\,.
\end{align}
Analogously, the third term in the sum in \eqn{eqn:tauDerivApp} is
\begin{align}
&\sum_{k=1}^q\sum_{j=0}^{k-1}s_{b_k,b_j} \left(f^{(1)}_{b_k,b_j}\left(\partial_{\eta_{b_k}}-  \theta_{j\geq 1} 
 \partial_{\eta_{b_{j}}}\right)+f^{(2)}_{b_k,b_j}\right)\phiChain(1,A)\phiChain(0,B) \nonumber\\
&=-\sum_{k=1}^q\sum_{j=0}^{k-1} \frac{1}{2}\left(\partial_{\eta_{b_k}}-  \theta_{j\geq 1} 
 \partial_{\eta_{b_j}}\right)^2s_{b_k,b_j}\phiChain(1,A)\phiChain(0,B)\nonumber\\
&\phantom{=}+\sum_{l=1}^q\wp(\eta_{B_{l,p+1}})\phiChain(1,A)\phiChain(S[(0,B_{1,l}),B_{l,p+1}])\,,
\end{align}
which was the last term missing, such that the whole \eqn{eqn:tauDerivApp} is expressed in terms of the $S$-map leading to \eqn{eqn:tauDerivSMap}. This formula is similar to the $S$-map formula \eqref{eqn:derZ0SMap} for the $z_0$-derivative: the differences are the diagonal terms, i.e.\ the first sum proportional to $\Ztzn{n}((1,A),(0,B))$, the derivatives $\Omega^+_{01}(\eta_{B_{l,q+1}})$, $\Omega^-_{01}(-\eta_{A_l,p+1})$ instead of $\Omega_{01}(\eta_{B_{l,q+1}})$, $\Omega_{01}(-\eta_{A_l,p+1})$, respectively, and the appearance of the terms including the Weierstra\ss{} $\wp$-function.

\subsection{Closed formula}
Similarly, we obtain a closed formula for the $\tau$-derivative continuing from eq.\ \eqref{eqn:tauDerivApp}. The last sum can be rewritten using the results in eqs.\ \eqref{eqn:OmegaABCchainsClosedFormula} to \eqref{eqn:OmegaABCchainsClosedFormulaK} from the $z_0$-derivative
\begin{align}
&\sum_{k=0}^p\sum_{j=0}^{q}s_{a_k,b_j} \left(f^{(1)}_{a_k,b_j}\left(  \theta_{k\geq 1} 
 \partial_{\eta_{a_k}}-  \theta_{j\geq 1} 
  \partial_{\eta_{b_{j}}}\right)-f^{(2)}_{a_k,b_j} \right)\phiChain(1,A) \phiChain(0,B)\nonumber\\
&=\sum_{k=0}^p\sum_{j=0}^{q}s_{a_k,b_j} \left(  \theta_{k\geq 1} 
\partial_{\eta_{a_k}}-  \theta_{j\geq 1}  
 \partial_{\eta_{b_{j}}}-\partial_{\xi}\right) \Omega_{a_k,b_j}(\xi)\phiChain(1,A) \phiChain(0,B)|_{\xi^0}\nonumber\\
&=\left(\frac{1}{2}\sum_{k=0}^p\sum_{j=0}^q s_{a_k,b_j}\left(  \theta_{k\geq 1} 
 \partial_{\eta_{a_k}}-  \theta_{j\geq 1}  
  \partial_{\eta_{b_{j}}}\right)^2-s_{(1,A),(0,B)}f^{(2)}_{01}\right)\phiChain(1,A)\phiChain(0,B)\nonumber\\
&\phantom{=}-\sum_{k=0}^p\sum_{j=0}^{q}s_{a_k,b_j} \sum_{i=1 }^{k}\sum_{l=1 }^{j}(-1)^{k+j-i-l}\Omega_{01}^+(\eta_{B_{l,q+1}})\phiChain(0,B_{1l})\nonumber\\
&\phantom{=\sum_{i=1 }^{k}\sum_{l=1 }^{j}} \times \phiChain(1,A_{1i}\shuffle(a_k,(\tilde{A}_{i,k}\shuffle A_{k+1,p+1})\shuffle (b_j,\tilde{B}_{l,j}\shuffle B_{j+1,q+1})))\nonumber\\
&\phantom{=}-\sum_{k=0}^p\sum_{j=0}^{q}s_{a_k,b_j} \sum_{i=1 }^{k}\sum_{l=1 }^{j}(-1)^{k+j-i-l}\Omega^-_{01}(-\eta_{A_{i,p+1}})\phiChain(1,A_{1i})\nonumber\\
&\phantom{=\sum_{i=1 }^{k}\sum_{l=1 }^{j}}  \times \phiChain(0,B_{1l}\shuffle(b_j,(\tilde{B}_{l,j}\shuffle B_{j+1,q+1})\shuffle (a_k,\tilde{A}_{i,k}\shuffle A_{k+1,p+1})))\,.
\end{align}
The second sum can be expressed from eq.\ \eqref{eqn:SecondSumTau} using a similar calculation as in eq.\ \eqref{eqn:SecondSumTauCalculation} as follows:
\begin{align}
&\Big(\left(\partial_{\eta_{a_k}}-  \theta_{j\geq 1}  
 \partial_{\eta_{a_j}}+\partial_{\xi}\right)\Omega_{a_k,a_j}(\xi)\phiChain(1,A)\Big)|_{\xi^0}\nonumber\\
&=\phiChain(A_{0,j+1})|_{\eta_{j}\to \eta_{A_{j,p+1}}}\sum_{l=j+1}^k(-1)^{k-l}\phiChain(A_{j,l})\nonumber\\
&\phantom{=}\Big(\left(\partial_{\eta_{a_k}}+\partial_{\xi}\right)\Omega_{a_k,a_j}(\xi)\Omega_{a_j,a_k}(\eta_{A_{l,p+1}})\Big)|_{\xi^0}\phiChain(a_k,\tilde A_{l,k}\shuffle A_{k+1,p+1})\nonumber\\
&=\phiChain(A_{0,j+1})|_{\eta_{j}\to \eta_{A_{j,p+1}}}\sum_{l=j+1}^k(-1)^{k-l}\phiChain(A_{j,l})\nonumber\\
&\phantom{=}\left(\wp(\eta_{A_{l,p+1}})-\frac{1}{2}\partial_{\eta_{a_k}}^2\right)\Omega_{a_j,a_k}(\eta_{A_{l,p+1}})\phiChain(a_k,\tilde A_{l,k}\shuffle A_{k+1,p+1})
\nonumber\\
&=-\frac{1}{2}\left(\partial_{\eta_{a_k}}-  \theta_{j\geq 1} 
 \partial_{\eta_{a_j}}\right)^2\phiChain(1,A)\nonumber\\
&\phantom{=}+\sum_{l=j+1}^k\wp(\eta_{A_{l,p+1}})(-1)^{k-l}\phiChain(A_{0,j+1})|_{\eta_{a_j}\to \eta_{A_{j,p+1}}}\phiChain(a_j, A_{j+1,l}\shuffle (a_k,\tilde A_{l,k}\shuffle A_{k+1,p+1}))\nonumber\\
&=-\frac{1}{2}\left(\partial_{\eta_{a_k}}-  \theta_{j\geq 1} 
 \partial_{\eta_{a_j}}\right)^2\phiChain(1,A)\nonumber\\
&\phantom{=}+\sum_{l=j+1}^k\wp(\eta_{A_{l,p+1}})(-1)^{k-l}\phiChain(1,A_{1,j},a_j, A_{j+1,l}\shuffle (a_k,\tilde A_{l,k}\shuffle A_{k+1,p+1}))\,,
\end{align}
while the third sum in eq.\ \eqref{eqn:tauDerivApp} is given by 
\begin{align}
&\Big(\left(\partial_{\eta_{b_k}} -   \theta_{j\geq 1}  
 \partial_{\eta_{b_j}}+\partial_{\xi}\right)\Omega_{b_k,b_j}(\xi)\phiChain(0,B)\Big)|_{\xi^0}
=-\frac{1}{2}\left(\partial_{\eta_{b_k}}-  \theta_{j\geq 1} 
 \partial_{\eta_{b_j}}\right)^2\phiChain(0,B)\nonumber\\
&\phantom{=}+\sum_{l=j+1}^k\wp(\eta_{B_{l,q+1}})(-1)^{k-l}\phiChain(0,B_{1,j},b_j, B_{j+1,l}\shuffle (b_k,\tilde B_{l,k}\shuffle B_{k+1,q+1}))\,.
\end{align}
Altogether, this leads to the closed expression of the $\tau$-derivative as given in \eqn{eqn:tauDerivClosed}. Note that as for the $S$-map formula, the only difference compared to the closed $z_0$-derivative \eqref{eqn:derZ0Closed} are the diagonal terms (first sum), the derivatives $\Omega^+_{01}(\eta_{B_{l,q+1}})$, $\Omega^-_{01}(-\eta_{A_l,p+1})$ instead of $\Omega_{01}(\eta_{B_{l,q+1}})$, $\Omega_{01}(-\eta_{A_l,p+1})$, respectively, and the appearance of the terms including the Weierstra\ss{} $\wp$-function (the second and third sum).

\subsection{Three-point example}
\label{app:tau3}

This appendix complements the discussion of the three-point $\tau$-derivative in section \ref{sec:tau3} by expressing $\partial_\tau \BZtzn{3}$ in terms of Weierstra\ss{} $\wp$-functions and Kronecker--Eisenstein series. The $\tau$-derivative following from \eqns{eqn:tauDerivSMap}{eqn:tauDerivClosed} is given by
\begin{align}
2\pi i \partial_{\tau}\BZtzn{3}&= \left(\frac{1}{2}(s_{02}+s_{12})\partial_{\eta_2}^2
+\frac{1}{2}(s_{03}+s_{13})\partial_{\eta_3}^2 + \frac{1}{2} s_{23} (\partial_{\eta_2}{-}\partial_{\eta_3})^2
 -2\zeta_2 s_{0123}\right) \BZtzn{3} \label{bigeq}\\
&\phantom{=}-
 \te{diag} \big( s_{0,123} ,\, s_{0,123} ,\,  s_{03,12} ,\,  s_{02,13} ,\,  s_{023,1}  ,\, s_{023,1}  \big)
 f^{(2)}_{01} \BZtzn{3} +  \begin{pmatrix}
d_{11}^\tau &d_{12}^\tau &d_{13}^\tau \\
d_{21}^\tau &d_{22}^\tau &d_{23}^\tau \\
d_{31}^\tau &d_{32}^\tau &d_{33}^\tau 
\end{pmatrix}\BZtzn{3} \, , \notag
\end{align}
where $d_{ij}^\tau$ denote $2{\times} 2$ blocks: while the diagonal blocks depend on $\eta_2,\eta_3$
via Weierstra\ss{} functions,
\begin{align}
d_{11}^\tau  &=
\ccb -s_{12} \wp(\eta_{23}) - (s_{13}{+}s_{23}) \wp(\eta_3) & s_{13} \big[ \wp(\eta_{23}) - \wp(\eta_3) \big]  \\
s_{12} \big[ \wp(\eta_{23}) - \wp(\eta_2) \big] &-s_{13} \wp(\eta_{23}) - (s_{12}{+}s_{23}) \wp(\eta_2) \cce\,,
\notag \\
d_{22}^\tau &=
\ccb -s_{12} \wp(\eta_2) - s_{03}\wp(\eta_3) &0 \\
0 &-s_{13} \wp(\eta_3) - s_{02} \wp(\eta_2)  \cce\,, \label{eq2.18} \\
d_{33}^\tau  &=
\ccb -s_{03} \wp(\eta_{23}) - (s_{02}{+}s_{23}) \wp(\eta_2) & s_{02} \big[ \wp(\eta_{23}) - \wp(\eta_2) \big] \\
s_{03} \big[ \wp(\eta_{23}) - \wp(\eta_3) \big] &-s_{02} \wp(\eta_{23}) - (s_{03}{+}s_{23}) \wp(\eta_3) \cce\, ,
\notag
\end{align}
the off-diagonal ones involve the derivatives (\ref{delKE}) of the Kronecker--Eisenstein series
\begin{align}
d_{12}^\tau &= \ccb s_{03} \Omega^{-}_{01}({-}\eta_3) &0 \\
0 &s_{02} \Omega^{-}_{01}({-}\eta_2) \cce \,, \notag \\
d_{13}^\tau &=  \Omega^-_{01}({-}\eta_{23}) \ccb
-s_{03}  &s_{02} \\ s_{03} &-s_{02}
\cce \,,\notag \\
d_{21}^\tau &= \ccb
(s_{13} {+} s_{23}) \Omega^+_{01}(\eta_{3}) &s_{13} \Omega^+_{01}(\eta_{3}) \\
s_{12}  \Omega^+_{01}(\eta_{2}) &(s_{12}{+}s_{23}) \Omega^+_{01}(\eta_{2})
\cce \,,\label{eq2.19} \\
d_{23}^\tau &= \ccb
(s_{02}{+}s_{23}) \Omega^-_{01}({-}\eta_{2}) &s_{02}  \Omega^-_{01}({-}\eta_{2}) \\
s_{03}  \Omega^-_{01}({-}\eta_{3}) &(s_{03}{+}s_{23}) \Omega^-_{01}({-}\eta_{3})
\cce\,, \notag \\
d_{31}^\tau  &=  \Omega^+_{01}(\eta_{23}) \ccb
-s_{12} &s_{13} \\ s_{12} &-s_{13}
\cce\,, \notag \\
d_{32}^\tau   &= \ccb
s_{12}  \Omega^+_{01}(\eta_{2}) &0 \\
0 &s_{13}\Omega^+_{01}(\eta_{3})
\cce \, .
\notag
\end{align}
Based on (\ref{bigeq}) and the $\eta$-expansions of (\ref{eq2.18}) and (\ref{eq2.19}),
one arrives at the expressions for $r_{0,3}(\eps_{k}) $ and $r_{0,3}(b_{k}) $
given in section \ref{sec:tau3}.

\section{Recovering genus-one Selberg integrals}\label{app:recoverSelbergInt}
In this appendix, we relate the language used in \secref{sec:derivatives} to the genus-one Selberg integrals
(\ref{eqn:Selberg}) employed in \rcite{Broedel:2019gba}.

\subsection{Recovering genus-one Selberg integrals}\label{app:ZtzTogenSelberg}
Let us begin with defining the generating series of genus-one Selberg integrals
with expansion variables $\xi_\ell$: 
\begin{align}
\label{defTintegral}
T^{\tau}_{0,n} \Big[
\begin{smallmatrix}
\xi_2 &\xi_3 &\dots &\xi_n \\
i_2 &i_3 &\dots  &i_n
\end{smallmatrix}
\Big]&= \! \! \! \! \! \! \! \! \!  \int\limits_{0<z_2<z_{3}<\ldots < z_n<z_0}  \phantom{}
\dd z_2\cdots \dd z_n \, {\rm KN}^\tau_{012\ldots n} \, \Omega_{i_{2}2}(\xi_{2})  \Omega_{i_{3}3}(\xi_{3})
\ldots  \Omega_{i_{n}n}(\xi_{n})
 \nonumber\\
&=e^{s_{01\dots n}\omega(1,0|\tau)}\sum_{k_2,\dots,k_n\geq 0}(-1)^{k_2+\dots+k_n}\xi_{2}^{k_2-1}\dots \xi_{n}^{k_n-1} \SIE{k_{n},\dots,k_2}{i_{n},\dots,i_2}(z_0)\,,
\end{align}
where the $i_\ell$ satisfy the admissibility condition \eqref{eqn:admissibility}, i.e.\ $i_\ell\in\{0,1,\ell{+}1,\ell{+}2,\dots,n\}$.  The integrals $T^{\tau}_{0,n} $ are related to the augmented $Z$-integrals $\Ztzn{n}$ by Fay identities.

This connection can be understood from the graphical approach in section \ref{graphsec} since the integrand $\Omega_{i_{2}2}(\xi_{2})  \ldots  \Omega_{i_{n}n}(\xi_{n})$ of \eqn{defTintegral} with admissible $i_\ell$ agrees precisely with the one in \eqn{eqn:Ztaug}. Once the $\xi_\ell$ are identified as suitable linear combinations of $\eta_2,\eta_3,\ldots,\eta_n$ to be denoted by $\eta_{C_{\ell}(\vec{i})}$, one can use Fay identities to expand
\begin{align}
T^{\tau}_{0,n} \Big[
\begin{smallmatrix}
\eta_{C_{2}(\vec{i})} &\eta_{C_{3}(\vec{i})} &\ldots &\eta_{C_{n}(\vec{i})} \\
i_2 &i_3 &\ldots &i_n
\end{smallmatrix}
\Big]&=\sum_{A,B}m^{\vec{i}}_{A,B}\,\Ztzn{n}((1,A),(0,B))
\label{fromTtoZ}
\end{align}
with coefficients $m^{\vec{i}}_{A,B} \in \ZZ$, and the sum runs over all disjoint sequences $A=(a_1,a_2,\dots,a_p)$ and $B=(b_1,b_2,\dots,b_q)$ such that $A\cup B = \{2,3,\dots,n\}$. The linear combinations $\xi_\ell = \eta_{C_{\ell}(\vec{i})}$ are fixed from the discussion in section \ref{sec:411}: the labels $\vec{i}=(i_2,i_3,\ldots,i_n)$ of the integrand $\Omega_{i_{2}2}(\xi_{2})  \ldots  \Omega_{i_{n}n}(\xi_{n})$ in \eqn{defTintegral} define a graph which in turn identifies $\xi_\ell=\eta_{C_{\ell}(\vec{i})}$ with a linear combination of  $\eta_2,\eta_3,\ldots,\eta_n$ through the weights of its edges as explained below  \eqn{eqn:Ztaug}.

In the same way as the $\Ztzn{n}((1,A),(0,B))$ are gathered in the $n!$-component vector $\BZ_{0,n}^{\tau}$ in \eqn{eqn:Ztzbasis}, we introduce an $n!$-component vector of admissible integrals \eqn{defTintegral},
\begin{align}\label{T0n}
\BT^{\tau}_{0,n}&=\begin{pmatrix}
T^{\tau}_{0,n} \Big[
\begin{smallmatrix}
\eta_{C_{2}(\vec{i})} &\eta_{C_{3}(\vec{i})} &\ldots &\eta_{C_{n}(\vec{i})} \\
i_2 &i_3 &\ldots &i_n
\end{smallmatrix}
\Big]
\end{pmatrix} \text{ for } i_\ell\in\{0,1,\ell{+}1,\ell{+}2,\dots,n\} \, .
\end{align}
In this setting, \eqn{fromTtoZ} defines an $n!{\times} n!$ transformation matrix $M_n$ with integer entries,  
\begin{align}\label{eqn:basisTrafoTZ}
\BT^{\tau}_{0,n}&= M_n\, \BZtzn{n}\,.
\end{align}
This basis transformation relates the component integrals of the augmented $Z$-integrals to the genus-one Selberg integrals by means of eq.\ \eqref{defTintegral}.

\subsubsection{Two-point example}

Let us illustrate the above definitions and the calculation of the transformation matrix $M_n$ on the two- and three-point examples. In the two-point case, the two integrals are
\begin{align}
T^{\tau}_{0,n} \Big[
\begin{smallmatrix}
\xi_2 \\
i_2
\end{smallmatrix}
\Big] &= \int_0^{z_0} \dd z_2  \, {\rm KN}^\tau_{012}
\Omega_{i_22}(\xi_{2})
\end{align}
whose $\xi_2$-variable is identified with $\eta_2$ for both of
$i_2=0,1$. The two vector components then
trivially agree with $\Ztzn{2}(1,2)$ and $\Ztzn{2}(0,2)$, respectively,
\begin{align}\label{eqn:twoPointM}
\BT^{\tau}_{0,2}&=\begin{pmatrix}
T^{\tau}_{0,2} \Big[
\begin{smallmatrix}
\xi_2 \\
1
\end{smallmatrix}
\Big]\\
T^{\tau}_{0,2} \Big[
\begin{smallmatrix}
\xi_2 \\
0
\end{smallmatrix}
\Big]
\end{pmatrix}=\begin{pmatrix}
1&0\\0&1
\end{pmatrix}
\BZtzn{2}\,.
\end{align}

\subsubsection{Three-point example}

In the case of $n=3$, the generating series \eqn{defTintegral} specializes to
\begin{align}
T^{\tau}_{0,3} \Big[
\begin{smallmatrix}
\xi_2& \xi_3 \\
i_2 &i_3
\end{smallmatrix}
\Big] &= \int_{0}^{z_0}\dd z_3\,\Omega_{i_33}(\xi_3)\int_0^{z_3}\dd z_2\,
\Omega_{i_22}(\xi_2){\rm KN}^\tau_{0123}\,,
\label{3ptTs}
\end{align}
and features six components $(i_2,i_3)\in \{0,1,3\}\times \{0,1\}$ compatible with the admissibility condition.  Those with $(i_2,i_3)\in\{(3,1),(3,0),(0,1),(1,0)\}$ immediately yield the Kronecker-Eisenstein series seen in a component of the $\BZtzn{3}$-vector \eqn{eqn:3PointExampleIntegral} which settles the respective identifications of $\xi_2, \xi_3$ with the $\eta_j$-variables:
\begin{align}
T^{\tau}_{0,3} \Big[
\begin{smallmatrix}
\eta_2 &\eta_{2,3} \\
3 &1
\end{smallmatrix}
\Big]&=\Ztzn{3}(1,3,2)\,,\nonumber\\
T^{\tau}_{0,3} \Big[
\begin{smallmatrix}
\eta_2 &\eta_{2,3} \\
3 &0
\end{smallmatrix}
\Big]&=\Ztzn{3}(0,3,2)\,,\nonumber\\
T^{\tau}_{0,3}\Big[
\begin{smallmatrix}
\eta_2 &\eta_{3} \\
0 &1
\end{smallmatrix}
\Big]&=\Ztzn{3}((1,3),(0,2))\,,\nonumber\\
T^{\tau}_{0,3} \Big[
\begin{smallmatrix}
\eta_2 &\eta_{3} \\
1 &0
\end{smallmatrix}
\Big]&=\Ztzn{3}((1,2),(0,3))\,.
\end{align}
The remaining entries of \eqn{3ptTs} with $(i_2,i_3)=(1,1)$ and $(i_2,i_3)=(0,0)$ 
yield Kronecker-Eisenstein series $\Omega_{12}(\xi_2)\Omega_{13}(\xi_3)$
and $\Omega_{02}(\xi_2)\Omega_{03}(\xi_3)$, respectively. Their expansion in
terms of entries of $\Ztzn{3}((1,A),(0,B))$ via Fay relations uniquely selects 
$\xi_2 = \eta_2$ and $\xi_3 = \eta_3$,
\begin{align}
T^{\tau}_{0,3}  \Big[
\begin{smallmatrix}
\eta_2 &\eta_{3} \\
1 &1
\end{smallmatrix}
\Big]&=\Ztzn{3}(1,2,3)+\Ztzn{3}(1,3,2) \\
T^{\tau}_{0,3}  \Big[
\begin{smallmatrix}
\eta_2 &\eta_{3} \\
0 &0
\end{smallmatrix}
\Big]&=\Ztzn{3}(0,2,3)+\Ztzn{3}(0,3,2)\,,
\end{align}
such that the full basis transformation is given by 
\begin{align}
\BT^{\tau}_{0,3}&=\begin{pmatrix}
T^{\tau}_{0,3}  \Big[
\begin{smallmatrix}
\eta_2 &\eta_{3} \\
1 &1
\end{smallmatrix}
\Big]\\
T^{\tau}_{0,3} \Big[
\begin{smallmatrix}
\eta_2 &\eta_{2,3} \\
3 &1
\end{smallmatrix}
\Big]\\
T^{\tau}_{0,3} \Big[
\begin{smallmatrix}
\eta_2 &\eta_{3} \\
1 &0
\end{smallmatrix}
\Big]\\
T^{\tau}_{0,3} \Big[
\begin{smallmatrix}
\eta_2 &\eta_{3} \\
0 &1
\end{smallmatrix}
\Big]\\
T^{\tau}_{0,3}  \Big[
\begin{smallmatrix}
\eta_2 &\eta_{2,3} \\
3 &0
\end{smallmatrix}
\Big]\\
T^{\tau}_{0,3}  \Big[
\begin{smallmatrix}
\eta_2 &\eta_{3} \\
0 &0
\end{smallmatrix}
\Big]
\end{pmatrix}=\begin{pmatrix}
1&1&0&0&0&0\\0&1&0&0&0&0\\
0&0&1&0&0&0\\
0&0&0&1&0&0\\
0&0&0&0&1&0\\
0&0&0&0&1&1
\end{pmatrix}\BZtzn{3}\,.
\end{align}

\subsection{Recovering the representation of \texorpdfstring{$x_{n,w}$}{x(n,w)}}
In \rcite{Broedel:2019gba}, it was shown that the differential equation of the vector of all admissible genus-one Selberg integrals \eqref{eqn:Selberg} of a certain weight $w=k_2+k_3+\dots+ k_n$, i.e.\
\begin{align}\label{eqn:defSnw}
\Selbld^{\E}_{n,w}(z_0|\tau)&=\begin{pmatrix}
\SIE{k_{n},\dots,k_2}{i_{n},\dots,i_2}(z_0)
\end{pmatrix} \text{ for } \bigg\{ \begin{array}{c} k_2+k_3+\dots+ k_n=w  \\ i_\ell\in\{0,1,\ell{+}1,\ell{+}2,\dots,n\} \end{array}
\end{align}
satisfies a differential equation of the form 
\begin{align}\label{eqn:degHatS}
\partial_0 \Selbld^{\E}_{n,w}(z_0|\tau)&=\sum_{k=0}^{w+1}f^{(k)}_{01}r^{\E}_{0,n}(x_{k,w})\Selbld^{\E}_{n,w+1-k}(z_0|\tau)\,,
\end{align}
where the matrices $r^{\E}_{0,n}(x_{k,w})$ are linear in the Mandelstam variables and play a crucial role in the associator construction proposed therein. These matrices are the finite, non-vanishing blocks of the matrices $r^{\E}_{0,n}(x_{k})$ in \eqn{eqn:KZBSelE}. In this subsection, we show how these matrices can be related to the matrices $r_{0,n}(x_k)$ in the KZB equation \eqref{eqn:KZBforZ}.

The basis transformation \eqref{eqn:basisTrafoTZ} along with \eqn{eqn:KZBforZ} imply that the vector of integrals $\BT^{\tau}_{0,n}$ satisfies the KZB equation 
\begin{align}
\partial_0 \BT^{\tau}_{0,n} &=\sum_{k= 0}^{\infty} f^{(k)}_{01} M_n r_{0,n}(x_k)M_n^{-1} \BT^{\tau}_{0,n} \,.
\end{align}
On the one hand, according to the definition \eqref{defTintegral} the vector $\BT^{\tau}_{0,n}$ can be expressed as a linear combination of vectors $\Selbld^{\E}_w(z_0|\tau)$ as follows
\begin{align}\label{eqn:trafoSelToT}
\BT^{\tau}_{0,n}&=e^{s_{01\dots n}\omega(1,0|\tau)}  \sum_{k_2,\dots,k_n\geq 0}   (-1)^{k_2+\dots+k_n}\xi_{2}^{k_2-1}\dots \xi_{n}^{k_n-1} \notag \\
&\ \ \ \ \times  \begin{pmatrix}
\SIE{k_{n},\dots,k_2}{i_{n},\dots,i_2}(z_0)
\end{pmatrix} \text{ for } i_\ell\in\{0,1,\ell{+}1,\ell{+}2,\dots,n\} \nonumber\\
&=e^{s_{01\dots n}\omega(1,0|\tau)}\sum_{w= 0}^{\infty}(-1)^{w}M_{n,w} \Selbld^{\E}_{n,w}(z_0|\tau)\,,
\end{align}
where $M_{n,w}$ is a matrix of homogeneous degree $w{+}1{-}n$ in the variables $\xi$. The entries of
$M_{n,w}$ are engineered to reproduce the linear combination of the length-$n!$ vector of admissible genus-one Selberg integrals in the first line of \eqn{eqn:trafoSelToT} for a certain sequence $(k_2,\dots,k_n)$:
\begin{align}\label{eqn:defMnw}
M_{n,w} \Selbld^{\E}_{n,w}(z_0|\tau)&=\sum_{\begin{smallmatrix}
	k_2,\dots,k_n\geq 0\\
	k_2+\dots+k_n=w
	\end{smallmatrix}}\xi_{2}^{k_2-1}\dots \xi_{n}^{k_n-1} \begin{pmatrix}
\SIE{k_{n},\dots,k_2}{i_{n},\dots,i_2}(z_0)
\end{pmatrix}   \text{ for } i_\ell\in\{0,1,\ell{+}1,\ell{+}2,\dots,n\}\,.
\end{align}
Thus, we obtain
\begin{align}\label{eqn:KZBforT}
\partial_0 \BT^{\tau}_{0,n}&=e^{s_{01\dots n}\omega(1,0|\tau)}\sum_{k= 0}^{\infty}\sum_{w= 0}^{\infty}(-1)^{w} f^{(k)}_{01} M_n r^{\E}_{0,n}(x_k)M^{-1}_n M_{n,w} \Selbld^{\E}_{n,w}(z_0|\tau)\,.
\end{align}
On the other hand, the derivative $\partial_0 \BT^{\tau}_{0,n}$ can also be calculated using eq.\ \eqref{eqn:degHatS}, which leads to
\begin{align}
\partial_0 \BT^{\tau}_{0,n}
&=e^{s_{01\dots n}\omega(1,0|\tau)}\sum_{w= 0}^{\infty}(-1)^{w}M_{n,w} \partial_0\Selbld^{\E}_{n,w}(z_0|\tau)\nonumber\\
&=e^{s_{01\dots n}\omega(1,0|\tau)}\sum_{w= 0}^{\infty}(-1)^{w}M_{n,w} \sum_{k=0}^{w+1}f^{(k)}_{01}r^{\E}_{0,n}(x_{k,w})\Selbld^{\E}_{n,w+1-k}(z_0|\tau)\\
&=e^{s_{01\dots n}\omega(1,0|\tau)}\sum_{k= 0}^{\infty}\sum_{w\geq \max(0,k-1)}(-1)^{w}f^{(k)}_{01}M_{n,w} r^{\E}_{0,n}(x_{k,w})\Selbld^{\E}_{n,w+1-k}(z_0|\tau)\nonumber\\
&=e^{s_{01\dots n}\omega(1,0|\tau)}\sum_{k= 0}^{\infty}\sum_{v=0}^{\infty}(-1)^{v+k-1}f^{(k)}_{01} (1-\delta_{v,0}\delta_{k,0})M_{n,v+k-1} r^{\E}_{0,n}(x_{k,v+k-1})\Selbld^{\E}_{n,v}(z_0|\tau)\,,\nonumber
\end{align}
where we have used the change of variables $v=w{+}1{-}k$. Comparing this with eq.\ \eqref{eqn:KZBforT} and using the independence of $f^{(k)}_{01}$ and $\Selbld^{\E}_{n,w}(z_0|\tau)$ for different $k$ and $w$, respectively, we can conclude that for any $k,w\geq 0$
\begin{align}\label{eqn:eAndx}
M_n r_{0,n}(x_k)M^{-1}_n M_{n,w}
&=(-1)^{k-1}(1-\delta_{w,0}\delta_{k,0})M_{n,w+k-1} r^{\E}_{0,n}(x_{k,w+k-1})\,.
\end{align}
This equation expresses the relation between the matrices $r_{0,n}(x_k)$ from section \ref{sec:z0Deriv} and the submatrices $r^{\E}_{0,n}(x_{k,w})$ of $r^{\E}_{0,n}(x_{k})$ appearing in the KZB equation in \rcite{Broedel:2019gba}.

\subsubsection{Two-point example}

Let us illustrate and check the formula \eqref{eqn:eAndx} for the two-point example and the block-matrices given in \eqns{eqn:2ptExamplex01}{eqn:2ptExamplex2}. The latter encode the matrices $r^{\E}_{0,n}(x_{k,w})$ for $k=0,1$ at weight $w=0$
\begin{align}\label{eqn:x0w}
r^{\E}_{0,2}(x_{0,0})&=
-s_{12}
\,,\quad r^{\E}_{0,2}(x_{1,0})=
-s_{01}\,,
\end{align}
for $k=0,1,2$ at weight $w=1$
\begin{align}\label{eqn:x1w}
r^{\E}_{0,2}(x_{0,1})&=\begin{pmatrix}
s_{02}&s_{02}
\end{pmatrix}\,,\quad r^{\E}_{0,2}(x_{1,1})=
-s_{012}\,,\quad r^{\E}_{0,2}(x_{2,1})=
s_{02}
\end{align}
and for $k=1,2$ at weight $w=2$
\begin{align}\label{eqn:x2w}
r^{\E}_{0,2}(x_{1,2})&=\begin{pmatrix}
-(s_{01}+s_{02})&s_{02}\\s_{12}&-(s_{01}+s_{12})
\end{pmatrix}\,,\quad r^{\E}_{0,2}(x_{2,2})=\begin{pmatrix}
-s_{12}\\
-s_{12}
\end{pmatrix}\,.
\end{align}
Moreover, comparing the first four entries of the two-point vector $\Selbld^\El_2(z_0)$ in \eqn{eqn:2PtExSelVec} with the defining \eqn{eqn:defSnw} of its constant-weight subvectors $\Selbld^{\E}_{n,w}(z_0|\tau)$, we find that 
\begin{align}
\Selbld^{\E}_{2,0}(z_0|\tau)&=\SIEzwei{0}{1}(z_0)\,,\quad \Selbld^{\E}_{2,1}(z_0|\tau)=\SIEzwei{1}{1}(z_0)\,,\quad \Selbld^{\E}_{2,2}(z_0|\tau)=\begin{pmatrix}\SIEzwei{2}{1}(z_0)\\ \SIEzwei{2}{0}(z_0)\end{pmatrix}\,.
\end{align}
According to \eqn{eqn:defMnw}, for $n=2$ the matrices $M_{2,w}$ are defined by the equation
\begin{align}
M_{2,w} \Selbld^{\E}_{2,w}(z_0|\tau)&=\xi^{w-1} \begin{pmatrix}
\SIE{w}{1}(z_0)\\
\SIE{w}{0}(z_0)
\end{pmatrix}
\end{align}
and, thus, for $w=0,1,2$ they are given by 
\begin{align}\label{eqn:M2w}
M_{2,0}&=\eta^{-1}\begin{pmatrix}
1\\1
\end{pmatrix}\,,\quad M_{2,1}=\begin{pmatrix}
1\\-\frac{s_{12}}{s_{02}}
\end{pmatrix}\,,\quad M_{2,2}=\eta \begin{pmatrix}
1&0\\0&1
\end{pmatrix}\,.
\end{align}
Now, we have all we need to check \eqn{eqn:eAndx} for some configurations of $k$ and $w$. We begin with $(k,w)=(0,0)$ and $r_{0,2}(x_k)$ (for $k=0,1,2$) from \eqn{eqn:2PointEMatrices}, which leads to the equation
\begin{align}
r_{0,2}(x_0)M_{2,0}&=\begin{pmatrix}
0\\0
\end{pmatrix}
\Leftrightarrow\begin{pmatrix}
-s_{02}\partial_{\eta}&-s_{02}/\eta\\s_{12}/\eta&s_{12}\partial_{\eta}
\end{pmatrix}\eta^{-1}\begin{pmatrix}
1\\1
\end{pmatrix}=\begin{pmatrix}
0\\0
\end{pmatrix} \, ,
\end{align}
which is indeed satisfied since $(\partial_\eta + \eta^{-1})\eta^{-1}=0$. For $(k,w)=(0,1)$ the right-hand side of \eqn{eqn:eAndx} is not vanishing and the full equation takes the form 
\begin{align}
r_{0,2}(x_0) M_{2,1}
&=- M_{2,0}r^{\E}_{0,2}(x_{0,0})
\Leftrightarrow \begin{pmatrix}
-s_{02}\partial_{\eta}&-s_{02}/\eta\\s_{12}/\eta&s_{12}\partial_{\eta}
\end{pmatrix}
\begin{pmatrix}
1\\-\frac{s_{12}}{s_{02}}
\end{pmatrix}=\eta^{-1}\begin{pmatrix}
1\\1
\end{pmatrix} s_{12}\, ,
\end{align}
while for $(k,w)=(0,2)$
\begin{align}
r_{0,2}(x_0) M_{2,2}
&=-M_{2,1} r^{\E}_{0,2}(x_{0,1})
\Leftrightarrow \begin{pmatrix}
-s_{02}\partial_{\eta}&-s_{02}/\eta\\s_{12}/\eta&s_{12}\partial_{\eta}
\end{pmatrix}\eta \begin{pmatrix}
1&0\\0&1
\end{pmatrix}
=\begin{pmatrix}
-1\\\frac{s_{12}}{s_{02}}
\end{pmatrix}\begin{pmatrix}
s_{02}&s_{02}
\end{pmatrix} \, .
\end{align}
The lowest-weight configurations for $k=1$ are $(k,w)=(1,0)$ leading to 
\begin{align}
r_{0,2}(x_1) M_{2,0}
&=M_{2,0} r^{\E}_{0,2}(x_{1,0})
\Leftrightarrow 
\begin{pmatrix}
-(s_{01}{+}s_{02})&s_{02}\\s_{12}&-(s_{01}{+}s_{12}) 
\end{pmatrix} \eta^{-1}\begin{pmatrix}
1\\1
\end{pmatrix}
&=-\eta^{-1}\begin{pmatrix}
1\\1
\end{pmatrix} s_{01} \, ,
\end{align}
and $(k,w)=(1,1)$ such that
\begin{align}
r_{0,2}(x_1)M_{2,1}
&=M_{2,1} r^{\E}_{0,2}(x_{1,1})
\Leftrightarrow 
\begin{pmatrix}
-(s_{01}{+}s_{02})&s_{02}\\s_{12}&-(s_{01}{+}s_{12}) 
\end{pmatrix}
\begin{pmatrix}
1\\-\frac{s_{12}}{s_{02}}
\end{pmatrix}
&=-\begin{pmatrix}
1\\-\frac{s_{12}}{s_{02}}
\end{pmatrix} s_{012} \, .
\end{align}
Finally, $(k,w)=(1,2)$ yields the equation 
\begin{align}
r_{0,2}(x_1) M_{2,2}
&=M_{2,2} r^{\E}_{0,2}(x_{1,2})
\end{align}
which holds by $r_{0,2}(x_1)= r^{\E}_{0,2}(x_{1,2})$ and $M_{2,2} = \eta \mathds{1}_{2{\times} 2}$.

For $k=2$, we can check the following two configurations with the explicit examples given above: the weight-zero case $(k,w)=(2,0)$
\begin{align}
r_{0,2}(x_2)M_{2,0}
&=-M_{2,1} r^{\E}_{0,2}(x_{2,1})
\Leftrightarrow \eta\begin{pmatrix}
0&-s_{02}\\s_{12}&0
\end{pmatrix}\eta^{-1}\begin{pmatrix}
1\\1
\end{pmatrix}
=-\begin{pmatrix}
1\\-\frac{s_{12}}{s_{02}}
\end{pmatrix} s_{02}
\end{align}
and weight one $(k,w)=(2,1)$
\begin{align}
r_{0,2}(x_2) M_{2,1}
&=-M_{2,2} r^{\E}_{0,2}(x_{2,2})
\Leftrightarrow \eta\begin{pmatrix}
0&-s_{02}\\s_{12}&0
\end{pmatrix} \begin{pmatrix}
1\\-\frac{s_{12}}{s_{02}}
\end{pmatrix}
=-\eta \begin{pmatrix}
1&0\\0&1
\end{pmatrix} \begin{pmatrix}
-s_{12}\\
-s_{12}
\end{pmatrix}\, .
\end{align}
Thus, we have explicitly approved \eqn{eqn:eAndx} for the lowest-weight configurations involving the objects \eqref{eqn:x0w} to \eqref{eqn:x2w} and \eqref{eqn:M2w} at two points.

\section{Subleading terms in \texorpdfstring{$\BC_{1,n}^{\tau}$}{C(1,n)tau}}\label{restTerms}
In this section, we address the claim above \eqn{eqn:finiteEntriesC1}. We argue that subleading terms appear in the limit $z_0\to 1$ of $\Ztzn{n}((1,A),(0,B))$ and estimate their order in $1{-}z_0$ by first giving the two-point example and afterwards generalizing to $n$ points.

\subsection{Two-point example}

As seen in \eqn{eqn:KNdegeneration1}, the integrand of
\begin{align}\label{eqn:app2pt}
\Ztzn{2}(1,2)&=\int_0^{z_0}\dd z_2 \, e^{-s_{01}\CG^{\tau}_{01}-s_{02}\CG^{\tau}_{02}-s_{12}\CG^{\tau}_{12}}\Omega_{12}(\eta)
\end{align}
has a leading singularity of order $(1{-}z_0)^{-s_{01}}$ as $z_0\to 1$:
\begin{align}
e^{-s_{01}\CG^{\tau}_{01}-s_{02}\CG^{\tau}_{02}-s_{12}\CG^{\tau}_{12}}\Omega_{12}(\eta)&=(-2\pi i (1{-}z_0))^{-s_{01}}e^{s_{01}\omega(1,0|\tau)}e^{-\tilde s_{12}\CG^{\tau}_{12}}\Omega_{12}(\eta)+\mathcal{O}((1{-}z_0)^{-s_{01}+1})\,.
\end{align}
This singularity appears due to the merging of the punctures $z_0\to 1\cong 0=z_1$. However, in order to calculate the limit $z_0\to 1$ of the integral, we can not simply set $z_0=1$ in \eqn{eqn:app2pt}, which can be deduced using the following integral decomposition 
\begin{align}
\Ztzn{2}(1,2)&=\int_0^{1}\dd z_2 \, e^{-s_{01}\CG^{\tau}_{01}-s_{02}\CG^{\tau}_{02}-s_{12}\CG^{\tau}_{12}}\Omega_{12}(\eta)-\int_{z_0}^{1}\dd z_2 \, e^{-s_{01}\CG^{\tau}_{01}-s_{02}\CG^{\tau}_{02}-s_{12}\CG^{\tau}_{12}}\Omega_{12}(\eta)\,.
\label{dec2pt}
\end{align}
Thus, taking $z_0\to 1$ involves a subleading term given by the latter integral above, where the three punctures $z_0<z_2<1\cong z_1$ merge. Similar to the estimate \eqref{eqn:KNlowerlimit} for the lower boundary value, this leads to a term of order $(1{-}z_0)^{-s_{012}}$,
\begin{align}
\int_{z_0}^{1}\dd z_2 \, e^{-s_{01}\CG^{\tau}_{01}-s_{02}\CG^{\tau}_{02}-s_{12}\CG^{\tau}_{12}}\Omega_{12}(\eta)&=\mathcal{O}((-2\pi i (1{-}z_0))^{-s_{012}})\,,
\label{scal2pt}
\end{align}
as can for instance be checked through a change of variables
$1{-}z_2 = x_2(1{-}z_0)$ with $x_2 \in (0,1)$. Altogether, we find the estimate that for $z_0\to 1$
\begin{align}
\Ztzn{2}(1,2)&=(-2\pi i (1{-}z_0))^{-s_{01}}e^{s_{01}\omega(1,0|\tau)}\Ztn{2}|_{\tilde s_{12}}+\mathcal{O}((1{-}z_0)^{-s_{01}+1},(1{-}z_0)^{-s_{012}})\,.
\end{align}
Therefore, if we follow the assumption \eqref{assumpGenusOne}, which implies $\Re(s_{012})<\Re(s_{01})<0$, one can extract the finite value 
\begin{align}
\lim_{z_0\to 1}(-2\pi i (1{-}z_0))^{s_{01}}\Ztzn{2}(1,2)&=e^{s_{01}\omega(1,0|\tau)}\Ztn{2}|_{\tilde s_{12}}\, ,
\end{align}
which reproduces the first component of \eqn{eqn:2ptC1}.

\subsection{\texorpdfstring{$n$}{n}-point generalization}

The above argument can be applied to the $n$-point case by generalizing
\eqn{dec2pt} to the decomposition of the integration domain of the $Z^\tau$ integrals 
without augmentations
\begin{align}
 \{z_2,z_3,\ldots,z_n \in \mathbb R\, |\, 0=z_1<z_2<z_3<\ldots <z_n<1 \}
= \gamma_{12\ldots n0} \cup \bigg( \bigcup^{n-1}_{k=1} \gamma_{12\ldots k,0,k+1\ldots n}  \bigg)
\label{decomnpt}
\end{align}
with the following generalization of the integration domain $\gamma_{12\ldots n0} $ in \eqn{shortdom}:
\beq
\gamma_{12\ldots k,0,k+1\ldots n} =  \{z_2,z_3,\ldots,z_n \in \mathbb R\, |\, 0=z_1<z_2<z_3<\ldots <z_k<z_0<z_{k+1} < \ldots <z_n<1 \}\,.
\label{dominterest}
\eeq
These domains $\gamma_{12\ldots k,0,k+1\ldots n} $ 
with $k=1,2,\ldots,n{-}1$ cause the $z_0 \rightarrow 1$ limit of $Z^\tau_{0,n}$ to deviate from 
the non-augmented $Z^\tau_{n}$ at arguments $\tilde s_{ij}$ in \eqn{shiftmands}, 
see e.g.\ the second term in \eqn{dec2pt} at $n=2$. The scaling behaviour of $\gamma_{12\ldots k,0,k+1\ldots n} $ 
integrals in the limit $z_0 \rightarrow 1$ can be conveniently extracted by substituting $z_0 = 1{-}w_0$ and
$z_j= 1{-}w_j$ for $j=k{+}1,\ldots,n$ followed by the rescaling $w_j=w_0 x_j$, see
\eqn{scal2pt} for the two-point result. For a generic function $F$ of $z_2,z_3,\ldots,z_n$ in the integrand, this
amounts to parametrizing the merging of $n{-}k{+}2$ consecutive punctures $z_0< z_{k+1}<\dots < z_{n}<1\cong z_1$
in the limit $z_0 \rightarrow 1$ or $w_0 \rightarrow 0$ via
\begin{align}
&\! \! \! \! \! \! \! \! \int \limits_{ \gamma_{12\ldots k,0,k+1\ldots n}  } \! \! \! \! \! \! \! \! \dd z_2\, \ldots \, \dd z_n \, F(z_2,\ldots,z_n)
= \! \! \! \! \! \! \! \! \! \! \int \limits_{0<z_2<\ldots<z_k<z_0}  \! \!  \! \! \! \! \! \! \! \!  \dd z_2\, \ldots \, \dd z_k 
\! \! \! \! \! \! \! \! \! \!  \int \limits_{z_0<z_{k+1} <\ldots < z_n<1} \! \!  \! \! \! \! \! \! \! \!  \dd z_{k+1} \, \ldots \, \dd z_{n} \, F(z_2,\ldots,z_n) 
\label{scalnpt} \\
&= w_0^{n-k}  \! \! \! \! \! \! \! \!  \int \limits_{0<z_2<\ldots<z_k<z_0} \! \! \! \! \! \! \! \!  \dd z_2\, \ldots \, \dd z_k 
\! \! \! \! \! \! \! \!  \int \limits_{0<x_n<\ldots < x_{k+1}<1} \! \! \! \! \! \! \! \!  \dd x_{k+1} \, \ldots \, \dd x_{n} \, F(z_2,\ldots,z_k,1{-}x_{k+1}w_0 , \ldots ,1{-}x_{n}w_0 ) \, . \notag
\end{align}
The integrands $F$ relevant to $\Ztzn{n}$ involve the augmented Koba-Nielsen factor
which scales as $w_0^{-s_{01(k+1)\dots n}}=(1{-}z_0)^{-s_{01(k+1)\dots n}}$ in this limit. This can be seen
by repeating the analysis of the $z_0 \rightarrow 0$ limit in \eqn{eqn:KNlowerlimit} with $w_0=1{-}z_0$ in
the place of $z_0$ and employing the arguments seen on the right-hand side of \eqn{scalnpt}. Then, as for the two-point example, the condition \eqref{assumpGenusOne}, such that $\Re(s_{01(k+1)\dots n})<\Re(s_{01})<0$,
ensures that \eqn{scalnpt} vanishes for $k=1,2,\ldots,n{-}1$ if the $z_0 \rightarrow 1$ limit is taken in presence of the regulating factor
$(-2\pi i (1{-}z_0))^{s_{01}}$ in front of $\Ztzn{n}((1,A),(0,B))$ in \eqn{eqn:finiteEntriesC1}. Hence, the $w_0^{-s_{01(k+1)\dots n}}$ scaling of the Koba-Nielsen factor entering the $F$ in \eqn{scalnpt} implies our claim
\eqref{eqn:finiteEntriesC1} for the regularized limit defining $\BC_{1,n}^{\tau}$.

\subsection{Further comments on subleading terms}

Note that the additional scaling $\sim w_0^{n-k}$ in \eqn{scalnpt} by integer powers does not affect
the derivation of \eqn{eqn:finiteEntriesC1} based on the scaling of the Koba-Nielsen factor: the 
prefactor $w_0^{n-k}$ is either
compensated by the $\Omega_{ij}$ in some of the vector components of $\Ztzn{n}((1,A),(0,B))$, or it
may suppress the contribution of  $\gamma_{12\ldots k,0,k+1\ldots n} $ to \eqn{decomnpt} in other vector components. At $n=3$ points, for instance, the integrand of $\Ztzn{3}$ in \eqn{eqn:3PointExampleIntegral}
involves Kronecker-Eisenstein series with the following scaling as $z_0 \rightarrow 1$
in the domains $\gamma_{1023}$ and $\gamma_{1203}$ of \eqn{dominterest}:
\begin{align}
\int \limits_{\gamma_{1023}} \! \! \! \dd z_2 \, \dd z_3\,\KN^{\tau}_{0123}\begin{pmatrix}
\Omega_{12}(\eta_{23}) \Omega_{23}(\eta_{3}) \\ 
\Omega_{13}(\eta_{23}) \Omega_{32}(\eta_{2}) \\ 
 \Omega_{12}(\eta_{2})  \Omega_{03}(\eta_{3}) \\
  \Omega_{13}(\eta_{3})  \Omega_{02}(\eta_{2}) \\
\Omega_{03}(\eta_{23}) \Omega_{32}(\eta_{2}) \\ 
\Omega_{02}(\eta_{23}) \Omega_{23}(\eta_{3})
\end{pmatrix} &\stackrel{z_0\rightarrow 1}{\longrightarrow} w_0^2 \! \! \! \! \! \! \! \! \int \limits_{0<x_3<x_2<1} \! \! \! \! \! \! \! \! \dd x_2 \, \dd x_3 \,
\KN^{\tau}_{0123}\begin{pmatrix}
(w_0^2 x_2 x_{32})^{-1} \\ 
(w_0^2 x_3 x_{23})^{-1} \\ 
(w_0^2 x_2(x_3{-}1))^{-1} \\
(w_0^2 x_3 (x_{2}{-}1))^{-1} \\
(w_0^2 x_{23}(x_3{-}1))^{-1} \\ 
(w_0^2 x_{32}(x_2{-}1))^{-1}
\end{pmatrix}  \\
\int \limits_{\gamma_{1203}} \! \! \! \dd z_2 \, \dd z_3\,\KN^{\tau}_{0123}\begin{pmatrix}
\Omega_{12}(\eta_{23}) \Omega_{23}(\eta_{3}) \\ 
\Omega_{13}(\eta_{23}) \Omega_{32}(\eta_{2}) \\ 
 \Omega_{12}(\eta_{2})  \Omega_{03}(\eta_{3}) \\
  \Omega_{13}(\eta_{3})  \Omega_{02}(\eta_{2}) \\
\Omega_{03}(\eta_{23}) \Omega_{32}(\eta_{2}) \\ 
\Omega_{02}(\eta_{23}) \Omega_{23}(\eta_{3})
\end{pmatrix} &\stackrel{z_0\rightarrow 1}{\longrightarrow} w_0 \int_0^{z_0} \dd z_2\, \int_{0}^1 \dd x_3 \,
\KN^{\tau}_{0123}\begin{pmatrix}
\Omega_{12}(\eta_{23}) \Omega_{23}(\eta_{3}) \\ 
(w_0 x_3)^{-1} \Omega_{32}(\eta_{2}) \\ 
(w_0 (x_3{-}1))^{-1}  \Omega_{12}(\eta_{2})   \\
(w_0 x_3)^{-1}   \Omega_{02}(\eta_{2}) \\
(w_0 (x_3{-}1))^{-1}  \Omega_{32}(\eta_{2}) \\ 
\Omega_{02}(\eta_{23}) \Omega_{23}(\eta_{3})
\end{pmatrix}  \notag 
\end{align}
In the first case of $\gamma_{1023}$, i.e.\ \eqn{scalnpt} at $k=1$, the Kronecker--Eisenstein series in all the six vector entries have a singular term that scales as $w_0^{-2}$ and compensates the prefactor from $w_0^{n-k}$.  In the second case of $\gamma_{1203}$, i.e.\ \eqn{scalnpt} at $k=2$, only the middle four vector components feature a Kronecker--Eisenstein series with most singular term $w_0^{-1}$. Accordingly, the first and last vector entry are suppressed by one additional power of $w_0$ as $w_0 \rightarrow 0$. 

Still, this example illustrates that the analysis of vector entries with additional suppression by powers of $w_0$ does not play any role for the above conclusion: all components of \eqn{scalnpt} vanish when the limit $w_0 \rightarrow 0$ is performed in presence of the regulating factor $(-2\pi i (1{-}z_0))^{s_{01}}$.

\vspace{-0.2cm}

\bibliographystyle{KZBTranslation}
\bibliography{KZBTranslation}

\def\cprime{$'$}
\begin{thebibliography}{10}
\ifx\href\asklfhas\newcommand{\href}[2]{#2}\fi
\ifx\arxivref\asklfhas\newcommand{\arxivref}[2]{\href{http://arxiv.org/abs/#1}{#2}}\fi
\ifx\doiref\asklfhas\newcommand{\doiref}[2]{\href{http://dx.doi.org/#1}{#2}}\fi
\raggedright
\small
\parskip 0pt

\bibitem{Mafra:2019xms}
C.~R.~Mafra and O.~Schlotterer,
\textit{``{One-loop open-string integrals from differential equations:
  all-order $\alpha$'-expansions at $n$ points}''},
\textsf{\doiref{10.1007/JHEP03(2020)007}{JHEP~2003,~007~(2020)}},
\texttt{\arxivref{1908.10830}{arxiv:1908.10830}}.

\bibitem{Mafra:2019ddf}
C.~R.~Mafra and O.~Schlotterer,
\textit{``{All-order alpha'-expansion of one-loop open-string integrals}''},
\textsf{\doiref{10.1103/PhysRevLett.124.101603}{Phys.~Rev.~Lett.~124,~101603~(2020)}},
\texttt{\arxivref{1908.09848}{arxiv:1908.09848}}.

\bibitem{Broedel:2019gba}
J.~Broedel and A.~Kaderli,
\textit{``{Amplitude recursions with an extra marked point}''},
\texttt{\arxivref{1912.09927}{arxiv:1912.09927}}.

\bibitem{Berends:1987me}
F.~A.~Berends and W.~T.~Giele,
\textit{``{Recursive Calculations for Processes with n Gluons}''},
\textsf{\doiref{10.1016/0550-3213(88)90442-7}{Nucl.~Phys.~B306,~759~(1988)}}.

\bibitem{Cachazo:2004kj}
F.~Cachazo, P.~Svrcek and E.~Witten,
\textit{``{MHV vertices and tree amplitudes in gauge theory}''},
\textsf{\doiref{10.1088/1126-6708/2004/09/006}{JHEP~0409,~006~(2004)}},
\texttt{\arxivref{hep-th/0403047}{hep-th/0403047}}.

\bibitem{Britto:2005fq}
R.~Britto, F.~Cachazo, B.~Feng and E.~Witten,
\textit{``{Direct proof of tree-level recursion relation in Yang-Mills
  theory}''},
\textsf{\doiref{10.1103/PhysRevLett.94.181602}{Phys.~Rev.~Lett.~94,~181602~(2005)}},
\texttt{\arxivref{hep-th/0501052}{hep-th/0501052}}.

\bibitem{CaronHuot:2011kk}
S.~Caron-Huot and S.~He,
\textit{``{Jumpstarting the All-Loop S-Matrix of Planar N=4 Super
  Yang-Mills}''},
\textsf{\doiref{10.1007/JHEP07(2012)174}{JHEP~1207,~174~(2012)}},
\texttt{\arxivref{1112.1060}{arxiv:1112.1060}}.

\bibitem{Baadsgaard:2015twa}
C.~Baadsgaard, N.~Bjerrum-Bohr, J.~L.~Bourjaily, S.~Caron-Huot, P.~H.~Damgaard
  and B.~Feng,
\textit{``{New Representations of the Perturbative S-Matrix}''},
\textsf{\doiref{10.1103/PhysRevLett.116.061601}{Phys.~Rev.~Lett.~116,~061601~(2016)}},
\texttt{\arxivref{1509.02169}{arxiv:1509.02169}}.

\bibitem{Cachazo:2015aol}
F.~Cachazo, S.~He and E.~Y.~Yuan,
\textit{``{One-Loop Corrections from Higher Dimensional Tree Amplitudes}''},
\textsf{\doiref{10.1007/JHEP08(2016)008}{JHEP~1608,~008~(2016)}},
\texttt{\arxivref{1512.05001}{arxiv:1512.05001}}.

\bibitem{Arkani-Hamed:2016byb}
N.~Arkani-Hamed, J.~L.~Bourjaily, F.~Cachazo, A.~B.~Goncharov, A.~Postnikov and
  J.~Trnka,
\textit{``{Grassmannian Geometry of Scattering Amplitudes}''},
Cambridge University Press (2016).

\bibitem{Jurco:2019yfd}
B.~Jur\v{c}o, T.~Macrelli, C.~Saemann and M.~Wolf,
\textit{``{Loop Amplitudes and Quantum Homotopy Algebras}''},
\textsf{\doiref{10.1007/JHEP07(2020)003}{JHEP~2007,~003~(2020)}},
\texttt{\arxivref{1912.06695}{arxiv:1912.06695}}.

\bibitem{Broedel:2013aza}
J.~Broedel, O.~Schlotterer, S.~Stieberger and T.~Terasoma,
\textit{``{All order $\alpha^{\prime}$-expansion of superstring trees from the
  Drinfeld associator}''},
\textsf{\doiref{10.1103/PhysRevD.89.066014}{Phys.Rev.~D89,~066014~(2014)}},
\texttt{\arxivref{1304.7304}{arxiv:1304.7304}}.

\bibitem{Aomoto}
K.~Aomoto,
\textit{``{Special values of hyperlogarithms and linear difference schemes}''},
\textsf{Illinois~J.~Math.~34,~191~(1990)}.

\bibitem{Terasoma}
T.~Terasoma,
\textit{``{Selberg Integrals and Multiple Zeta Values}''},
\textsf{Compositio~Mathematica~133,~1~(2002)}.

\bibitem{Mafra:2016mcc}
C.~R.~Mafra and O.~Schlotterer,
\textit{``{Non-abelian $Z$-theory: Berends-Giele recursion for the
  $\alpha'$-expansion of disk integrals}''},
\textsf{\doiref{10.1007/JHEP01(2017)031}{JHEP~1701,~031~(2017)}},
\texttt{\arxivref{1609.07078}{arxiv:1609.07078}}.

\bibitem{Puhlfuerst:2015gta}
G.~Puhlfürst and S.~Stieberger,
\textit{``{Differential Equations, Associators, and Recurrences for
  Amplitudes}''},
\textsf{\doiref{10.1016/j.nuclphysb.2015.11.005}{Nucl.~Phys.~B902,~186~(2016)}},
\texttt{\arxivref{1507.01582}{arxiv:1507.01582}}.

\bibitem{Levin}
A.~Levin,
\textit{``{Elliptic polylogarithms: An analytic theory}''},
\textsf{Compositio~Mathematica~106,~267~(1997)}.

\bibitem{BrownLev}
F.~Brown and A.~Levin,
\textit{``{Multiple elliptic polylogarithms}''},
\texttt{\arxivref{1110.6917v2}{arxiv:1110.6917v2}}.

\bibitem{Enriquez:Emzv}
B.~Enriquez,
\textit{``Analogues elliptiques des nombres multiz\'etas''},
\textsf{Bull.~Soc.~Math.~France~144,~395~(2016)},
\texttt{\arxivref{1301.3042}{arxiv:1301.3042}}.

\bibitem{Broedel:2015hia}
J.~Broedel, N.~Matthes and O.~Schlotterer,
\textit{``{Relations between elliptic multiple zeta values and a special
  derivation algebra}''},
\textsf{\doiref{10.1088/1751-8113/49/15/155203}{J.~Phys.~A49,~155203~(2016)}},
\texttt{\arxivref{1507.02254}{arxiv:1507.02254}}.

\bibitem{Broedel:2014vla}
J.~Broedel, C.~R.~Mafra, N.~Matthes and O.~Schlotterer,
\textit{``{Elliptic multiple zeta values and one-loop superstring
  amplitudes}''},
\textsf{\doiref{10.1088/1126-6708/2015/07/112}{JHEP~1507,~112~(2015)}},
\texttt{\arxivref{1412.5535}{arxiv:1412.5535}}.

\bibitem{Broedel:2017jdo}
J.~Broedel, N.~Matthes, G.~Richter and O.~Schlotterer,
\textit{``{Twisted elliptic multiple zeta values and non-planar one-loop
  open-string amplitudes}''},
\textsf{\doiref{10.1088/1751-8121/aac601}{J.~Phys.~A51,~285401~(2018)}},
\texttt{\arxivref{1704.03449}{arxiv:1704.03449}}.

\bibitem{Broedel:2019vjc}
J.~Broedel and O.~Schlotterer,
\textit{``{One-Loop String Scattering Amplitudes as Iterated Eisenstein
  Integrals}''},
in: \textit{``{KMPB Conference}: {Elliptic Integrals, Elliptic Functions and
  Modular Forms in Quantum Field Theory}''},
133--159p.

\bibitem{DHoker:2015wxz}
E.~D'Hoker, M.~B.~Green, O.~Gurdogan and P.~Vanhove,
\textit{``{Modular Graph Functions}''},
\textsf{\doiref{10.4310/CNTP.2017.v11.n1.a4}{Commun.~Num.~Theor.~Phys.~11,~165~(2017)}},
\texttt{\arxivref{1512.06779}{arxiv:1512.06779}}.

\bibitem{DHoker:2016mwo}
E.~D'Hoker and M.~B.~Green,
\textit{``{Identities between Modular Graph Forms}''},
\textsf{J.~Number~Theor.~189,~25~(2018)},
\texttt{\arxivref{1603.00839}{arxiv:1603.00839}}.

\bibitem{Zerbini:2015rss}
F.~Zerbini,
\textit{``{Single-valued multiple zeta values in genus 1 superstring
  amplitudes}''},
\textsf{\doiref{10.4310/CNTP.2016.v10.n4.a2}{Commun.~Num.~Theor.~Phys.~10,~703~(2016)}},
\texttt{\arxivref{1512.05689}{arxiv:1512.05689}}.

\bibitem{Brown:2017qwo}
F.~Brown,
\textit{``{A class of non-holomorphic modular forms I}''},
\textsf{Res.~Math.~Sci.~5,~5:7~(2018)},
\texttt{\arxivref{1707.01230}{arxiv:1707.01230}}.

\bibitem{Brown:2017qwo2}
F.~Brown,
\textit{``{A class of non-holomorphic modular forms II : equivariant iterated
  Eisenstein integrals}''},
\texttt{\arxivref{1708.03354}{arxiv:1708.03354}}.

\bibitem{DHoker:2019xef}
E.~D'Hoker and M.~Green,
\textit{``{Absence of irreducible multiple zeta-values in melon modular graph
  functions}''},
\textsf{\doiref{10.4310/CNTP.2020.v14.n2.a2}{Commun.~Num.~Theor.~Phys.~14,~315~(2020)}},
\texttt{\arxivref{1904.06603}{arxiv:1904.06603}}.

\bibitem{Zagier:2019eus}
D.~Zagier and F.~Zerbini,
\textit{``{Genus-zero and genus-one string amplitudes and special multiple zeta
  values}''},
\textsf{\doiref{10.4310/CNTP.2020.v14.n2.a4}{Commun.~Num.~Theor.~Phys.~14,~413~(2020)}},
\texttt{\arxivref{1906.12339}{arxiv:1906.12339}}.

\bibitem{Mizera:2019gea}
S.~Mizera,
\textit{``{Aspects of Scattering Amplitudes and Moduli Space Localization}''},
PhD thesis,
Perimeter Inst. Theor. Phys.,
2019.

\bibitem{Mandelstam:1974fq}
S.~Mandelstam,
\textit{``{Dual - Resonance Models}''},
\textsf{\doiref{10.1016/0370-1573(74)90034-9}{Phys.~Rept.~13,~259~(1974)}}.

\bibitem{Brown:2019wna}
F.~Brown and C.~Dupont,
\textit{``{Single-valued integration and superstring amplitudes in genus
  zero}''},
\texttt{\arxivref{1910.01107}{arxiv:1910.01107}}.

\bibitem{DHoker:1994gnm}
E.~D'Hoker and D.~Phong,
\textit{``{The Box graph in superstring theory}''},
\textsf{\doiref{10.1016/0550-3213(94)00526-K}{Nucl.~Phys.~B~440,~24~(1995)}},
\texttt{\arxivref{hep-th/9410152}{hep-th/9410152}}.

\bibitem{Mafra:2011nv}
C.~R.~Mafra, O.~Schlotterer and S.~Stieberger,
\textit{``{Complete N-Point Superstring Disk Amplitude I. Pure Spinor
  Computation}''},
\textsf{\doiref{10.1016/j.nuclphysb.2013.04.023}{Nucl.Phys.~B873,~419~(2013)}},
\texttt{\arxivref{1106.2645}{arxiv:1106.2645}}.

\bibitem{Broedel:2013tta}
J.~Broedel, O.~Schlotterer and S.~Stieberger,
\textit{``{Polylogarithms, Multiple Zeta Values and Superstring Amplitudes}''},
\textsf{\doiref{10.1002/prop.201300019}{Fortsch.Phys.~61,~812~(2013)}},
\texttt{\arxivref{1304.7267}{arxiv:1304.7267}}.

\bibitem{Azevedo:2018dgo}
T.~Azevedo, M.~Chiodaroli, H.~Johansson and O.~Schlotterer,
\textit{``{Heterotic and bosonic string amplitudes via field theory}''},
\textsf{\doiref{10.1007/JHEP10(2018)012}{JHEP~1810,~012~(2018)}},
\texttt{\arxivref{1803.05452}{arxiv:1803.05452}}.

\bibitem{Bern:2008qj}
Z.~Bern, J.~Carrasco and H.~Johansson,
\textit{``{New Relations for Gauge-Theory Amplitudes}''},
\textsf{\doiref{10.1103/PhysRevD.78.085011}{Phys.Rev.~D78,~085011~(2008)}},
\texttt{\arxivref{0805.3993}{arxiv:0805.3993}}.

\bibitem{aomoto1987}
K.~Aomoto,
\textit{``Gauss-Manin connection of integral of difference products''},
\textsf{\doiref{10.2969/jmsj/03920191}{J.~Math.~Soc.~Japan~39,~191~(1987)}}.

\bibitem{Mizera:2017cqs}
S.~Mizera,
\textit{``{Combinatorics and Topology of Kawai-Lewellen-Tye Relations}''},
\textsf{\doiref{10.1007/JHEP08(2017)097}{JHEP~1708,~097~(2017)}},
\texttt{\arxivref{1706.08527}{arxiv:1706.08527}}.

\bibitem{Mafra:2011nw}
C.~R.~Mafra, O.~Schlotterer and S.~Stieberger,
\textit{``{Complete N-Point Superstring Disk Amplitude II. Amplitude and
  Hypergeometric Function Structure}''},
\textsf{\doiref{10.1016/j.nuclphysb.2013.04.022}{Nucl.Phys.~B873,~461~(2013)}},
\texttt{\arxivref{1106.2646}{arxiv:1106.2646}}.

\bibitem{Goncharov:2001iea}
A.~Goncharov,
\textit{``{Multiple polylogarithms and mixed Tate motives}''},
\texttt{\arxivref{math/0103059}{math/0103059}}.

\bibitem{Duhr:2011zq}
C.~Duhr, H.~Gangl and J.~R.~Rhodes,
\textit{``{From polygons and symbols to polylogarithmic functions}''},
\textsf{\doiref{10.1088/1126-6708/2012/10/075}{JHEP~1210,~075~(2012)}},
\texttt{\arxivref{1110.0458}{arxiv:1110.0458}}.

\bibitem{Panzer:2015ida}
E.~Panzer,
\textit{``{Feynman integrals and hyperlogarithms}''},
PhD thesis,
Humboldt U.,
2015.

\bibitem{Kaderli:2019dny}
A.~Kaderli,
\textit{``{A note on the Drinfeld associator for genus-zero superstring
  amplitudes in twisted de Rham theory}''},
\texttt{\arxivref{1912.09406}{arxiv:1912.09406}},
accepted for publication in JPhysA.

\bibitem{MZVWebsite}
J.~Broedel, O.~Schlotterer and S.~Stieberger,
\href{http://mzv.mpp.mpg.de}{\texttt{http://mzv.mpp.mpg.de}}.

\bibitem{gitrep}
C.~Mafra and O.~Schlotterer,
\href{https://repo.or.cz/BGap.git}{\texttt{https://repo.or.cz/BGap.git}}.

\bibitem{Oprisa:2005wu}
D.~Oprisa and S.~Stieberger,
\textit{``{Six gluon open superstring disk amplitude, multiple hypergeometric
  series and Euler-Zagier sums}''},
\texttt{\arxivref{hep-th/0509042}{hep-th/0509042}}.

\bibitem{Stieberger:2006te}
S.~Stieberger and T.~R.~Taylor,
\textit{``{Multi-Gluon Scattering in Open Superstring Theory}''},
\textsf{\doiref{10.1103/PhysRevD.74.126007}{Phys.Rev.~D74,~126007~(2006)}},
\texttt{\arxivref{hep-th/0609175}{hep-th/0609175}}.

\bibitem{Stieberger:2009rr}
S.~Stieberger,
\textit{``{Constraints on Tree-Level Higher Order Gravitational Couplings in
  Superstring Theory}''},
\textsf{\doiref{10.1103/PhysRevLett.106.111601}{Phys.Rev.Lett.~106,~111601~(2011)}},
\texttt{\arxivref{0910.0180}{arxiv:0910.0180}}.

\bibitem{Boels:2013jua}
R.~H.~Boels,
\textit{``{On the field theory expansion of superstring five point
  amplitudes}''},
\textsf{\doiref{10.1016/j.nuclphysb.2013.08.009}{Nucl.~Phys.~B876,~215~(2013)}},
\texttt{\arxivref{1304.7918}{arxiv:1304.7918}}.

\bibitem{Mafra:2016ltu}
C.~R.~Mafra,
\textit{``{Berends-Giele recursion for double-color-ordered amplitudes}''},
\textsf{\doiref{10.1007/JHEP07(2016)080}{JHEP~1607,~080~(2016)}},
\texttt{\arxivref{1603.09731}{arxiv:1603.09731}}.

\bibitem{Polchinski:1998rq}
J.~Polchinski,
\textit{``{String theory. Vol. 1: An introduction to the bosonic string}''},
Cambridge University Press (2007).

\bibitem{Green:1987mn}
M.~B.~Green, J.~Schwarz and E.~Witten,
\textit{``{Superstring Theory. Vol. 2: Loop amplitudes, anomalies and
  phenomenology}''},
Cambridge, UK: Univ. Pr. (1987) (Cambridge Monographs on Mathematical Physics)
  (1987).

\bibitem{CalaqueGonz}
D.~Calaque and M.~Gonzalez,
\textit{``{On the universal ellipsitomic KZB connection}''},
\texttt{\arxivref{1908.03887}{arxiv:1908.03887}}.

\bibitem{Broedel:2018izr}
J.~Broedel, O.~Schlotterer and F.~Zerbini,
\textit{``{From elliptic multiple zeta values to modular graph functions: open
  and closed strings at one loop}''},
\textsf{\doiref{10.1007/JHEP01(2019)155}{JHEP~1901,~155~(2019)}},
\texttt{\arxivref{1803.00527}{arxiv:1803.00527}}.

\bibitem{Zerbini:2018sox}
F.~Zerbini,
\textit{``{Elliptic multiple zeta values, modular graph functions and genus 1
  superstring scattering amplitudes}''},
PhD thesis,
Bonn U.,
2017.

\bibitem{Zerbini:2018hgs}
F.~Zerbini,
\textit{``{Modular and holomorphic graph function from superstring
  amplitudes}''},
\texttt{\arxivref{1807.04506}{arxiv:1807.04506}},
in: \textit{``{KMPB Conference: Elliptic Integrals, Elliptic Functions and
  Modular Forms in Quantum Field Theory Zeuthen, Germany, October 23-26,
  2017}''}.

\bibitem{Kronecker}
L.~Kronecker,
\textit{``{Zur Theorie der elliptischen Funktionen}''},
\textsf{Mathematische~Werke~IV,~313~(1881)}.

\bibitem{mumford1984tata}
D.~Mumford, M.~Nori and P.~Norman,
\textit{``Tata Lectures on Theta I, II''},
Birkh{\"a}user (1983, 1984).

\bibitem{Brown:mmv}
F.~Brown,
\textit{``Multiple modular values and the relative completion of the
  fundamental group of $M_{1,1}$''},
\texttt{\arxivref{1407.5167v4}{arxiv:1407.5167v4}}.

\bibitem{Matthes:edzv}
N.~Matthes,
\textit{``Elliptic double zeta values''},
\textsf{\doiref{10.1016/j.jnt.2016.07.010}{J.~Number~Theory~171,~227~(2017)}}.

\bibitem{Matthes:Thesis}
N.~Matthes,
\textit{``{Elliptic multiple zeta values}''},
PhD thesis,
Universit\"at Hamburg,
2016.

\bibitem{Henn:2013pwa}
J.~M.~Henn,
\textit{``{Multiloop integrals in dimensional regularization made simple}''},
\textsf{\doiref{10.1103/PhysRevLett.110.251601}{Phys.~Rev.~Lett.~110,~251601~(2013)}},
\texttt{\arxivref{1304.1806}{arxiv:1304.1806}}.

\bibitem{Adams:2018yfj}
L.~Adams and S.~Weinzierl,
\textit{``{The $\varepsilon$-form of the differential equations for Feynman
  integrals in the elliptic case}''},
\textsf{\doiref{10.1016/j.physletb.2018.04.002}{Phys.~Lett.~B~781,~270~(2018)}},
\texttt{\arxivref{1802.05020}{arxiv:1802.05020}}.

\bibitem{Drinfeld:1989st}
V.~G.~Drinfeld,
\textit{``Quasi-{H}opf algebras''},
\textsf{Algebra~i~Analiz~1,~114~(1989)}.

\bibitem{Drinfeld2}
V.~Drinfeld,
\textit{``{On quasitriangular quasi-Hopf algebras and on a group that is
  closely connected with $\text{Gal}(\bar{\mathbb Q}/ \mathbb Q)$}''},
\textsf{Leningrad~Math.~J. 2 (4),~829~(1991)}.

\bibitem{KZB}
D.~Calaque, B.~Enriquez and P.~Etingof,
\textit{``Universal {KZB} equations: the elliptic case''},
in: \textit{``Algebra, arithmetic, and geometry: in honor of {Y}u. {I}.
  {M}anin. {V}ol. {I}''},
Birkh\"auser Boston, Inc., Boston, MA (2009),
165--266p.

\bibitem{EnriquezEllAss}
B.~Enriquez,
\textit{``Elliptic associators''},
\textsf{\doiref{10.1007/s00029-013-0137-3}{Selecta~Math.~(N.S.)~20,~491~(2014)}}.

\bibitem{Hain}
R.~Hain,
\textit{``{Notes on the universal elliptic KZB equation}''},
\texttt{\arxivref{1309.0580}{arxiv:1309.0580}}.

\bibitem{DHoker:2019blr}
E.~D'Hoker and M.~B.~Green,
\textit{``{Exploring transcendentality in superstring amplitudes}''},
\textsf{\doiref{10.1007/JHEP07(2019)149}{JHEP~1907,~149~(2019)}},
\texttt{\arxivref{1906.01652}{arxiv:1906.01652}}.

\bibitem{Gerken:2020yii}
J.~E.~Gerken, A.~Kleinschmidt and O.~Schlotterer,
\textit{``{Generating series of all modular graph forms from iterated
  Eisenstein integrals}''},
\texttt{\arxivref{2004.05156}{arxiv:2004.05156}}.

\bibitem{Kotikov:1990kg}
A.~Kotikov,
\textit{``{Differential equations method: New technique for massive Feynman
  diagrams calculation}''},
\textsf{\doiref{10.1016/0370-2693(91)90413-K}{Phys.~Lett.~B~254,~158~(1991)}}.

\bibitem{ArkaniHamed:2010gh}
N.~Arkani-Hamed, J.~L.~Bourjaily, F.~Cachazo and J.~Trnka,
\textit{``{Local Integrals for Planar Scattering Amplitudes}''},
\textsf{\doiref{10.1007/JHEP06(2012)125}{JHEP~1206,~125~(2012)}},
\texttt{\arxivref{1012.6032}{arxiv:1012.6032}}.

\bibitem{Broedel:2018qkq}
J.~Broedel, C.~Duhr, F.~Dulat, B.~Penante and L.~Tancredi,
\textit{``{Elliptic Feynman integrals and pure functions}''},
\textsf{\doiref{10.1007/JHEP01(2019)023}{JHEP~1901,~023~(2019)}},
\texttt{\arxivref{1809.10698}{arxiv:1809.10698}}.

\bibitem{LeMura}
T.~Le and J.~Murakami,
\textit{``{Kontsevich's integral for the Kauffman polynomial}''},
\textsf{Nagoya~Math~J.~142,~93~(1996)}.

\bibitem{Tsunogai}
H.~Tsunogai,
\textit{``On some derivations of {L}ie algebras related to {G}alois
  representations''},
\textsf{\doiref{10.2977/prims/1195164794}{Publ.~Res.~Inst.~Math.~Sci.~31,~113~(1995)}}.

\bibitem{LNT}
J.-G.~Luque, J.-C.~Novelli and J.-Y.~Thibon,
\textit{``{Period polynomials and Ihara brackets}''},
\texttt{\arxivref{math/0606301}{math/0606301}}.

\bibitem{Pollack}
A.~Pollack,
\textit{``{Relations between derivations arising from modular forms}''},
Undergraduate thesis, Duke University.

\bibitem{Gerken:2019cxz}
J.~E.~Gerken, A.~Kleinschmidt and O.~Schlotterer,
\textit{``{All-order differential equations for one-loop closed-string
  integrals and modular graph forms}''},
\textsf{\doiref{10.1007/JHEP01(2020)064}{JHEP~2001,~064~(2020)}},
\texttt{\arxivref{1911.03476}{arxiv:1911.03476}}.

\bibitem{Enriquez:EllAss}
B.~Enriquez,
\textit{``Elliptic associators''},
\textsf{\doiref{10.1007/s00029-013-0137-3}{Selecta~Math.~(N.S.)~20,~491~(2014)}}.

\bibitem{Mafra:2014oia}
C.~R.~Mafra and O.~Schlotterer,
\textit{``{Multiparticle SYM equations of motion and pure spinor BRST
  blocks}''},
\textsf{\doiref{10.1007/JHEP07(2014)153}{JHEP~1407,~153~(2014)}},
\texttt{\arxivref{1404.4986}{arxiv:1404.4986}}.

\bibitem{Kleiss:1988ne}
R.~Kleiss and H.~Kuijf,
\textit{``{Multi - Gluon Cross-sections and Five Jet Production at Hadron
  Colliders}''},
\textsf{\doiref{10.1016/0550-3213(89)90574-9}{Nucl.~Phys.~B312,~616~(1989)}}.

\bibitem{Schocker}
M.~Schocker,
\textit{``{Lie elements and Knuth relations}''},
\textsf{Can.~J.~Math~56,~871~(2004)},
\texttt{\arxivref{math/0209327}{math/0209327}}.

\bibitem{Broedel:2018iwv}
J.~Broedel, C.~Duhr, F.~Dulat, B.~Penante and L.~Tancredi,
\textit{``{Elliptic symbol calculus: from elliptic polylogarithms to iterated
  integrals of Eisenstein series}''},
\textsf{\doiref{10.1007/JHEP08(2018)014}{JHEP~1808,~014~(2018)}},
\texttt{\arxivref{1803.10256}{arxiv:1803.10256}}.

\end{thebibliography}

\end{document}